\documentclass[twocolumn]{aa}
\usepackage{graphicx}
\usepackage{natbib}

\usepackage{txfonts}
\begin{document}
   \title{The XMM-Newton Extended Survey of the Taurus Molecular Cloud (XEST)}
   \titlerunning{XMM-Newton Extended Survey of Taurus}
   \authorrunning{G\"udel et al.}

   \author{Manuel G\"udel
          \inst{1}
          \and
	  Kevin R. Briggs
	  \inst{1}
	  \and
	  Kaspar Arzner
	  \inst{1}
	  \and
	  Marc Audard
	  \inst{2}
	  \and
	  Jer\^ome Bouvier
	  \inst{3}
	  \and
	  Eric D. Feigelson
	  \inst{4}
	  \and
	  Elena Franciosini
	  \inst{5}
	  \and
	  Adrian Glauser
	  \inst{1}
	  \and
	  Nicolas Grosso
	  \inst{3}
	  \and
	  Giusi Micela
	  \inst{5}
	  \and
	  Jean-Louis Monin
	  \inst{3}
	  \and
	  Thierry Montmerle
	  \inst{3}
	  \and
	  Deborah L. Padgett
	  \inst{6}
	  \and
	  Francesco Palla
	  \inst{7}
	  \and
	  Ignazio Pillitteri
	  \inst{8}
	  \and
	  Luisa Rebull
	  \inst{6}
	  \and
	  Luigi Scelsi
	  \inst{8}
	  \and
	  Bruno Silva
	  \inst{9, 10}
	  \and
	  Stephen L. Skinner
	  \inst{11}
	  \and
	  Beate Stelzer
	  \inst{5}
	  \and
	  Alessandra Telleschi
	  \inst{1}
          }

   \offprints{M. G\"udel}

   \institute{Paul Scherrer Institut, W\"urenlingen and Villigen,
              CH-5232 Villigen PSI, Switzerland\\
              \email{guedel@astro.phys.ethz.ch}
	 \and
	      Columbia Astrophysics Laboratory,
              Mail Code 5247, 550 West 120th Street,
              New York, NY 10027,
              USA     
         \and
             Laboratoire d'Astrophysique de Grenoble,
              Universit\'e Joseph Fourier - CNRS,
             BP 53,
             F-38041 Grenoble Cedex,
             France 
	 \and 
             Department of Astronomy \& Astrophysics,
             Penn State University,
             525 Davey Lab,
             University Park, PA 16802,
             USA 
	 \and 
	     INAF - Osservatorio Astronomico di Palermo,
             Piazza del Parlamento 1,
             I-90134 Palermo,
             Italy    
	 \and
	     Spitzer Science Center, 
	     California Institute of Technology, 
	     Mail Code 220-6, Pasadena, CA 91125,
	     USA   
	 \and
	     INAF - Osservatorio Astrofisico di Arcetri,
             Largo Enrico Fermi, 5,
             I-50125 Firenze,
             Italy     
	 \and 
             Dipartimento di Scienze Fisiche ed Astronomiche, 
	     Universit\`a di Palermo, Piazza del Parlamento 1,
             I-90134 Palermo,
	     Italy	 
	 \and 
             Centro de Astrof\'{\i}sica da Universidade do Porto, Rua das Estrelas, 4150
             Porto, Portugal
         \and
             Departamento de Matem\'atica Aplicada, Faculdade de Ci\^ecias da
             Universidade do Porto, 4169 Porto, Portugal	     
	 \and 
	     CASA, 389,
             University of Colorado,
             Boulder, CO 80309-0389,
             USA   
             }

   \date{Received 31 May 2006; accepted 5 August 2006}

  \abstract
   {The Taurus Molecular Cloud (TMC) is the nearest large star-forming region, prototypical for the
    distributed mode of low-mass star formation. Pre-main sequence stars are luminous X-ray sources,
    probably mostly owing to magnetic energy release. }
    {The {{\it XMM-Newton Extended Survey of the Taurus Molecular Cloud}} (XEST) presented in this paper
    surveys the most populated $\approx$5 square degrees of the TMC, using the {{\it XMM-Newton}}
    X-ray observatory to study the thermal structure, variability, and  long-term evolution of 
    hot plasma, to investigate the  magnetic dynamo, and to search for new potential members of the association.
    Many targets are also studied in the optical, and high-resolution X-ray grating spectroscopy has been
    obtained for selected bright sources.   }
    {The X-ray spectra have been coherently analyzed with two different thermal models (2-component thermal
    model, and a continuous emission measure distribution model). We present overall correlations with fundamental
    stellar parameters that were derived from the previous literature. A few detections from {{\it Chandra}}
    observations have been added.}
    {The present overview paper introduces the project and provides the basic results from the
    X-ray analysis of all sources detected in the XEST survey. Comprehensive tables summarize the
    stellar properties of all targets surveyed. The survey goes deeper than previous X-ray surveys
    of Taurus by about an order of magnitude and for the first time systematically accesses
    very faint and strongly absorbed TMC objects. We find a detection rate of 85\% and 98\% for
    classical and weak-line T Tau stars (CTTS resp. WTTS), and identify about half of the surveyed protostars and
    brown dwarfs. Overall, 136 out of 169 surveyed stellar systems are detected.  We describe an X-ray luminosity vs.
    mass correlation, discuss the distribution of X-ray-to-bolometric luminosity ratios, and show evidence for lower
    X-ray luminosities in CTTS compared to WTTS. Detailed analysis (e.g., variability, rotation-activity relations,
    influence of accretion on X-rays) will be discussed in a series of accompanying papers.  }
  {}

   \maketitle
%

\section{Introduction}

Studies of star-forming regions have drawn a picture in which cool, molecular gas,
contracting to protostars with surrounding accretion disks, co-exists with high-energy
radiation. The latter is emitted by plasma in the stellar environment that is continuously
heated to temperatures beyond one million degrees. This radiation is most prominently
seen in the soft (0.1--10~keV) X-ray range and is conventionally attributed to the
presence of magnetically trapped plasma in the outer stellar atmosphere. This plasma is 
reminiscent of, in the broadest sense, the solar coronal plasma. There is little doubt that
both lower-energy (extreme ultraviolet) and higher-energy (hard X-ray and gamma-ray) radiation
is amply present, but strong photoelectric absorption or, respectively, extremely
low flux precludes direct detection. Further, non-thermal radio emission from relativistic
electrons trapped in the magnetic fields gives unambiguous evidence for the presence
of an accelerated population of electrons similar to populations seen in the solar 
and in stellar coronae during flares \citep{guedel02}. Again, by implication from the solar  analogy, we
expect that high-energy, non-thermal particles (electrons and ions) play a crucial
role also outside the confines of the magnetic stellar atmosphere. 

But in contrast to the solar case, high-energy radiation from forming stars has a profound influence
on its environment as dense accretion disks and the larger-scale molecular 
envelope are irradiated, heated, and ionized by high-energy photons and particles. There are
numerous consequences of these processes. Ultraviolet and X-ray irradiation of molecular
disks heats their upper layers \citep{alexander04, glassgold04}, making them accessible 
to magnetic fields, thus inducing, for example, the magnetorotational instability \citep{balbus91}.
The latter is thought to be a prime source for increased disk viscosity, which in turn is the
main driver of the accretion process and thus the stellar formation process in the first place.
Complex chemical networks are also put in operation by heating and ionizing the 
molecular environment \citep{glassgold04}, thus fundamentally altering composition, opacity, 
and cooling behavior of the material. Magnetic fields are thought to attach to the inner
border of  the lightly ionized accretion disk, thus  transporting 
material to the stellar  surface while at the same time applying torques to the star that may regulate
its rotation period  \citep{montmerle00}. If the stellar magnetic fields are generated in a 
solar-like dynamo in which rotation is a key parameter, then the long-term evolution
of the magnetic field production and thus of the high-energy radiation is indeed directly
controlled by the accretion process in the feedback loop sketched above.  Further, magnetic
fields and ionization of the disk may be relevant for the launch of bipolar jets
and molecular outflows \citep{pudritz86}, a widely observed but poorly understood mechanism that seems
to regulate the mass buildup of the forming star. Alternatively, these jets may originate from
the magnetic fields spanning between the star and the disk \citep{shu94}, or from plasma clouds 
ejected in the process of magnetic reconnection of these same fields \citep{hayashi96}.
 
The origin of the X-ray emission is still not entirely clear. While the bulk of 
the X-rays from T Tau stars is thought to originate in a magnetically trapped
corona consisting of complex magnetic loops anchored in photospheric magnetic regions
like in the Sun, features unknown to the Sun may exist. Speculation has arisen that
some very soft X-rays may be formed in accretion shocks near the stellar surface  (e.g., 
\citealt{kastner02}). Herbig-Haro objects far from the forming stars have been detected
to be very soft X-ray emitters due to shock interactions with the interstellar medium
(e.g., \citealt{pravdo01, bally03, guedel05}), but new spectral phenomenology suggests
that appreciable X-ray emission may also be formed close to the stars, at the (spatially
unresolved) base of the jets otherwise seen in radio and in the optical \citep{guedel05}.
Whatever the analogy to the Sun, we also note that the level of the X-ray luminosity
(between a few times $10^{28}$~erg~s$^{-1}$ and $10^{31}$~erg~s$^{-1}$ for low-mass pre-main 
sequence stars) and the measured plasma temperatures of up to 100~MK are extreme and
may imply heating mechanisms related to continuous flaring \citep{guedel03, wolk05}.

Studies of X-ray radiation in pre-main sequence stars are thus crucial
for our further understanding of star formation, accretion processes, angular momentum
transport, and mass outflow mechanisms in the earliest stages of a star's life. For recent
reviews of X-rays from pre-main sequence stars, we refer the reader to \citet{feigelson99},
\citet{guedel04}, and \citet{feigelson06}. {\it {\it XMM-Newton}} and {\it Chandra} penetrate dense molecular envelopes 
providing unequaled information on high-energy processes. A recent, large survey of the Orion 
Nebula Cluster (ONC) with {\it Chandra} (the COUP project, \citealt{getman05}) has 
monitored approximately 1400 young stars in a $17\arcmin\times 17\arcmin$ field around the 
Trapezium. This study has provided unprecedented information on the ONC stellar population and 
the evolution of X-ray production. For example, \citet{preibisch05} concluded that a significant 
correlation between rotation and X-ray luminosity consistently found in main-sequence stars is 
absent in the ONC sample, arguing for a different magnetic dynamo process or a very different 
internal structure of these youngest stars. X-ray radiation was found to be largely variable, 
with numerous flares contributing significantly to the overall X-ray production \citep{wolk05}.
We refer the interested reader to a special volume of papers from this study (\citealt{getman05} and
associated papers), or to a summary provided by \citet{feigelson06}. 
 
The ONC  represents the so-called ``clustered mode'' of star formation in which hundreds to thousands
of low-mass stars form around massive O-type stars such as those in the Trapezium.
Such environments produce extreme stellar densities, and the presence of hot stars has
a decisive impact on star formation. UV radiation and the winds of the central O stars have a 
deteriorating effect on the molecular environment, the pre-stellar  cores and the circumstellar disks
by ionizing and dispersing the molecular material, prematurely altering the accretion
environment \citep{odell98} and eventually halting the star formation process altogether
\citep{dolan01}. The ionized H~II Orion Nebula is the most 
vivid expression of this environment.

Star formation occurs, however, also in relatively unspectacular low-mass star forming regions 
devoid of the influence of radiation and outflow-related effects from luminous O
stars. These so-called ``dark clouds'' typically form a smaller number of stars and
often do so in a ``dispersed'' or ``isolated mode'' in which stars or very small
groups of stars form in relative isolation. Stars form in a similar manner also in high-mass
star formation regions, outside the clusters and at larger distances from the central OB
stars \citep{carpenter00}.

The closest and best-studied  large, low-mass star formation region
is the  Taurus Molecular Cloud (TMC) complex that extends to 
the adjacent Auriga Clouds. TMC has served as a testbed for low-mass star-formation theory 
and has provided some of the best examples of phenomenology relevant to star formation. 
There are only a few Herbig Ae/Be stars in this vast region (V892 Tau and AB Aur in our survey). The
rest are predominantly   sub-solar in mass although there have been claims that a substantial
population of non-accreting B stars may be members of the Taurus population \citep{walter91}, in
agreement with expectations from the initial mass function.
Owing to the near-absence of massive stars and their mechanical/radiative effects, the molecular gas is 
still  present in large amounts ($M_{\rm cloud} \approx ~3\times 10^4~M_{\odot}$, \citealt{ungerechts87})
much in excess of the total mass in stars already formed 
($M_{\rm stars}$ of order  $2\times 10^2~M_{\odot}$).

The Taurus region is quite large (some 10--15 degrees in diameter,
corresponding to about 25--35~pc at a distance of 140~pc), making comprehensive studies
of the entire population difficult. We have  initiated a large X-ray study of 
the TMC with the {\it XMM-Newton} X-ray observatory, concentrating on the denser cloud areas that contain 
the majority of the TMC stellar population. The emphasis of this {\it XMM-Newton Extended Survey of the Taurus
Molecular Cloud} (XEST henceforth)  is on a wide-field sampling
rather than on long exposures. It goes systematically deeper by about an order of magnitude than
previous Taurus X-ray surveys. While it is less sensitive than the COUP survey, it surveys an area about 
60 times larger ($\approx 5$~square degrees), which is required because of the low surface density of Taurus members.
 The survey is accompanied by a deep, large-field optical
survey with the Canada-France-Hawaii Telescope (CFHT) 
and a mid-infrared survey of the entire cloud complex with the {\it Spitzer Space Telescope} 
\citep{padgett06}. These latter surveys will be presented separately (see summary of all three surveys in 
\citealt{guedel06a}).  All studies combined provide  an unsurpassed database for the nearest 
major star-forming cloud complex. 

The purpose of the present paper is to give
an introduction to the XEST  project, to present an overview of 
the surveyed stellar sample, and to  discuss the strategies followed during the data 
reduction and analysis.  The detailed results are described in a series of associated papers.

The outline of this paper is as follows. We discuss some pivotal properties of the TMC
in Sect.~\ref{TMC}, including  results from previous studies that support the key role
of this region for further star-formation studies. Sect.~\ref{XEST} introduces the TMC X-ray survey,
and Sect.~\ref{strategy} presents the data reduction and analysis strategies.
Sect.~\ref{results} contains an overview of the basic results, presenting detected and undetected 
stellar populations, accompanied by comprehensive tabulations of fundamental
properties of all surveyed objects and spectral X-ray results from the detections. 
We summarize overall statistical properties of XEST in Sect.~\ref{conclusions}.

The subsequent series of papers related to this survey discusses 
X-ray properties in the context of stellar class and accretion \citep{telleschi06a}, 
rotation-activity-age relations \citep{briggs06a}, 
a dedicated study of the L1495E subsample \citep{silva06}, 
correlated behavior of X-rays and optical/ultraviolet emission \citep{audard06},
X-ray properties in a sample of TMC protostars \citep{briggs06b}, 
X-rays from jet-driving stars in TMC \citep{guedel06b}, 
high-resolution X-ray spectroscopy of classical and weak-line T Tau stars \citep{telleschi06b}, 
X-rays from brown dwarfs \citep{grosso06a}, 
a $U$-band survey of brown dwarfs with the {\it XMM-Newton} Optical Monitor \citep{grosso06b}, 
light-curve variability studies \citep{stelzer06}, 
interpretation of X-ray flares \citep{franciosini06}, 
an investigation of statistical fluctuations in X-ray light curves \citep{arzner06a},
an investigation of spectral parameters in extremely faint sources \citep{arzner06b}, 
an analysis of the gas-to-dust ratio in the TMC region \citep{glauser06},  a case study
of the accreting, prototypical CTTS T Tau \citep{guedel06c},
a study of the supposedly single Herbig star in TMC, AB Aurigae \citep{telleschi06c},
and a search for new TMC members based on near-infrared and X-ray properties of field
sources detected in XEST \citep{scelsi06}.


\section{The Taurus Molecular Cloud Complex}\label{TMC}
At a distance around 140~pc (e.g., \citealt{loinard05, kenyon94}), the TMC is  the nearest large star formation
region and reveals characteristics that make it ideal for detailed physical studies.
One of the most notable properties of TMC in this regard is its structure in which several loosely
associated but otherwise rather isolated molecular cores each produce one or only a few low-mass
stars, different from the much denser cores in $\rho$ Oph or in Orion. TMC features a low stellar 
density of only 
1--10 stars~pc$^{-3}$ (e.g., \citealt{luhman00}). Strong mutual influence due
to outflows, jets, or gravitational effects are therefore minimized. Strong stellar winds and ionizing UV radiation
are mostly absent in TMC because there are no O stars and only very few B and A stars \citep{walter91}.
Further, most stars in
TMC are subject to relatively modest extinction, providing access to a broad spectrum of 
stars at all evolutionary stages from Class 0 sources to near-zero age main-sequence 
T Tau stars. TMC has also become of central interest for the study of substellar
objects, in particular brown dwarfs, with regard to their evolutionary history and their
spatial distribution and dispersal \citep{briceno02, guieu06}.

TMC has figured prominently in star-formation studies at all wavelengths. It has provided
the best-characterized sample of classical and weak-line T Tau stars (CTTS and WTTS, respectively, 
or ``Class II'' and ``Class III'' objects in the infrared classification - \citealt{kenyon95}); 
most of our current picture of low-density star formation is indeed based on IRAS studies of 
Taurus \citep{strom89, weaver92}. Among the key results from TMC studies as listed in \citet{kenyon95} 
figure the following: i) more than 50\% of the TMC objects have IR excess beyond the photospheric
contribution, correlating with other activity indicators (H$\alpha$, UV excess etc.) and indicating
the presence of warm circumstellar material predominantly in the form of envelopes for Class I
protostars and circumstellar disks for Class II stars. ii) Class III sources (mostly to be identified 
with WTTS) are distinctly different from Class I-II objects by not revealing optically thick disks or 
signatures of accretion. iii) Star formation has been ongoing at a similar level during the past
1-2~Myr, with the Class-I protostars having ages of typically 0.1--0.2~Myr. iv) There is clear support
for an evolutionary sequence Class I$\rightarrow$II$\rightarrow$III, although there is little 
luminosity evolution along this sequence, indicating different evolutionary
speeds for different objects (see also \citealt{hartmann02}). The infall time scale is a few times 
$10^5$~yrs, while the disk phase  amounts to a few times $10^6$~yrs. An evolutionary scenario is 
also suggested from the different spatial distribution of Class 
0/I stars vs CTTS/WTTS with respect to the gas distribution: The 
former classes are still within the boundaries of the high density gas, while 
the latter are found in regions of lower density \citep{palla02}.

TMC has also been well-studied at millimeter wavelengths. This
region has better high-resolution molecular line maps than
any other star-forming region \citep{onishi02}. Most of the
higher-mass CTTS  and Class I protostars have
been surveyed by millimeter interferometers for molecular line
emission from disks (\citealt{dutrey96, ohashi96}, etc.) and
many detailed studies of individual sources have been published
(e.g., \citealt{qi03, duchene03}). A variety of millimeter
continuum observations of these sources have enabled studies of cold
disk frequency into substellar mass ranges and raised the possibility
of grain growth within circumstellar disks (\citealt{scholz06, rodman06,
wolf03, kitamura02}, etc.).
Due to its proximity and easy accessibility to northern hemisphere
radio telescopes, the TMC has served as a template for
millimeter studies of young stellar objects.

Although TMC has been regarded, together with the $\rho$ Oph dark cloud, as the prototypical
low-mass star-forming region, a few apparent peculiarities deserve to be mentioned. TMC contains
an anomalous number of binaries \citep{ghez93, duchene99a}, compared with other SFRs 
(e.g., Orion) or with field stars.
In TMC, about two thirds of all members are bound in multiple systems, with an average
separation of about $0.3^{\prime\prime}$ (e.g., \citealt{leinert93, mathieu94,
simon95, duchene99b, white01, hartigan03}). \citet{petr98} reported, at the 96\% confidence
level,  a three times higher binary occurrence in TMC compared to Orion for component separations 
in the range of 63--225~AU.
Also, TMC cloud cores are comparatively small and of low
mass, at least when compared with cores in Orion or Perseus \citep{kun98}. 

Initially, TMC was also found to be deficient of lowest-mass stars and brown dwarfs, with a mass 
distribution significantly enriched in 0.5--1~M$_{\odot}$ stars, compared to Orion samples 
\citep{luhman00, briceno02}.  It was speculated that the formation of brown dwarfs could be different 
in the low-density environment of TMC compared to the dense packing of stars in Orion. However, a new wide-field
search for low-mass TMC members now indicates that there is no BD deficit in Taurus \citep{guieu06}.

In X-rays, Taurus has again played a key role in our understanding of high-energy processes
and circumstellar magnetic fields around pre-main sequence stars. Taurus X-ray studies provided the
first detailed view of the X-ray behavior of T Tauri stars \citep{feigelson81a, feigelson81b, walter81}.
 Among the key surveys of the entire region are those 
by \citet{feigelson87},  \citet{walter88}, \citet{bouvier90}, \citet{strom90}, 
\citet{damiani95a}, \citet{damiani95b},  based on {\it Einstein Observatory} observations,
and the work by \citet{strom94}, \citet{neuhaeuser95}, and \citet{stelzer01} based on {\it ROSAT}. 
These surveys have characterized the overall luminosity behavior of TTS and studied 
the dependence of X-ray activity on rotation. \citet{bally03} and \citet{favata03} reported,
respectively, {\it Chandra} and {\it XMM-Newton} studies of the L1551 cloud region (the longer
exposure from {\it XMM-Newton} will be included in our survey; the {\it Chandra} exposure contains
no additional sources).

But again, for reasons very poorly understood, TMC differs from other SFRs
significantly also with regard to X-ray properties. Whereas no X-ray activity-rotation correlation 
(analogous to that in main-sequence stars) is found for samples in the Orion star-forming regions,
perhaps suggesting that all stars are in a saturated 
state \citep{flaccomio03a, preibisch05}, the X-ray activity  
in TMC stars has been reported to decrease for increasing rotation period 
(e.g., \citealt{neuhaeuser95, damiani95a, stelzer01}). Also, claims have been made
that the X-ray behavior of TMC CTTS and WTTS 
is significantly different, CTTS being less luminous than WTTS \citep{strom94,
damiani95b, neuhaeuser95, stelzer01}. This  
contrasts with other star-forming regions \citep{flaccomio00, preibisch01},
but recent reports reveal a similar 
segregation also for Orion and some other SFRs \citep{flaccomio03b, preibisch05}.
Some of these discrepancies may be due to selection and detection bias 
(e.g., WTTS are predominantly identified in X-ray studies, in contrast to CTTS), but also
to incomplete samples due to high detection limits. Issues related to rotation and
selection biases will be discussed in detail in one of the companion papers, see \citet{briggs06a}.

\section{The XEST Survey}\label{XEST}

\subsection{Scientific goals}

The scientific goals of XEST are:
\begin{itemize}
\item To collect X-ray spectra and light curves from a statistically meaningful
           sample of TMC objects, and to characterize them in terms
	   of X-ray emission measure distributions,  temperatures, X-ray luminosities,
	   and variability.

\item To interpret X-ray emission in the context of other stellar properties
           such as rotation, mass, and radius.

\item  To investigate changes in the X-ray behavior as a young stellar object evolves.

\item To obtain a census of X-ray emitting objects at the stellar mass limit and
          in the substellar regime (brown dwarfs = BDs).

\item To study in what sense the stellar  environment (circumstellar disks, jets, 
            accretion) influences the X-ray production, and vice versa.

\item To study the gas-to-dust ratio in the circumstellar environment, making use
      of extinction measurements obtained in the optical or near-infrared.

\item To assess the role of flares in coronal heating, and to study flare 
      characteristics in their own right.

\item To search for new, hitherto unrecognized, TMC members. 

\end{itemize}
The outstanding characteristics of the survey are its sensitivity and its energy resolution,
covering a large fraction of the most densely populated regions of the Taurus clouds. It is
by far the largest-area Taurus survey at this sensitivity. It reaches sensitivities about 
ten times better than previous surveys such as
those conducted with ROSAT \citep{neuhaeuser95, stelzer01} and therefore permits
a systematic study of the lowest-mass Taurus members. Because of the harder band used for the XEST project
(compared to ROSAT's 0.1-2.4~keV band), XEST also detects several deeply embedded Taurus protostars 
systematically for  the first time, while they remained undetected in softer-band surveys. 
Given the achieved  
sensitivity, the XEST project accesses the known TMC population in the surveyed fields nearly completely,
thus suppressing potential bias that previous surveys may have been subject to.  
Energy resolution  permits a detailed description of the plasma properties together
with the measurement of the absorbing gas columns that are located predominantly in the
TMC clouds themselves, and even in the immediate circumstellar environment in the case
of strongly absorbed objects.

\subsection{Instruments and exposures}

XEST is a wide-field X-ray 
survey obtained  with the {\it XMM-Newton} X-ray observatory, principally based on
combined CCD camera exposures, but complemented by exposures
of a few bright targets with the reflection grating spectrometers and fields 
observed with the optical monitor. A few complementary observations obtained
with the {\it Chandra X-Ray Observatory} have also been included.

{\it XMM-Newton} \citep{jansen01} orbits the Earth in a 48~hr orbit, permitting long, 
uninterrupted  exposures. It carries three high-throughput telescopes feeding two
suites of X-ray instruments, and also features an optical telescope. A short description
of the instruments and the chosen setups follows.\footnote{For further details, see {\it XMM-Newton} User's Handbook
(http://xmm.vilspa.esa.es/)} 

i) The {\bf European Photon Imaging Cameras} (EPICs) are three CCD-based X-ray cameras,
one per telescope, that operate entirely independently. Two cameras are of the MOS type
\citep{turner01}, and one is of the PN type \citep{strueder01}.  The circular field of view of
each camera has a diameter of 30\arcmin,  and the three telescopes are nearly co-aligned. The PN camera 
provides
most  counts because the beams to the MOS cameras are intersected by the reflection gratings
that consume approximately half of the X-ray flux. The MOS PSF has a full-width-at-half-maximum
(FWHM) of about 4--5\arcsec, while the half-energy width (HEW) is larger, amounting to
some 13--15\arcsec. The cameras provide pixel sizes of  1.1\arcsec\ (MOS)
or 4\arcsec\ (PN). The energy resolution is approximately $E/\Delta E \approx 45$ at 6.7~keV, 
scaling as $E^{-1/2}$. To prevent the spectra from
being altered by optical load, various filters can be inserted. 

We usually 
applied the medium filter to all EPIC cameras, providing a combined on-axis effective area for
all three cameras of $\approx 1800$~cm$^2$ at 1.5~keV. Exceptions were exposure XEST-26 for
which the thick filter was used for the two MOS cameras to suppress optical load
from the Herbig star  AB Aur (the PN camera was not recording), and
exposures XEST-27 and XEST-28 for which the thick filter was used for each EPIC camera. 
All cameras were operated in full window mode, i.e., the entire field of view was exposed, 
confining the time resolution to 2.6~s for the MOS and to 73~ms for the PN camera, entirely 
appropriate for the moderate brightness of our sources. The only exceptions here were the fields
around V773 Tau (XEST-20) where MOS2 was operated in the small-window mode, and the field  
 of T Tau (XEST-01) for which MOS1 was operated in small-window mode, and MOS2 was operated in large-window
mode. These modes confine the central CCD to $100\times 100$ and $300\times 300$ pixels,
respectively, compared to $600\times 600$ pixels for the full window mode, but avoiding 
pile-up during potential large flares.

ii) Two {\bf Reflection Grating Spectrometers} (RGS, \citealt{denherder01}) are each fed by about
50\% of the light of one of the telescopes. The dispersed spectra comprise the wavelength
range from 5 to 35\AA, although in each spectrometer, one of the nine chips has failed
earlier in the mission, leaving a gap in the 10.6--13.8~\AA\  range for RGS1 and in the 
20.0--24.1~\AA\  range for RGS2. The first-order spectral resolution is approximately 
60--70~m\AA\ (FWHM) at any wavelength, corresponding to $\lambda/\Delta\lambda \approx 
300$ at $\lambda = 20$~\AA. The combined first-order effective area is $\approx 115$~cm$^{2}$ 
at 15~\AA.
 
iii) The {\bf Optical Monitor} (OM, \citealt{mason01}) is a microchannel-plate enhanced CCD camera
fed by a 30~cm mirror co-aligned with the EPIC cameras. Its square field of view is, however, 
somewhat smaller than the latters', amounting to 17\arcmin\ side-length, i.e., a radius
of 12\arcmin\ along the diagonal.  The OM can be used in combination with various filters.
Filter use may be dictated by the brightness of the stars in the field of view. We obtained
most of the exposures with the U band filter inserted, given the diagnostic importance of this band for
both chromospheric flare processes and accretion mechanisms. In a few cases, another
near-UV filter was used (UVW1: 2500-3500~\AA; UVW2: 1700-2500~\AA) to suppress excessive optical load. The window
mode we applied essentially observed the entire field of view. The detector area is split up into
a series of windows, some of which are sequentially exposed and read out. The exposure time for one
of these frames determines the time resolution for any source within the window. This time
resolution is of order 1000 - 2000~s but varies within any given XEST OM field, and is also not identical
between different XEST fields. A detailed presentation is given by \citet{audard06}. Additionally, if
a bright source was present on-axis, the central $10.5\arcsec\times 10.5\arcsec$ of the OM field were 
exposed within a high-time resolution window that collects photons  with a time resolution of 0.5~s.

Our initial project collected 19 fields in the TMC coherently with exposure 
times of about 30~ks each. The effective exposure time varied between 31.2~ks and 41.9~ks. 
Relevant information is given in Table~\ref{tab1}. The start and end times of the observations as listed in
Table~\ref{tab1} are the times of the earliest and latest recording of any of the three EPIC cameras,
and the quoted exposure time is  the difference between end and start times. The pointing coordinates
are the nominal boresight coordinates, which are not  identical with the coordinates
of the center of the EPIC fields of view, the latter being slightly misaligned (e.g., to avoid
the PN CCD edges crossing the exact field center). The last column gives the filter used
for the OM (U band, or one of the ultraviolet filters, UVW1 or UVW2). A letter {\it F} indicates
that the central star was recorded in fast mode (see \citealt{audard06}).

\begin{table*}
\centering
\caption{Observing log of XEST program}
\begin{tabular}{rlrrrrrr}
\hline
\hline
Exposure &ObsID$^a$ &RA(J2000.0)$^b$& $\delta$(J2000.0)$^b$        & Start time$^c$&  Stop time$^c$ & Exposure & OM mode,        \\
\#       &          & h\ \ m \ \ s &  $\deg\ \ \arcmin\ \ \arcsec$ & y-m-d h:m:s  &  y-m-d h:m:s    & time (s) & OM filter$^d$     \\
\hline
1$^e$    &0301500101& 04 21 59.4   & 19 32 06			   &2005-08-15\ 13:52:13 & 2005-08-16\ 12:55:22	&  82989   & F, UVW1	\\
2        &0203540201& 04 27 19.6   & 26 09 25			   &2004-08-17\ 06:08:10 & 2004-08-17\ 17:32:46 &  41076   & U	      \\
3        &0203540301& 04 32 18.9   & 24 22 28			   &2004-08-22\ 06:45:22 & 2004-08-22\ 16:37:04 &  35502   & F, U	      \\
4        &0203540401& 04 33 34.4   & 24 21 08			   &2005-02-21\ 01:18:21 & 2005-02-21\ 10:31:17 &  33176   & F, U	      \\
5        &0203540501& 04 39 34.9   & 25 41 46		           &2005-02-21\ 11:25:51 & 2005-02-21\ 20:13:47 &  31676   & U	      \\
6        &0203540601& 04 04 42.9   & 26 18 56			   &2004-08-25\ 20:53:54 & 2004-08-26\ 05:41:50 &  31676   & U	      \\
7        &0203540701& 04 41 12.5   & 25 46 37			   &2005-02-24\ 20:28:00 & 2005-02-25\ 05:15:58 &  31678   & U		\\
8        &0203540801& 04 35 52.9   & 22 54 23	     	           &2004-08-26\ 06:36:23 & 2004-08-26\ 18:10:59 &  41676   & F, U		\\
9        &0203540901& 04 35 55.1   & 22 39 24	  	           &2005-02-25\ 08:38:53 & 2005-02-25\ 17:26:50 &  31677   & U		\\
10       &0203542201& 04 42 20.9   & 25 20 35			   &2005-03-05\ 05:56:38 & 2005-03-05\ 14:44:30 &  31672   & F, U		\\
11       &0203541101& 04 21 51.1   & 26 57 33			   &2004-08-18\ 06:44:13 & 2004-08-18\ 18:21:54 &  41861   & U		\\
12       &0203542101& 04 35 17.4   & 24 15 00			   &2005-03-04\ 20:22:29 & 2005-03-05\ 05:01:54 &  31165   & F, U		\\
13       &0203541301& 04 29 52.0   & 24 36 47	  	           &2004-08-25\ 11:11:23 & 2004-08-25\ 19:59:19 &  31676   & U		\\
14       &0203541401& 04 30 30.6   & 26 02 14			   &2005-02-09\ 03:01:02 & 2005-02-09\ 12:30:38 &  34176   & U		\\
15       &0203541501& 04 29 42.4   & 26 32 51		           &2005-02-09\ 13:12:40 & 2005-02-09\ 22:38:18 &  33938   & F, U		\\
16       &0203541601& 04 19 43.0   & 27 13 34			   &2004-08-21\ 05:53:24 & 2004-08-21\ 17:11:22 &  40678   & -		\\
17       &0203541701& 04 33 21.2   & 22 52 41			   &2005-02-11\ 01:19:56 & 2005-02-11\ 10:07:50 &  31674   & U		\\
18       &0203541801& 04 33 54.7   & 26 13 28	   	           &2004-08-13\ 21:48:05 & 2004-08-14\ 06:36:03 &  31678   & F, U		\\
19       &0203541901& 04 32 43.0   & 25 52 32			   &2004-08-14\ 07:18:05 & 2004-08-14\ 17:50:41 &  37956   & F, U		\\
20       &0203542001& 04 14 12.9   & 28 12 12			   &2004-09-12\ 07:04:43 & 2004-09-12\ 15:52:37 &  31674   & F, U		\\
21$^e$   &0101440701& 04 21 59.0   & 28 18 08			   &2000-09-05\ 02:57:44 & 2000-09-05\ 15:47:55 &  46211   & -		\\
22$^e$   &0109060301& 04 31 39.0   & 18 10 00			   &2000-09-09\ 18:29:39 & 2000-09-10\ 10:18:12 &  56913   & UVW2		\\
23$^e$   &0086360301& 04 18 31.2   & 28 27 16			   &2001-03-11\ 12:46:45 & 2001-03-12\ 09:13:24 &  73599   & (UVW2)$^f$		\\
24$^e$   &0086360401& 04 18 31.2   & 28 27 16			   &2001-03-12\ 09:29:38 & 2001-03-12\ 21:47:51 &  44293   & (UVW2)$^f$		\\
25$^e$   &0152680201$^g$& 04 34 55.5 &  24 28 54  		   &2003-02-14\ 02:18:48 & 2003-02-14\ 07:01:32 &  16964   & UVW2  	     \\
26$^e$   &0101440801& 04 55 59.0   & 30 34 02			   &2001-09-21\ 01:34:17 & 2001-09-22\ 13:34:31 & 129614   & -  	\\
27$^e$   &0201550201& 03 54 07.9   & 31 53 01			   &2004-02-13\ 21:45:10 & 2004-02-14\ 09:48:07 &  43377   & -  	\\
28$^e$   &0200370101& 04 19 15.8   & 29 06 27			   &2004-08-15\ 06:14:30 & 2004-08-16\ 18:42:57 & 131307   & UVW1		\\
\hline
\multicolumn{8}{l}{$^a$ {\it XMM-Newton} observation identification number }\\
\multicolumn{8}{l}{$^b$ Nominal boresight coordinates}\\
\multicolumn{8}{l}{$^c$ Earliest respectively latest time of recording for any EPIC camera is reported}\\
\multicolumn{8}{l}{$^d$ F = fast + imaging  mode (only imaging otherwise); for filter name, see {{\it XMM-Newton}} User's 
                        Handbook (http://xmm.vilspa.esa.es/) }\\
\multicolumn{8}{l}{$^e$ Observation from separate projects}\\
\multicolumn{8}{l}{$^f$ The OM was recording only for short time intervals}\\
\multicolumn{8}{l}{$^g$ Eight different observations of $\approx 16-19$~ks (PN) each around AA Tau, 2003-02-14 -- 2003-02-28, ObsID = 0152680201-901}\\
\end{tabular}
\label{tab1}
\normalsize
\end{table*}

\begin{table*}
\centering
\caption{Complementary {\it Chandra} observations}
\begin{tabular}{rlrrrrrrr}
\hline
\hline
Exposure	   &Main      &Obs &RA(J2000.0)$^b$& $\delta$(J2000.0)$^b$        & Start time	  &  Stop time  	 & Exposure &  Instr.	      \\
\#		   &Target&ID$^a$  & h\ \ m \ \ s &  $\deg\ \ \arcmin\ \ \arcsec$ & y-m-d h:m:s	  &  y-m-d h:m:s	 & time (s) &         \\
\hline
C1		   & V410 Tau &3364& 04 18 34.6   & 28 22 47		    &2002-03-07\ 06:16:32 & 2002-03-07\ 11:45:24 & 17734    & ACIS-S	 \\
C2		   & FS Tau   &4488& 04 22 00.2   & 26 58 07		    &2003-11-08\ 12:57:58 & 2003-11-08\ 21:56:34 & 29674   & ACIS-S	  \\
C3		   & DG Tau   &4487& 04 27 02.3   & 26 04 56		    &2004-01-11\ 02:58:51 & 2004-01-11\ 11:52:21 & 29717   & ACIS-S	  \\
C4		   & GV Tau   &4498& 04 29 23.3   & 24 32 44		    &2003-12-28\ 08:56:57 & 2003-12-28\ 16:37:59 & 24650   & ACIS-I	  \\
C5		   & HR~1442  & 612& 04 33 33.1   & 18 01 15		    &2000-09-11\ 17:53:33 & 2000-09-11\ 19:41:57 &  4679   & HRC-I	    \\
C6		   & L1527    &2563& 04 39 52.7   & 26 03 05		    &2002-12-06\ 08:30:12 & 2002-12-06\ 14:21:15 & 19317   & ACIS-I	\\
\hline
\multicolumn{8}{l}{$^a$ {\it Chandra} observation identification number}\\
\multicolumn{8}{l}{$^b$ Nominal boresight coordinates}\\
\end{tabular}
\label{tab2}
\normalsize
\end{table*}

\begin{figure*}[t!]
\centerline{\resizebox{0.99\hsize}{!}{\includegraphics{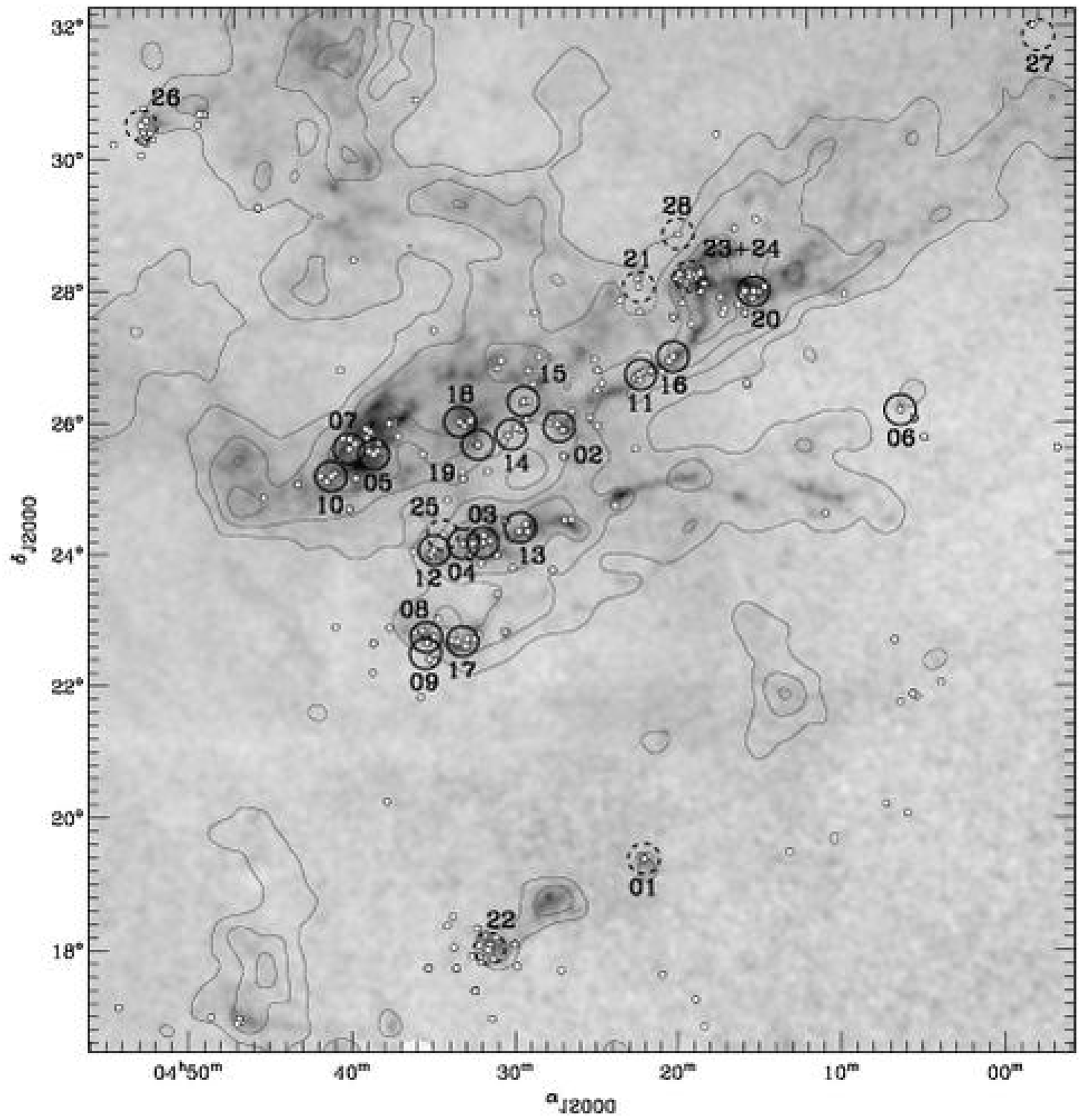}}}
\caption{\small Map of the TMC region (oriented in RA - dec, north is up, east to the left). The grayscale background 
map is an extinction ($A_{\rm V}$) map from \citet{dobashi05}. Contours show the CO emission \citep{dame87}.
The large (0.5 degrees diameter) circles show the fields of
view of the {\it XMM-Newton} survey (the dashed circles marking fields from separate projects also
used for the survey). Small white dots mark the positions of individual young TMC members.
The  labels correspond to the XEST exposure numbers in Table~\ref{tab1}. Note the outlying
{\it XMM-Newton} fields XEST-26 around SU Aur (NE corner), XEST-27 around $\zeta$ Per (NW corner),  XEST-01 around T Tau, and
XEST-22 around L1551 (the latter two at the southern border of the figure).  
  \label{fig-1}} 
\end{figure*}

The fields of view were selected such that they cover the densest 
concentrations of CO gas, which also show the strongest accumulations of TMC stellar and substellar 
members. We complemented this sample with  exposures that were obtained  as part of separate
projects, partly retrieved from the {\it XMM-Newton} archive (marked in Table~\ref{tab1}); 
in all of these observations very similar instrument setups were used,
although in most cases the exposure times were longer and the strategies for the use of the OM were 
different. The eight additional fields are:
1) L1495 centered at V410 Tau (XEST exposures 23 and 24, see Table~\ref{tab1}; two exposures totaling 117.9~ks; PI F. Walter); 
2) L1551, centered near L1551 IRS-5 (XEST exposure 22, 56.9~ks, PI F. Favata); 
3) a field centered at AA Tau (eight exposures of 16-19.5~ks each of which we use only one [XEST exposure 25]  
   that reveals the strongly variable AA Tau X-ray source in an average state - details in 
   Grosso et al., in preparation; PI J. Bouvier); 
4) a field centered at HD~283572 (XEST exposure 21,  46.2~ks, PI R. Pallavicini); 
5) a field around BP Tau (XEST exposure 28, 131.3~ks, PI J. Schmitt); 
6) a field around SU Aur (XEST exposure 26, 129.6~ks, PI R. Pallavicini); 
7) a field around $\zeta$ Per (XEST exposure 27, 43.4~ks, PI W. Waldron); 
8) a field around T Tau (XEST exposure 1, 83.0~ks - a special discussion of this observation  
   will be provided in a separate, future paper; PI M. G\"udel).
The entire survey includes approximately 5 sq. degrees in total, with a total exposure time
of about 1.3~Ms. The spatial coverage is  illustrated in Figure~\ref{fig-1}. 

We have included a few exposures obtained by {\it Chandra} that contain important objects
outside the {\it XMM-Newton} coverage, or faint objects not detected by {\it XMM-Newton} but
detected by {\it Chandra}. Objects that were covered by the XEST program (regardless of
their detection status) and were {\it not} detected in these {\it Chandra} exposures will 
not be reported in our survey tables.
The additional fields are defined in Table~\ref{tab2}. Note that different detectors were used
for these observations. The entire fields of view were scanned for TMC objects that add to the project, but
no systematic survey of all  X-ray sources in the fields was performed.
The exposure around GV Tau does not contain new sources beyond those detected in the {\it XMM-Newton} field 
but  provides better angular resolution  for the study of the origin of the X-rays in this binary system 
\citep{guedel06b}.  We also add the strong X-ray detection coincident with HD 28867 = HR~1442 that has
been suspected to be a late-B (non-Herbig) binary member of the TMC (e.g., \citealt{walter91}). This object
has been extensively studied by \citet{walter03} who found evidence for a G-type companion to 
one of the B-type components.   

The XEST work presented here and in the accompanying papers addresses X-ray properties of the commonly
known stellar and substellar population. A considerable number of hitherto unrecognized Taurus members may be present 
in the survey, e.g., deeply embedded sources or  extincted stars on the far side of the clouds. An attempt
to identify some objects of this population has been undertaken by \citet{scelsi06}.

\section{Strategy and analysis of the TMC survey}\label{strategy}

Our {\it XMM-Newton} data analysis procedure predominantly uses tasks provided in the
{\it XMM-Newton} Science Analysis Software (SAS) v6.1.0, augmented by
procedures in FTOOLS, and is tied together in Perl scripts developed
at the Paul Scherrer Institut.
Source identification was based on procedures involving maximum-likelihood 
algorithms and wavelet analysis. In order to optimize detection of faint sources,
periods of high background radiation levels due to local particles were cut out. 
All programs named below can be assumed
to be SAS tasks unless otherwise noted.

\subsection{Creation of event lists and event selection} 

An event list was produced for each active EPIC instrument from the
original data files using the standard SAS procedures EPCHAIN and
EMCHAIN. MOS event lists were filtered to include only events flagged as
`good' and with patterns 0--12 \citep{kirsch06}. PN event lists were
filtered using the selection expression (FLAG \& 0xfb0825)$=0$, which
rejects events with patterns $> 12$, close to CCD windows or on-board
bad pixels, in spoiled frames, or outside the field of view, but
retains events on or next to offset columns, close to bright pixels or
dead pixels. Remaining bright pixels, columns and partial columns
observed in images at energies $> 300$~eV were further added to the
bad pixel list and rejected. 

Source detection was performed in three energy bands: soft 
(500--2000~eV), hard (2001--7300~eV) and full (500--7300~eV). These
were chosen with consideration of the energy-dependent sensitivity of
the EPIC instruments, the expected spectra of our sources of interest,
the energies of strong fluorescent spectral features from the detectors
(particularly Ni, Cu and Zn K$\alpha$ at energies 7.3--9~keV), and
compatibility of count-rates and fluxes derived therefrom with those
quoted using previous and current instruments for X-ray astronomy. 
We give an outline of the source detection procedure here and describe
each stage in more detail below. 

\noindent For each EPIC instrument and energy band:
\begin{itemize}
\item Define good time intervals that reject times of high background
  rate.
\item Extract an image using good time intervals. 
\item Generate an exposure map.
\item Create a background map. 
\end{itemize}
\noindent For each band:
\begin{itemize}
\item Mosaic the images of the (up to) three instruments to use the
  full sensitivity of the combined EPIC detectors. 
\item Mosaic the background maps.
\item Scale the exposure maps to account for the differences in
  sensitivity and mosaic.
\item Locate candidate sources using a wavelet transform algorithm.
\end{itemize}
\noindent Finally:
\begin{itemize}
\item Match candidate sources from all bands and concatenate into a
  single list.
\item Parameterize candidate sources in all bands simultaneously via
  maximum-likelihood fitting of the point-spread function and reject
  those with low detection significance.
\item Verify by eye and reject obvious spurious detections.
\item Calculate upper limits in bands where each source is undetected.
\item Calculate and correct any systematic offset of the X-ray source
  positions from the positions of near-infrared counterparts.
\end{itemize}

\subsection{Definition of good time intervals}

In almost every observation the background count-rate was highly variable,
by factors up to several hundred\footnote{The EPIC background is described at
  http://xmm.vilspa.esa.es/ external/xmm\_sw\_cal/background/index.shtml;
  see also \citet{read03}.}. To optimize sensitivity to
the detection of faint sources, it was necessary to exclude intervals
of highest background count-rate. Good time intervals (GTIs) were defined
independently for each instrument and energy band as follows. 

\begin{figure*}
\centering
\includegraphics[width=0.45\textwidth]{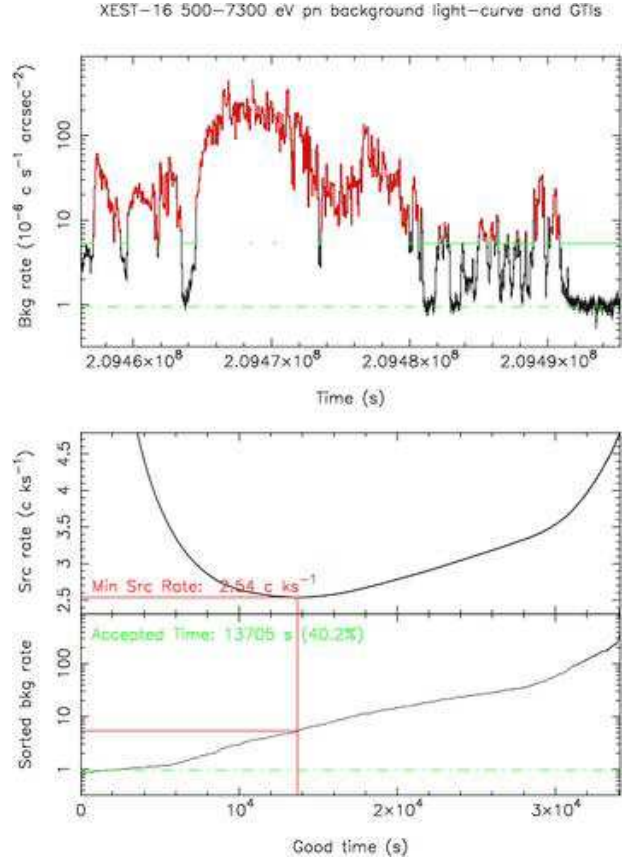}
\caption{The selection of good time intervals to reject times affected by 
high background and so optimize sensitivity to the detection of faint sources. 
The background count rate per unit area in the 0.5--7.3~keV energy band is 
shown in the top panel. The expected quiescent background level in this band is
 marked by the green dot-dashed lines. The bottom panel shows the background 
 count-rate sorted in ascending order. The middle panel shows as a function of 
 accumulated good exposure time the minimum source count-rate that will give a 
 4$\sigma$ detection against the accumulated background in a circular source 
 extraction region of radius 15\arcsec, containing 67 per cent of the source counts. 
 The minimum defines the amount of accepted good time, noted in the bottom panel 
 in green in seconds and as a percentage of the total exposure time, and the maximum 
 acceptable background rate, marked in the upper panel by a green solid line. The time 
 bins marked in red in the upper panel are rejected. XEST-16, shown here, was one of the 
 observations worst-affected by high background, but time intervals with background levels 
 more than 100 times the quiescent level occurred in most observations.}\label{fig_gtis}
\end{figure*}

A light curve was extracted of the whole active detector area (excluding bright
sources, as described in the creation of background maps, below) with
a binning of 52~s, scaled to a 1 sq.~arcsec area, and ordered by increasing
count-rate. Its cumulative distribution was made and the source
count-rate required to give a $4\sigma$ detection in a circular region
of 15~arcsec\footnote{This encloses approximately 68 per cent of
  source counts and is the area used by the SAS source
  parameterization task EMLDETECT.} was
calculated for each bin, and the minimum was found. This defined the
maximum background count-rate accepted, and hence the good time
intervals (see Fig.~\ref{fig_gtis}). In some observations, the source
count-rate required for a $4\sigma$ detection is a factor three higher
than would be expected for a 30-ks observation performed with
background at the quiescent level. The total accepted good exposure
times for the full energy band for each instrument in each observation
are listed in Table~\ref{tab3}.

\begin{table*}
\caption{Accepted good exposure time and fraction of total exposure time remaining after filtering out 
intervals of high background in the full energy band (0.5--7.3~keV). Calculated for each EPIC instrument 
in each {\it XMM-Newton} observation used in XEST. The penultimate column lists the approximate lowest 
detectable count-rate for a 4$\sigma$ detection in the PN; for a 30~ks exposure in purely quiescent background, 
this would be 1.1~c\,ks$^{-1}$. For XEST-26 only the MOS cameras were used and the MOS1 value has been multiplied 
by a factor 2.7 to calculate a PN-equivalent source count-rate. The final column lists the total number of 
detections made in the summed EPIC images.}
\label{tab3}
\centering
\begin{tabular}{crrrrrrcr}
\hline\hline
Field & \multicolumn{2}{c}{M1} & \multicolumn{2}{c}{M2} & \multicolumn{3}{c}{PN} & EPIC\\
      & $t_{\rm good}$ & Per cent & $t_{\rm good}$ & Per cent & $t_{\rm good}$ & Per cent & $C_{\rm 4\sigma}$ & $N_{\rm det}$\\
      & s & & s & & s & & c\,ks$^{-1}$ \\
\hline
01 & 50215 & 62.9 & 50944 & 63.9 & 41756 & 57.2 & 1.06 & 104\\
02 & 39078 & 95.4 & 39638 & 96.8 & 27895 & 71.0 & 1.63 &  76\\
03 & 22814 & 65.2 & 23908 & 68.3 & 15647 & 47.1 & 1.81 &  37\\
04 & 29722 & 89.9 & 29710 & 89.9 & 29131 & 95.0 & 1.19 &  70\\
05 & 24125 & 76.3 & 24229 & 76.6 & 15963 & 53.9 & 1.86 &  36\\
06 & 30777 & 97.4 & 30894 & 97.7 & 27858 & 92.9 & 1.30 & 103\\
07 & 30013 & 95.1 & 30183 & 95.6 & 25338 & 84.2 & 1.29 &  52\\
08 & 35577 & 85.6 & 35841 & 86.2 & 26699 & 71.7 & 1.66 &  82\\
09 & 25009 & 79.0 & 25170 & 79.5 & 19272 & 64.3 & 1.82 &  81\\
10 & 28597 & 90.3 & 28649 & 90.5 & 25476 & 84.8 & 1.31 &  66\\
11 & 34245 & 82.5 & 35075 & 84.2 & 30972 & 80.7 & 1.14 & 104\\
12 & 30398 & 97.9 & 30451 & 98.1 & 28673 & 97.4 & 1.17 &  90\\
13 & 24561 & 77.8 & 24979 & 79.1 & 20918 & 70.5 & 3.11 &  37\\
14 & 28832 & 84.6 & 28942 & 84.9 & 24173 & 74.9 & 1.65 &  78\\
15 & 23372 & 75.5 & 23475 & 75.9 & 19166 & 63.1 & 1.67 &  84\\
16 & 23864 & 58.9 & 25683 & 63.3 & 13705 & 40.2 & 2.54 &  46\\
17 & 26750 & 86.7 & 26804 & 86.9 & 22898 & 78.3 & 1.59 &  68\\
18 & 28540 & 90.1 & 28654 & 90.5 & 26500 & 89.8 & 1.24 &  66\\
19 & 30235 & 80.0 & 30544 & 80.8 & 26086 & 75.4 & 1.29 &  90\\
20 & 31494 &100.0 & 31479 &100.0 & 29877 &100.0 & 1.18 &  76\\
21 & 29060 & 74.9 & 29094 & 75.4 & 18386 & 40.4 & 2.03 &  75\\
22 & 55031 & 96.8 & 54931 & 96.6 & 50484 & 92.7 & 0.76 & 116\\
23 & 57519 & 79.9 & 57496 & 79.8 & 41190 & 59.5 & 1.11 &  76\\
24 & 32069 & 72.6 & 31991 & 72.4 & 24253 & 60.2 & 1.57 &  63\\
25 & 16411 & 97.8 & 16466 & 98.1 & 14399 & 94.9 & 1.84 &  54\\
26 &124995 & 97.4 &125270 & 97.6 &   -   &  -   & 0.81 & 165\\
27 & 37770 & 87.3 & 27864 & 87.5 & 35416 & 85.2 & 1.16 & 153\\
28 &124833 & 95.4 &124951 & 95.5 &120693 & 93.1 & 0.52 & 199\\
\hline
\end{tabular}
\end{table*}

\subsection{Extraction of images}

An image was created for each instrument and each
energy band, in sky coordinates with pixel size $2\arcsec\times 2\arcsec$,
using data in good time intervals only. The intrinsic pixel sizes
of the PN and MOS detectors are 4$\arcsec$ and 1.1$\arcsec$, respectively. In
processing, a randomization of the event position within a pixel is
performed to avoid fixed pattern noise. 2\arcsec\ was simply found to 
provide much better separation of close sources than 4\arcsec.  

For each energy band, the images from the three detectors were
mosaicked to use the full sensitivity of the combined EPIC cameras. 
The relative astrometry of the EPIC detectors is less than
1.5\arcsec, so we applied no offset between the images. 

An exposure map was generated for each image using 
EEXPMAP. A summed exposure map was also produced for each energy
band, but first MOS exposure maps were scaled to account for the lower
sensitivity compared with the PN. The scaling factors used were 0.370,
0.345 and 0.385 in the full, soft and hard bands, respectively: these
factors are strictly dependent on the individual source spectrum; the
values used are median values found for all sources detected in the
first 12 analyzed observations. The count rates resulting from our
source detection are therefore approximate PN-equivalent count rates.

\subsection{Creation of background maps}

A background map was created for each image: each instrument and energy
band. The basic method was to mask out regions of the image affected
by source counts and apply a 2-D Gaussian smoothing filter of $\sigma
= 60\arcsec$. 
The background map was produced in three iterations which successively 
improved the source parameterization and hence the mask. We first used 
the SAS wavelet transform algorithm EWAVELET to locate and approximately 
parameterize sources. The resultant source list and background map were 
input into the maximum-likelihood fitting algorithm EMLDETECT to produce 
an improved background which was input into a second 
run of EMLDETECT to produce the final background map.
The mask, produced using REGION, masked out each source with
an elliptical area, whose eccentricity reflected the shape of the
point-spread function at the source position and whose size reflected
the number of source counts.
Smoothing was performed using ASMOOTH. 
EMLDETECT outputs an image of the parameterized sources, which was
masked and smoothed in the same way as the image and subtracted from
the background to remove the effect of remaining source counts in the
background map. 
Out-of-time events (OOTEs) in PN images were modeled into the background
map. An image of OOTEs was created\footnote{see http://xmm.vilspa.esa.es/sas/new/documentation/threads/ EPIC\_OoT.html} 
and smoothed only along
the CCD readout direction, subtracted from the true image before
source detection and smoothing, and added to the resulting background map.
Where a central MOS CCD observed a bright source in a window mode such that no reliable background
level could be assessed from this CCD, the
background in that CCD was estimated as 1.15 times the mean background
in the other CCDs, as measured from MOS observations in full frame mode
lacking a bright central source.
For each energy band, we mosaicked the background maps from all EPIC detectors.

\subsection{Location of source candidates}

Source candidates in the mosaicked EPIC image in each energy band were
located using two wavelet transform algorithms: the SAS task EWAVELET, and
PWXDETECT developed at the Osservatorio Astronomico di Palermo \citep{damiani97}. 
In PWXDETECT the wavelet transform (WT) is obtained from the rate image (image$/$exposure map) $r(x,y)$ as:
\begin{equation}
w(x,y;a) = \int\int\,g\left(\frac{x-x'}{a},\frac{y-y'}{a}\right)\,r(x',y')\,dx'\,dy'
\end{equation}
The WT is evaluated at different values of the scale parameter, $a$, of the
generating wavelet\footnote{The wavelet function $g$ adopted in this method is a
two-dimensional 'Mexican hat': $g(x/a,y/a) = (2-r^2/a^2)\,e^{-r^2/2\,a^2}$
\hspace{0.2cm}$(r^2=x^2+y^2)$. }, going from $2.8''$ to $16''$ with a logarithmic
step of $\sqrt 2$.  This property  makes this multiscale WT
particularly suited when the PSF varies across the detector image, as well as
very effective in detecting extended sources: for the case of the EPIC images,
point-like sources are detected typically at $a=8''$ (although faint sources
with undersampled PSF may be detected either at low and at high detection
scale), while extended sources typically result with $a=16''$.

Since sources are defined as local maxima in the WT space, thresholds for
detection at a given confidence level must be assessed for each field and each
band. This is obtained by means of extensive simulations on summed PN and MOS
images containing only background photons. The number of background photons in
each simulated PN or MOS image is evaluated from the observed image relevant to
that band, instrument and field. To avoid rejecting possible source
candidates, unconservative thresholds were adopted so as to give about 150 spurious
detections due to 
background fluctuations, per band and per field. These simulations were also used to 
assess the number of false detections expected after maximum likelihood fitting of the 
candidate sources (see below). EWAVELET uses a similar methodology and we used a 
similarly low detection significance threshold of $3.5\sigma$. 
For each algorithm the source lists in the three bands were merged to
create a single list input to EMLDETECT.

\subsection{Parameterization of detected sources}

Source candidates were parameterized in the three energy bands
simultaneously via maximum likelihood fitting of the spatially-varying
point spread function (PSF) using EMLDETECT.
The free fit parameters were the source location, fixed to be the same
in all energy bands, and the count rate in each energy band. 

We retained detections with maximum likelihood $ML > 8$ in
at least one band. Extensive tests of simulated blank-field EPIC
images have demonstrated that we expect approximately 1.5 false detections
per full-band image due to background fluctuations alone.

Detections were overplotted on the image and examined by eye. Obvious
false detections, e.g. in the wings of the point spread function
of bright sources, were removed and the maximum likelihood fitting
repeated.

This procedure was run using the EWAVELET candidate list and the
PWXDETECT candidate list. The two source lists were merged, adding
0--5 extra sources per observation, and the maximum likelihood fitting
repeated.

For each source, the parameters fitted by EMLDETECT were retained in
each band where the source was detected with $ML > 5$, else upper
limits were calculated at the 95 per cent confidence level using the
prescription of \citet{kraft91}.

\subsection{Boresight correction}

2MASS counterpart positions were used to correct X-ray source
positions for systematic shifts; this boresight correction was made
iteratively. First, we cross-correlated the  XEST catalog and the 
2MASS All-Sky Catalogue of Point Sources \citep{cutri03},
using a 5\arcsec\ correlation radius and keeping only the
minimum-distance match to compute the XEST-2MASS position
offsets. The XEST positions were  corrected by subtracting the median
position offset. Then, we repeated this correction process, using a
3.5\arcsec~correlation radius, until the median position offset became
zero, which was obtained in a few steps. The boresight corrections are small
(0.4\arcsec--3.8\arcsec), and the final residual registration error between
2MASS and XEST sources is 0.9\arcsec-- 2.2\arcsec.

The resultant boresight correction was
applied to the X-ray source positions. The RMS error on the boresight
shift of each field was calculated and added in quadrature to the
statistical positional error calculated for each source in that field
by EMLDETECT to give a $1\sigma$ positional error circle.  

We applied the same
method to OM sources using 10\arcsec~and 3\arcsec~radii. We found larger
boresight corrections (1\arcsec--8\arcsec) but with residual registration
errors lower than 1.6\arcsec.

\subsection{Candidate TMC members}

To assess membership in TMC of the detected X-ray sources, we compiled a comprehensive
membership catalog mostly based on the following references:
\citet{cohen79}, \citet{kenyon95}, \citet{briceno98}, \citet{white01}, \citet{briceno02}, \citet{hartmann02}, 
\citet{hartigan03}, \citet{luhman03}, \citet{luhman04}, \citet{luhman06}, \citet{white04}, \citet{andrews05}, and \citet{guieu06}.
The catalog was subsequently used to decide which X-ray sources to use for the physical interpretation, and
to assess which TMC members remained undetected. The adopted coordinates (mostly from 2MASS) are
described in more detail below (Sect.~\ref{coord}).

\subsection{Extraction of source spectra and light curves}

Source events were extracted from circular regions centered on the
EMLDETECT source positions. Although the off-axis PSF is
obviously elongated, there is currently no available parameterization
of an elliptical form of the PSF and the enclosed energy correction
calculated by the SAS task ARFGEN in creating the ancillary response
file is not energy dependent unless a circular source extraction
region is used. We calculated the source extraction radius that
maximized the source signal-to-noise ratio in the full energy band
using: the total number of source counts derived by EMLDETECT for the full-band 
mosaicked EPIC image, the expected number of background counts per unit area
in the source region, calculated from the full-band EPIC background
map, and the circularly-symmetric EPIC PSF formulation of
\citet{ghizzardi02}. The SAS task REGION was used to
calculate smaller extraction radii in cases where nearby contaminating
sources existed. The source extraction radius was constrained to be
between 8\arcsec\ and 60\arcsec. 

Annular background extraction regions were calculated by the REGION
task, with inner radii calculated such that the source count density
(counts per sq. arcsec) had fallen to 30 per cent of the local background
count density. The task excluded neighboring sources using the same
criterion. The outer radius was simply 3 times the inner radius.
For each case where a source of interest was clearly contaminated by
counts from a neighboring source, the background region was
individually defined to account for these, as an annulus around the
contaminating source with inner and outer radii exactly enclosing the
extraction region of the source of interest (with the source itself
excluded with a larger circle), or as a number of circles within this
annulus of identical size to the extraction region of the source of
interest, chosen to avoid other nearby sources. 

The same source and background regions were used for the extraction of
PN and MOS products. Background products were scaled by the ratio of
active detector area in the source region to that in the background region.

Light curves and spectra were extracted for all sources
identified as members of Taurus--Auriga, for all other sources with more than
150 EPIC counts (see e.g. Scelsi et~al. 2006), and all other sources
detected by the Optical Monitor \citep{audard06}. 

\subsubsection{Light curves}

Light curves were extracted for each instrument in the full-band
only and used all available exposure time. Time intervals affected by high 
background were not excluded as 
intrinsic source variability can still be detected in these times.
 Also, the background flares provide a sensitive test of
the reliability of the background subtraction. 

The active exposure time in each bin was calculated, accounting for intervals
when the instrument (or CCD) was not live, which are frequent when the
instrument experiences high event rates during background
flares. The background light curve was corrected for differences in
active exposure 

The time bin size was calculated such that the average number of EPIC
source counts in each bin was 20. However, the minimum bin size was fixed to
100~s to produce easily-viewed plots and the maximum bin size was fixed to
1000 s, the timescale on which flares are typically observed, to avoid
short bright flares on faint sources being smoothed out. The same bin
size was used for PN and MOS light curves to enable simple summation to
form an EPIC light curve. 

The light curves often have few counts per bin, in which case the 
traditional approach -- calculating the source counts as $T - B$, where 
$T$ is the total number of counts in the source extraction region and $B$ 
is the number of expected background counts, with a symmetric $1\sigma$ 
uncertainty calculated by propagating $\sqrt(N)$ errors -- breaks down.
We used the Bayesian approach of \citet{kraft91} to 
calculate the most-likely source count-rate and the 68 percent confidence 
interval in each bin with fewer than 20 net source counts or where the 
traditional $1\sigma$ uncertainty extended below zero.

\begin{figure}
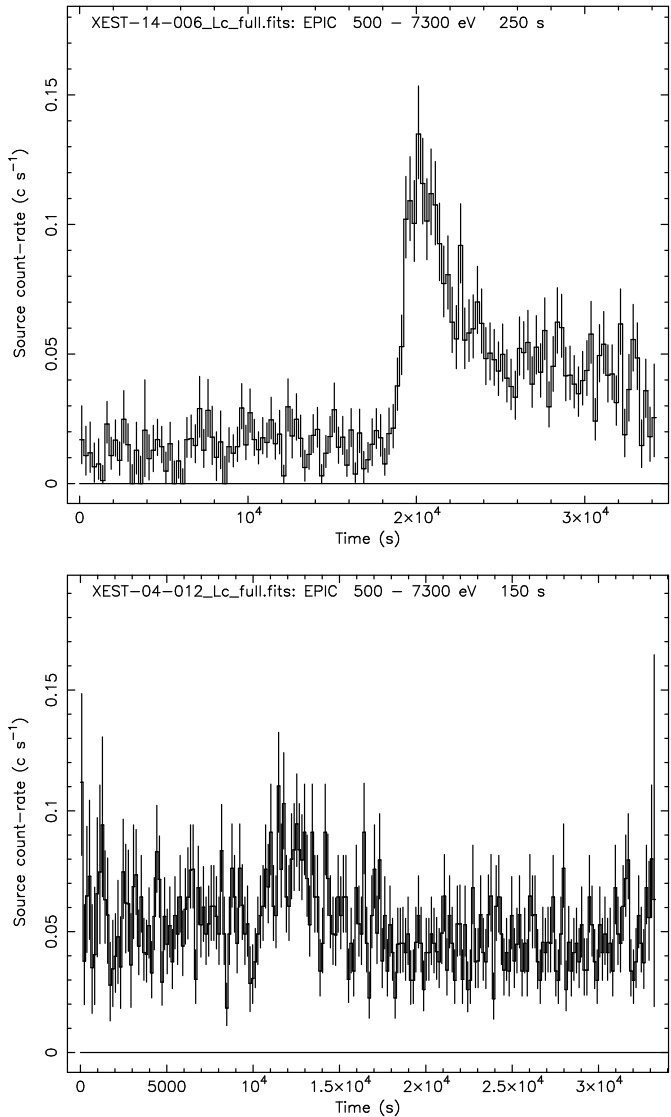

\centering
\includegraphics[angle=-90,width=0.49\textwidth]{guedel_f3a.ps}
\vskip 0.3truecm
\includegraphics[angle=-90,width=0.49\textwidth]{guedel_f3b.ps}
\caption{Two X-ray light curves, summed from all three EPIC detectors in the energy range 0.5-7.3~keV. 
The total count rate has been re-normalized to the effective area
of the PN. The upper panel shows IQ Tau = XEST-14-006 that flared strongly
in the second half of the observation. The low-level episode used for spectral fitting 
was defined to cover the first 18~ks (bin size: 250~s). The lower panel shows V807 Tau = XEST-04-012,
exhibiting low-level flaring variability seen in most XEST objects (bin size: 150~s). Here, all data were used for the spectral fits.
}\label{lightcurve}
\end{figure}

We also calculated a summed EPIC light curve. The counts in the PN,
MOS1 and MOS2 source light curves were summed, as were those in the
background light curves, and the net source light curve was calculated
as described above. A PN-effective count rate was calculated for
each bin by scaling the MOS active exposure time by the factor 0.370
(see above) and dividing the summed net counts by the sum of active
exposure times in each bin. Two example light curves are shown in Fig.~\ref{lightcurve},
one containing a strong flare (IQ Tau = XEST-14-006), and one containing 
some low-level activity (V807 Tau = XEST-04-012) seen in most stellar light curves in 
XEST.

\subsubsection{Spectra}

Source and background spectra were produced for each instrument
using data in the GTIs used for source detection. This may be too
conservative for some bright sources, but in trying to extract
products automatically for sources with a wide range of count-rates,
this is most straightforward. For sources whose light curves showed
clear flares we also excluded times of flaring emission where
possible, as we aimed to avoid measuring properties of a source in a
temporary peculiar state. 

PN spectra were extracted using events with patterns 0--4, as the
spectral response is not calibrated for higher patterns.
Ancillary responses files (ARFs) were produced for each source
spectrum using the SAS task ARFGEN. We used the canned response
matrix files (RMFs) appropriate for SAS v6.1.0. For each source
spectrum we used the PN RMF appropriate for patterns 0--4, the
observing mode and the source position on the CCD, and MOS RMFs
appropriate for Imaging mode, patterns 0--12, and the orbit (date) of
observation.

Three examples of EPIC PN spectra are shown in Fig.~\ref{epicspectra} together with
the best DEM fits (see below).

\begin{figure}
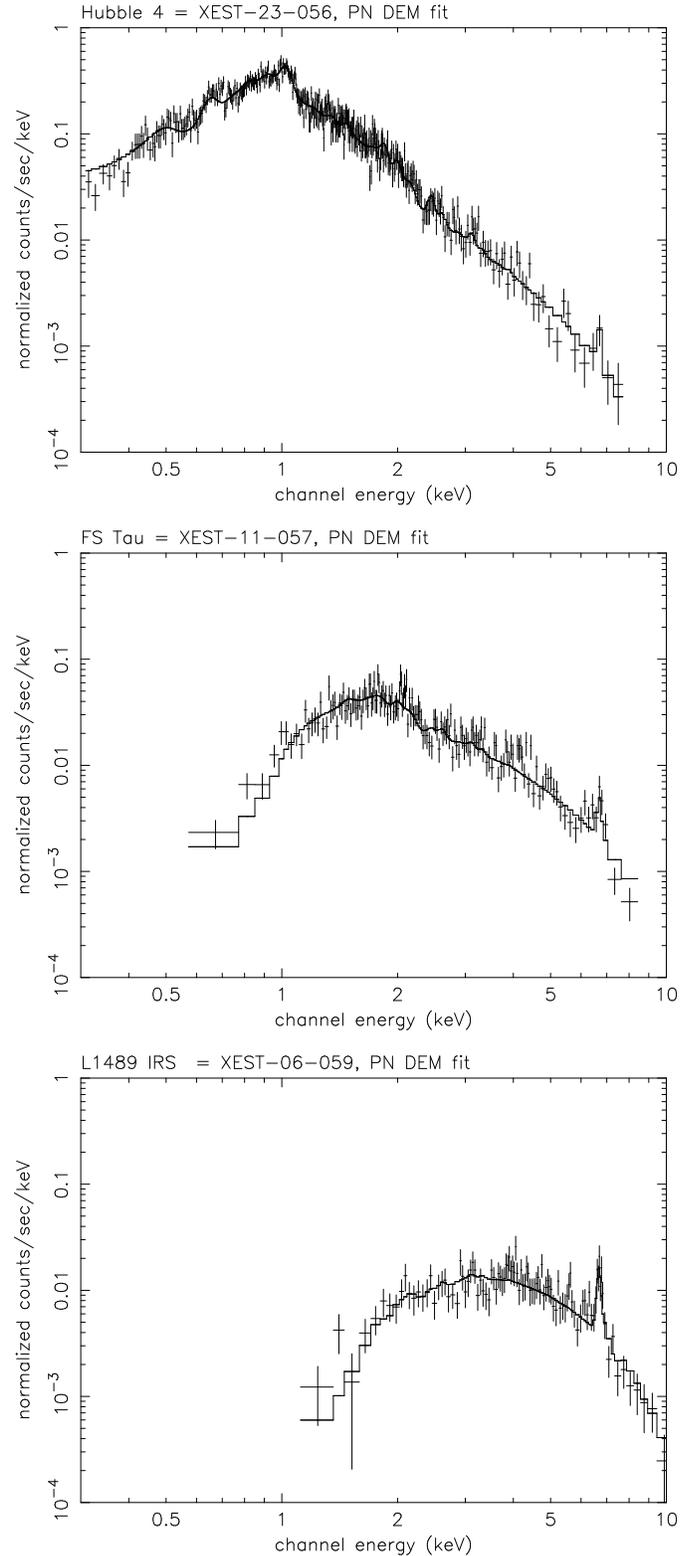

\centering
\includegraphics[angle=-90,width=0.49\textwidth]{guedel_f4a.ps}
\vskip 0.3truecm
\includegraphics[angle=-90,width=0.49\textwidth]{guedel_f4b.ps}
\vskip 0.3truecm
\includegraphics[angle=-90,width=0.49\textwidth]{guedel_f4c.ps}
\caption{Three example CCD spectra from the EPIC PN camera. All spectra extracted from the source position were binned to a minimum
of 15 cts per bin. The histograms show the best fits derived from DEM fits.
Top: Hubble 4 = XEST-23-056, an X-ray strong WTTS with modest absorption (total of 27648~cts in EPIC, $N_{\rm H} = 3.1\times 10^{21}$~cm$^{-3}$).
Middle: FS Tau = XEST-11-057, an intermediately X-ray strong  CTTS with strong absorption (total of 7398~cts in EPIC, $N_{\rm H} = 1.4\times 10^{22}$~cm$^{-3}$).
Bottom: L1489 IRS = IRAS 04016+2610 = XEST-06-059, an X-ray bright protostar with very strong absorption (total of 3336~cts in EPIC, $N_{\rm H} = 6.6\times 10^{22}$~cm$^{-3}$).
}\label{epicspectra}
\end{figure}


\subsection{Overall sensitivity}

\begin{figure}[t!]
\centerline{\resizebox{0.98\hsize}{!}{\includegraphics{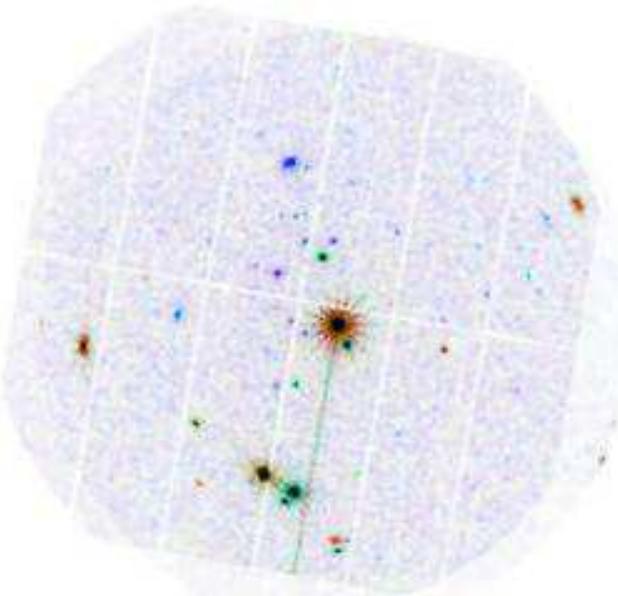}}}
\caption{\small Example of an XMM EPIC image (the field around V410 Tau = L1495).
         The field diameter is 30$^{\prime}$, and the point-spread function has
	 an FWHM of about 5$^{\prime\prime}$. North is up. The colors indicate X-ray hardness,
	 the red-orange sources being softest and the blue sources being hardest. This exposure 
	 was longer than average (74~ks compared
	 to the more typical 30-40~ks). It contains 76 X-ray detections, of which 20 are 
	 TMC members.
  \label{fig-2}} 
\end{figure}

An example of a rich EPIC field in L1495, centered at V410 Tau, is displayed in Figure~\ref{fig-2},
showing a particularly deep observation within our survey.
A typical exposure of $\approx 30$~ks duration with an average background contamination level reached a detection
threshold of $\approx 9\times 10^{27}$~erg~s$^{-1}$ on-axis and  $\approx 1.3\times 
10^{28}$~erg~s$^{-1}$ at 10$^{\prime}$ off-axis for an X-ray 
source with a thermal spectrum characteristic of T Tau stars (see below) subject to
a hydrogen  absorption column density of $N_{\rm H} = 3\times 10^{21}$~cm$^{-2}$. This 
threshold turns out to be appropriate to detect essentially every
T Tau star  in the surveyed fields. The precise detection limits depend on the level of
particle background contamination (hence the exposure time obtained during low background
radiation), and the off-axis angle. The effective area drops to about 50\% at 10\arcmin\
distance from the center and to 35\% at the border of the detectors (15\arcmin\ distance).

A typical XEST exposure contained between 50 and 100 X-ray point sources (Table~\ref{tab3}) of which, however, a minority
(of order 10) are known TMC stellar members. The remainder are probably mostly extragalactic sources,
but a non-negligible contribution may be defined by as yet unidentified new TMC members that
may be embedded or be located behind the molecular cloud complex (see \citealt{scelsi06} for
a detailed study).

\subsection{Chandra data}

The few exposures obtained with {\it Chandra} were reduced using standard strategies for ACIS,
while we used the standard, reduced events file for the HRC-I observation around HR~1442.
The ACIS-S observations of the fields around DG Tau and FS Tau used the ``Very Faint Mode'' of ACIS, while those around
 the fields of GV  Tau, V410 Tau, L1527  used the ``Faint Mode''. 
The data were reduced in CIAO vers. 3.3.0.1 following the standard analysis 
threads\footnote{http://asc.harvard.edu/ciao/threads/}. These procedures 
included, for ACIS, corrections for charge transfer inefficiency and afterglow, and selection
of good time intervals.  Spectra were extracted from level 2 event
files with  DMEXTRACXT, and response and ancillary response files were generated with the 
MKRMF and MKARF tasks, respectively. We re-emphasize that these data are
added here as they provide information on some  objects undetected by {\it XMM-Newton},
but the observations have been obtained with  rather different exposure times, and the 
detectors in use vary (ACIS-S, ACIS-I, HRC-I). Some of the objects were at large off-axis 
angles and were thus subject to severe distortion (PSF FWHM  up to several arcseconds). The data 
quality, e.g., positional accuracy or spectral information, is therefore very inhomogeneous for 
this sample.

\subsection{Spectral analysis} \label{analysis}

\subsubsection{Multi-thermal fits} 

For the spectral analysis, the MOS spectra provide little or no additional constraints for the 
many fainter sources, with a total count rate smaller than the count rate of the PN. To 
keep our analysis as uniform as possible, we therefore used only the PN spectrum for
our spectral fits. Exceptions are those sources for which the PN data were not
available, e.g., for sources whose image falls into a PN CCD gap, or for all sources
in XEST exposure 26 for which the PN was not operational.
To derive basic X-ray properties from the spectra, we followed two strategies. 
First, conventional one- or two-component spectral fits (1-$T$ or 2-$T$ henceforth) were performed. 
The two components define two plasmas  with different temperatures $T_{1,2}$ and emission measures EM$_{1,2}$. 
The latter parameters were fitted in XSPEC \citep{arnaud96} using the {\sl vapec} thermal collisional 
ionization equilibrium model that includes emission lines and continua. The abundances were fixed 
at values typical for pre-main sequence or extremely active zero-age main-sequence stars (based on 
results from \citealt{telleschi05}, \citealt{argiroffi04}, \citealt{garcia05}, and \citealt{scelsi05}).\footnote{The
adopted abundances are, with respect to the solar photospheric abundances given by \citet{anders89} = AG89:
C = 0.45, N = 0.788, O = 0.426, Ne = 0.832, Mg = 0.263, Al = 0.5, Si = 0.309, S = 0.417, Ar = 0.55, Ca = 0.195,
Fe = 0.195, Ni =0.195.}  The spectral components were 
subject to  photoelectric absorption  based on the \citet{morrison83} cross sections. 
The equivalent hydrogen column density, $N_{\rm H}$, was treated as a further fit parameter. For the
faintest sources, a 2-$T$ fit provided too many free parameters for the information 
contained in the spectrum. A 1-$T$ fit was performed in these cases.

\subsubsection{Emission-measure distributions} 

For highly absorbed spectra,  the multi-thermal fit method  can become problematic because the softer component
is strongly suppressed or may not be detected at all, biasing the fits toward high 
temperatures. This may further bias the total X-ray luminosity and $N_{\rm H}$. We therefore, in an alternative approach,
combined several thermal components in such a way
that they describe the shape of the differential emission measure distribution (DEM) in a similar way as previously 
found for nearby pre-main sequence stars or extremely active zero-age main sequence 
stars \citep{telleschi05, argiroffi04, garcia05, scelsi05}. In short, we have adopted a model in which the DEM
shows one peak located at the temperature  $T_0$,
and two power-law distributions toward lower and higher temperatures, characterized by
their power-law indices $\alpha$ and $\beta$, respectively. Given the poor temperature discrimination 
of CCD spectra in the softer X-ray range (in particular in the presence of significant photoelectric absorption), 
the power-law index on the low-temperature side was kept 
fixed at a value often found for magnetically active main-sequence and pre-main sequence stars, namely
$\alpha = 2$, while $\beta$ was treated as a free parameter. The DEM was cut off at $\log T = 6.0$ below which
EPIC is not sensitive to photons,  and at $\log T = 7.95$ above which EPIC is insensitive to temperature. 
Within this range, the DEM was binned to $d\log T = 0.1$, i.e., 20 bins (centered at $\log T = $ 6.0, 6.1, ...7.9)
were used.  The element abundances were fixed 
at the same values as those adopted for the 1- or 2-$T$ fits. Again, the absorbing hydrogen column density 
$N_{\rm H}$ was also fitted to the spectrum. The final fit parameter was the normalization, defined as the EM 
in the temperature bin at $T_0$. This method thus assumes that the DEM can be described by four fit parameters, namely $T_0$, $\beta$,
EM$_0$, and $N_{\rm H}$. We summed the EM over all EM bins used for the calculation to obtain the total EM$_t$. 

We set the following
hard limits to the parameters: $-3 \le \beta\le 1$, $2~{\rm MK} \le T_0 \le 32$~MK. In exceptional cases (very hard, absorbed
spectra), we extended the range for $T_0$ to slightly higher values, although we emphasize that at such temperatures,
the dominant contribution to the spectrum is a featureless bremsstrahlung continuum, apart from the Fe~K complex
at 6.7~keV. 

For very faint sources, such fits provide too many degrees of freedom. We thus fixed one or more
of the following parameters: $\beta$, usually fixed at a value of $-1$ which is often found
for brighter sources, or $N_{\rm H}$, fixed at a value corresponding to the tabulated $A_{\rm V}$ assuming a standard
gas-to-dust ratio (after \citealt{vuong03}), or both. In a few  cases, it was necessary to fix 
$T_0$ as well, usually  at $\log T_0 = 7$. Finally, for non-detections with reasonably well-known $A_{\rm V}$, we derived 
upper limits to $L_{\rm X}$ as follows: We used the 95\% confidence  upper limits to the count rates derived at the optical or 
infrared position of the star, and then adopted an $N_{\rm H}$ derived from $A_{\rm V}$. We thus estimated the upper limit using an
average thermal model.
For brown dwarfs and very-low mass stars, we adopted $\beta = -3$ and $\log T_0 = 6.9$, characteristic values found
for the detected low-mass objects. For more massive T Tau stars, we adopted $\beta = -1$ and $\log T_0 = 7$, again characteristic
values found for the detected stars. These upper limits should only be used as rough estimates. For embedded
sources, uncertain visual or infrared extinctions and the strong influence of $N_{\rm H}$ on the measured count rate
make such estimates unreliable, which will therefore not be given. As we will predominantly use the DEM-fit results,
we have not derived upper limits based on 1-$T$ plasmas.

For both models, the total X-ray luminosity was computed for the energy range 0.3-10~keV based on
the integration of the best-fit model. We adopted a distance of 140~pc for all objects.

The same spectral-fit procedures were applied to the {\it Chandra} ACIS data if feasible 
(most of the faint sources were interpreted only with a 1-$T$ model). For the HRC data, we assumed 
a standard electron temperature of 10~MK and estimated $N_{\rm H}$ from $A_{\rm V}$ using standard interstellar 
conversion ratios (\citealt{vuong03} and references therein). The unabsorbed source flux  and the 
corresponding luminosity  were then estimated based on the PIMMS 
software\footnote{http://heasarc.gsfc.nasa.gov/Tools/w3pimms.html}.

\section{Results}\label{results}

Tables~\ref{tab4} -- \ref{tab11} summarize basic X-ray parameters and  fundamental properties of all observed 
TMC members. We reproduce the first ten entries per table for illustration. The entire tables with 
comments and references are available in the electronic version of this paper. 

All tables are of identical length, each line number referring to the same observation of the
same target across all tables. The tables are sorted in increasing right ascension. The first column of 
each table gives the XEST catalog entry. This number is composed of the two-digit code corresponding
to the exposure number listed in Table~\ref{tab1}, followed by the X-ray source number for this exposure
(including every X-ray source regardless of TMC membership). This source number is `000' for X-ray 
non-detections of known TMC members. The second column gives a conventional name 
frequently used in the literature. Preference was mostly given to variable-star names if available. Note that
several objects were observed twice within the principal XEST project 
(in particular in exposures 23 and 24). They are listed as separate entries. XEST-25 is one of a series of
exposures containing the TMC member AA Tau. This star is highly X-ray variable probably due to variable
absorption by its disk. We use only one observation to report the typical X-ray properties of this star.
A more detailed analysis of the temporal behavior of AA Tau is given by Grosso et al. (in preparation). There are
no other known TMC stellar members in this field. 

The final portion of each table refers to results from the complementary {\it Chandra}
observations. Here, we list  TMC objects that were either not covered by {\it XMM-Newton}, regardless of
their detection status  with {\it Chandra}, or objects that were detected by {\it Chandra} but not by
{\it XMM-Newton}. We also add FS Tau because {\it XMM-Newton} observed this star during a slow flux decrease, presumably
from a flare, and Haro 6-5~B which is slightly contaminated by the wings of FS Tau A in the
{\it XMM-Newton} observation. GV Tau is added because of the higher resolution of {\it Chandra} that
attributes the flux to GV Tau A \citep{guedel06b}, while the embedded object Haro 6-10 B at $1\farcs 3$ remains undetected.

\begin{table*}
\caption{X-ray parameters of targets in XEST (1): Positions and count rates [first ten entries]}
\begin{tabular}{rlrrrrrrrrr}
\hline
\hline
XEST & Name & RA$_{\rm X}$  & Dec$_{\rm X}$               &  Poserr    & Offset     & ML$_{\rm det}^a$ & Scts   & $T_{\rm exp}$  &  Rate  & Var$^b$           \\
     &      & h\ \ m \ \ s  & $\deg\ \ \arcmin\ \ \arcsec$   & ($\arcsec$)&($\arcsec$) &                  &        &  (s)           &  (ct~s$^{-1}$)  &   \\
\hline
27-115 &  HBC 352           &  3 54 29.54 & 32 03 02.2 & 1.04 & 0.89 &   9395 &      3832 &  15776 &    0.2429 & 0 \\
27-000 &  HBC 353           &  3 54 30.17 & 32 03 04.3 & 0.00 & 0.00 &      0 & $<$   183 &  14469 & $<$0.0127 & 0 \\
06-005 &  HBC 358 AB        &  4 03 49.27 & 26 10 53.1 & 1.48 & 1.22 &   1073 &       879 &   7361 &    0.1194 & 0 \\
06-007 &  HBC 359           &  4 03 50.82 & 26 10 53.0 & 1.46 & 0.34 &   2921 &      1605 &   7664 &    0.2094 & 0 \\
06-059 &  L1489 IRS         &  4 04 43.07 & 26 18 56.3 & 1.46 & 0.10 &   7632 &      3336 &  46853 &    0.0712 & 0 \\
20-001 &  LkCa 1            &  4 13 14.03 & 28 19 09.9 & 1.58 & 1.71 &    455 &       336 &   7813 &    0.0430 & 0 \\
20-005 &  Anon 1            &  4 13 27.28 & 28 16 23.3 & 1.53 & 1.64 &  24658 &      8299 &  21634 &    0.3836 & 0 \\
20-000 &  IRAS 04108+2803 A &  4 13 53.29 & 28 11 23.4 & 0.00 & 0.00 &      0 & $<$    26 &  29957 & $<$0.0009 & 0 \\
20-022 &  IRAS 04108+2803 B &  4 13 54.72 & 28 11 32.2 & 1.55 & 0.70 &   1145 &       696 &  30660 &	0.0227 & 1 \\
20-000 &  2M J04141188+28   &  4 14 11.88 & 28 11 53.5 & 0.00 & 0.00 &      0 & $<$    54 &  48645 & $<$0.0011 & 0 \\
\hline
\end{tabular}
\begin{minipage}{0.87\textwidth}
\footnotetext{
\hskip -0.5truecm $^a$  Maximum likelihood for detection for {\it XMM-Newton} data, CIAO WAVDETECT 'Significance' for {\it Chandra} data\\
$^b$  Variability flag: 0 = no or only low-level variability; 1 = clear flaring, flare intervals removed in spectral fit;  
     2 = clear flaring observed but flare intervals not removed; 3 = slow decay of flare throughout observation, all data used\\
}
\end{minipage}
\vskip 0.6truecm
\caption{X-ray parameters of targets in XEST (2): Plasma parameters from the DEM fits [first ten entries]}
\begin{tabular}{lllllrlrrrr}
\hline
\hline
XEST & Name & $N_{\rm H}$ \hfill{\scriptsize (1$\sigma$ range)}                 &  $T_0$ \hfill{\scriptsize (1$\sigma$ range)}  &  $\beta$ \hfill{\scriptsize (1$\sigma$ range)} & EM$_t^a$                & $L_{\rm X}^b$ \hfill{\scriptsize (range)}                  & log         & $T_{\rm av}$ & $\chi^2_{\rm red}$ & dof       \\
   &      & ($10^{22}$~cm$^{-2}$) &  (MK)   &          & ($10^{52})$             & ($10^{30}$~erg~s$^{-1}$)               & $L_{\rm X}/L_*$   & (MK)   &              &           \\
\hline
27-115 &  HBC 352           & 0.22\hfill{~\scriptsize(0.19,0.25)} &  6.5\hfill{~\scriptsize( 5.0, 8.2)} & -0.60\hfill{~\scriptsize(-0.79,-0.47)} &  25.21 &     2.657\hfill{~\scriptsize(2.43, 2.87)} &    -3.03 & 12.00 &  0.85 & 129  \\
27-000 &  HBC 353           & 0.17\hfill                          & 10.0\hfill                          & -1.00\hfill                            &    ... & $<$ 0.176                                 & $<$-4.04 &   ... &   ... &  ... \\
06-005 &  HBC 358 AB        & 0.01\hfill{~\scriptsize(0.00,0.04)} &  4.7\hfill{~\scriptsize( 3.0, 7.5)} & -0.60\hfill{~\scriptsize(-1.24,-0.34)} &   3.79 &     0.383\hfill{~\scriptsize(0.37, 0.44)} &    -3.45 &  9.42 &  0.74 & 21   \\
06-007 &  HBC 359           & 0.01\hfill{~\scriptsize(0.00,0.02)} &  7.2\hfill{~\scriptsize( 5.8, 8.9)} & -1.25\hfill{~\scriptsize(-1.84,-0.91)} &   6.65 &     0.663\hfill{~\scriptsize(0.64, 0.69)} &    -3.18 &  9.20 &  1.05 & 57   \\
06-059 &  L1489 IRS         & 6.63\hfill{~\scriptsize(6.18,7.09)} & 50.9\hfill{~\scriptsize(30.9,60.5)} & -3.00\hfill{~\scriptsize(-3.00, 0.02)} &  30.39 &     4.471\hfill{~\scriptsize(4.07, 4.91)} &    -3.63 & 40.42 &  0.94 & 99   \\
20-001 &  LkCa 1            & 0.07\hfill{~\scriptsize(0.04,0.15)} &  4.9\hfill{~\scriptsize( 3.7, 5.7)} & -3.00\hfill{~\scriptsize(-3.00,-1.94)} &   2.66 &     0.232\hfill{~\scriptsize(0.20, 0.37)} &    -3.80 &  4.35 &  0.57 & 18   \\
20-005 &  Anon 1            & 0.33\hfill{~\scriptsize(0.30,0.35)} &  7.6\hfill{~\scriptsize( 6.4, 9.1)} & -1.01\hfill{~\scriptsize(-1.26,-0.88)} &  40.27 &     4.139\hfill{~\scriptsize(3.84, 4.42)} &    -3.38 & 10.72 &  0.92 & 228  \\
20-000 &  IRAS 04108+2803 A &                                 ... &                                 ... &                                    ... &    ... &       ...                                 &      ... &   ... &   ... &  ... \\
20-022 &  IRAS 04108+2803 B & 5.68\hfill{~\scriptsize(2.99,8.19)} &  4.5\hfill{~\scriptsize( 2.0,31.6)} & -1.00\hfill                            &   4.39 &     0.417\hfill{~\scriptsize(0.07, 0.57)} &    -3.57 &  6.93 &  0.68 & 8    \\
20-000 &  2M J04141188+28   & 0.18\hfill                          &  7.9\hfill                          & -3.00\hfill                            &    ... & $<$ 0.028                                 & $<$-3.31 &   ... &   ... &  ... \\
\hline
\end{tabular}
\begin{minipage}{0.95\textwidth}
\footnotetext{
\hskip -0.5truecm $^a$ EM$_t$ is sum of EM over all DEM bins from log$T$ = 6.0 to log$T$ = 7.9; given in units of $10^{52}$~cm$^{-3}$\\
$^b$ $L_{\rm X}$ for [0.3,10]~keV, in units of $10^{30}$~erg~s$^{-1}$\\
}
\end{minipage}
\vskip 0.6truecm
\caption{X-ray parameters of targets in XEST (3): Plasma parameters from the 1-$T$ and 2-$T$ fits [first ten entries]}
\begin{tabular}{rllrrrrrrrrr}
\hline
\hline
XEST & Name & $N_{\rm H}$  {\scriptsize \hfill(1$\sigma$ range)}              &  $T_1^a$  &  $T_2$    & EM$_1^b$               & EM$_2^b$             & $L_{\rm X}^c$                     & log      & $T_{\rm av}$ & $\chi^{2\ d}_{\rm red}$ & dof     \\
   &      & ($10^{22}$~cm$^{-2}$) &  (MK)   &  (MK)     & ($10^{52})$ & ($10^{52})$       & ($10^{30})$                & $L_{\rm X}/L_*$ & (MK)         &                      &          \\
\hline
27-115 &  HBC 352           & 0.19\hfill{\scriptsize~(0.17,0.21)} &   7.54 &   23.77 &  10.35 &  11.52 &  2.307 & -3.09 &  13.81 &  0.87  &  128 \\
27-000 &  HBC 353           &                                 ... &    ... &     ... &    ... &    ... &    ... &   ... &    ... & ...    &  ... \\
06-005 &  HBC 358 AB        & 0.00\hfill{\scriptsize~(0.00,0.03)} &   4.29 &   14.26 &   1.41 &   2.12 &  0.346 & -3.49 &   8.82 &  0.73  &   20 \\
06-007 &  HBC 359           & 0.00\hfill{\scriptsize~(0.00,0.02)} &   5.33 &   15.19 &   3.06 &   3.53 &  0.628 & -3.20 &   9.34 &  1.09  &   56 \\
06-059 &  L1489 IRS         & 6.54\hfill{\scriptsize~(6.08,6.88)} &    ... &   50.20 &    ... &  26.57 &  4.003 & -3.67 &  50.20 &  1.02  &  100 \\
20-001 &  LkCa 1            & 0.07\hfill{\scriptsize~(0.00,0.21)} &   1.39 &    8.35 &   1.76 &   0.82 &  0.205 & -3.85 &   2.46 &  1.27  &   15 \\
20-005 &  Anon 1            & 0.28\hfill{\scriptsize~(0.27,0.29)} &   8.58 &   24.12 &  19.05 &  14.11 &  3.473 & -3.46 &  13.32 &  1.02  &  228 \\
20-000 &  IRAS 04108+2803 A &                                 ... &    ... &     ... &    ... &    ... &    ... &   ... &    ... & ...    &  ... \\
20-022 &  IRAS 04108+2803 B & 7.86\hfill{\scriptsize~(3.40,28.0)} &    ... &   11.59 &    ... &   5.95 &  0.588 & -3.42 &  11.59 &  0.61  &    8 \\
20-000 &  2M J04141188+28   &				      ... &    ... &	 ... &    ... &    ... &    ... &   ... &    ... & ...    &  ... \\
\hline
\end{tabular}
\begin{minipage}{0.93\textwidth}
\footnotetext{
\hskip -0.5truecm $^a$ '=' sign before number indicates that parameter was fixed\\
$^b$ EM in units of $10^{52}$~cm$^{-3}$\\
$^c$ $L_{\rm X}$ for [0.3,10]~keV, in units of $10^{30}$~erg~s$^{-1}$\\
$^d$ Numbers followed by 'C' denote C statistic (for low-background {\it Chandra} data)\\
}
\end{minipage}
\normalsize
\end{table*}

\setcounter{table}{6}
\begin{table*}[t!]
\caption{Fundamental parameters of targets in XEST (1): Names and coordinates [first ten entries]}
\begin{tabular}{rlcrlll}
\hline
\hline
XEST & Name & 2MASS$^a$ & IRAS$^b$ & Alternative names  &  RA(J2000.0)$^c$   & Dec(J2000.0)$^c$                 \\
     &      &           &          &      & h\ \ m \ \ s  & $\deg\ \ \arcmin\ \ \arcsec$ \\
\hline
27-115 &  HBC 352           &  03542950+3203013  & ...          & NTTS 035120+3154SW                    & 3 54 29.51 & 32 03 01.4     \\
27-000 &  HBC 353           &  03543017+3203043  & ...          & NTTS 035120+3154NE                    & 3 54 30.17 & 32 03 04.3     \\
06-005 &  HBC 358 AB        &  04034930+2610520  & ...          & NTTS 040047+2603W                     & 4 03 49.31 & 26 10 52.0     \\
06-007 &  HBC 359           &  04035084+2610531  & ...          & TTS 040047+2603                       & 4 03 50.84 & 26 10 53.2     \\
06-059 &  L1489 IRS         &  04044307+2618563  & 04016+2610   & ...                                   & 4 04 43.07 & 26 18 56.4     \\
20-001 &  LkCa 1            &  04131414+2819108  & ...          & HBC 365, V1095 Tau, JH 141            & 4 13 14.14 & 28 19 10.8     \\
20-005 &  Anon 1            &  04132722+2816247  & ...          & HBC 366, V1096 Tau                    & 4 13 27.23 & 28 16 24.8     \\
20-000 &  IRAS 04108+2803 A &  04135328+2811233  & 04108+2803A  & L1495N IRS                            & 4 13 53.29 & 28 11 23.4     \\
20-022 &  IRAS 04108+2803 B &  04135471+2811328  & 04108+2803	& ...					& 4 13 54.72 & 28 11 32.9     \\
20-000 &  2M J04141188+28   &  04141188+2811535  & ...  	& ...					& 4 14 11.88 & 28 11 53.5     \\
\hline
\end{tabular}
\begin{minipage}{0.96\textwidth}
\footnotetext{
\hskip -0.5truecm $^a$ Nearest 2MASS entry within 5\arcsec\ to coordinates given by references 5, 18, 33, or SIMBAD. Unlikely identifications in parentheses \\
$^b$ Nearest IRAS catalog entry, within 10\arcsec\\
$^c$ 2MASS coordinates. For unlikely identifications, SIMBAD (S) or reference 5 (B)\\
}
\end{minipage}
\vskip 1.truecm
\caption{Fundamental parameters of targets in XEST (2): Multiplicity [first ten entries]}
\begin{tabular}{rlrrr}
\hline
\hline
XEST & Name & Comp & Separations	  & Refs\\
     &      &      & (\arcsec) &	 \\
\hline
27-115 &  HBC 352           & 1 & ...             & ...                                  \\
27-000 &  HBC 353           & 1 & ...             & ...                                  \\
06-005 &  HBC 358 AB        & 3 & 0.15, 1.55      &  11, 20, 31, 40                      \\
06-007 &  HBC 359           & 1 & ...             & ...                                  \\
06-059 &  L1489 IRS         & 1 & ...             & ...                                  \\
20-001 &  LkCa 1            & 1 & ...             & ...                                  \\
20-005 &  Anon 1            & 1 & ...             & ...                                  \\
20-000 &  IRAS 04108+2803 A & 1 & ...             & ...                                  \\
20-022 &  IRAS 04108+2803 B & 1 & ...             & ...                                  \\
20-000 &  2M J04141188+28   & 1 & ...             & ...                                  \\
\hline
\end{tabular}
\vskip 1.6truecm
\caption{Fundamental parameters of targets in XEST (3): Photometry and spectroscopy [first ten entries]\hfill }
\begin{tabular}{rlrrrrrrr}
\hline
\hline
XEST & Name & Spec$^a$  & Refs & $A_{\rm V}^a$ & $A_{\rm J}$  & $T_{\rm eff}^a$ &  $L_*^b$        & Refs	\\
     &      &		 &	& (mag)        & (mag)   &  (K) 	   & ($L_{\odot})$     &	\\
\hline
27-115 &  HBC 352	    & G0	&   27      &	     0.87 &  0.25 &	   6030 &		0.740 &   27	       \\
27-000 &  HBC 353	    & G5	&   27      &	     0.97 &  0.28 &	   5770 &		0.500 &   27	       \\
06-005 &  HBC 358 AB	    & M2	&   27      &	     0.21 &  0.06 &	   3580 &		0.280 &   27	       \\
06-007 &  HBC 359	    & M2	&   27      &	     0.49 &  0.14 &	   3580 &		0.260 &   27	       \\
06-059 &  L1489 IRS	    & K4	&   59      &	    10.20 &   ... &	   4500 &		4.900 &   10, 59       \\
20-001 &  LkCa 1	    & M4	&    5      &	     0.00 &  0.00 &	   3270 &		0.380 &    5, 27       \\
20-005 &  Anon 1	    & M0	&    5      &	     1.32 &  1.03 &	   3850 &		2.600 &    5, 27       \\
20-000 &  IRAS 04108+2803 A & ...	&  ...      &	      ... &   ... &	    ... &		  ... &   ...	       \\
20-022 &  IRAS 04108+2803 B & ...	&  ...      &	      ... &   ... &	   3500 &		0.400 &   10	       \\
20-000 &  2M J04141188+28   & M6.25	&   33      &	      ... &  0.28 &	   2962 &		0.015 &   33	       \\
\hline
\end{tabular}  \\
\begin{minipage}{0.89\textwidth}
\footnotetext{
\hskip -0.5truecm $^a$ For multiples, first number or spectral type refers to primary, second to secondary component\\
$^b$ For multiples, three numbers give primary/secondary/total system luminosity\\
$^c$ Referring to $L_{\rm bol}$ as derived from integration of the optical and infrared spectrum\\
}
\end{minipage}
\vfill
\normalsize
\end{table*}

\setcounter{table}{9}
\begin{table*}
\caption{Fundamental parameters of targets in XEST (4): Age, mass, radius, rotation [first ten entries]}
\begin{tabular}{rlrlrlrlrl}
\hline
\hline
XEST & Name & Age$^{a,b}$   &  Mass$^{a,c}$       & Refs & Radius$^d$    &  $P$ & Refs & $v\sin i$         & Refs  \\
    &       & (Myr)         & ($M_{\odot}$)       &      & ($R_{\odot}$) &  (d) &      & (km~s$^{-1}$)     &        \\
\hline
27-115 &  HBC 352           &       ... &        1.05 &         2 &  0.79 & $<$ 0.53 & C          & $>$ 75.00 &   48       \\
27-000 &  HBC 353           &       ... &         ... &       ... &  0.71 & $<$ 4.08 & C          &      8.80 &    2       \\
06-005 &  HBC 358 AB        &      3.26 &        0.41 &        27 &  1.38 &      ... & C          & $<$ 10.00 &    2       \\
06-007 &  HBC 359           &      3.54 &        0.41 &        27 &  1.33 &      ... & C          & $<$ 10.00 &   48       \\
06-059 &  L1489 IRS         &      0.80 &        1.45 &        10 &  3.65 & $<$ 4.02 & C          &     46.00 &   10       \\
20-001 &  LkCa 1            &      0.87 &        0.27 &         5 &  1.93 & $<$ 4.18 & C          &     23.30 &   48       \\
20-005 &  Anon 1            &      0.50 &        0.56 &         5 &  3.63 &      ... & ...        &       ... &   ...      \\
20-000 &  IRAS 04108+2803 A &       ... &         ... &       ... &   ... &      ... & ...        &       ... &   ...      \\
20-022 &  IRAS 04108+2803 B &      1.60 &        0.36 &        10 &  1.72 & $<$ 6.23 & C          &     14.00 &   10       \\
20-000 &  2M J04141188+28   &       ... &        0.08 &        46 &  0.47 &      ... & ...        &       ... &   ...      \\
\hline
\end{tabular}
\begin{minipage}{0.79\textwidth}
\footnotetext{
\hskip -0.5truecm $^a$ For binaries, first number refers to primary, second to secondary component (calculated from $L_*$ and $T_{\rm eff}$).\\
$^b$ Ages derived after \citet{siess00} using the same principal parameters as for masses, quoted in Table~\ref{tab9})\\
$^c$ Masses derived after \citet{siess00} using principal parameters quoted in Table~\ref{tab9}.\\
$^d$ For multiples, radius is given only for primary if luminosity of primary is explicitly known\\
}
\end{minipage}
\vskip 0.2truecm
\caption{Fundamental parameters of targets in XEST (5): Accretion and evolution [first ten entries]}
\begin{tabular}{rlrlrllrll}
\hline
\hline
XEST & Name & $\dot{M}$ (min/max)$^a$  &  Refs  &  EW(H$\alpha$)$^b$  & TTS    & Refs$^c$  & IR$^d$& Refs$^d$ & Type  \\
   &      & ($M_{\odot}$yr$^{-1}$) &        & (\AA)                   & type   &  	     & class  &          &         \\
\hline
27-115 &  HBC 352           &                 ... & ...       &          0 & W     &  29            & III     &  1, 27     & 3 \\
27-000 &  HBC 353           &                 ... & ...       &          0 & W     &  29            & III     & 27         & 3 \\
06-005 &  HBC 358 AB        & $<$  8.97           &   20      &      4- 10 & W/W   &  37, 29        & III     &  1, 27     & 3 \\
06-007 &  HBC 359           &                 ... & ...       &      2-  9 & W     &  56, 37        & III     & 27         & 3 \\
06-059 &  L1489 IRS         &     -7.15           &   59      &     41- 56 & C*    &  59, 29        & I       & 1, 59, 27  & 1 \\
20-001 &  LkCa 1            & $<$ -9.72           &   58      &      3-  4 & W     &  29, 45        & III     &  1, 27     & 3 \\
20-005 &  Anon 1            & $<$ -8.94           &   58      &      1-  3 & W     &  37, 29        & III     &  1, 27     & 3 \\
20-000 &  IRAS 04108+2803 A &                 ... &  ...      &         37 & C     &  29            & II      & 27         & 2 \\
20-022 &  IRAS 04108+2803 B &                 ... &  ...      &        ... & ...   &  ...           & I       & 59, 27     & 1 \\
20-000 &  2M J04141188+28   &    -10.00           &   46      &        250 & C     &  46            & ...     &  ...       & 4 \\
\hline
\end{tabular}
\begin{minipage}{0.84\textwidth}
\footnotetext{
\hskip -0.5truecm $^a$ Range of $\dot{M}$ reported in literature given. For multiple systems, numbers refer to primary or integrated system\\
$^b$ Range of EW reported in literature given. For multiple systems, numbers refer to primary or integrated system\\
$^c$ For EW range, first reference for minimum, second for maximum reported.\\
$^d$ Infrared classification; double entries: '/' for transition objects, '+' for components, ';' for different types, ',' for  different references. FS = flat-spectrum source\\
}
\end{minipage}
\normalsize
\end{table*}

\subsection{The XEST X-ray results}

The individual X-ray results from our survey will be discussed in depth in the series of accompanying 
papers. We collect here the basic, statistical X-ray results for the stellar TMC sources. Our 
results refer to episodes during the observations that were not affected by outstanding stellar
flares (i.e., flares in excess of the typical slow modulation within factors of $\approx 2$ of the average 
level), although low-level variability is common to pre-main sequence X-ray sources. Exceptions are
CFHT-BD-Tau 1 = XEST-17-068 that showed a flare with a slow decay, dominating most of the exposure
time, and FS Tau = XEST-11-057, DH Tau = XEST-15-040 and V830 Tau = XEST-04-016   that showed slowly decreasing
light curves, probably following a strong flare starting before our exposures.
Specific discussions of the time-dependent behavior of all
observed sources will be given by \citet{stelzer06}, \citet{franciosini06}, \citet{arzner06a},
and \citet{audard06}. 

Table~\ref{tab4} provides a summary of detection results for each XEST TMC member. The table lists
the detection coordinates with statistical errors and the offsets from the expected position as given 
later in Table~\ref{tab7}. For non-detections, the expected coordinates of the source is given (see below), which
were used to calculate upper limits to the count rates.
Then, the maximum likelihood (ML) is given for each detection, referring to the full (0.5-7.3~keV) band. Non-detections
show ML = 0. For the  {\it Chandra} observations in Table~\ref{tab4}, we list the ``significance'' of the detection as
provided by the CIAO WAVDETECT algorithm instead of ML$_{\rm det}$.
Then, we list the total number of detected EPIC counts, the effective PN-equivalent 
on-axis exposure time that corrects for vignetting and also takes the (lower-sensitivity) MOS data into account,
as described above. For non-detections, the 95\% upper limits are reported. 
Next, the table gives the average count rate (on-axis equivalent rate for PN). Finally,
we report a variability flag from inspection of the light curves; the codes have the following meaning:
0 = no or only low-level variability, no flares visible that may dominate the X-ray spectrum;
1 = clear flaring observed and flare intervals removed in spectral fit;  
2 = clear flaring observed but flare intervals not removed;
3 = slow decay of flare throughout the observation, all data used. Examples for light curves with
flag 1 (IQ Tau) and flag 0  (V807 Tau) 
are shown in Fig.~\ref{lightcurve}. In the former case, spectral fits were obtained using only
data during the low-level episode, defined by the first 18~ks of the light curve.  A more quantitative
analysis of light curve variability and an interpretation of the role of flares in coronal heating
will be given by \citet{stelzer06}.

Table~\ref{tab5} summarizes the basic spectral-fit results from our DEM method.
The abundances were kept fixed at characteristic values as described in Sect.~\ref{analysis}.
The columns provide $N_{\rm H}$, the temperature of the DEM peak, $T_0$, the power-law index $\beta$ of
the high-temperature DEM slope, the total EM integrated over all bins centered at values of
log$T = 6.0$  to log$T = 7.9$ in steps of $d\log T = 0.1$, and the inferred $L_{\rm X}$ in the 0.3--10~keV range. 
We list 68\% (1$\sigma$) confidence ranges for $N_{\rm H}$, $\beta$, and $T_0$ in parentheses. 
 For fixed parameters, no error range is given. For non-detections,
the spectral parameters used to estimate upper limits to $L_{\rm X}$ are also given as fixed parameters.
The next columns list the logarithms of the normalized luminosities  $L_{\rm X}/L_*$ where the stellar bolometric
luminosity $L_*$ is adopted from 
Table~\ref{tab9}, and the average electron temperature that is obtained as a mean of $\log T$ over the entire DEM, 
where  DEM($T$) is used as a weight function, calculated analytically as follows:
\begin{eqnarray}
&A& = \left. {T^{\alpha}\over \alpha T_0^{\alpha}}\left(\log T - {1\over c\alpha} \right)\right|_{T_l}^{T_0}
    + \left. {T^{\beta}\over \beta T_0^{\beta}} \left(\log T - {1\over c\beta} \right)\right|_{T_0}^{T_h}\\
&B& = \left. {T^{\alpha}\over \alpha T_0^{\alpha}}\right|_{T_l}^{T_0}
    + \left. {T^{\beta}\over \beta T_0^{\beta}}\right|_{T_0}^{T_h}\\
&\log T_{\rm av}& = {A \over B}    
\end{eqnarray}
where $ c = {\rm ln}10 \approx 2.3$ (note that we use decadic logarithms throughout), and we define the limits 
of the integration as $T_l = 10^6$~K and $T_h = 10^8$~K. 
Finally, we list the reduced $\chi_{\rm red}^2$ achieved by the fit together with the number of degrees of 
freedom (dof).

Table~\ref{tab6} equivalently reports the results from 2-$T$ or, for faint sources, 
1-$T$ fits. The columns give $N_{\rm H}$ (with errors if fitted), $T_1$, $T_2$, and the corresponding EM$_1$, EM$_2$. If only one component was fitted, 
its temperature and EM values are listed in the columns for $T_2$ and EM$_2$ if the temperature exceeds
10~MK, and in the columns for $T_1$ and EM$_1$ otherwise. If a temperature was held fixed, its value is preceded by a ``='' sign.
Further, the table gives $L_{\rm X}$, $\log (L_{\rm X}/L_*)$, and $T_{\rm av}$, the latter defined as the EM-weighted logarithmic average
of the temperature in the case of two non-vanishing components:
\begin{equation}
\log T_{\rm av} = { {\rm EM}_1 \log T_1 + {\rm EM}_2 \log T_2 \over {\rm EM}_1 + {\rm EM}_2 }. 
\end{equation} 
Finally, the table lists the $\chi_{\rm red}^2$ values and the number of degrees of freedom (dof).

\begin{figure}[t!]
\centerline{\resizebox{1\hsize}{!}{\includegraphics{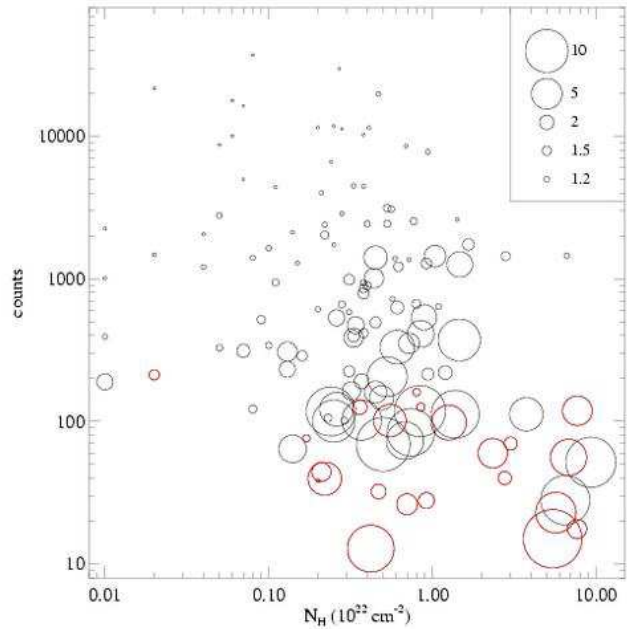}}}
\caption{\small Magnitude of the error ranges for $L_{\rm X}$, estimated from maximum variations of $N_{\rm H}$
(see text for details), as a function of the best-fit $N_{\rm H}$ and the number of counts used in the spectral
fit. Results from the DEM fit method are shown. The circle size scales with the logarithm of the ratio between 
the upper and the lower bounds of $L_{\rm X}$. Red circles denote sources for which the $\beta$ parameter was held fixed.
  \label{fig-err}} 
\end{figure}

Errors of EM and $L_{\rm X}$ are difficult to compute, and their meaning is not unequivocal. 
Our EM models are rather crude approximations to true emission measure distributions.
Formal error ranges for the EM values may therefore bear little relevance. Further, $L_{\rm X}$ is a complicated
function of the thermal emission model. Although resolved spectral features can usually be fitted sufficiently well
in a $\chi^2$ sense using a number of thermal components that describe the absorbed, observed spectral flux density,
the major uncertainty in $L_{\rm X}$ is introduced  by fitting (and correcting for) $N_{\rm H}$. An instability
discussed below makes determinations of $L_{\rm X}$ rather uncertain once cool plasma components are subject
to large absorption while the spectrum is of modest quality. To quantify this latter effect that dominates
the uncertainty of $L_{\rm X}$ in many cases, we proceeded as follows. We adopted the upper and lower 1$\sigma$ bounds
of $N_{\rm H}$ (if $N_{\rm H}$ was a fit parameter), kept these values fixed and refitted the spectrum. The $L_{\rm X}$ values 
thus determined, $L_{\rm X,1}$ and $L_{\rm X,2}$ for the lower and the upper bound, are taken to bracket the error range 
of the tabulated $L_{\rm X}$. We performed this error analysis for the DEM fit method. We graphically show the magnitude 
of the ratio $L_{\rm X,2}/L_{\rm X,1}$ in Fig.~\ref{fig-err} as a function of the best-fit $N_{\rm H}$ (set to $0.01\times 10^{22}$~cm$^{-2}$
for sources with $N_{\rm H}$ smaller than this value) and the number of 
counts used in the spectral fit. Expectedly, the error range is largest for faint and strongly absorbed
sources, and smallest for bright and weakly absorbed sources. For 83 sources, $L_{\rm X,2}/L_{\rm X,1} \le 2$ and for 
48 sources,  $L_{\rm X,2}/L_{\rm X,1} > 2$. For 99 sources, $L_{\rm X,2}/L_{\rm X,1} \le 3$ and for 
32 sources,  $L_{\rm X,2}/L_{\rm X,1} > 3$. \citet{gagne04} described a similar analysis performed for errors in $L_{\rm X}$
for {\it Chandra} observations of the $\rho$ Ophiuchus Cloud, with similar conclusions.

We note, however, that T Tau stars are manifestly variable on timescales of hours, typically within a factor of at
least  two, while most of the EM errors are formally much smaller. A detailed variability analysis of all XEST sources
will be presented by \citet{stelzer06}. Therefore, for the majority of our sources, the formal uncertainties 
from the spectral fits underestimate the true uncertainty. Realistically, thus,
standard error bars of (at least) 0.3~dex should be adopted  for all values of $L_{\rm X}$ (and EM). Larger statistical
errors from spectral fits are typically found for $N_{\rm H} \ga 2\times 10^{21}$~cm$^{-2}$ and  $\la 600$~cts in the spectrum,
or in regions where cts$/(N_{\rm H}/10^{22}~{\rm cm}^{-2}) \la 500$.

\subsection{The XEST catalog}

Tables~\ref{tab7}-\ref{tab11} identify the X-ray sources and provide a summary of the fundamental properties
of the stellar systems. The parameters were extracted from the existing  literature, in particular from \citet{cohen79},
\citet{kenyon95}, \citet{briceno98}, \citet{white01}, \citet{briceno02}, \citet{hartmann02}, \citet{hartigan03}, 
\citet{luhman03}, \citet{luhman04},  \citet{white04}, \citet{andrews05}, and \citet{guieu06}. Several further references 
are listed in the bibliography summary in the table section. For the suspected late-B member of
TMC, HD 28867, all basic information except for the 2MASS identification comes from \citet{walter03} and references 
therein. A short description of the table entries follows.

\subsubsection{Names, coordinates, and multiplicity}\label{coord}

Table~\ref{tab7}  lists the 2MASS catalog entry likely to be identified with the X-ray source.
For this identification, we first adopted object coordinates mostly from 
\citet{briceno02}, \citet{luhman03},  \citet{luhman04}, and \citet{guieu06} if available, or 
from SIMBAD in other cases, and then identified the closest object in the most recent 2MASS catalog version \citep{cutri03} 
within a circle of radius 5\arcsec\ around this position. In three  cases, no 2MASS object was found within 5\arcsec. 
The next columns of Table~\ref{tab7} give the nearest IRAS catalog entry, confined to a search radius of 10\arcsec\
or directly identified from SIMBAD, and a selection of commonly used alternative names.
Finally, the J2000.0 coordinates refer to the adopted 2MASS identifications. If no
2MASS identification is given, the coordinates were taken from SIMBAD,
marked with  '(S)' after the coordinates.

Table~\ref{tab8} provides binary information, i.e., 
the number of components, the approximate component
separations, and references. Note that binary components with a separation of $\la 5\arcsec$\ 
are unresolved in {\it XMM-Newton} observations, and pairs at 5--10\arcsec\ are difficult
to separate. Typically, binaries with separations up to about $10\arcsec$\ have been recorded
as a single X-ray source (e.g., V819 Tau, but note the case of HP Tau/G2 and G3, two stars with a separation of 
10\arcsec\ for which a joint spectral model was derived, while the count rates and thus the luminosities
were estimated separately based on PSF fitting). 

\subsubsection{Photometry and spectroscopy}

Table~\ref{tab9} lists spectral types, visual and $J$-band extinctions ($A_{\rm V}$ and $A_{\rm J}$, respectively, the latter only
for the primary in multiple systems),
effective temperatures ($T_{\rm eff}$; for binaries, separately given for primary and secondary if available), 
and bolometric luminosities ($L_*$) of the stars. The latter are bolometric luminosities of the 
stellar photospheres derived from the optical or near-IR except in cases where only the IR bolometric
luminosity, possibly including  contributions from the disk and/or the envelope, was available (these cases
are marked with `Lb' in the reference column). In the case of binaries, $L_*$
is given for the primary and the secondary if available, and the entire system. The component and the system
values may originate from different references, which accounts for some discrepancies with regard
to the sum of the component  $L_*$. Because spectral types, $A_{\rm V}$, $A_{\rm J}$, and
$L_*$ sensitively depend on the interpretation of the measured $T_{\rm eff}$ and the optical
and near-infrared photometry, we attempted to use the same references for these parameters for a given star, 
and opted predominantly for \citet{briceno02}, \citet{luhman03} and \citet{luhman04}, and
\citet{kenyon95} (in this sequence) for single stars or integrated multiple systems, and for \citet{white01} 
and \citet{hartigan03} for binary components. Significant information for deeply embedded sources was 
adopted from \citet{white04}, and for
brown dwarfs from \citet{guieu06}, \citet{luhman03}, and \citet{luhman04}. \citet{grosso06a} rederived
 $T_{\rm eff}$ of some BDs from the spectral types, while we report here values as originally
published. Strict consistency
was not possible, in particular for $A_{\rm V}$ and $A_{\rm J}$ that had to be adopted from different authors
and that are inconsistent in several cases (the second references in the last column mostly refer to
$A_{\rm V}$).

\subsubsection{Ages, masses, and rotation}

Table~\ref{tab10} lists ages and masses, $M$, that we derived from $T_{\rm eff}$ and $L_*$ as listed
in Table~\ref{tab9} using the \citet{siess00} isochrones. For binaries, component ages and masses
are given if the fundamental parameters were available separately. In a few cases, we did not
find a solution from the \citet{siess00} tracks; if available, some masses (and ages) were therefore
taken directly from the literature (see footnotes to Table~\ref{tab10}). The references specifically
used for $T_{\rm eff}$ and $L_*$ are listed. Assessing the uncertainties of these parameters is
difficult because the uncertainties in $T_{\rm eff}$ and $L_*$ are usually not reported in
the literature. If multiple measurements have been reported for an object, then the agreement
was (for the TMC) within 100~K for $T_{\rm eff}$ for about half of the objects, and within 150~K for two thirds.
We conservatively adopt an error of $\pm 150$~K for $T_{\rm eff}$.
Using the Siess et al. calculations for the regions of densest population in the HRD 
($T_{\rm eff} = 4700, 4000,$ and 3500~K with $L_* = 2, 1,$ and 0.1$L_{\odot}$, respectively),
we find mass uncertainties of $\pm$(0.07--0.16)$M_{\odot}$ or a fractional mass uncertainty of about $\pm$(10--20)\% 
(higher for cooler stars). An uncertainty in $T_{\rm eff}$ also affects the age estimate, resulting in 
typical deviations of $\pm (0.1-0.18)$~dex (factors of 1.3--1.5) from the best-fit value. The age
estimate is further affected by uncertainties in $L_*$ (while $L_*$ has little influence on the mass
estimate in the region where the evolutionary tracks are nearly vertical).  We found that in the
region of $T_{\rm eff} \approx 3500-4000$~K, a shift of $d\log L_*$  in $\log L_*$ results in a shift
of approximately $-d\log L_*$ in age. For $\approx 80$\% of the TMC objects with multiple 
reports of $L_*$ in the literature, deviations are up to a factor of 2 (0.3~dex), but often much smaller. 
We thus adopt a conservative, characteristic uncertainty of factors of 2.6--3.0 for the ages, noting that this 
is a gross overestimate for many objects, while for a minority of objects, in particular for embedded 
protostars with poor $L_*$ determinations, ages may become entirely unreliable.

The radius, $R$,  was calculated from $L_* = 4\pi R^2 \sigma T_{\rm eff}^4$, 
where $\sigma = 5.67\times 10^{-5}$~erg~cm$^{-2}$~s$^{-1}$~K$^{-4}$ is the Stefan-Boltzmann
constant. Rotation periods $P$ are given as measured by various authors. The values were taken
from \citet{rebull04} who give a comprehensive list of references (not reproduced here). That work includes periods
of the lower-mass sample. We have added further rotation periods (Rebull, private communication)
extracted from the same bibliography. If not available from the literature, rotation periods
were estimated as upper 
limits if spectroscopic $v\sin i$ measurements were available (listed in column 9, with references in
column 10, many again referring to the tabulations and the bibliography given by \citealt{rebull04}). 
In those cases, $P \le 2\pi R/(v\sin i)$, and the reference column for these calculated
periods contains a `C'. 

\subsubsection{Accretion and evolutionary stage}

Table~\ref{tab11} provides parameters relevant for the interpretation of the evolutionary status. Column 3 lists mass accretion rates
$\dot{M}$. Accretion rates may be variable, and different indirect methods have been used by previous studies
to estimate $\dot{M}$, hence significantly discrepant values may be found in 
the published literature. In those cases, we provide ranges, giving a reference for each the
minimum and maximum values reported. The same applies to the equivalent width of the H$\alpha$ line
[EW(H$\alpha$)], reported in column 5. In multiple systems, both of the above values refer to the primary star 
even if, in a few cases, a separate measurement was available for a companion. We adopted this strategy 
because the primary star is most likely also to be the dominant X-ray source. 

We assessed the ``T Tauri''  type (classical or weak-line) based on EW as follows. For  spectral types 
G and K, stars with  EW(H$\alpha) \ge 5$~\AA\ are defined as classical T Tauri stars (C), all other stars are weak-line T Tauri
stars (W). For early-M stars, the separation 
was adopted at EW(H$\alpha$) = 10~\AA, and for mid-M type stars, we adopted EW(H$\alpha$) = 20~\AA\ for 
discrimination. For late-M stars, i.e., typically BDs, the low continuum makes a sensible definition 
difficult. We will therefore not include BDs when studying differences between CTTS and WTTS, although we have
adopted the criterion described by \citet{barrado03} that defines accreting stars and BDs as those that show 
$L({\rm H}\alpha)/L_*$ greater than the chromospheric saturation limit. This criterion also supports our
classifications at earlier spectral types. Borderline cases are marked with a question mark. In such cases,
we also included other supporting evidence. For example, KPNO-Tau 14 (XEST-18-004) was classified as `W?' (and subsequently
as `type 3', see below) 
despite one published measurement of EW(H$\alpha$) somewhat higher than the saturation limit, because there is one report
with much lower EW(H$\alpha$), a very low upper limit to $\dot{M}$, and no K band excess \citep{luhman03, muzerolle05}.
`C*' denotes sources for which accretion signatures are evident, but that are embedded and share characteristics 
with protostars. We have added some `C` and `W' designations from \citet{luhman04} even if the explicit EW values 
were not given.

Column 8 lists the ``Young Stellar Object'' (YSO) infrared class as derived from the infrared spectral energy 
distribution, which is thought to provide information on the presence of disks and molecular envelopes. Class 0
and I objects are protostars, Class II objects are disk-surrounded T Tau stars, whereas disk signatures are weak
or absent in Class III stars. `FS' designates ``flat spectrum'' sources, intermediate between 
Class I and II.
Most entries refer to IRAS measurements (and were mostly adopted from \citealt{kenyon95}), but some recent
information from {\it Spitzer} is also available \citep{hartmann05}. If discrepant entries are
given, they are separated by a semi-colon, and so are the corresponding references in the subsequent 
column. References referring to the same classification are separated by a comma. Transition objects
are designated by two types separated by a '/', while for multiple systems with different component
classifications, we give the two classes separated by '+'.

There is a rather good correlation between TTS type indicating presence or absence of
accretion (CTTS and WTTS) and YSO infrared class (Classes II and III, respectively) indicating the presence or absence of
circumstellar disks. There are, however, exceptions that may be borderline cases or in which  
IR companions may bias the YSO classification. We therefore adopted a final classification scheme that 
is predominantly based on the accretion signatures (WTTS vs. CTTS) and takes the IR classification as supporting 
evidence in borderline cases, except for protostars for which the IR classification is the 
only relevant parameter. Hence, in col. 10, '0' and '1' stand, respectively, for a protostar of Class 0 or Class I, `2' for an
accreting (classical) T Tau star that usually shows a Class-II IR spectrum; `3' corresponds to 
a weak-line or Class-III object; `4' designates a brown dwarf (spectral class equal to or later
than M6.25) irrespective of its accretion signatures; and `5' marks the Herbig Ae/Be stars in TMC. For 
uncertain classifications or other object types, we use the code `9'.

\section{Discussion and conclusions}\label{conclusions}

\subsection{Spectral interpretation: Quality of results}

\begin{figure}[t!]
\centerline{\resizebox{0.95\hsize}{!}{\includegraphics{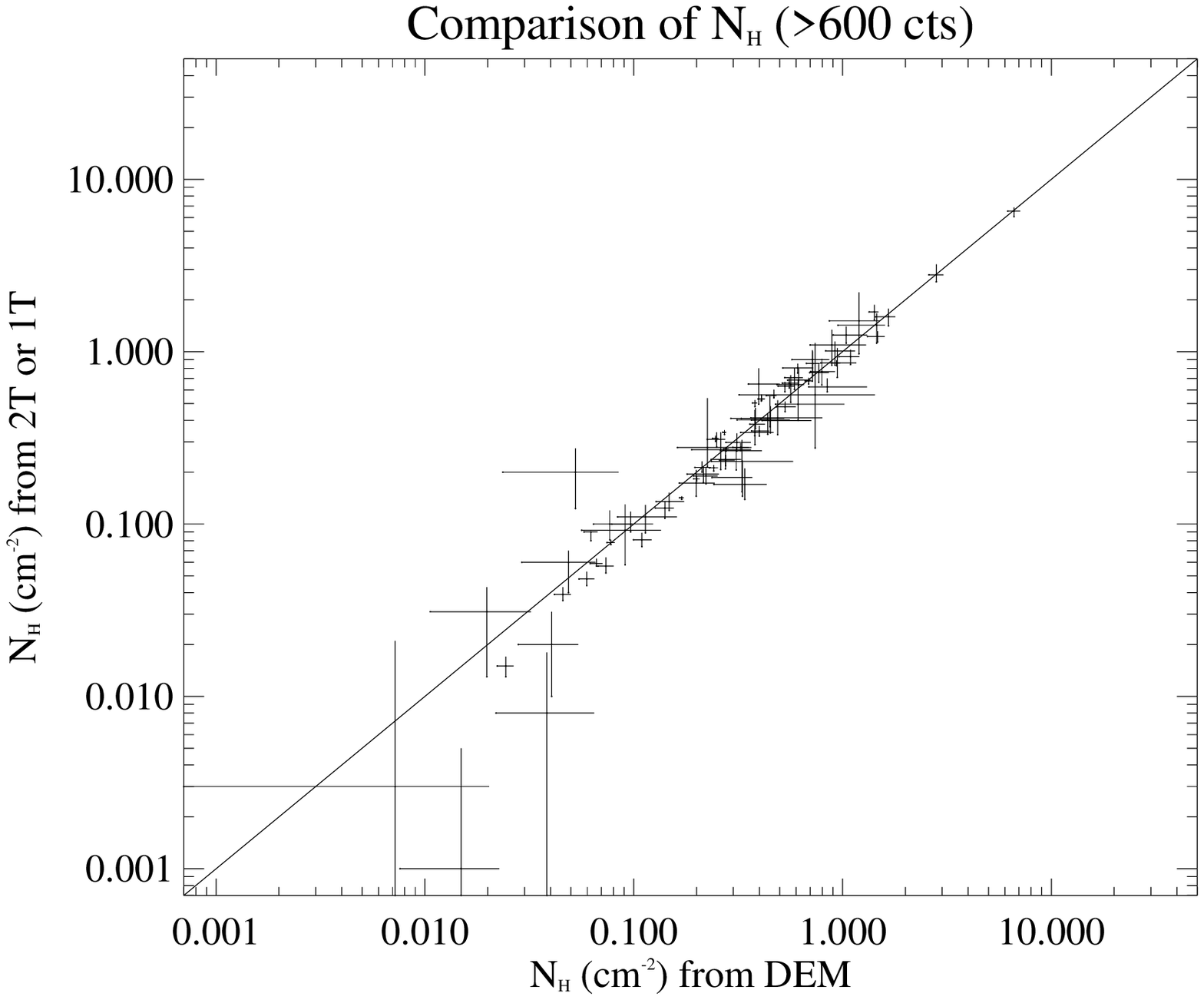}}}
\centerline{\resizebox{0.95\hsize}{!}{\includegraphics{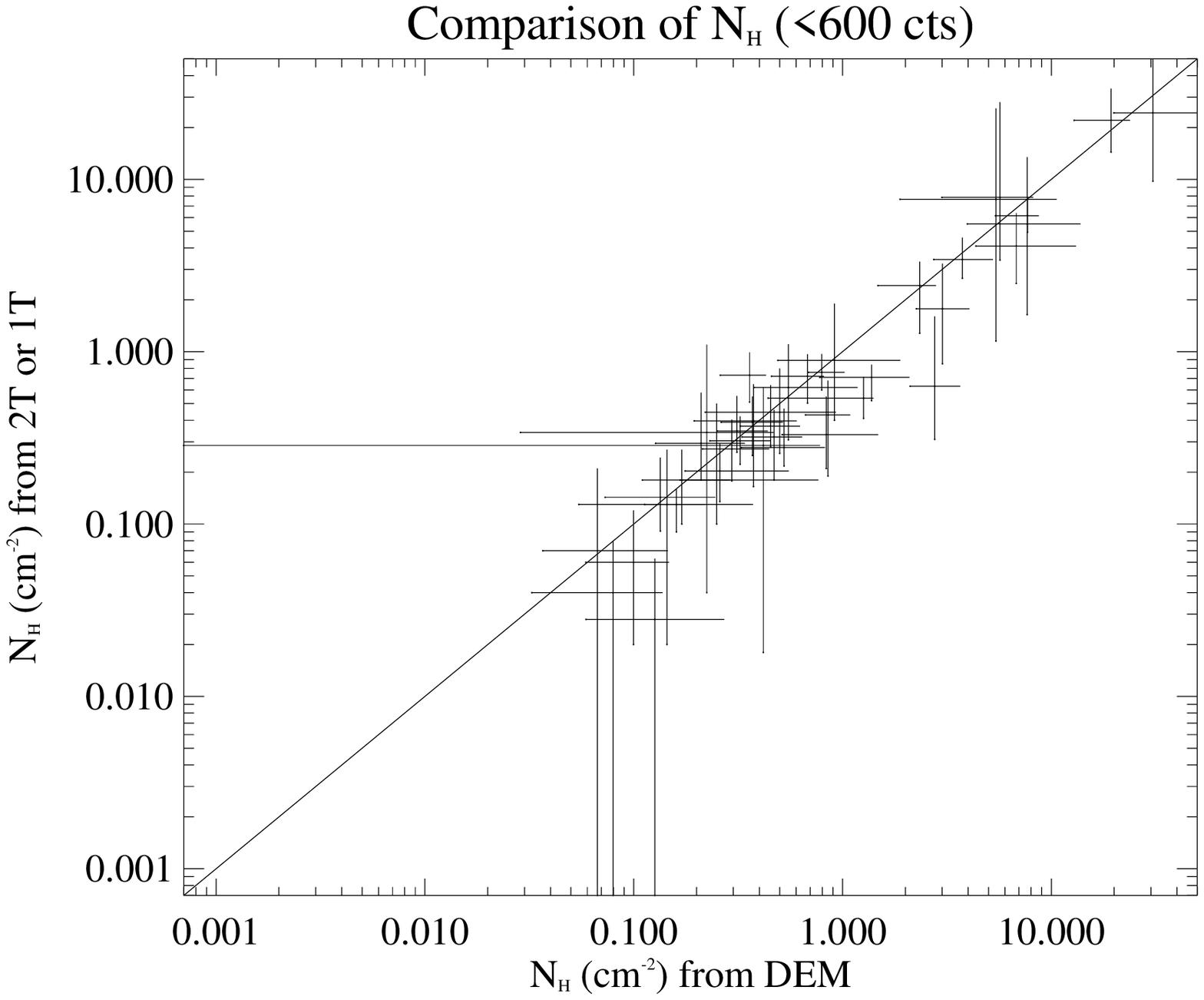}}}
\caption{Comparison of $N_{\rm H}$ values derived from the DEM method (x axis) and from 1-$T$ or 2-$T$ fits (y axis; Tables~\ref{tab5} and
\ref{tab6}). Only results from {\it XMM-Newton} data are reported.
Upper plot is for sources with more than 600 counts in total (after Table~\ref{tab4}), lower plot
for sources with fewer than 600 source counts. The solid lines indicate equal $N_{\rm H}$.}\label{NHcomp}
\end{figure} 

To assess the quality of our spectral interpretations, we now compare results derived
from the two fit methods (DEM vs 1- or 2-$T$ plasma).
Fig.~\ref{NHcomp} compares $N_{\rm H}$ values derived from the DEM fits (shown along the abscissa)
and from the 1- or 2-$T$ fits (along the ordinate). We show results for bright spectra 
($> 600$~cts for the source counts parameter in Table~\ref{tab4} of which typically about half 
are found in the PN camera) and for faint spectra ($< 600$~cts) separately. The agreement is
excellent. The brighter sources show a trend toward  $N_{\rm H}$ being slightly lower for the 
1-$T$ or 2-$T$ fit than for the DEM fit. This can be  understood as follows. The DEM model assumes
the presence of cool material down to $\log T = 6$. This plasma accounts for some of the soft emission
where the 1- or 2-$T$ model may be subject to an  EM deficit. To account for the same spectral flux,
the trend is to lower $N_{\rm H}$ for the spectrum with deficient cool EM. The scatter obviously becomes
larger for fainter sources, but the error bars grow accordingly so that the agreement is of similar 
quality (lower plot).

\begin{figure}
\centerline{\resizebox{0.95\hsize}{!}{\includegraphics{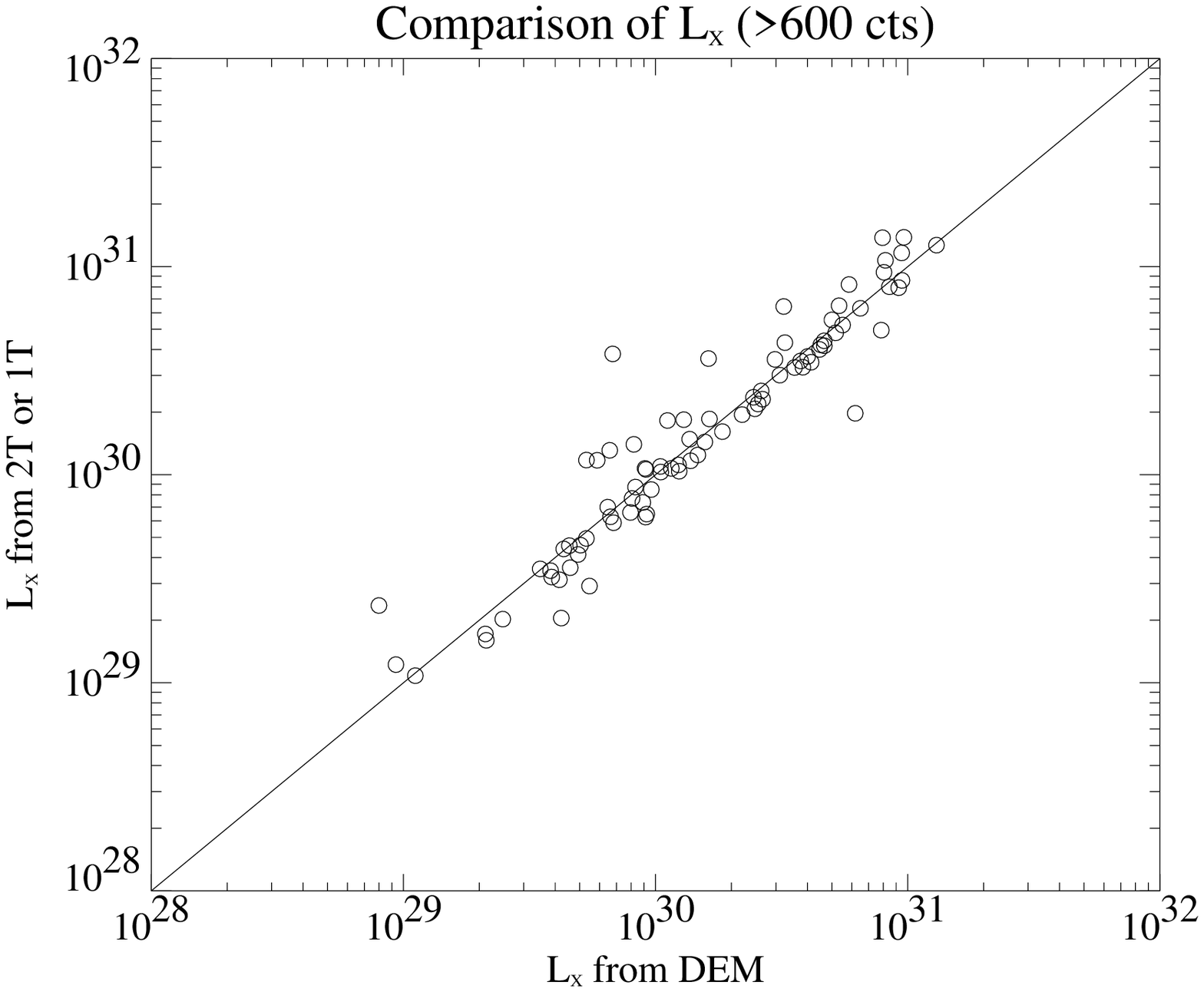}}}
\centerline{\resizebox{0.95\hsize}{!}{\includegraphics{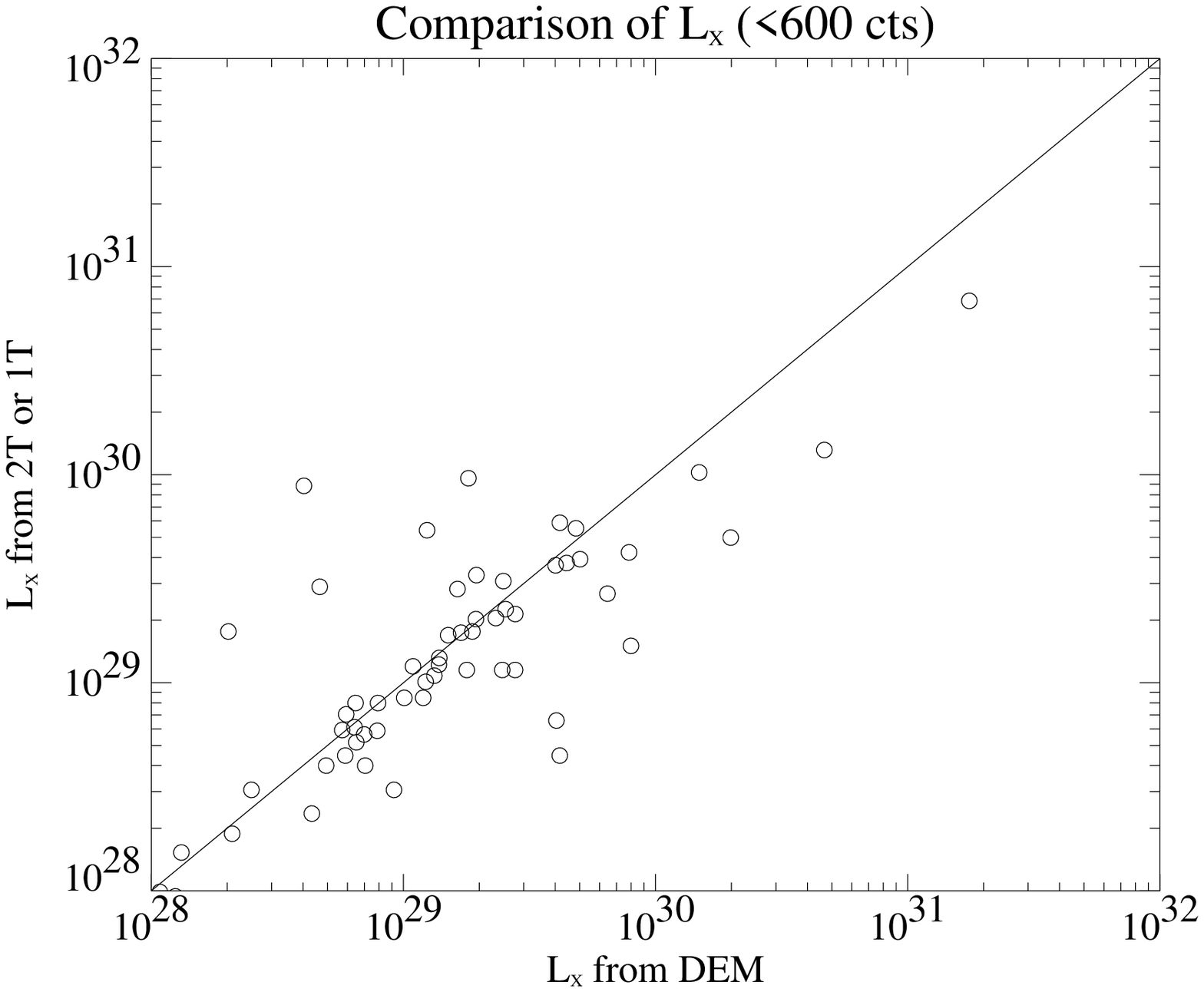}}}
\caption{Comparison of $L_{\rm X}$ values derived from the DEM method (x axis) and from 1-$T$ or 2-$T$ fits (y axis; Tables~\ref{tab5} and
\ref{tab6}). Only results from {\it XMM-Newton} data are reported.
Upper plot is for sources with more than 600  counts in total (after Table~\ref{tab4}), lower plot
for sources with fewer than 600 source counts. The solid lines indicate equal $L_{\rm X}$.}\label{LXcomp}
\end{figure}

\begin{figure}
\centerline{\resizebox{0.95\hsize}{!}{\includegraphics{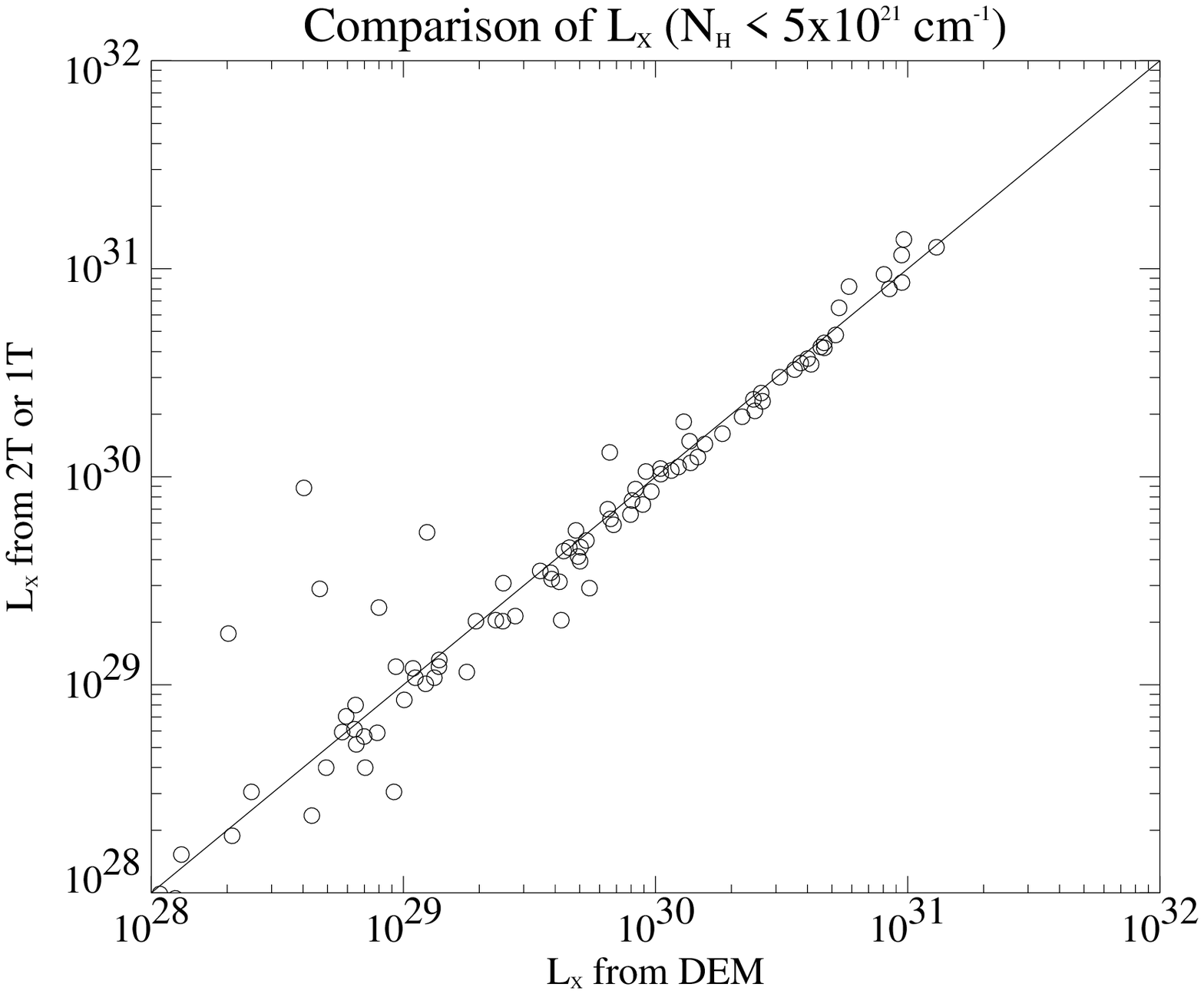}}}
\centerline{\resizebox{0.95\hsize}{!}{\includegraphics{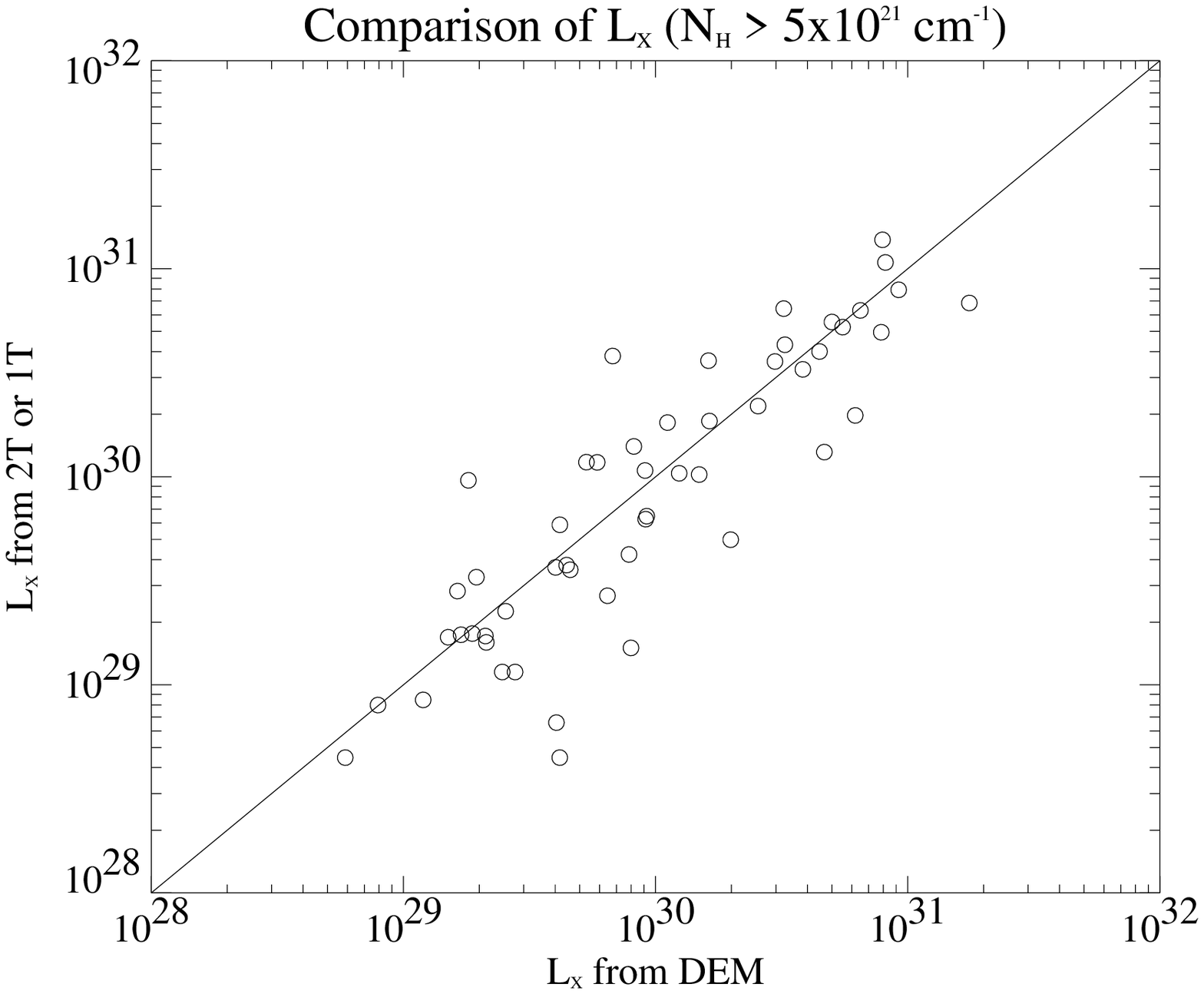}}}
\caption{Comparison of $L_{\rm X}$ values derived from the DEM method (x axis) and from 1-$T$ or 2-$T$ fits (y axis; Tables~\ref{tab5} and
\ref{tab6}). Only results from {\it XMM-Newton} data are reported.
Upper plot is for sources with $N_{\rm H} < 5\times 10^{21}$~cm$^{-2}$, lower plot for $N_{\rm H} > 5\times 10^{21}$~cm$^{-2}$  
(based on the DEM fits in Table~\ref{tab5}). The solid lines indicate equal $L_{\rm X}$.}\label{LXcomp2}
\end{figure}

Fig.~\ref{LXcomp} analogously compares $L_{\rm X}$. For most of the brighter stars, the agreement is
excellent although there are a few cases for which the 1- or 2-$T$ fit produces significantly higher $L_{\rm X}$.
The reason for this deviation is the following. Soft components can be added with relatively little constraints
if they are sufficiently absorbed. The temperature then becomes uncertain. A cooler plasma subject to photoelectric absorption
will, however, require a much larger EM to produce the same observed flux. If the cool EM is a free fit parameter,
an instability may occur with $T$ converging to low values while $N_{\rm H}$ and EM grow excessively. 
This becomes more evident for fainter sources (lower plot). In several cases, we
therefore fixed $N_{\rm H}$ at the value expected from $A_{\rm V}$ for the DEM fit, whereas the 1-$T$ fits, requiring one parameter less,
were performed without this constraint. The sources that likely suffer from this numerical instability in the 1-$T$ fit are
CIDA 7 (XEST-10-034), HO Tau (XEST-09-010), V410 X6 (XEST-23-061),  CFHT-Tau 7 (XEST-03-017), MHO 9 (XEST-22-013), and
V410 A25 (XEST-23-029) that all show $L_{\rm X}$ from the 1-$T$ or 2-$T$ fits at least 3 times higher compared to the DEM fit.
In the former, they show $N_{\rm H}$ considerably higher than in the latter, or much higher than expected from $A_{\rm V}$, while the dominant 
electron temperature was very low, between 1~MK and 3.3~MK. The first four of those six stars  are faint, while MHO 9  
is at the limit between bright and faint as defined for the plots in Fig.~\ref{LXcomp}. 

Given that the DEM model is more rigid and avoids biases toward  strongly absorbed soft emission
from large amounts of cool EM, we will base most of our statistical investigations in the accompanying papers
on the DEM results. This model is also more physically meaningful because work on bright spectra of
nearby stars has clearly shown continuous emission measure distributions of this kind (e.g., \citealt{telleschi05}).
As shown here and below, however, the results for $L_{\rm X}$, $N_{\rm H}$, and $T_{\rm av}$ from the two methods 
agree well for the brighter spectra. Caution is in order essentially only for spectra defined as ``faint'' here.

\begin{figure}
\centerline{\resizebox{0.95\hsize}{!}{\includegraphics{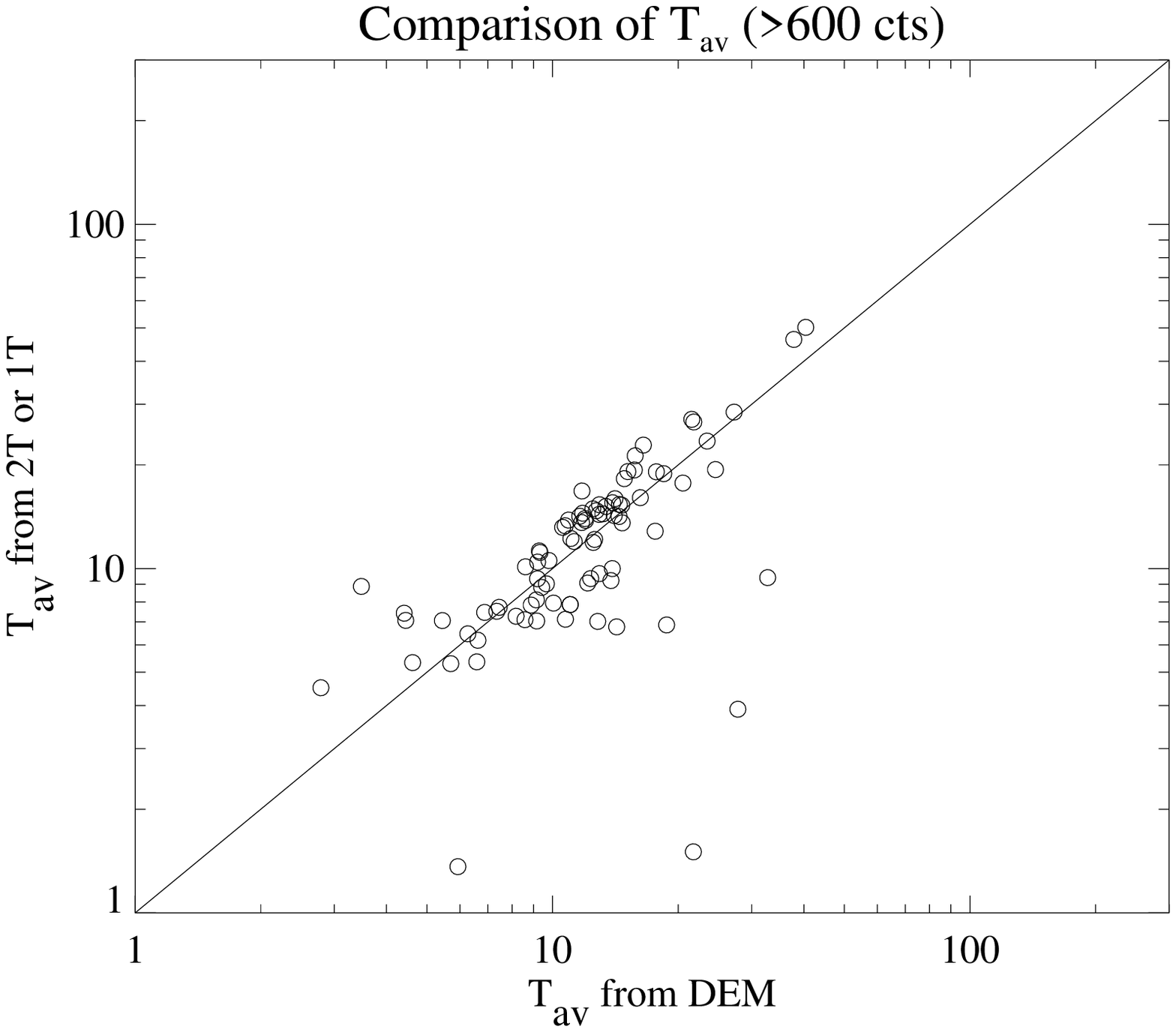}}}
\centerline{\resizebox{0.95\hsize}{!}{\includegraphics{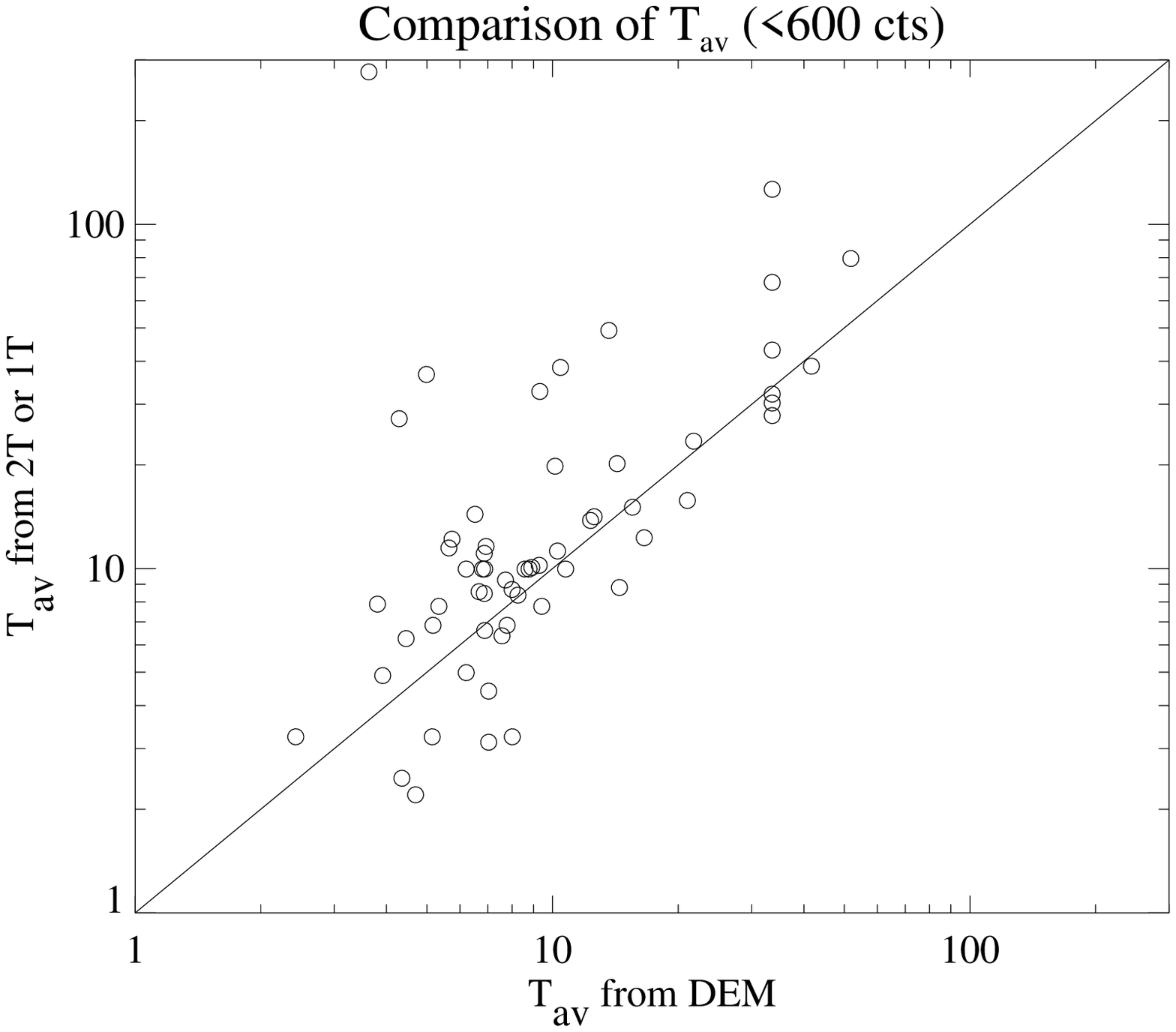}}}
\caption{Comparison of $T_{\rm av}$ values derived from the DEM method (x axis) and from 1-$T$ or 2-$T$ fits (y axis; Tables~\ref{tab5} and
\ref{tab6}). Only results from {\it XMM-Newton} data are reported.
Upper plot is for sources with more than 600 counts in total (after Table~\ref{tab4}), lower plot
for sources with fewer than 600 source counts. The solid lines indicate equal $T_{\rm av}$.}\label{Tavcomp}
\end{figure}

Fig.~\ref{LXcomp2} shows the same comparison for weakly absorbed sources 
(upper plot, for $N_{\rm H} < 5\times 10^{21}$~cm$^{-2}$) and for more strongly absorbed sources 
(lower plot, for $N_{\rm H} > 5\times 10^{21}$~cm$^{-2}$).  It is evident that higher absorption introduces
more uncertainty to the precise determination of $L_{\rm X}$. Also, these sources obviously 
tend to be fainter (in terms of count rates).

Fig.~\ref{Tavcomp} compares the average electron temperature, $T_{\rm av}$, derived from 
the DEM and the 2-$T$ or 1-$T$ fits. Given the rather different methodology, the
agreement is satisfactory for the brighter sources, while the
scatter increases for faint sources, making accurate statements on electron temperatures difficult.
The few outliers among the bright sources show very cool average temperatures in 1-$T$ or 2-$T$ fits 
probably due to the same numerical instability described above. The worst agreement is found for the very faint spectrum
of LR~1 (XEST-23-048) for which $T_0$ of the DEM fit is unconstrained despite fixed $\beta = -1$. In the presence
of  severe absorption, the DEM fit finds a low best-fit $T_0$, while the 1-$T$ fit converges to a very high temperature. Both
$L_{\rm X}$ are reasonable, while the DEM fit converges to a higher $N_{\rm H}$ which is preferred in the light of the very high $A_{\rm J}$ of
this star.

Concluding from this comparison, we are confident that $N_{\rm H}$ values are meaningful within factors of about 1.5
for the  brighter sample and mostly within factors of 2 for the fainter sample. X-ray luminosities appear
to be reasonably constrained within a factor of two (although with exceptions), which is a typical 
range of variability for most sources in any case.

\subsection{Population statistics}

\begin{figure*}
\centerline{\resizebox{0.9\hsize}{!}{\includegraphics{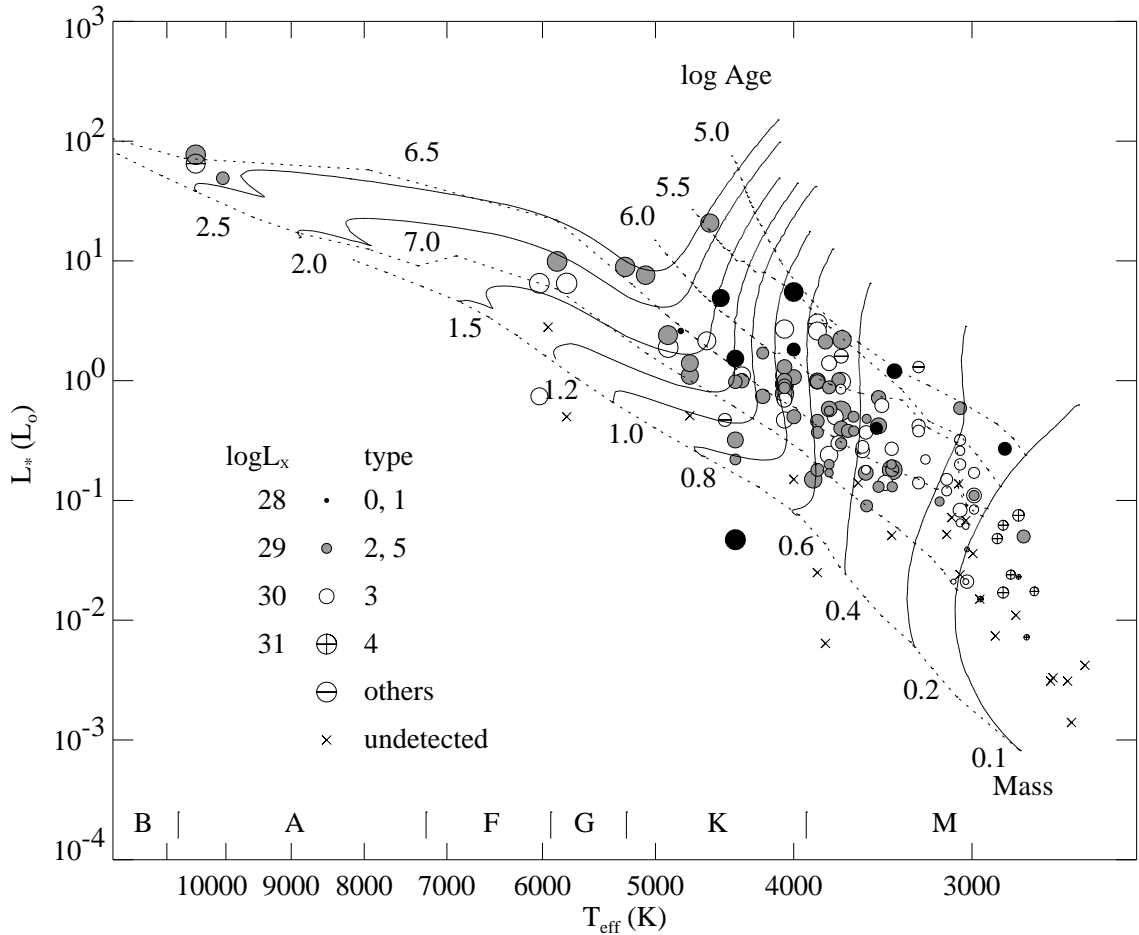}}}
\caption{HRD of all observed TMC objects for which $L_*$ and $T_{\rm eff}$ are known. Symbol indicates object type 
(see panel in lower left corner for the key), and symbol size scales with $\log L_{\rm X}$,  crosses marking 
undetected sources. Additional sources from {\it Chandra} have been included, but note that the L1551 IRS5 protostar is
marked here as a non-detection. Solid lines show evolutionary tracks 
toward the zero-age main sequence (ZAMS) for masses as given, in units of the solar mass (ZAMS). Dotted curves indicate 
isochrones at ages given (after \citealt{siess00}). The lowest of these curves marks the ZAMS. The spectral class ranges given at the
bottom of the figure refer to main-sequence stars.}\label{hrd}
\end{figure*}

Fig.~\ref{hrd} shows an Hertzsprung-Russell Diagram (HRD)  of all sources observed in the framework of this survey 
with isochrones and evolutionary tracks (after \citealt{siess00}) overplotted. The source symbols indicate 
the source type, and the symbol size  corresponds to the X-ray luminosity. 

The following 
characteristics are noteworthy: i) The surveyed sources cover a broad range of masses, from the 
substellar range up to 2.5-3$M_{\odot}$, the three most massive stars being HD~28867 and the
Herbig stars AB Aur and V928 Tau (the latter being a multiple system). Most objects are concentrated
at ages between $3\times 10^5 - 3\times 10^6$~yr, and very few at ages around $10^7$~yr.  
ii) As pointed out by \citet{kenyon95}, CTTS and WTTS (Class II and III, shown in gray and white, 
respectively) occupy the same region in the HRD. In fact, this is also true for the detected Class-I protostars. 
The low-mass and substellar HRD will be more specifically discussed by \citet{grosso06a}. 
iii) Most  undetected sources are in the 
very low-mass/substellar regimes and/or have the largest isochronal ages, although for some of the
latter objects, the fundamental parameters are poorly determined, as detailed below.

A few objects are apparently located below the zero-age main sequence (ZAMS) which is likely to  be a consequence of
poor photometry. This is particularly true for embedded sources
where scattered light distorts photometric measurements. The stars below the ZAMS are, in order of 
increasing $T_{\rm eff}$: HH~30 (XEST-22-000, an edge-on star-disk system, \citealt{burrows96}), 
IRAS~S04301+261 (XEST-18-000, a highly reddened object probably primarily detected in scattered light,
\citealt{briceno02}),  
Haro 6-5~B = FS Tau B (XEST-11-054 = C2-1, a near-edge-on star-disk system, \citealt{padgett99}, with 
$A_{\rm V} \approx 10$~mag), HBC 353 (XEST-27-000), and HBC 352 (XEST-27-115).  These 
stars will not be considered in statistical studies whenever $L_*$ is required (e.g., in $L_{\rm X}/L_*$).  A few 
stars are located between the 10~Myr isochrone and the ZAMS, making them much older than the bulk
population in Taurus. Several of these objects are relatively poorly studied and may suffer from
inaccurate photometry. For most of them, no $A_{\rm V}$ has been given, or the extinction is large, and indeed several
of them have not been detected (crosses in Fig.~\ref{hrd}). These objects
are, in order of  increasing $T_{\rm eff}$ above the 0.1$M_{\odot}$ track: ITG~33A (XEST-07-000, suggested to
be an  edge-on  star-disk system, \citealt{martin00}: $A_{\rm V} = 3.5$~mag), FS Tau  (XEST-11-057, a strongly extincted
close binary of CTTS, $A_{\rm V} \approx 5$~mag),
CoKu Tau 1 (XEST-23-000, an edge-on star-disk system, \citealt{padgett99}, $A_{\rm V} = 6.8$~mag), 
HN Tau (C5-2), LR 1 (XEST-23-048, with a large $A_{\rm J} = 6.4$~mag), V955 Tau (XEST-10-020, $A_{\rm V} = 3.7$~mag),
 V410 A20 (XEST-23-000, with a large $A_{\rm J} = 6.57$~mag), 
and V410 A24 (XEST-23-000, with a large $A_{\rm J} = 6.73$~mag - see Table~\ref{tab9} for $A_{\rm V}$ and $A_{\rm J}$ values).

The detection statistics of our X-ray survey is summarized in Table~\ref{tab12}
(considering only the {\it XMM-Newton} observations),
based on our final classification scheme in Table~\ref{tab11}. Here, the protostar L1551 IRS5 has been
treated as a non-detection as the source closely associated with this binary has been interpreted as
X-ray emission from the jets \citep{favata02, bally03}. We also list the statistics (in parentheses)
if {\it Chandra} detections and non-detections are counted (note that DG Tau B and FV Tau/c have
been counted as non-detections in the {\it XMM-Newton} sample, while they are detections with {\it Chandra}).
An important point for further statistical studies is that the X-ray sample
of detected CTTS and WTTS is nearly complete for the surveyed fields (as
far as the population is known). Most of the few remaining, undetected objects are
considerably extincted and by implication X-ray absorbed: 
IRAS S04301+261 = XEST-18-000  with $A_{\rm J} \approx 1.8$, and CoKu Tau 1 = XEST-23-000, CFHT-Tau 19 = XEST-11-000, 
FV Tau/c = XEST-02-000, CFHT-Tau 20 = XEST-13-000, ITG33 A = XEST-07-000,  CFHT-Tau 8 = XEST-07-000,
and HH~30 = XEST-22-000  with  $A_{\rm V} = $ 6.8, 7.3, 3.25, 3.6, 3.5, 1.8, and $\approx 3$~mag and 
CFHT-Tau 12 = XEST-17-000 with $A_{\rm J} = 3.44$~mag,
where the $A_{\rm V}$ of  IRAS S04301+261, CoKu Tau 1, HH~30, ITG 33A  may have been severely underestimated owing 
to scattered light and strong extinction by near-edge-on disks (see above). FV Tau/c was, however, detected 
in a {\it Chandra} observation (C3-1) reported here, with an unexpectedly high photoelectric absorption 
corresponding to  $N_{\rm H} \approx 10^{23}$~cm$^{-2}$. Some of the non-detections are very-low mass
stars with late spectral types (CFHT-Tau 19 = XEST-11-000, CFHT-Tau 20 = XEST-13-000, CFHT-Tau 12 = XEST-17-000,
and CFHT-Tau 8 = XEST-07-000 with spectral types of  M5.25, M5.5, M6, and M5.5, respectively). Very little is known
about IRAS 04108+2803~A. Its companion, a Class~I object, was detected, its light curve showing
a strong flare and its spectrum revealing very strong absorption. Perhaps,  both objects are
subject to similar absorption. The only formally undetected WTTS in the {\it XMM-Newton} sample is HBC 353, but
it is located on the wings of HBC 352 = XEST-27-115, falls into a CCD gap of the
PN detector, and is located far off-axis, elevating the corresponding detection limit to 
$ \approx 180$~cts. From Table~\ref{tab8}, we derive that only 3 out of 45 multiple systems remain 
undetected  (2/47 including {\it Chandra} results),
while  29 out of 114 single stars are undetected  (31/122 including {\it Chandra} results).
In total,  45 of the 159 systems surveyed by {\it XMM-Newton}  are multiple 
(47/169 including {\it Chandra} results).

\setcounter{table}{11}

\begin{table}[t]
\caption{XEST X-ray detection statistics\label{tab12} }
\begin{tabular}{rlrrl}
\hline
\hline
\multicolumn{2}{l}{Object type}  & Members     &  Detections & Detection \\  
\multicolumn{2}{l}{ }            & surveyed    &  	     & fraction  \\
\hline
0, 1& Protostars  	  & 20 \hfill (21)         &     8 \hfill (10)     & 40\% \hfill (48\%)	\\
2& CTTS                   & 65 \hfill (70)         &    55 \hfill (60)     & 85\% \hfill (86\%)	\\
3& WTTS                   & 50 \hfill (52)         &    49 \hfill (50)     & 98\% \hfill (96\%)	\\
4& BDs    		  & 16 \hfill (17)         &     8 \hfill (9)      & 50\% \hfill (53\%)	\\
5& Herbig		  & 2  \hfill {\ }         &    2  \hfill {\ }     & 100\% \hfill {\ } 	\\
9& others/unident.        & 6  \hfill (7)          &    4  \hfill (5)      & 67\% \hfill (71\%)	\\
\hline
& Total		          & 159 \hfill (169)       &    126 \hfill (136)   & 79\% \hfill (80\%)	\\
\hline
\multicolumn{5}{l}{NOTES:}\\
\multicolumn{5}{l}{Numbers in parentheses include {\it Chandra} observations. Source}\\
\multicolumn{5}{l}{near L1551 IRS5 not considered (non-detection for XEST-22-040)}\\ 
\end{tabular}
\end{table}

Previous X-ray surveys of TMC did not detect the  intrinsically fainter TTS population. Those previous
surveys covered a wider field in Taurus (e.g., the ROSAT surveys by \citealt{neuhaeuser95, stelzer01})
but were about ten times less sensitive and were confined 
to soft sources. Stellar classes that were thus inaccessible by these earlier surveys but are
systematically (albeit not completely) detected in XEST are: i) strongly absorbed, embedded protostars
or T Tau stars absorbed by their own disks seen nearly edge-on; their non-detection in previous surveys 
is little surprising given the strong photoelectric absorption. Most of the detected protostars 
show no X-ray counts below 
2~keV \citep{briggs06b}. ii) Low-$L_{\rm X}$ brown dwarfs; in XEST, the detection rate of BDs
(50\% for the {\it XMM-Newton} sample, 53\% if one detection from {\it Chandra} is added) 
is  high; the remaining objects of this class are
likely to be intrinsically fainter than our detection limit rather than being
excessively absorbed by gas ($A_{\rm V}$ of those objects typically being no more than 
a few magnitudes; see \citealt{grosso06a} for further discussion). 
iii) Several T Tau stars with double-peaked X-ray spectra in which a strongly absorbed hard component 
is accompanied by a soft excess \citep{guedel06b}. 

The incomplete statistics in previous surveys potentially introduces bias into statistical 
correlations and population studies. On the other hand, we emphasize that the so far cataloged
population of Taurus members may not be complete either. In particular, WTTS are predominantly 
detected by their X-rays (e.g., \citealt{neuhaeuser95}). The WTTS statistics in Taurus 
may thus be incomplete in particular with regard to faint X-ray sources or WTTS located
behind the TMC. Several dozen new TMC member candidates are indeed reported from XEST \citep{scelsi06}
but will require spectroscopic follow-up observations to confirm membership.

\subsection{Statistical properties and correlations}

Table~\ref{tab13} summarizes statistical properties of the surveyed sources, providing 
median, mean, and standard deviation of the sample for various parameters. The sample sizes vary (given in the last
column, N) because not all parameters are available for all sources (see details in the
footnotes of the table). The salient features of 
this compilation are the following: i) The agreement between the sample X-ray parameters derived from the
DEM model (``DEM'' in column 2)  and the 1-$T$ or 2-$T$ model (``2-$T$'' in column 2) is excellent.
ii) Characteristic electron temperatures are around 6-8~MK as characterized by $T_0$ in the DEM 
model and by $T_1$ in the 1-$T$ or 2-$T$ model. The hotter component shows a mean and median around 23~MK. 
iii) The large majority of the sample shows $-4 \la \log L_{\rm X}/L_* \la -3$. iv) Protostars are,
as expected, significantly more extincted in the visual band ($A_{\rm V}$) in our sample than T Tau stars; 
among the latter, objects of type 2 (CTTS, Class-II objects) are more strongly extincted than objects
of type 3 (WTTS, Class III). The median extinction of the detected sources is $\approx 1.4$~mag,
corresponding to a hydrogen column density of $\log N_{\rm H} \approx 21.44$, close to the measured median of
$\log N_{\rm H} \approx 21.5-21.6$. iv) As pointed out on the HRD, the $L_*$ 
distributions of objects of type 1, 2, and 3 strongly overlap, although a marginal trend 
is seen toward a decreasing sample {\it mean} of $L_*$ with increasing type, perhaps reflecting some overall 
evolutionary trend. v) A wide range of accretion rates is covered (for protostars and CTTS), with a mean and
 median around $10^{-8}~M_{\odot}$~yr$^{-1}$.

\begin{table}[t!]
\caption{XEST parameter statistics\label{tab13} }
\begin{tabular}{lrrrrr}
\hline
\hline
Parameter                        &Sample$^a$& Median      & Mean     & Std.        &  N \\  
                                     &      &             &          & dev.$^b$    &    \\
\hline
\multicolumn{6}{c}{X-ray parameters} \\
\hline
$\beta^c$                            &  DEM & -1.25       & -1.49    & 0.95        &  99 \\
$\log T_0^c$ [K]                     &  DEM & 6.90        & 6.90     & 0.28        & 119 \\
$\log T_1^{c,d}$ [K]                 & 2-$T$ & 6.79       & 6.76     & 0.21        &  94 \\
$\log T_2^{c,d}$ [K]                 & 2-$T$ & 7.37       & 7.38     & 0.25        & 105 \\
$\log T_{\rm av}^{e}$ [K]            & DEM  & 7.04        & 7.03     & 0.24        & 123 \\
                                     & 2-$T$ & 7.05       & 7.08     & 0.32        & 125 \\
$\log N_{\rm H}^d$ [cm$^{-2}$]       &  DEM & 21.57       & 21.52    & 0.67        & 123 \\
                                     & 2-$T$ & 21.53      & 21.50    & 0.65        & 123 \\
$\log L_{\rm X}^e$ [erg~s$^{-1}$]    & DEM  & 29.82       & 29.76    & 0.77        & 123 \\
                                     & 2-$T$ & 29.80      & 29.72    & 0.78        & 125 \\
$\log L_{\rm X}/L_*^{e,f}$           & DEM  & -3.56       & -3.55    & 0.50        & 116 \\
                                     & 2-$T$ & -3.56      & -3.58    & 0.56        & 118 \\
 \hline
\multicolumn{6}{c}{Fundamental parameters} \\
\hline
$A_{\rm V}$ [cm$^{-2}$]              & all  & 1.46        & 2.80     & 3.38        & 116    \\
		                     & 1    & 10.20	  & 9.52     & 1.93	   & 6	 \\
		                     & 2    & 1.95	  & 3.30     & 3.61	   & 57    \\
		                     & 3    & 0.87	  & 1.36     & 1.56	   & 38   \\
                                     & det  & 1.35        & 2.60     & 3.36        & 97    \\
$\log L_*^{f,g}$ [$L_{\odot}$]       & all  & -0.34       & -0.46    & 0.82        & 146    \\
		                     & 1    & 0.08        & -0.01    & 0.49        & 16    \\
		                     & 2    & -0.25	  & -0.33    & 0.61        & 60    \\
 		                     & 3    & -0.43	  & -0.44    & 0.59        & 46   \\
                                     & det  & -0.30       & -0.33    & 0.71        & 119    \\
$\log({\rm age})^{f,h}$ [Myr]        & all  & 0.39        & 0.38     & 0.46        & 112    \\
		                     & 1    & 0.05        & 0.23     & 0.54        & 9    \\
		                     & 2    & 0.43	  & 0.44     & 0.42        & 52    \\
 		                     & 3    & 0.39	  & 0.32     & 0.35        & 43   \\
$\log({\rm mass})^{f,g}$ [$M_{\odot}$]& all & -0.29       & -0.37    & 0.43        & 123    \\
		                     & 1    & -0.17       & -0.23    & 0.22        & 10   \\
		                     & 2    & -0.25	  & -0.27    & 0.31        & 54    \\
 		                     & 3    & -0.39	  & -0.40    & 0.34        & 43  \\
$\log(P)^{i}$ [d]                    & all  & 0.69        &  0.64    & 0.28        & 69    \\
		                     & 1    & 0.63        & 0.54     & 0.32        & 10   \\
		                     & 2    & 0.77	  & 0.73     & 0.21        & 35    \\
 		                     & 3    & 0.58	  & 0.56     & 0.33        & 22  \\
$\log(\dot{M})^{j}$ [$M_{\odot}{\rm yr}^{-1}$]  & all  & -8.08       &  -8.20      & 1.21        & 57    \\
\hline
\multicolumn{6}{l}{}\\ 
\multicolumn{6}{l}{NOTES: Only results from {\it XMM-Newton} have been considered.}\\ 
\multicolumn{6}{l}{For multiple detections, parameters were averaged}\\ 
\multicolumn{6}{l}{$^a$ DEM: from DEM fit; 2-$T$: from 1-$T$ or 2-$T$ fit; all; all objects;}\\ 
\multicolumn{6}{l}{ 1, 2, 3; object types; det = X-ray detected objects}\\ 
\multicolumn{6}{l}{$^b$ Standard deviation of distribution}\\ 
\multicolumn{6}{l}{$^c$ Only non-fixed fit parameters for detections considered}\\ 
\multicolumn{6}{l}{$^d$ Only non-zero values for detections considered}\\ 
\multicolumn{6}{l}{$^e$ Only detections considered}\\ 
\multicolumn{6}{l}{$^f$ Only for known $L_*$ above the ZAMS}\\ 
\multicolumn{6}{l}{$^g$ For multiples, value for primary if available}\\ 
\multicolumn{6}{l}{$^h$ For multiples, logarithmic average of components if available}\\ 
\multicolumn{6}{l}{$^i$ Upper limits are adopted as measured values}\\ 
\multicolumn{6}{l}{$^j$ Average of reported range if at least one non-upper limit given}\\ 
\end{tabular}
\end{table}	

\begin{figure}[t!]
\centerline{\resizebox{1.05\hsize}{!}{\includegraphics{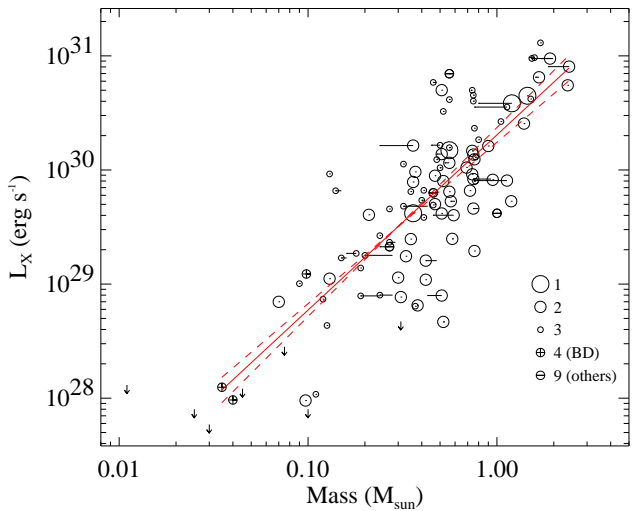}}}
\caption{X-ray luminosity $L_{\rm X}$ (based on DEM fits) vs. stellar mass $M$ for all detected XEST sources ({\it Chandra} sources and Herbig stars are not included),
or upper limits (arrows). For multiple systems, the primary mass has been used if available. Symbols defining the object type
are given in the lower
right corner. The horizontal bars for some objects show the ranges of masses derived from literature $L_*$ and $T_{\rm eff}$, 
while the circle is centered at the adopted $M$. The straight line gives a linear regression using the
logarithmic values for the X-ray detections, and the 1$\sigma$ ranges for the slopes: $\log L_{\rm X} = (1.54\pm 0.12) \log M +30.31 \pm 0.06$.}\label{lxm}
\end{figure}

In Fig.~\ref{lxm} we characterize the detected stellar population by plotting $L_{\rm X}$ (from the DEM
models) as a function of mass 
$M$  for all observed stars and substellar objects with known masses, although we exclude
the following objects: DG Tau A (XEST-02-022), GV Tau (XEST-13-004), DP Tau (XEST-10-045), and CW Tau (XEST-20-046)
have composite X-ray spectra that may originate from two unrelated sources \citep{guedel06b}. L1551 IRS5 is a deeply
embedded protostar, while the X-ray source seen close to it is only lightly absorbed and may be related to the jet
\citep{favata02, bally03}. Finally, the light curves of DH Tau (XEST-15-040), FS Tau (XEST-11-057), and V830 Tau (XEST-04-016) 
showed a decay presumably from large flares.  For objects that were observed and detected
twice, the logarithmic average of $L_{\rm X}$ has been adopted. Although for pre-main sequence
stars there is no strict correlation between $L_{*}$ and $M$, it is interesting
that we find a rather well-developed correlation between $L_{\rm X}$ and $M$. The correlation coefficient for the $\log L_{\rm X}$ vs $\log M$
sample is $C = 0.79$ for 99 data points. For the subsample of  type 2 (``CTTS'')  and type 3 objects (``WTTS''), we find 
$C = 0.75$ and $C = 0.85$ for 45 and 43  data points, respectively. All values point to a very significant correlation. 
We find a linear regression,   $\log L_{\rm X} = ( 1.54\pm 0.12) \log M +  30.31 \pm 0.06$. The slopes for 
type 2 and type 3 objects are, respectively, $1.52\pm 0.21$ and $1.78\pm 0.17$. While these regressions are not significantly
different, we note the larger errors for type 2 objects, and also their lower correlation coefficient, which is due
to a larger scatter of $L_{\rm X}$ at a given mass. Similar trends have been
noted  in Orion \citep{preibisch05}, with $\log L_{\rm X} = (1.44\pm 0.10) \log M + 30.37 \pm 0.06$ for the entire sample. 
This coincidence between XEST and COUP reveals that in TMC and ONC, 
{\it the basic X-ray production mechanism scales precisely the same way 
with the most basic property of the stellar objects, namely their mass}, regardless of any differences
in age distribution, star-formation mode, or the presence or {\nolinebreak (near-)\linebreak} absence of  high-mass stars.
Part of this correlation might be explained by higher-mass stars being larger, i.e., 
providing more surface area for coronal active regions. Assuming similar average internal stellar density, the 
correlation between surface area and $L_{\rm X}$ is, however, considerably weaker  than the trend shown 
in Fig.~\ref{lxm}, so that additional effects may play a role (convection zone depth, convective mass, etc).

The figure shows another important feature of our survey. Almost all sources plotted are at $L_{\rm X}$ considerably
higher than our approximate detection limit of  $\approx 10^{28}$~erg~s$^{-1}$. There are hardly any objects
in the range of $(1-5)\times 10^{28}$~erg~s$^{-1}$ although they would have been detected. Exceptions are 
the brown dwarfs. This again testifies to the near-completeness of our survey within the observed fields.

\begin{figure}[t!]
\centerline{\resizebox{1.05\hsize}{!}{\includegraphics{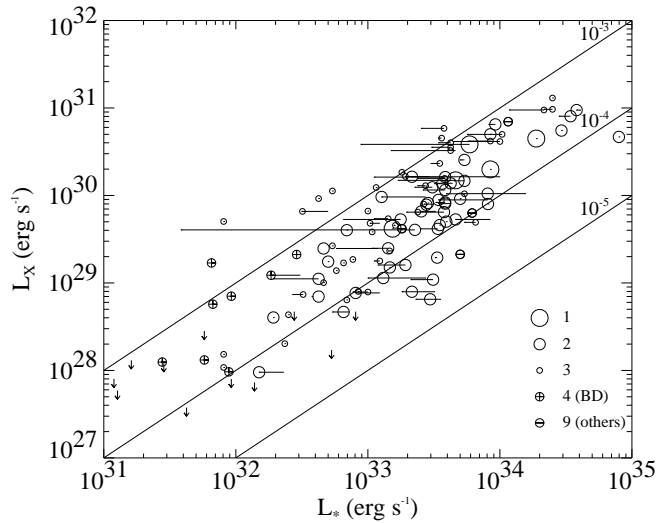}}}
\caption{X-ray luminosity $L_{\rm X}$ (based on DEM fits) vs. stellar luminosity $L_*$ for all X-ray detected XEST  sources 
({\it Chandra} sources, Herbig stars, and objects ``below'' the 
ZAMS not included), or upper limits (arrows). For multiple systems, total (``system'')  luminosity $L_*$ has been used. 
Symbols defining the object type are given in the lower
right corner. The horizontal bars for some objects show the ranges of literature values for $L_*$, 
while the circle marks the adopted $L_*$. }\label{lxlb}
\end{figure}

Fig.~\ref{lxlb} shows the distribution of the  $L_{\rm X}$ (derived from the DEM models)
as a function of the (stellar, photospheric)
bolometric luminosity $L_{*}$ for all spectrally 
modeled TTS and protostars,  and also including  BDs. We again exclude the peculiar 
spectral sources with two absorbers, the protostar L1551 IRS5, 
the three stars with gradually decaying light curves (DH Tau = XEST-15-040, FS Tau = XEST-11-057,  V830 Tau = XEST-04-016),
and also all sources located below the ZAMS in the HRD. 
For stars observed and detected twice, we again plot the logarithmic average of $L_{\rm X}$.
Most stars cluster between $L_{\rm X}/L_{*} = 10^{-4} - 10^{-3}$ as is often found in star-forming regions 
(see \citealt{preibisch05} for a similar analysis for  the Orion sample). The value $L_{\rm X}/L_{*} = 10^{-3}$ 
corresponds to the saturation value for rapidly rotating main-sequence stars. One key parameter that drives 
the X-ray output is obviously $L_{*}$. 

\begin{figure}[t!]
\centerline{\resizebox{1.05\hsize}{!}{\includegraphics{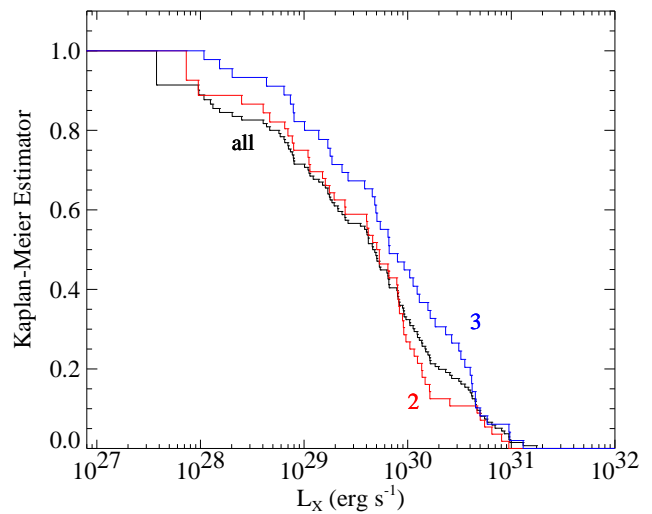}}}
\caption{X-ray luminosity function for the entire sample (black), type 2 objects (red), and
type 3 objects (blue).  }\label{xlf}
\end{figure}

We finally present the X-ray luminosity function (XLF) of our Taurus sample in Fig.~\ref{xlf}. The XLF has been calculated using 
the Kaplan-Meier estimator in  the ASURV software package \citep{lavalley92} that considers upper limits to $L_{\rm X}$ also. Non-detections for which no
reasonable upper limit can be given were not included. We also dropped the same non-standard X-ray sources excluded above
(double-absorber spectra, L1551 IRS 5, and sources with gradual flare decays), but include BDs and Herbig stars, thus using 136 X-ray 
luminosity values, including 17 upper limits. We also plot 
separately the XLF for type 2 (CTTS, 54 values, including 6 upper limits) and type 3 (WTTS, 49 values, 1 upper limit). We note 
the shift of the CTTS distribution toward lower luminosities compared to WTTS, by  a factor of about two. This is also evident for 
$L_{\rm X}/L_{\rm bol}$  \citep{telleschi06a}. A two-sample test based on the Wilcoxon test  and the logrank test performed in ASURV
indicates a probability of only 6-9\% that the two distributions are drawn from the same parent population. A more detailed discussion
on distinctions between CTTS and WTTS is given by \citet{telleschi06a}.

\section{Outlook}

The XEST project gives near-complete access to the entire pre-main sequence population
of the Taurus Molecular Clouds within the $\approx 5$~sq. degrees covered by this survey.
Essentially all WTTS and nearly all CTTS have been detected, exceptions being a few low-mass,
strongly absorbed sources. About half of the embedded protostellar sample is detected,
and a similar fraction of the BDs. This survey thus goes deeper than previous X-ray 
surveys of Taurus (e.g., \citealt{neuhaeuser95, stelzer01})
by about an order of magnitude for low-extinction objects and provides systematic
access to the protostellar and the brown-dwarf samples for the first time. The survey forms 
the basis for more specialized studies as discussed in the accompanying series of papers.

Furthermore, the survey will provide a useful database for correlation studies at other wavelengths.
In particular, gas-to-dust ratio studies and investigations of the
effects of accretion disks on magnetic fields in pre-main sequence systems
 will be undertaken in conjunction with a wide-field survey 
available from the {\it Spitzer Space Telescope} \citep{padgett06}.

\section*{Appendix A: The XEST source catalog}

The complete catalog of all X-ray sources detected in the XEST fields  (not including the complementary
fields from {\it Chandra}) is
available in electronic form. The data reduction and source identification procedures are described in 
Sect.~\ref{strategy}.  The catalog also provides errors of the source counts and the
count rates (not given in Table~\ref{tab4} for the ``Scts'' and ``Rate'' columns). The numbering scheme 
is as defined in Sect.~\ref{results}, all TMC members being included. The catalog is sorted in right 
ascension and contains a total of 2347 identified X-ray sources. If two XEST fields overlap, the same sources may
have been identified twice, with different XEST IDs assigned.  The first ten entries of the catalog
are given in the Table~\ref{tab14} below for illustration. Note that  if ML $< 5$ in any band the number of 
counts and the count rate are 95\% upper 
limits in that band. The HR is the upper (or lower) limit as appropriate in these situations 
with the low and high values set to -1 and to the HR upper limit, respectively (or to the HR 
lower limit and to +1, respectively).

\begin{table*}
\caption{XEST Catalog (first ten entries)\label{tab14} }
\begin{tabular}{rrrrrrrrrrr}
\hline
\hline
XEST   &  RA$_{\rm X}$  & Dec$_{\rm X}$  & RA     &Dec          &Poserr  &ML$_{\rm F}$&N$_{\rm F}$&  T$_{\rm exp, F}$  &  Rate$_{\rm F}$   &Err$_{\rm F}$    \\
1      &  2             & 3	         & 4      &  5          &	  6    &     7   &    8 &    9    &	10    &  11         \\
\hline 
27-001 &  03 53 03.92  &+31 53 02.8  & 58.266348 & +31.884122	  & 1.17    &  337.9   &  250 & 10386	&  0.0241 & 0.0017     \\	      
27-002 &  03 53 05.45  &+31 54 19.8  & 58.272721 & +31.905495	  & 1.47    &	20.0   &   41 & 11432	&  0.0037 & 0.0008     \\	      
27-003 &  03 53 12.71  &+31 55 12.7  & 58.302955 & +31.920193	  & 1.28    &	51.7   &  102 & 27319	&  0.0038 & 0.0005     \\	      
27-004 &  03 53 15.59  &+31 53 01.3  & 58.314976 & +31.883708	  & 1.20    &  138.4   &  180 & 29916	&  0.0060 & 0.0006     \\	      
27-005 &  03 53 19.30  &+31 47 39.1  & 58.330416 & +31.794186	  & 2.01    &	11.1   &   42 & 28822	&  0.0015 & 0.0004     \\	      
27-006 &  03 53 19.73  &+31 59 21.4  & 58.332216 & +31.989273	  & 1.95    &	 9.2   &   38 & 25935	&  0.0015 & 0.0004     \\	      
27-007 &  03 53 22.62  &+31 56 24.9  & 58.344257 & +31.940258	  & 1.66    &	11.1   &   38 & 31200	&  0.0012 & 0.0003     \\	      
27-008 &  03 53 24.35  &+31 43 20.5  & 58.351468 & +31.722347	  & 1.83    &	 8.0   &   24 & 10810	&  0.0022 & 0.0007     \\	      
27-009 &  03 53 25.01  &+31 42 13.4  & 58.354220 & +31.703724	  & 1.60    &	 8.8   &   24 &  9042	&  0.0027 & 0.0008     \\	      
27-010 &  03 53 25.25  &+32 03 26.5  & 58.355189 & +32.057353	  & 1.21    &	80.4   &  129 & 10663	&  0.0121 & 0.0014    \\    
\hline
\end{tabular}
\vskip 0.5truecm 
\begin{tabular}{rrrrrrrrrrrrrr}
\hline
\hline
 ML$_{\rm S}$   &  N$_{\rm S}$ &  T$_{\rm exp,S}$  & Rate$_{\rm S}$  & Err$_{\rm S}$  &  ML$_{\rm H}$ &  N$_{\rm H}$ &  T$_{\rm exp,H}$  &  Rate$_{\rm H}$   & Err$_{\rm H}$  &  HR & HR$_{\rm low}$   & HR$_{\rm high}$   & Verif. \\
 12    &	13 &   14  &   15   &   16     &	 17    &    18 &    19   &    20   &   21    &    22 &    23	&   24  &  25	 \\
\hline 
   75.5  &    69 & 10471  &   0.0066 & 0.0009	&   271.6  &   181 & 10440   &  0.0174 & 0.0015  & +0.448& +0.370  &+0.526 &	    \\  	 
   17.8  &    29 & 11517  &   0.0026 & 0.0006	&     3.0  &	21 & 11503   &  0.0019 & 0.0000  & -0.160& -1.000  &-0.160 &	    \\  	 
   49.9  &    73 & 28589  &   0.0026 & 0.0004	&     9.1  &	32 & 27099   &  0.0012 & 0.0003  & -0.372& -0.519  &-0.224 &	    \\  	 
  106.7  &   115 & 31188  &   0.0037 & 0.0004	&    34.4  &	62 & 29754   &  0.0021 & 0.0004  & -0.265& -0.361  &-0.169 &	    \\  	 
    4.1  &    41 & 30079  &   0.0014 & 0.0000	&     6.4  &	24 & 28636   &  0.0009 & 0.0003  & -0.226& -0.226  &+1.000 &	    \\  	 
    8.4  &    27 & 27223  &   0.0010 & 0.0003	&     1.7  &	34 & 25697   &  0.0013 & 0.0000  & +0.142& -1.000  &+0.142 &	    \\  	 
    3.5  &    34 & 32427  &   0.0011 & 0.0000	&     5.8  &	22 & 31072   &  0.0007 & 0.0003  & -0.178& -0.178  &+1.000 &	    \\  	 
    0.2  &    11 & 10896  &   0.0011 & 0.0000	&    10.5  &	21 & 10894   &  0.0020 & 0.0005  & +0.297& +0.297  &+1.000 &	    \\  	 
   11.7  &    19 &  9131  &   0.0021 & 0.0006	&     0.3  &	19 &  9101   &  0.0022 & 0.0000  & +0.006& -1.000  &+0.006 &	    \\  	 
  112.0  &   132 & 11298  &   0.0117 & 0.0012	&     0.1  &	11 & 10463   &  0.0011 & 0.0000  & -0.825& -1.000  &-0.825 &	    \\  	 
\hline
\end{tabular} 
\end{table*}

A description of the columns follows:
\begin{description}
\item[1]	XEST Id (field-srcnr)
\item[2]	RA (J2000, X-ray, corrected for boresight shift)
\item[3]	Dec (J2000, X-ray, corrected for boresight shift)
\item[4]	RA as above, in decimal degrees
\item[5]	Dec as above, in decimal degrees
\item[6]	1$\sigma$ positional error (including RMS of boresight shift)
\item[7]	Maximum likelihood of detection in the mosaicked EPIC full band (0.5-7.3~keV) image
\item[8]	Number of detected counts in the mosaicked EPIC full band image. If ML$_{\rm F} < 5$, then 95\% upper limit is given.
\item[9]	Total PN-equivalent effective exposure time in s at source position in in the mosaicked EPIC full band image
\item[10]	PN-equivalent source count-rate in the mosaicked EPIC full band image. If ML$_{\rm F} < 5$, then 95\% upper limit is given.
\item[11]	1$\sigma$ statistical error in PN-equivalent source count-rate in the mosaicked EPIC full band image (note: does not include uncertainty in spectrally-dependent PN/MOS sensitivity ratio)
\item[12]	Maximum likelihood of detection in the mosaicked EPIC soft band (0.5-2~keV) image
\item[13]	Number of detected counts in the mosaicked EPIC soft band image. If ML$_{\rm S} < 5$, then 95\% upper limit is given.
\item[14]	Total PN-equivalent effective exposure time in s at source position in in the mosaicked EPIC soft band image
\item[15]	PN-equivalent source count-rate in the mosaicked EPIC soft band image. If ML$_{\rm S} < 5$, then 95\% upper limit is given.
\item[16]	1$\sigma$ statistical error in PN-equivalent source count-rate in the mosaicked EPIC soft band image (note: does not include uncertainty in spectrally-dependent PN/MOS sensitivity ratio)
\item[17]	Maximum likelihood of detection in the mosaicked EPIC hard band (2-7.3~keV)  image
\item[18]	Number of detected counts in the mosaicked EPIC hard band image. If ML$_{\rm H} < 5$, then 95\% upper limit is given.
\item[19]	Total PN-equivalent effective exposure time in s at source position in in the mosaicked EPIC hard band image
\item[20]	PN-equivalent source count-rate in the mosaicked EPIC hard band image. If ML$_{\rm H} < 5$, then 95\% upper limit is given.
\item[21]	1$\sigma$ statistical error in PN-equivalent source count-rate in the mosaicked EPIC hard band image (note: does not include uncertainty in spectrally-dependent PN/MOS sensitivity ratio)
\item[22]	Hardness ratio (Rate$_{\rm H}$-Rate$_{\rm S}$)/Rate$_{\rm F}$. If object is not detected in soft (hard) band, then HR is upper (lower) limit. If object is detected only in full band,
                then HR = 0.
\item[23]	1$\sigma$ lower limit to hardness ratio
\item[24]	1$\sigma$ upper limit to hardness ratio
\item[25]	Comment concerning by-eye verification of source
\end{description}

\section*{Appendix B: The XEST atlas}

\begin{figure*}
\hbox{
{\resizebox{0.45\hsize}{!}{\includegraphics{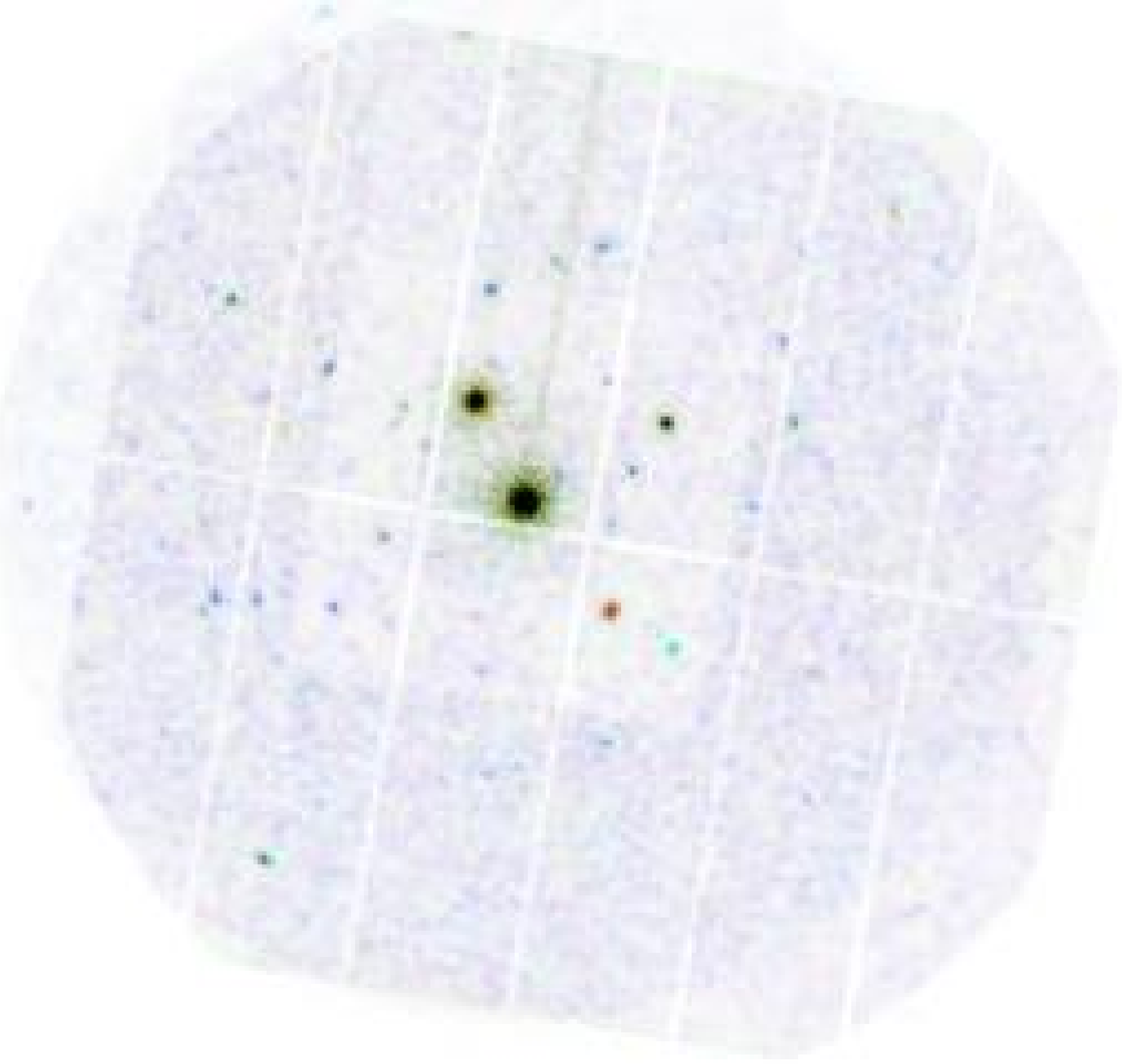}}}
{\resizebox{0.45\hsize}{!}{\includegraphics{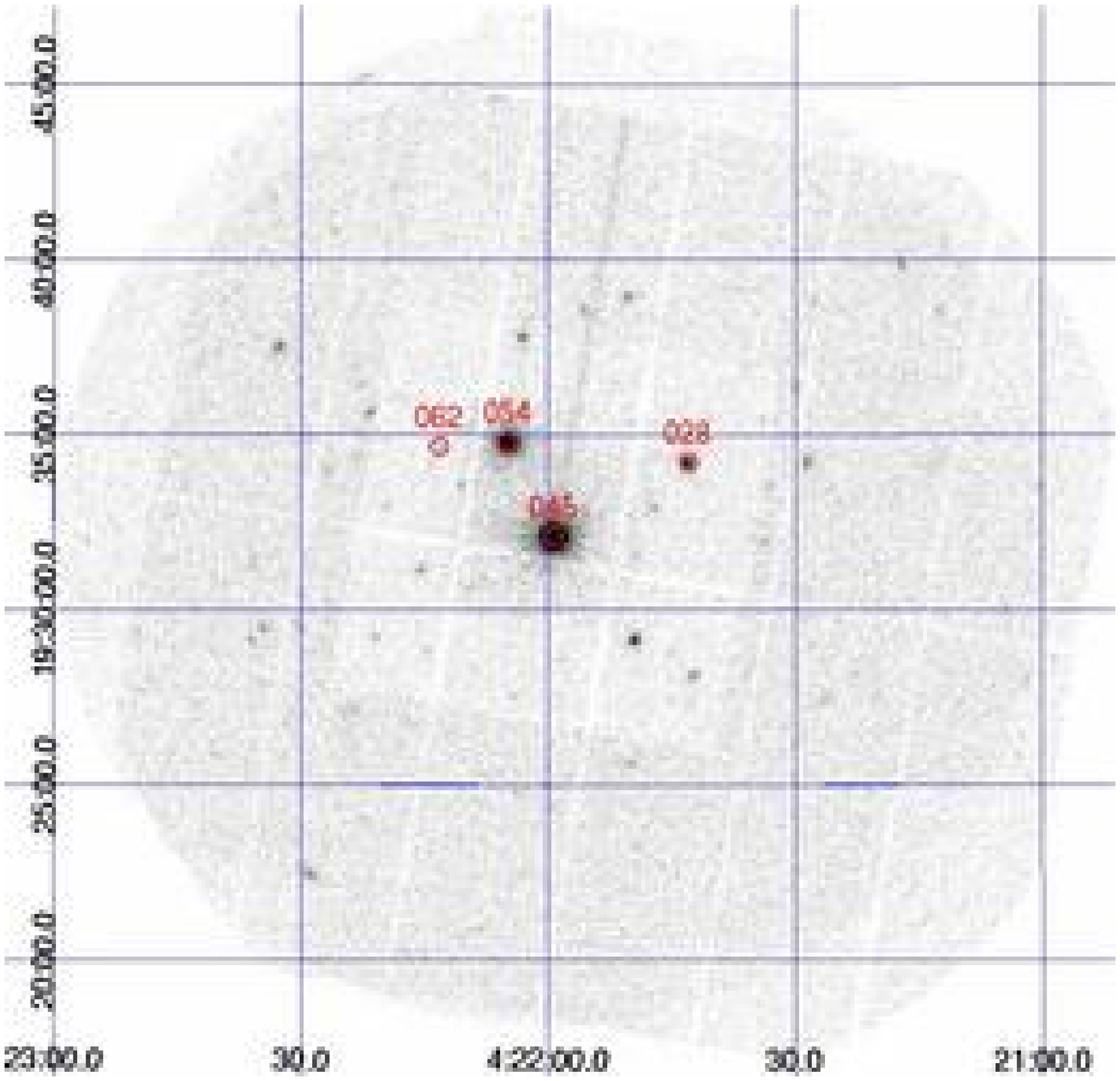}}}
}
\hbox{
{\resizebox{0.45\hsize}{!}{\includegraphics{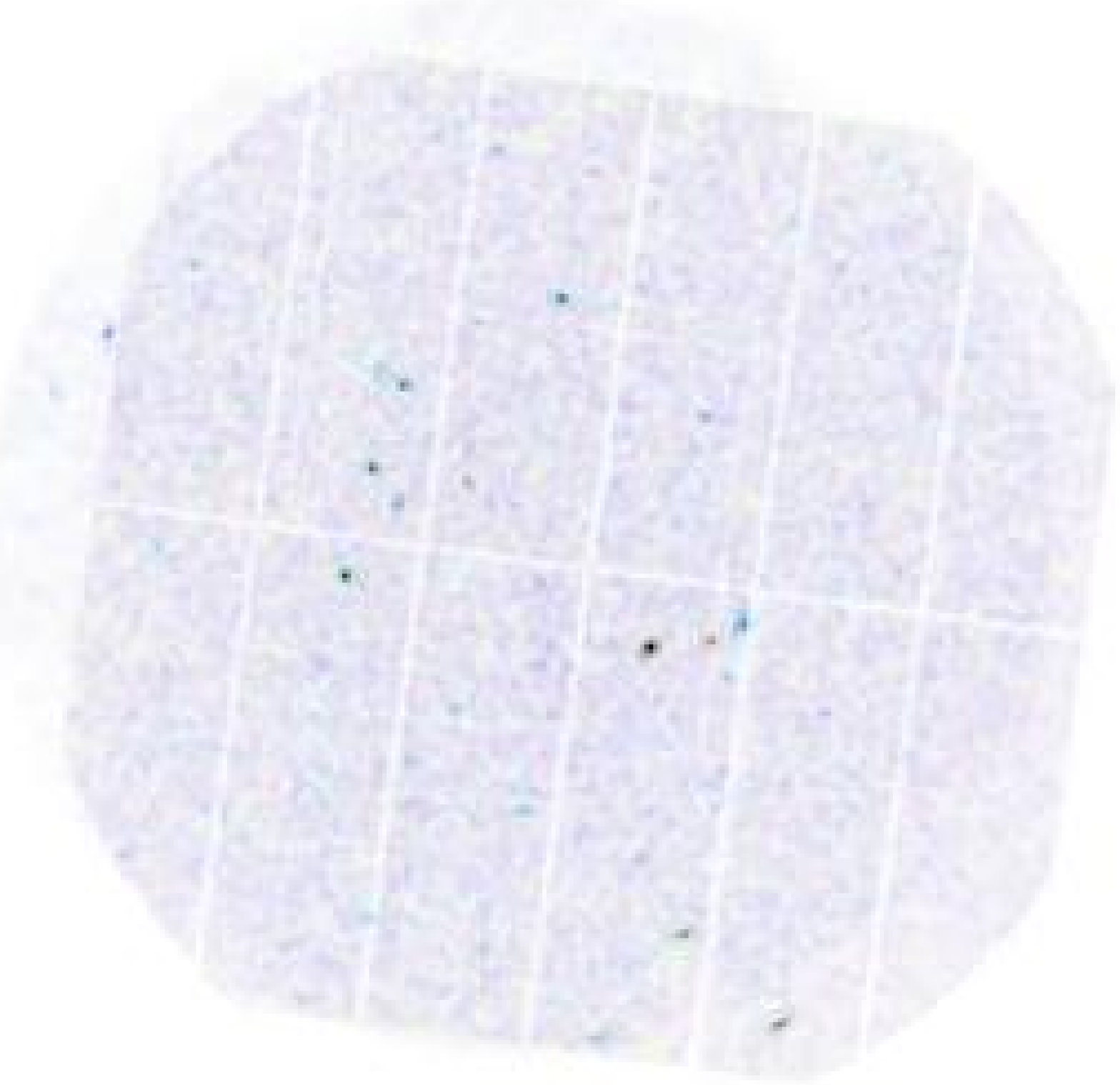}}}
{\resizebox{0.45\hsize}{!}{\includegraphics{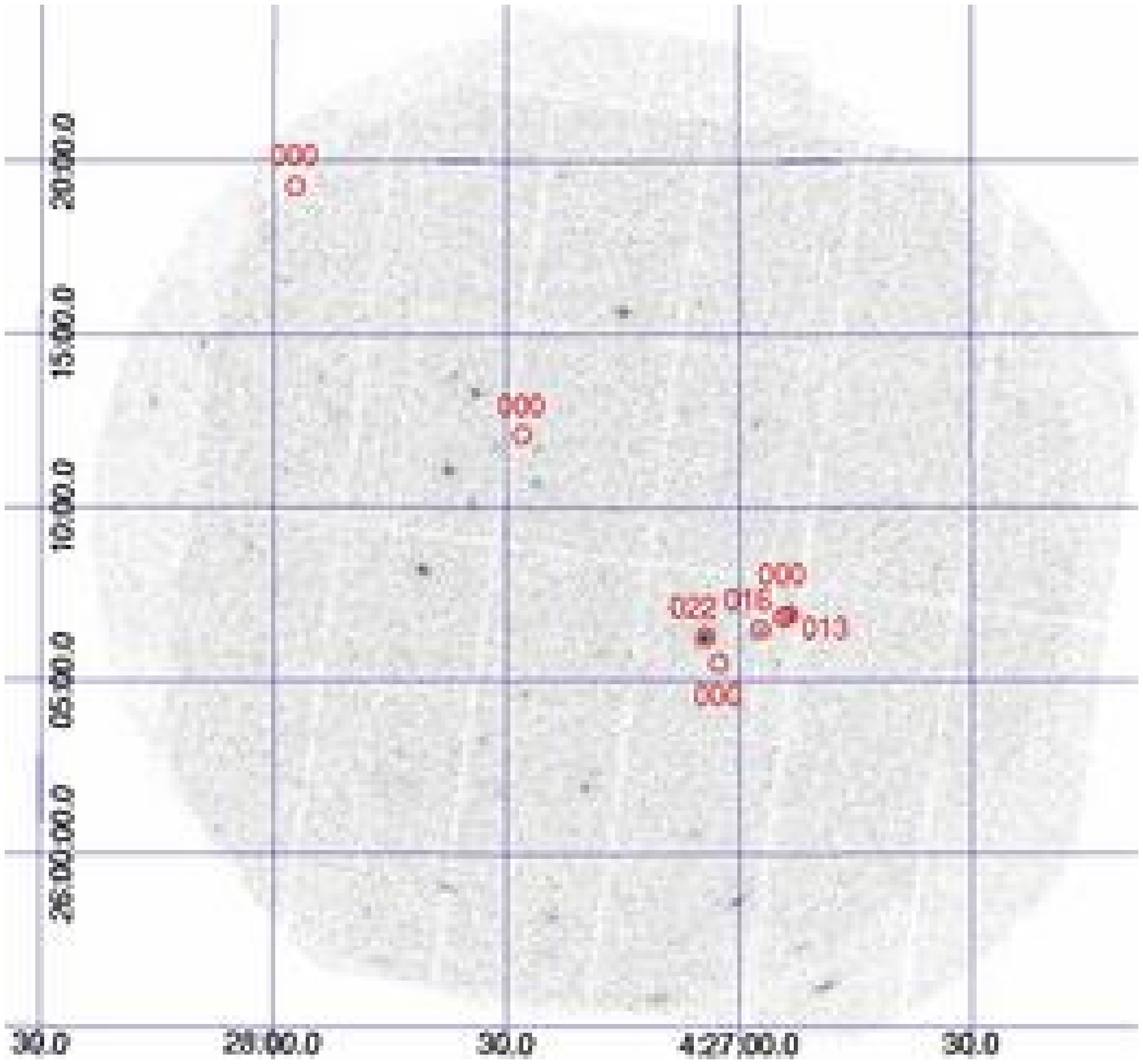}}}
}
\hbox{
{\resizebox{0.45\hsize}{!}{\includegraphics{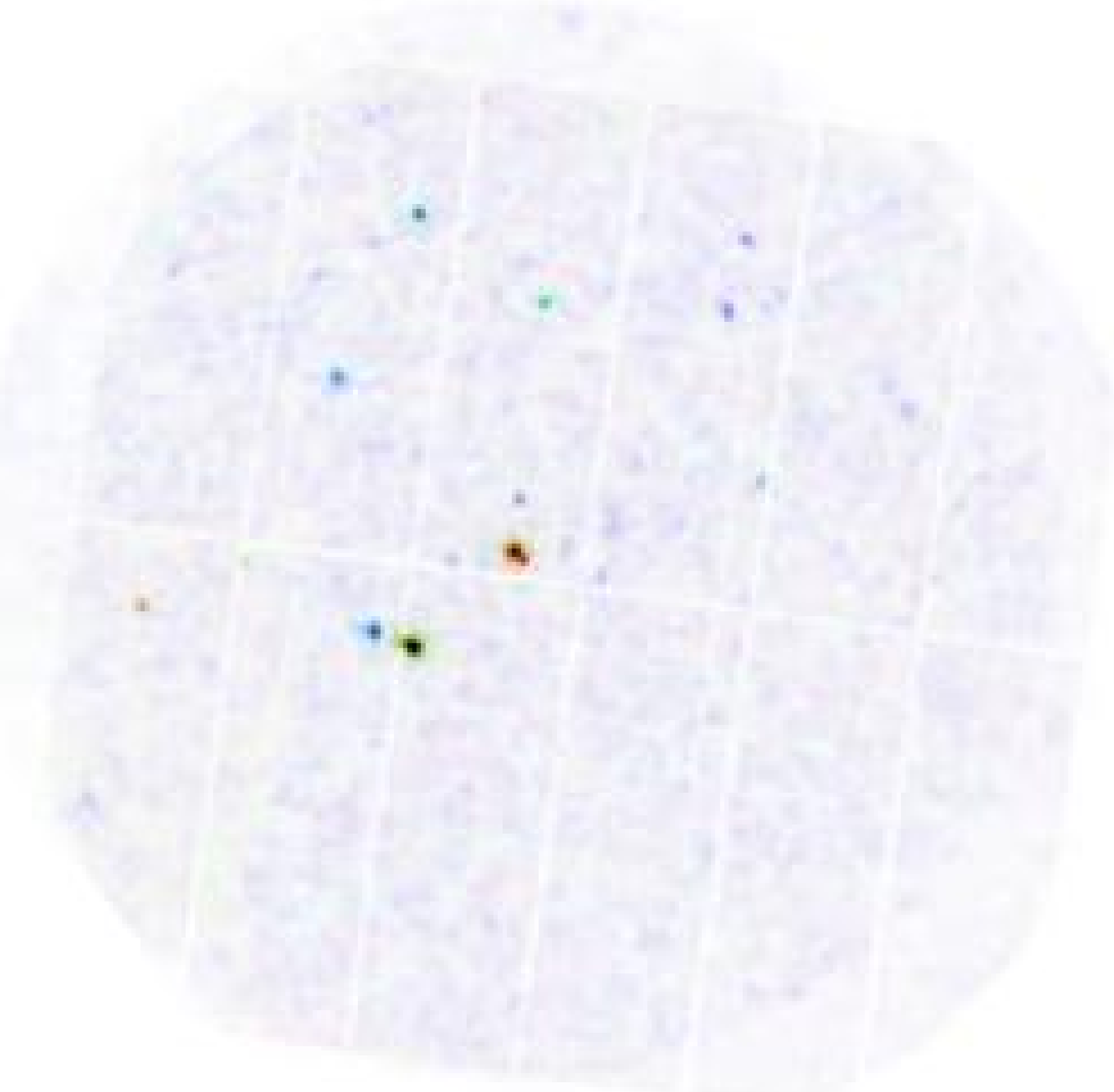}}}
{\resizebox{0.45\hsize}{!}{\includegraphics{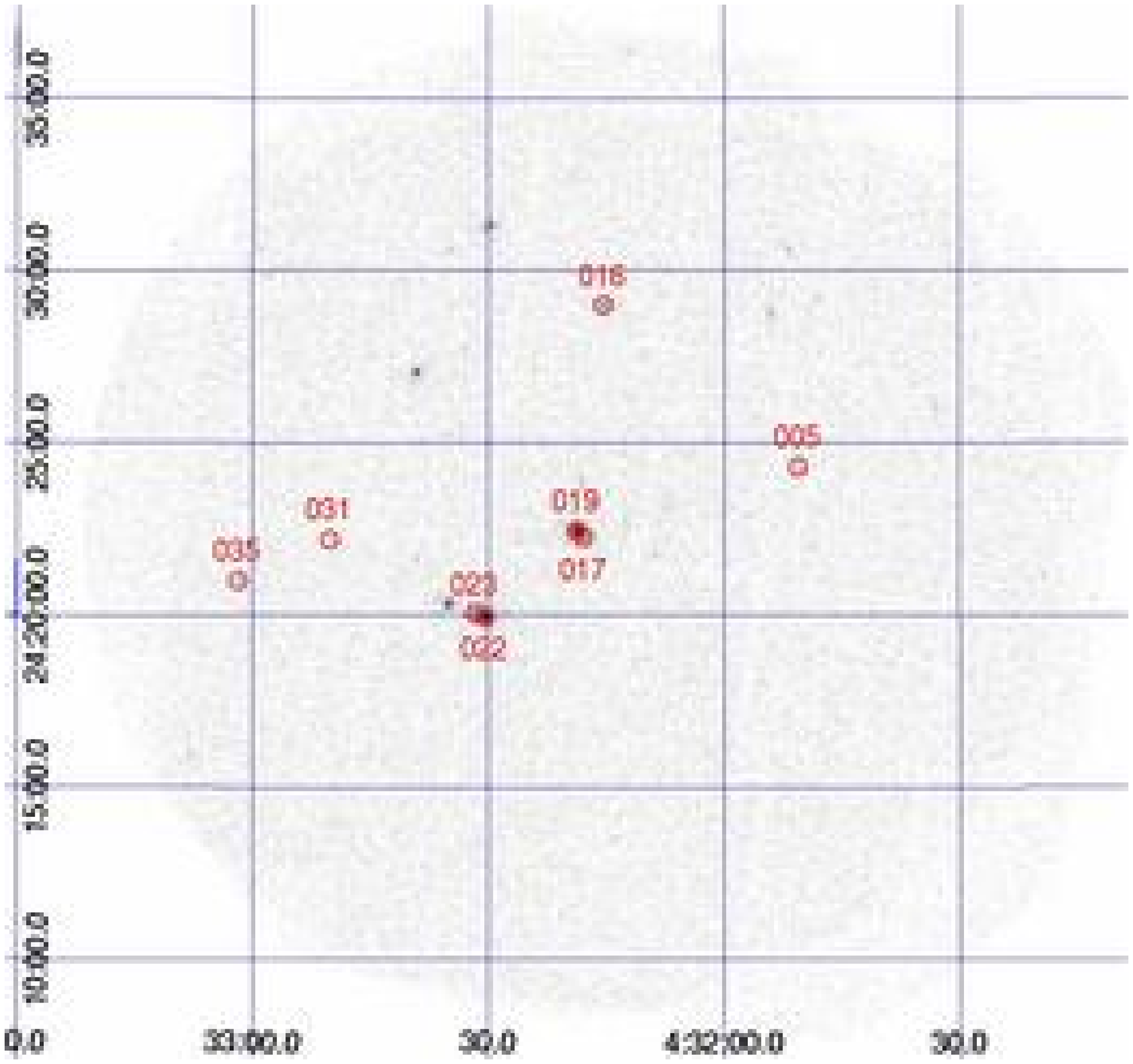}}}
}
\caption{Co-added EPIC images of field XEST-01, XEST-02, and XEST-03 (from top to bottom). Left: Smoothed images, color coded for hardness; right: 
      coordinate grid and TMC identifications included.\label{atlas1}} 
\end{figure*}

\begin{figure*}
\hbox{
{\resizebox{0.45\hsize}{!}{\includegraphics{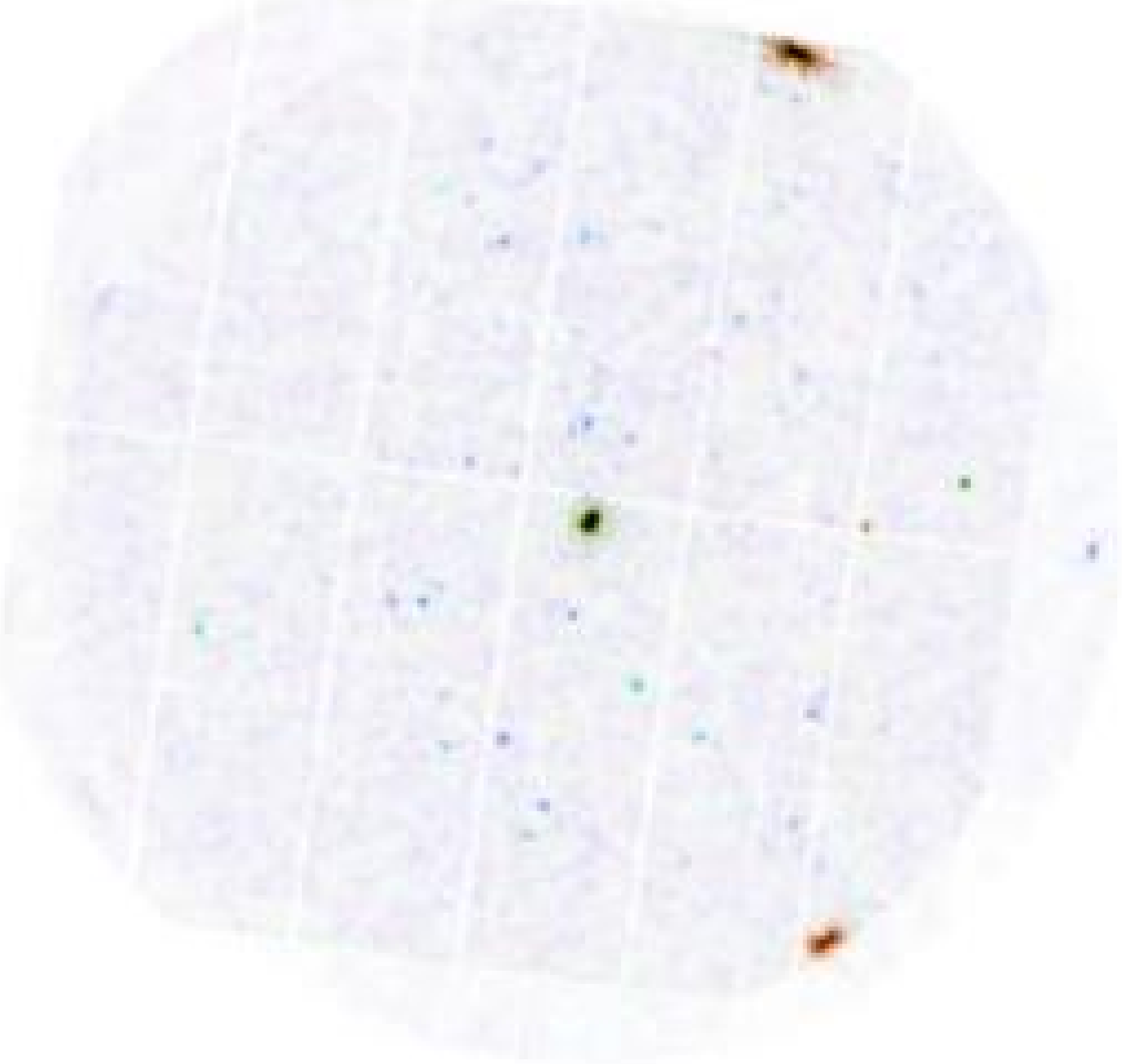}}}
{\resizebox{0.45\hsize}{!}{\includegraphics{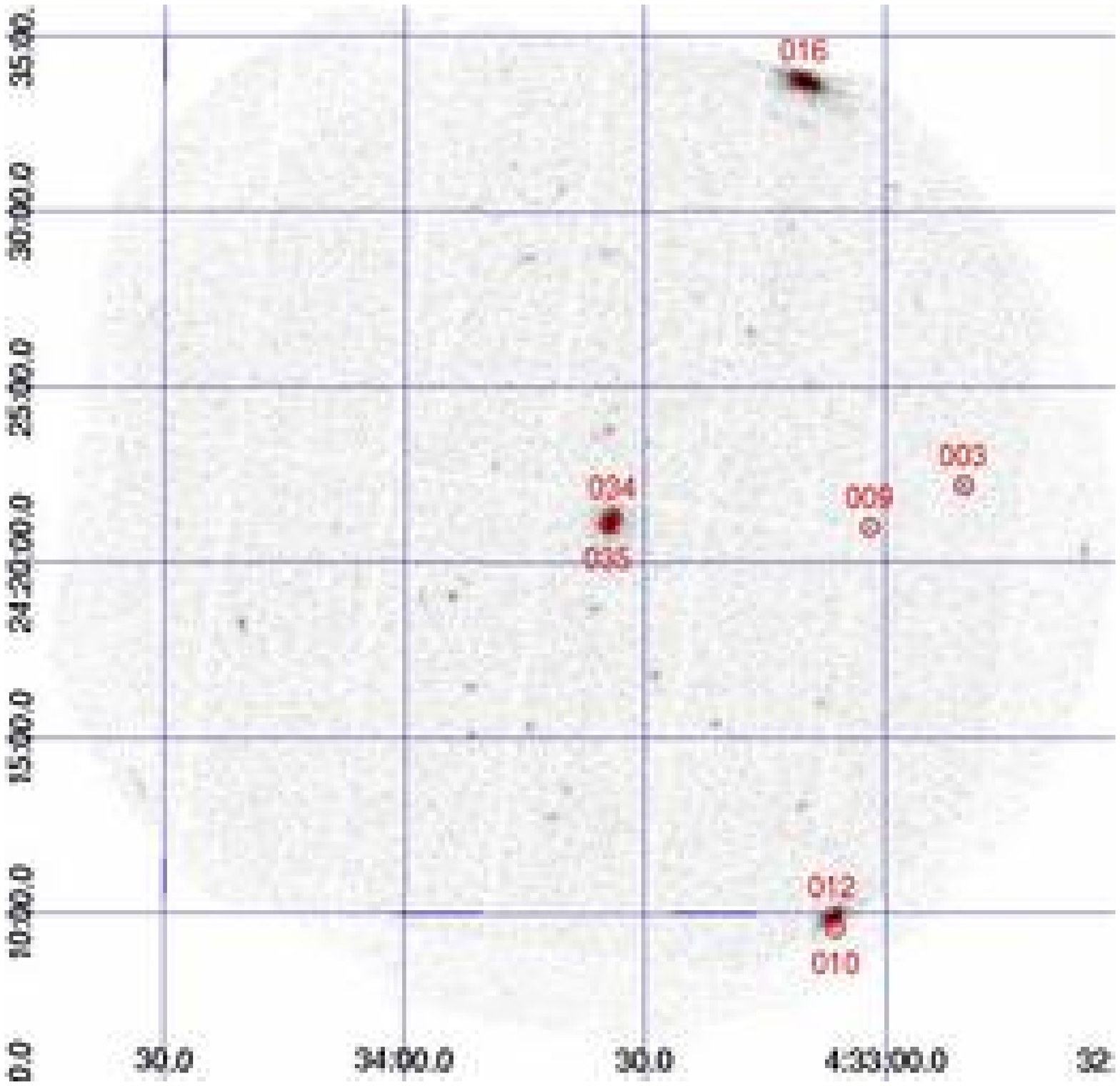}}}
}
\hbox{
{\resizebox{0.45\hsize}{!}{\includegraphics{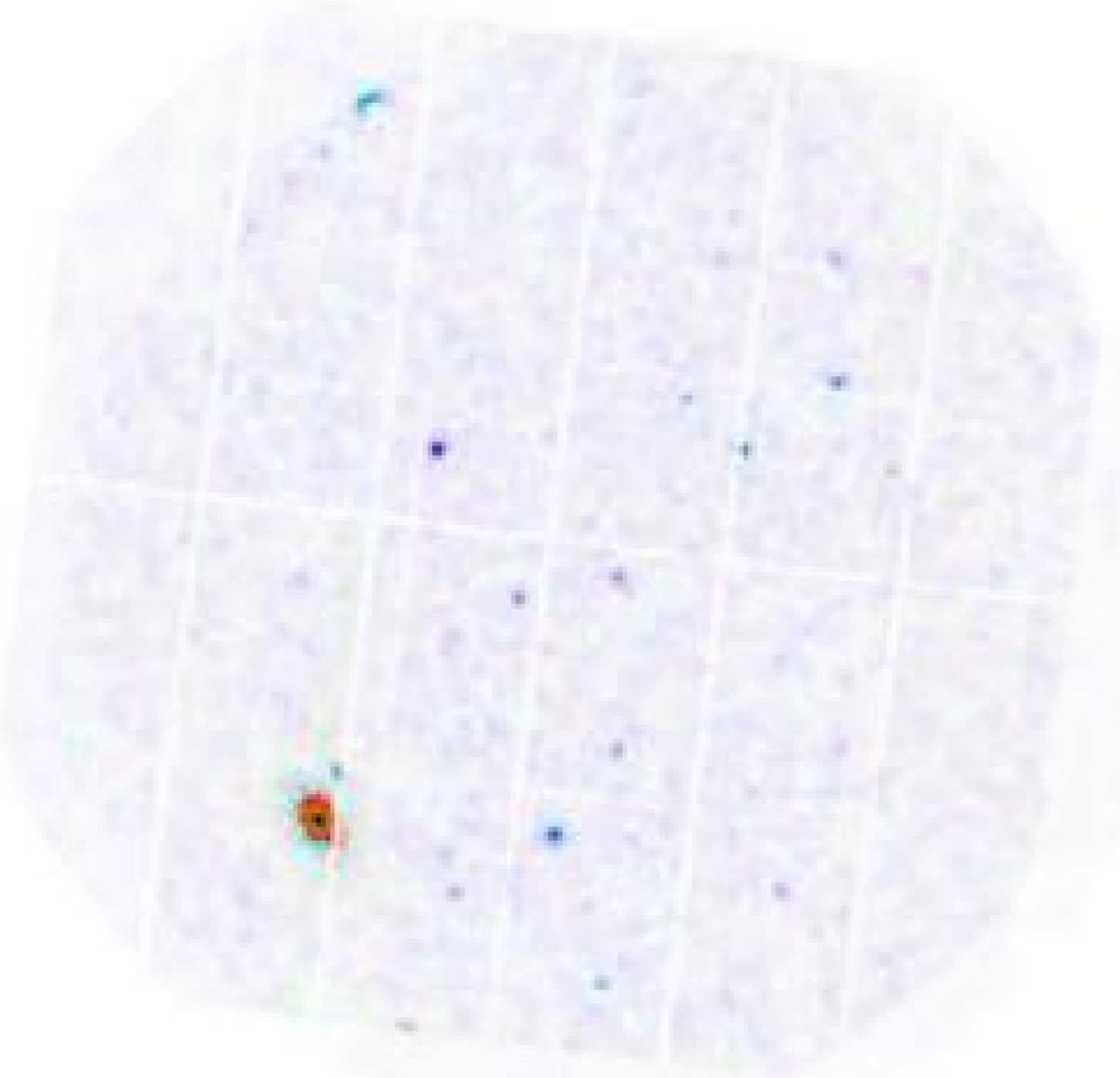}}}
{\resizebox{0.45\hsize}{!}{\includegraphics{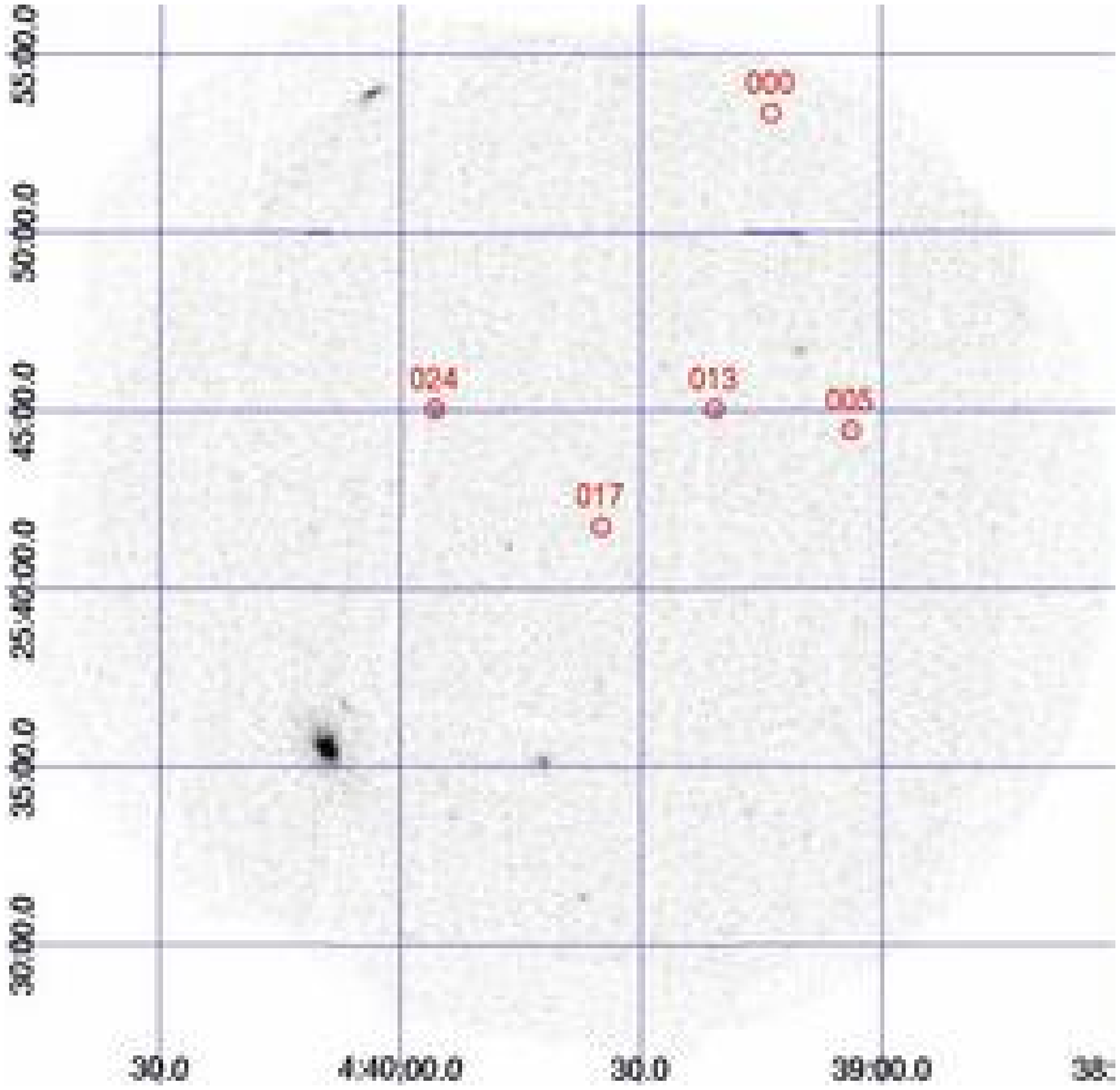}}}
}
\hbox{
{\resizebox{0.45\hsize}{!}{\includegraphics{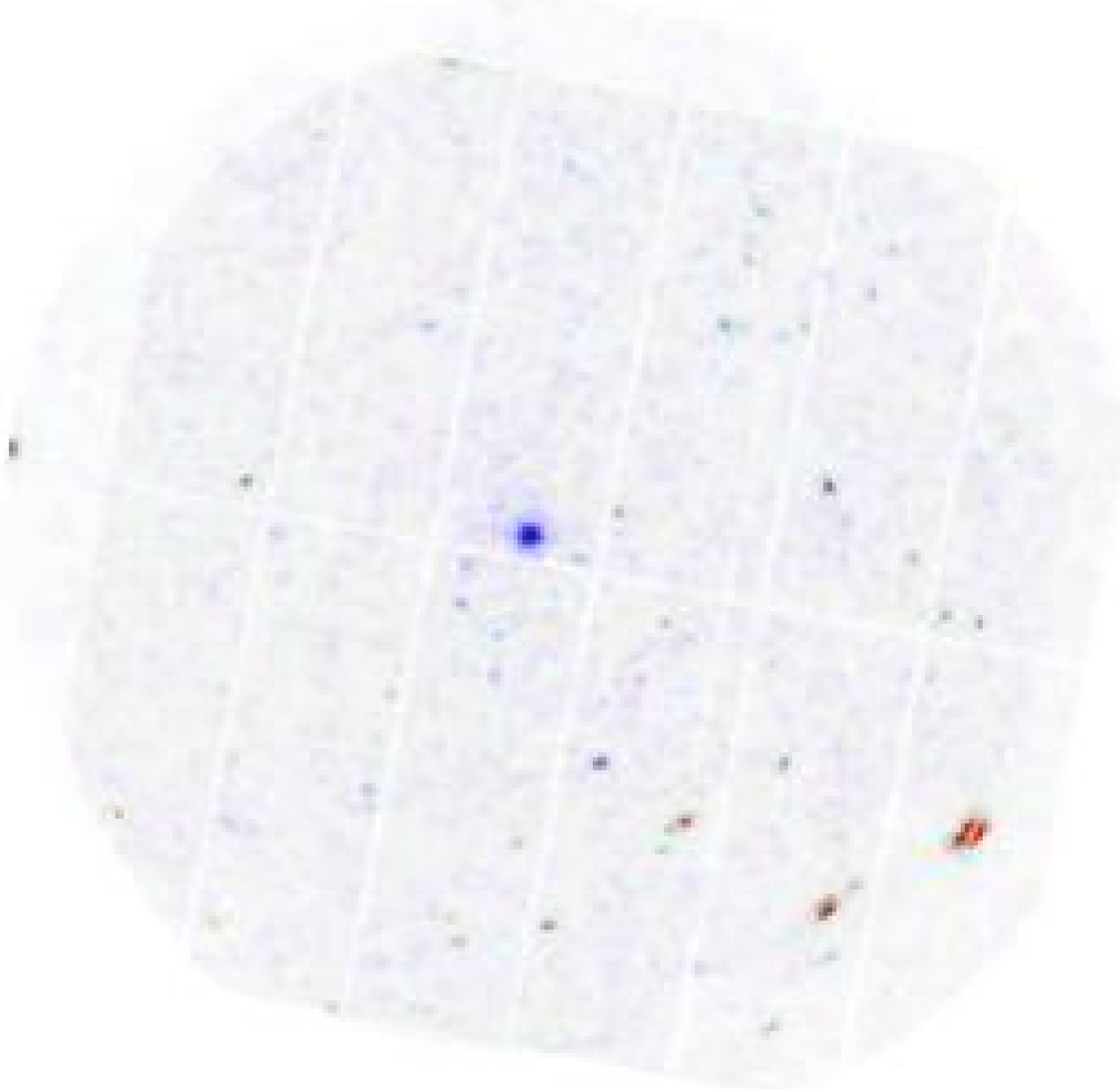}}}
{\resizebox{0.45\hsize}{!}{\includegraphics{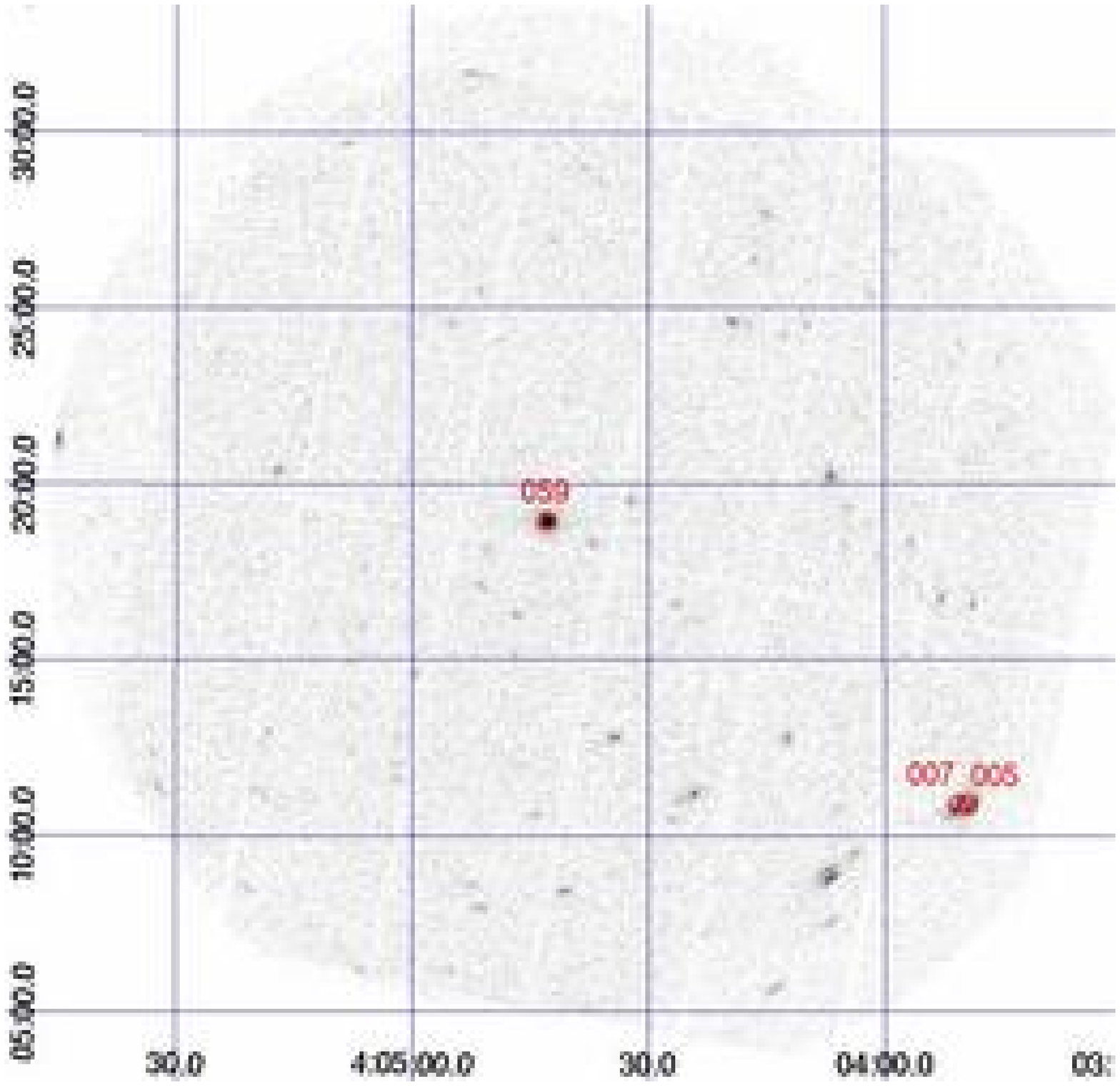}}}
}
\caption{Co-added EPIC images of field XEST-04, XEST-05, and XEST-06 (from top to bottom). Left: Smoothed images, color coded for hardness; right: 
      coordinate grid and TMC identifications included.\label{atlas2}} 
\end{figure*}

\begin{figure*}
\hbox{
{\resizebox{0.45\hsize}{!}{\includegraphics{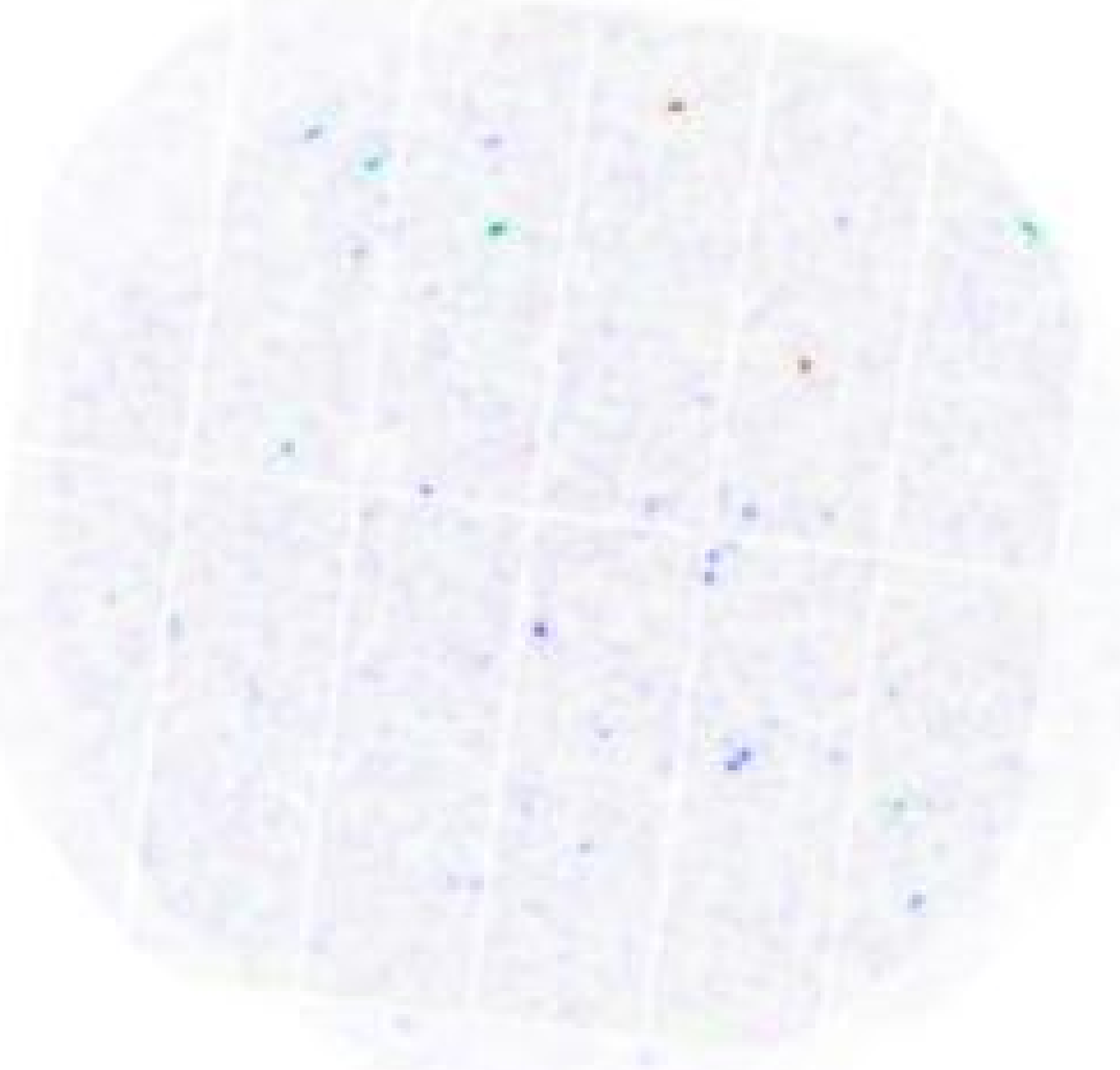}}}
{\resizebox{0.45\hsize}{!}{\includegraphics{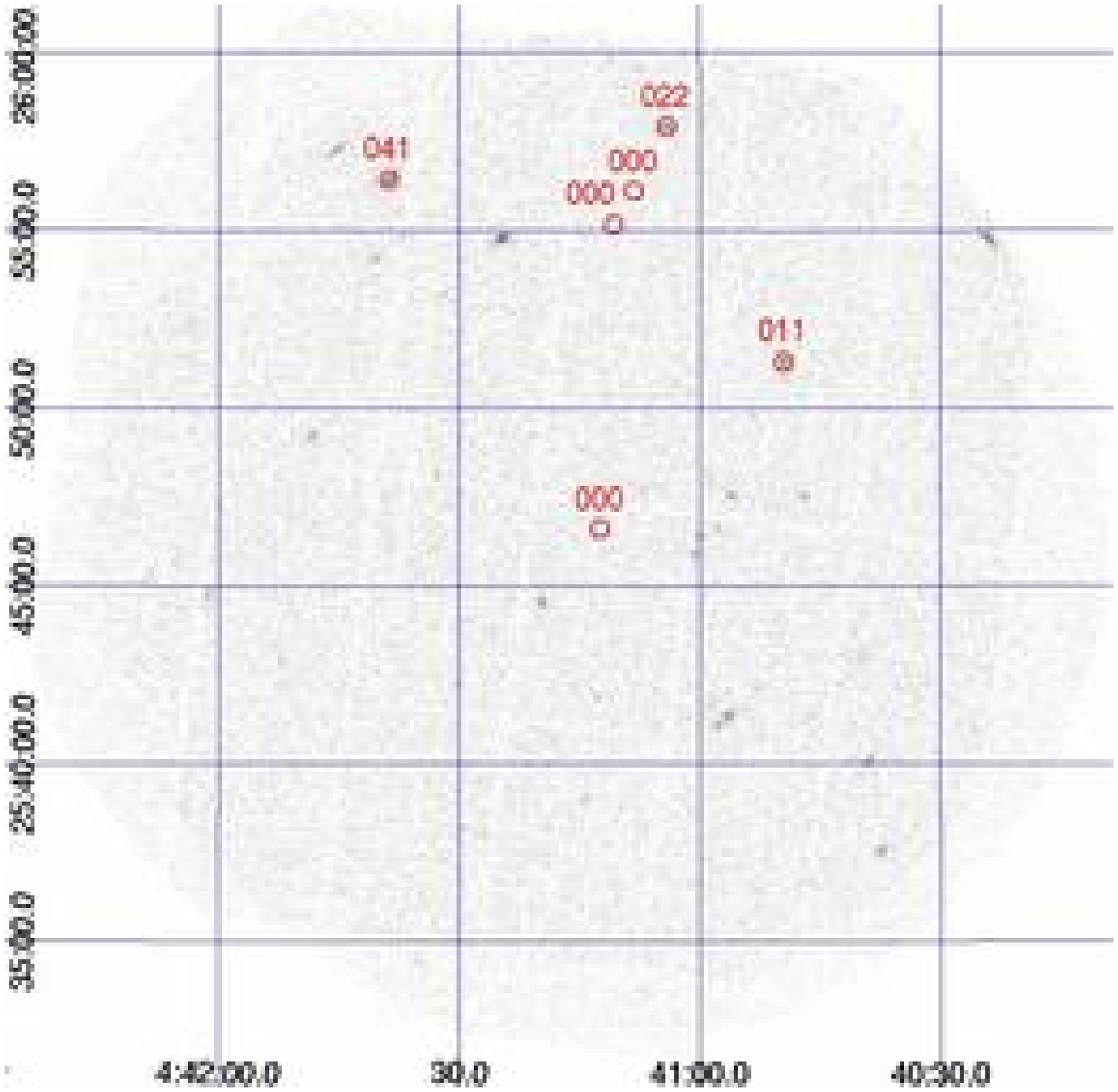}}}
}
\hbox{
{\resizebox{0.45\hsize}{!}{\includegraphics{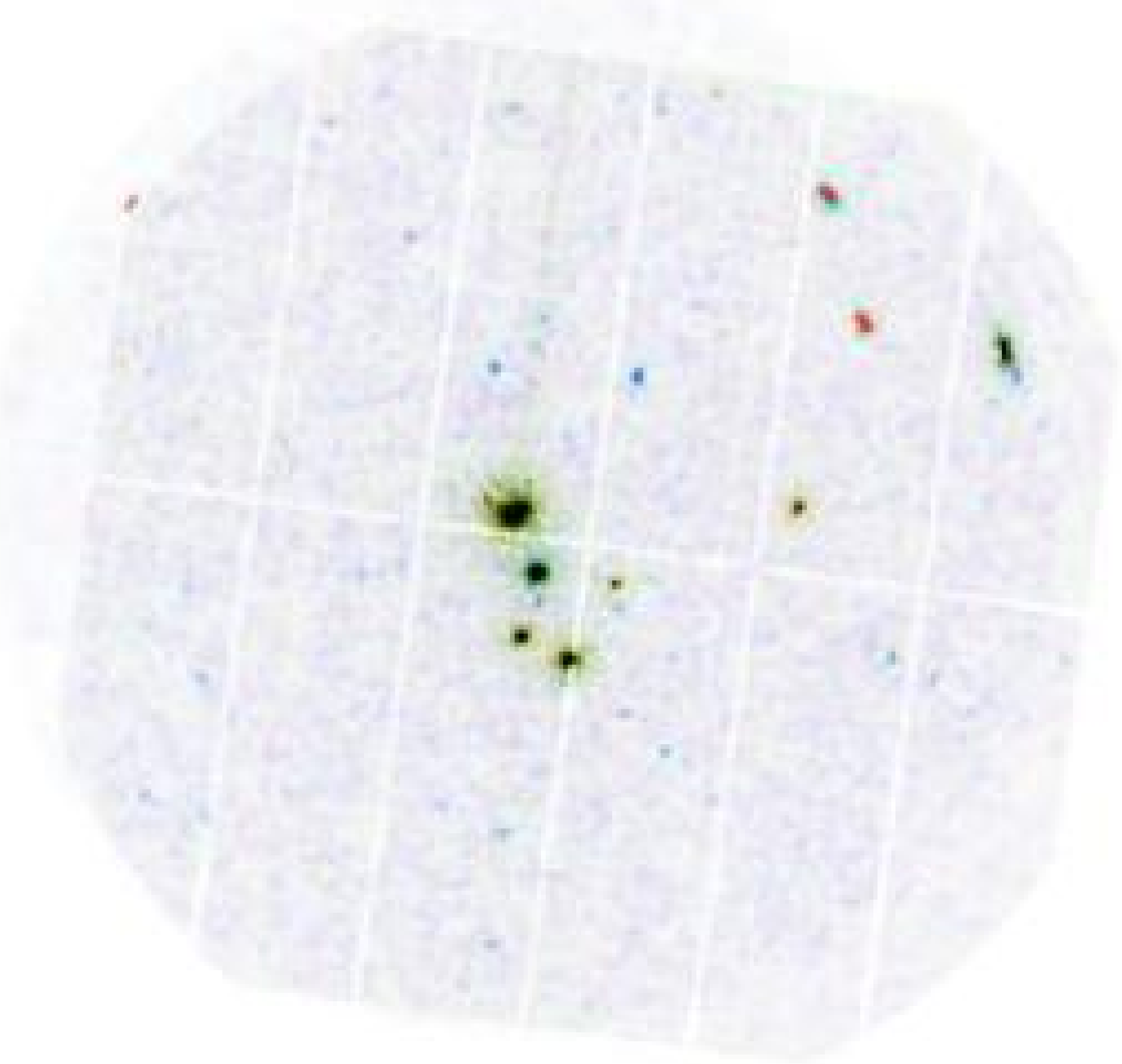}}}
{\resizebox{0.45\hsize}{!}{\includegraphics{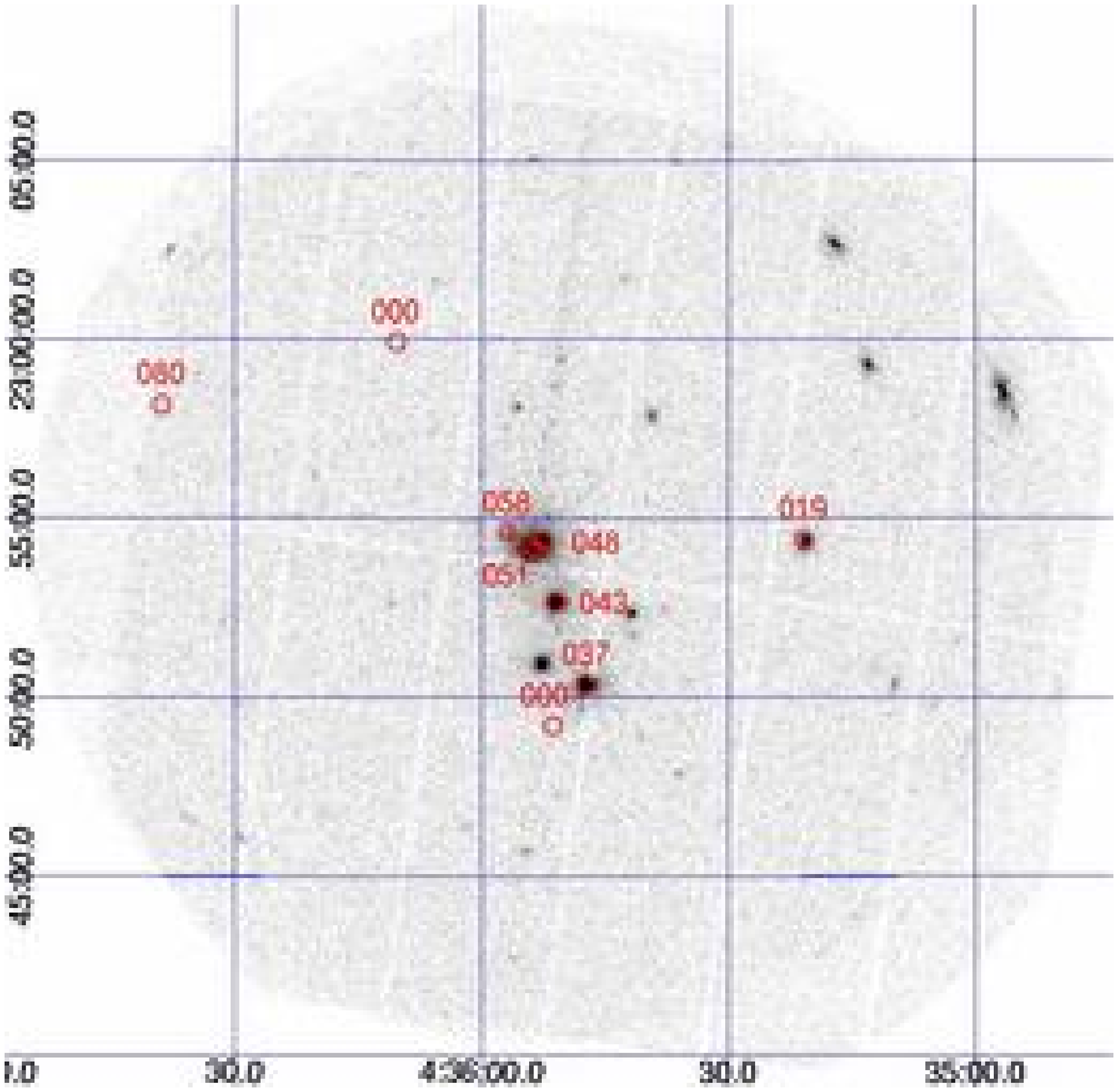}}}
}
\hbox{
{\resizebox{0.45\hsize}{!}{\includegraphics{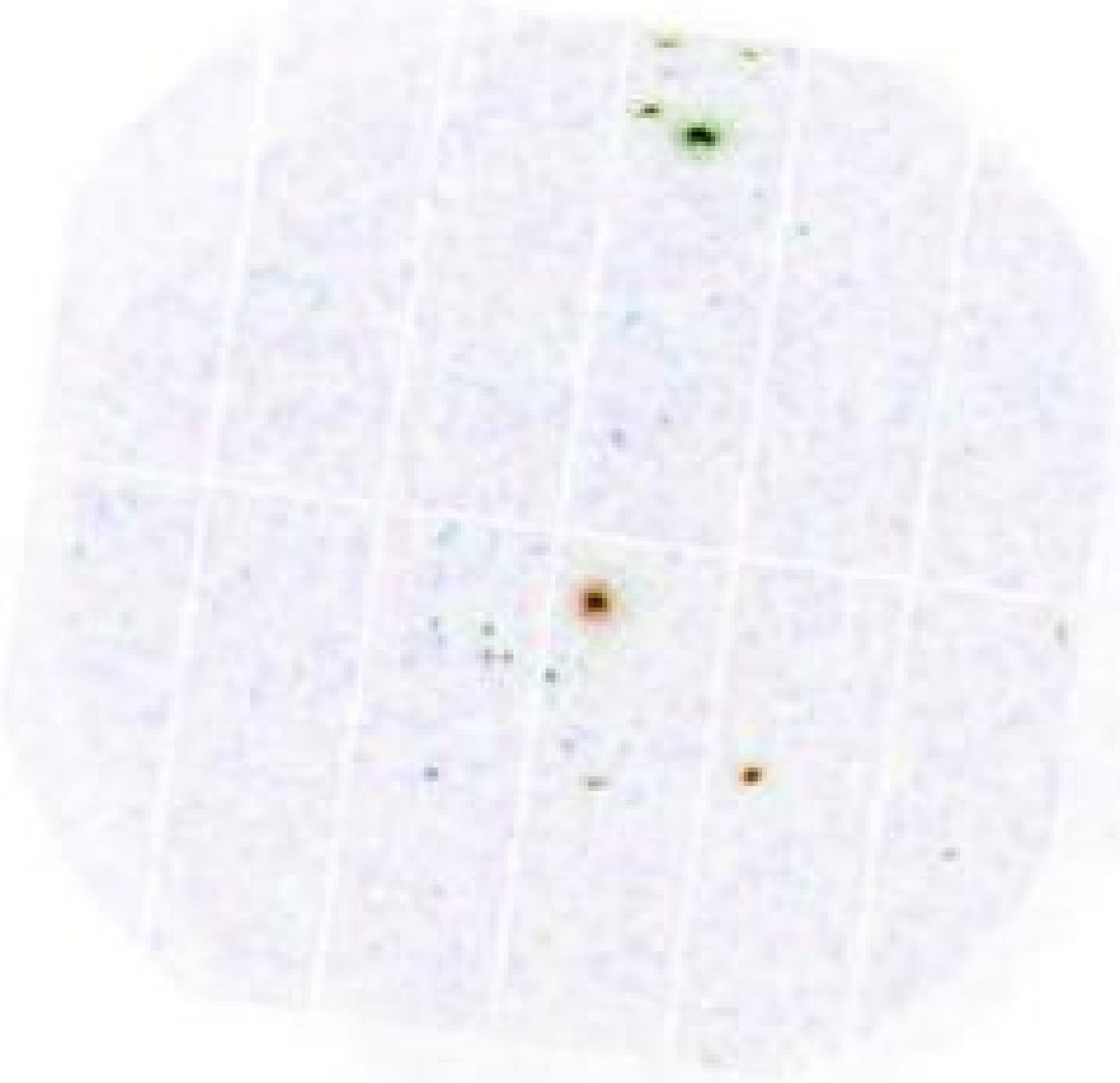}}}
{\resizebox{0.45\hsize}{!}{\includegraphics{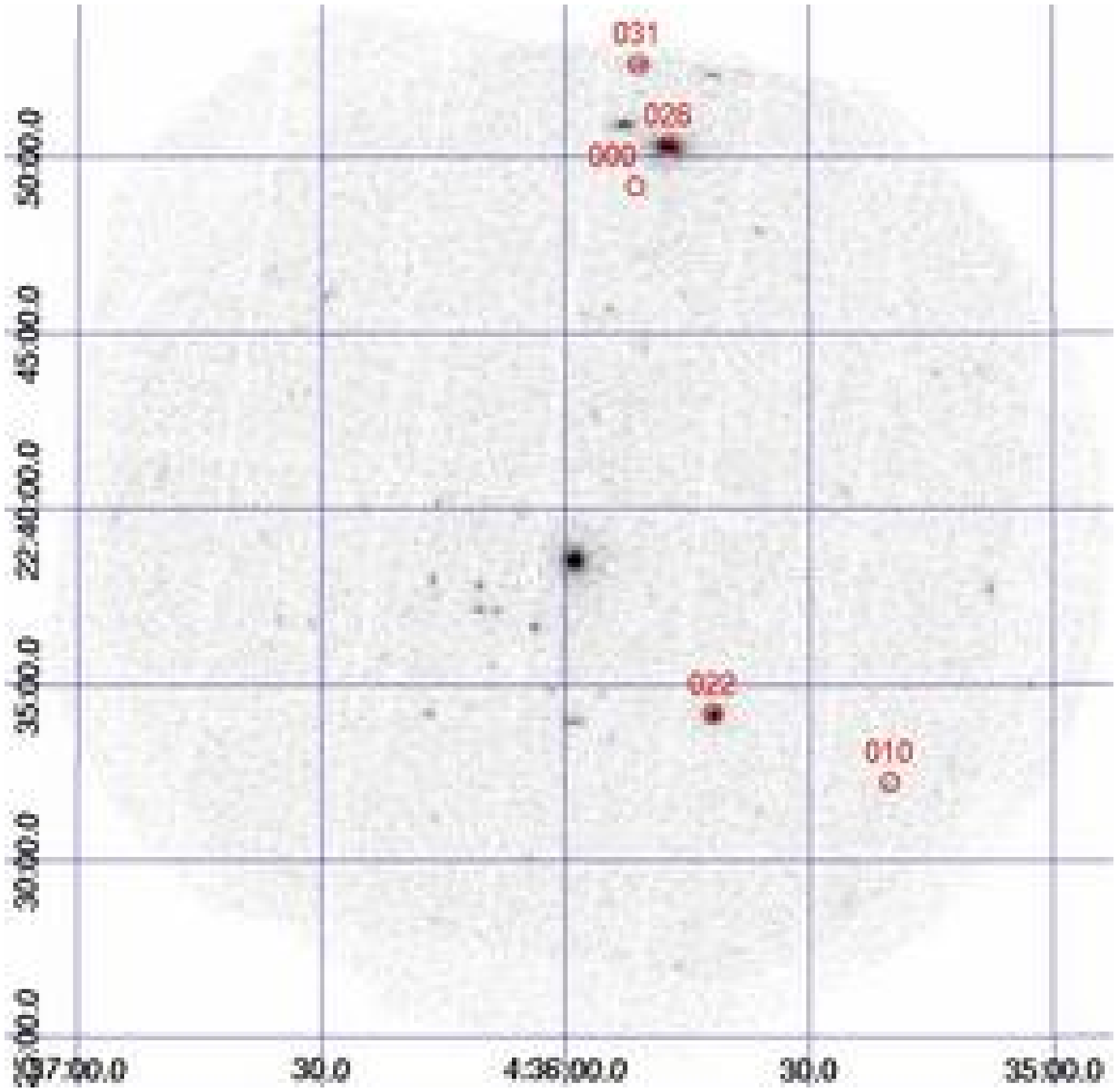}}}
}
\caption{Co-added EPIC images of field XEST-07, XEST-08, and XEST-09 (from top to bottom). Left: Smoothed images, color coded for hardness; right: 
      coordinate grid and TMC identifications included.\label{atlas3}} 
\end{figure*}

\begin{figure*}
\hbox{
{\resizebox{0.45\hsize}{!}{\includegraphics{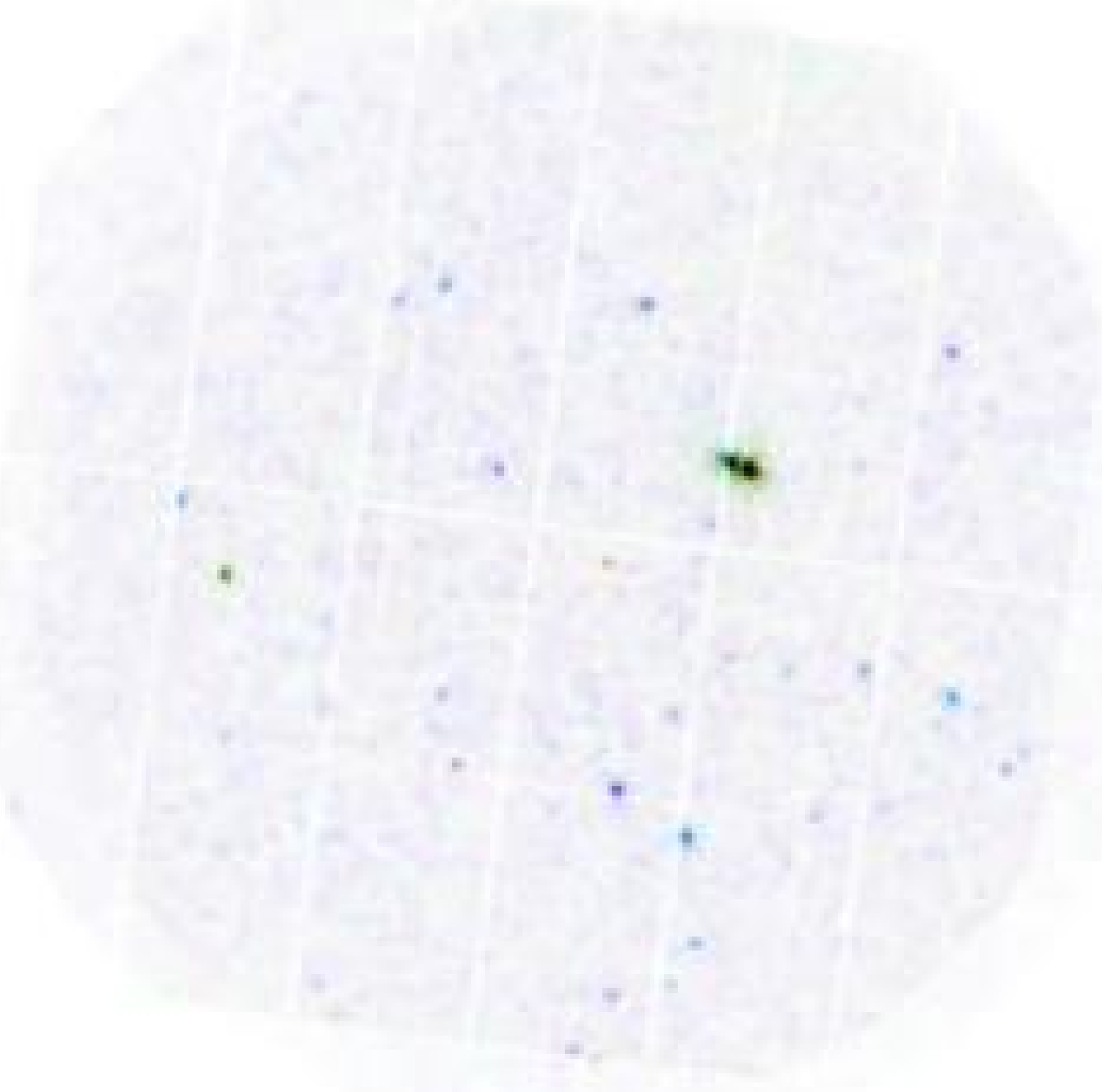}}}
{\resizebox{0.45\hsize}{!}{\includegraphics{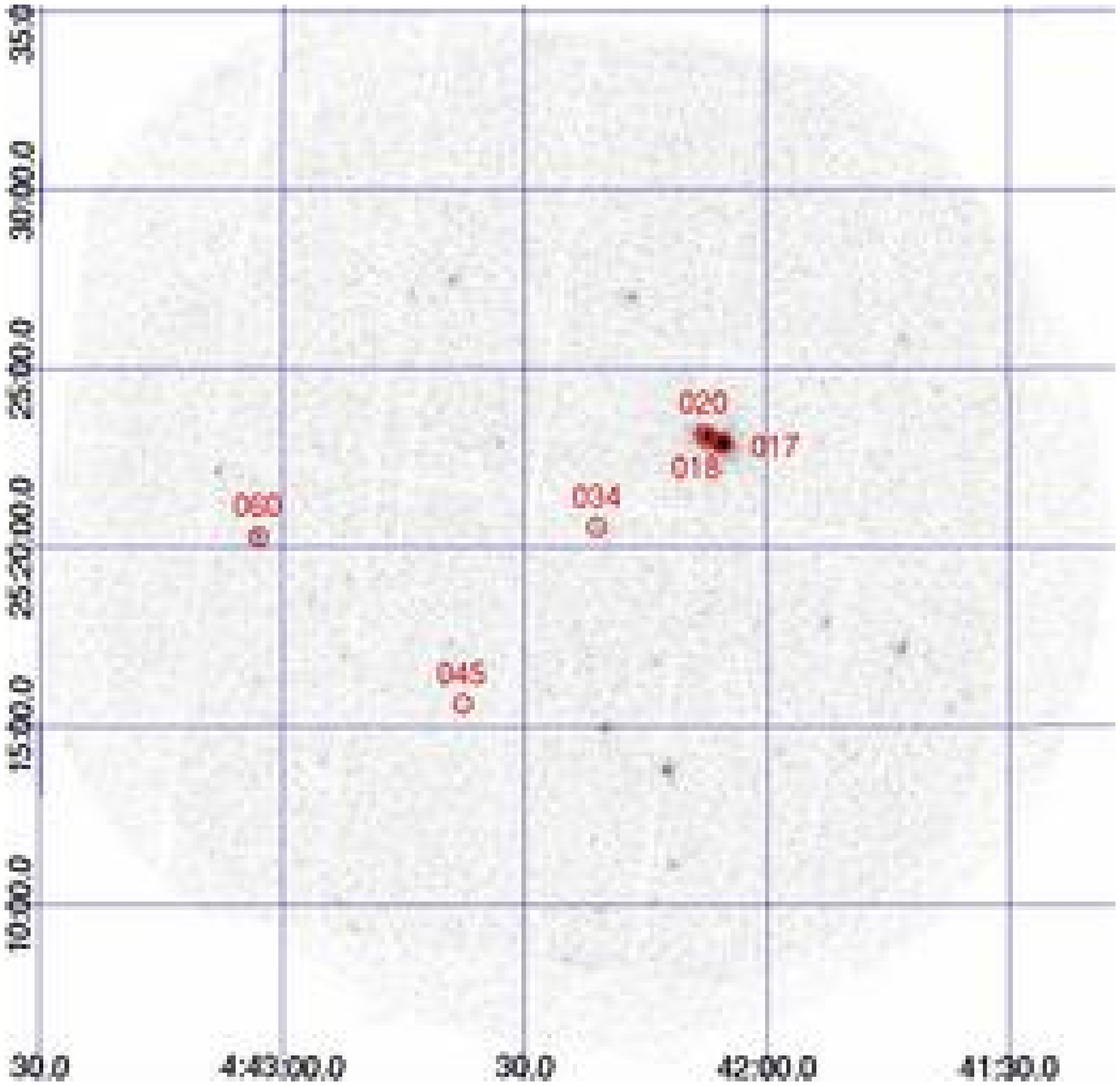}}}
}
\hbox{
{\resizebox{0.45\hsize}{!}{\includegraphics{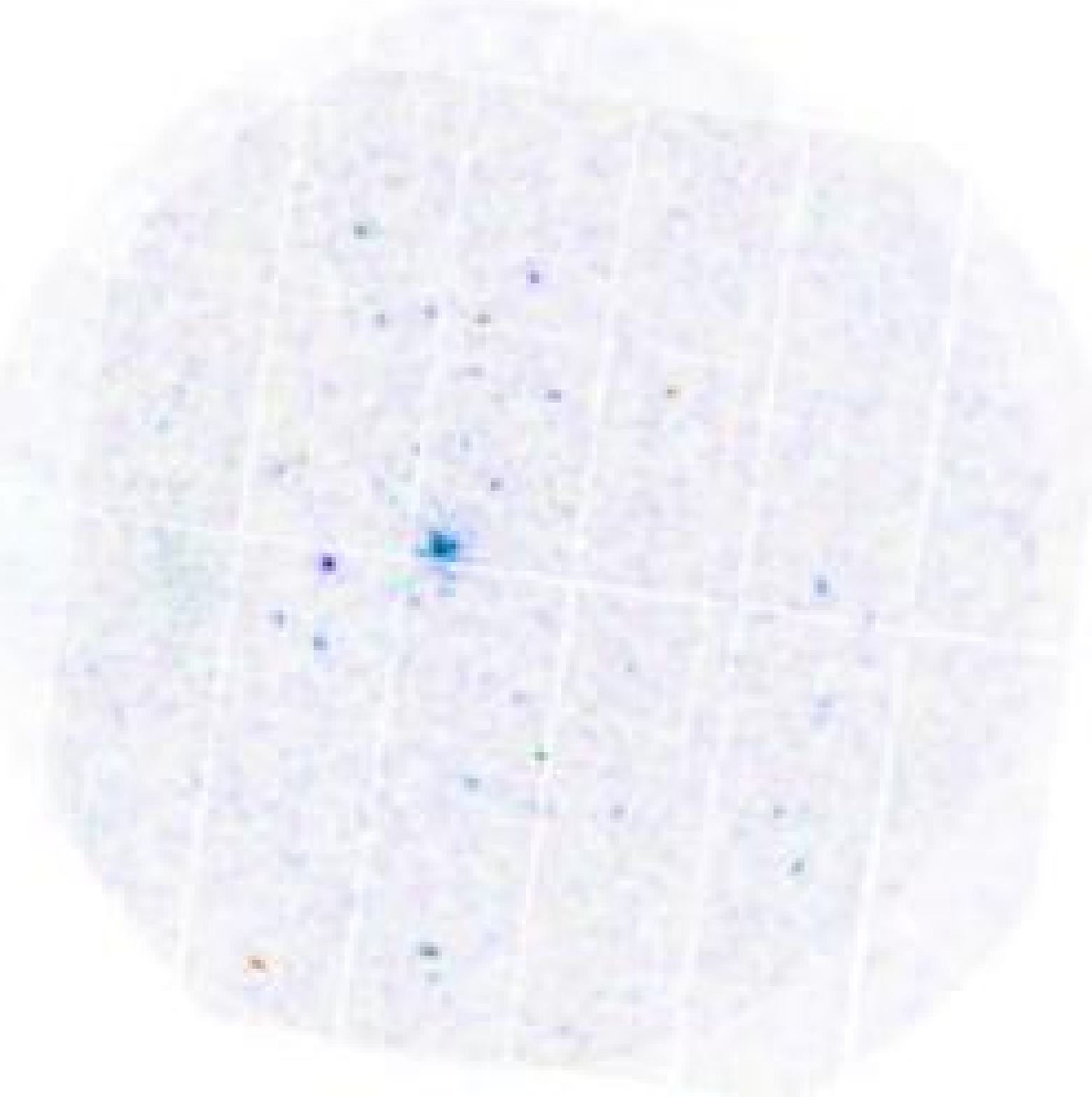}}}
{\resizebox{0.45\hsize}{!}{\includegraphics{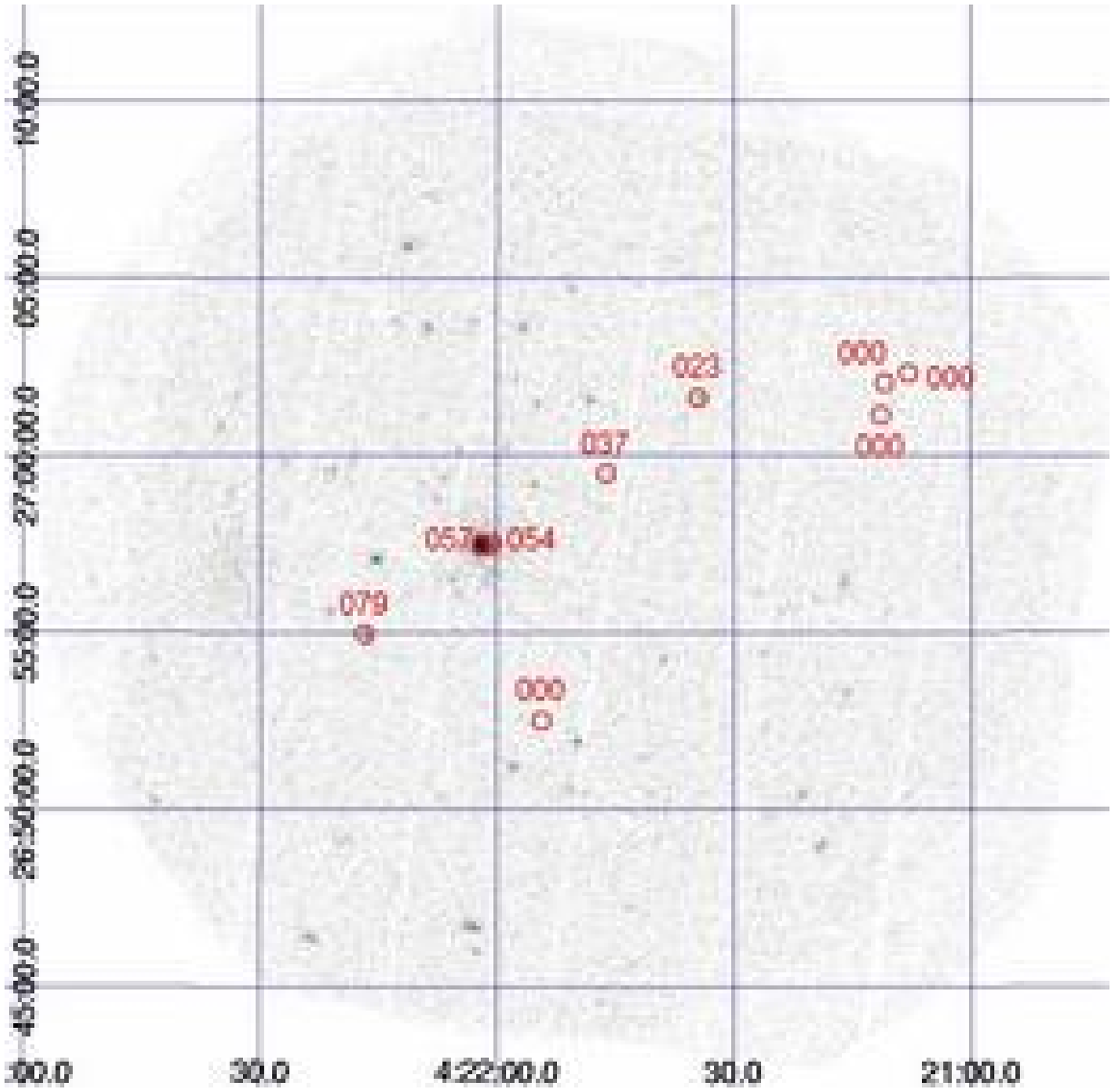}}}
}
\hbox{
{\resizebox{0.45\hsize}{!}{\includegraphics{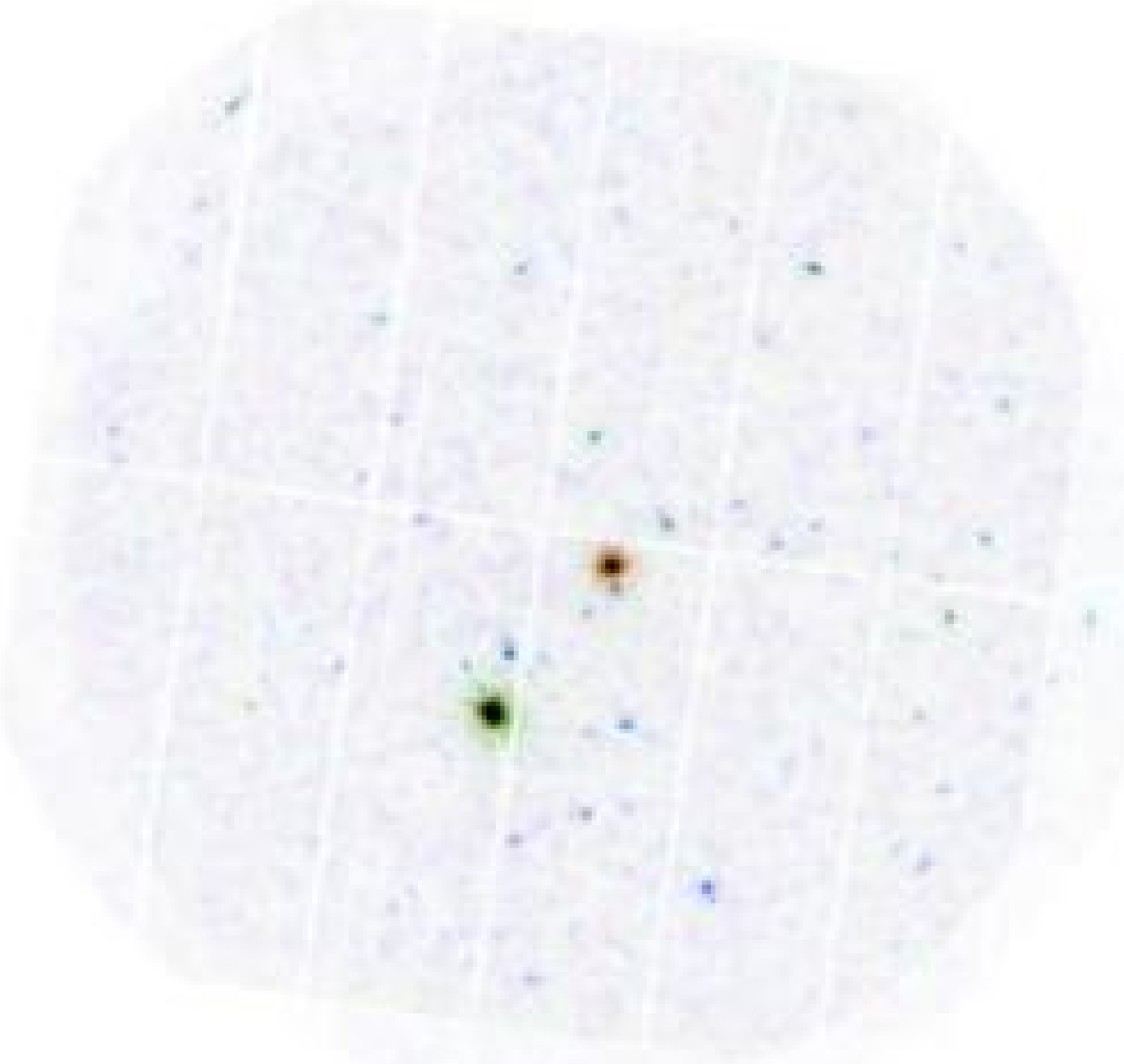}}}
{\resizebox{0.45\hsize}{!}{\includegraphics{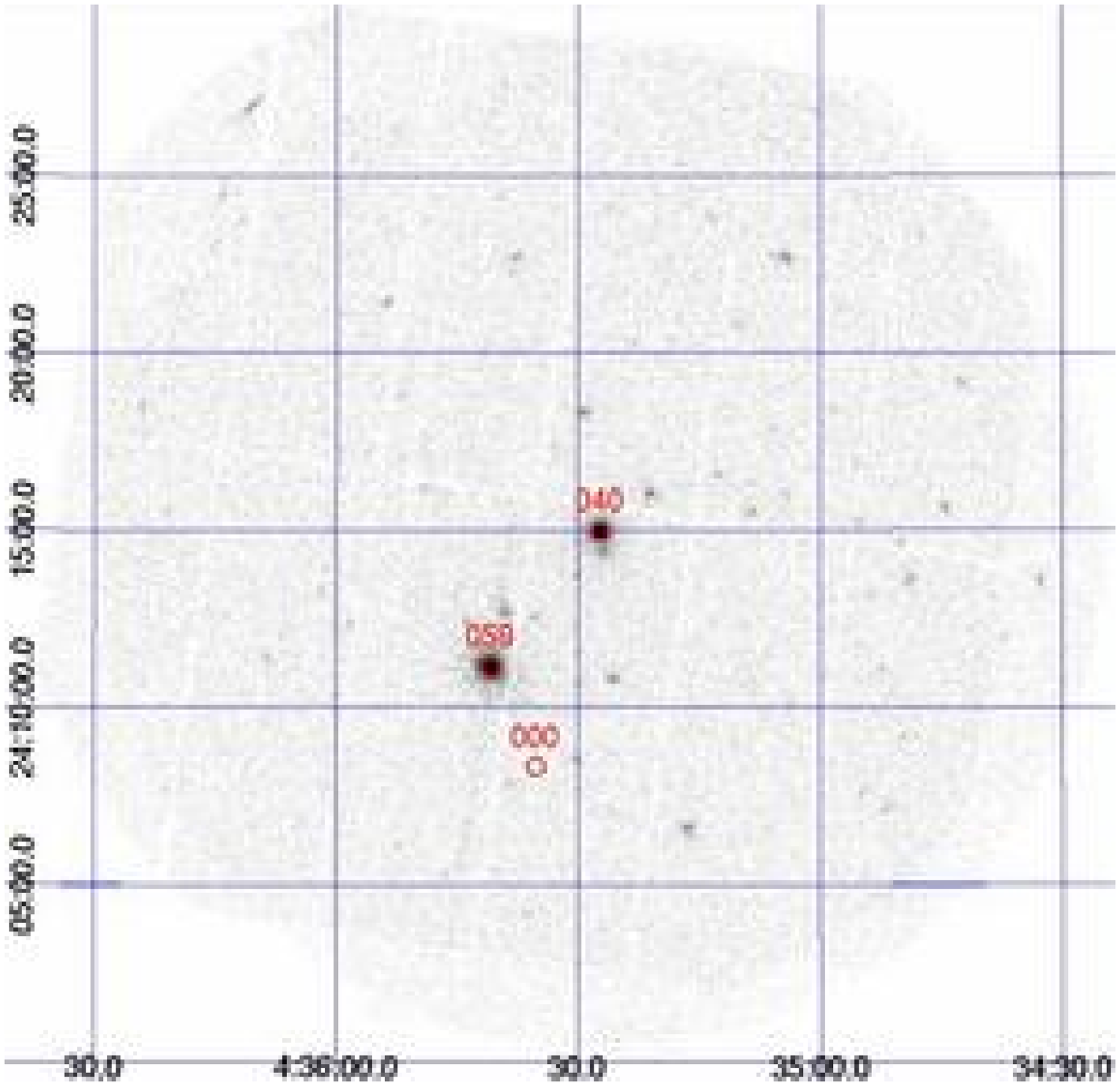}}}
}
\caption{Co-added EPIC images of field XEST-10, XEST-11, and XEST-12 (from top to bottom). Left: Smoothed images, color coded for hardness; right: 
      coordinate grid and TMC identifications included.\label{atlas4}} 
\end{figure*}

\begin{figure*}
\hbox{
{\resizebox{0.45\hsize}{!}{\includegraphics{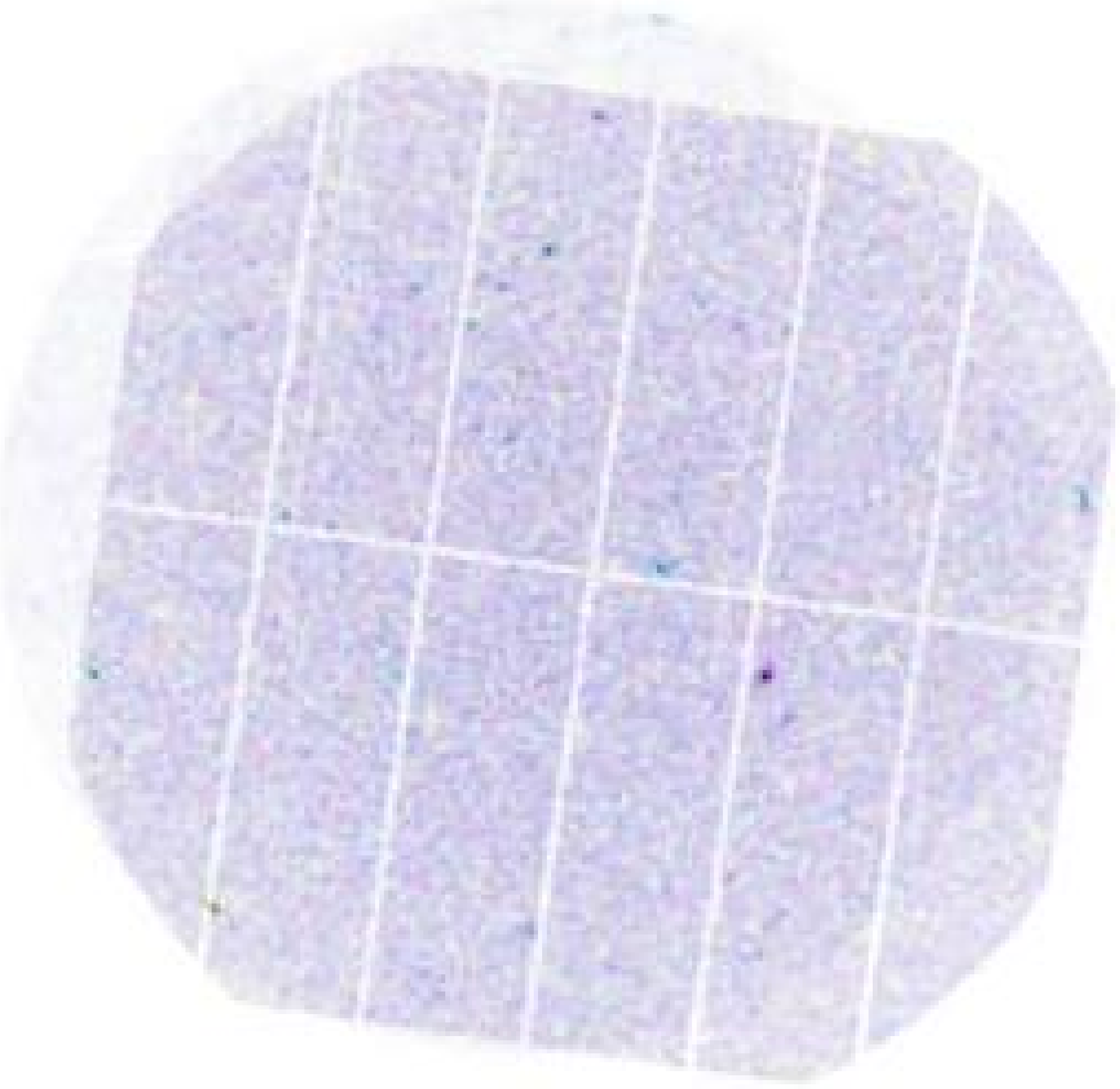}}}
{\resizebox{0.45\hsize}{!}{\includegraphics{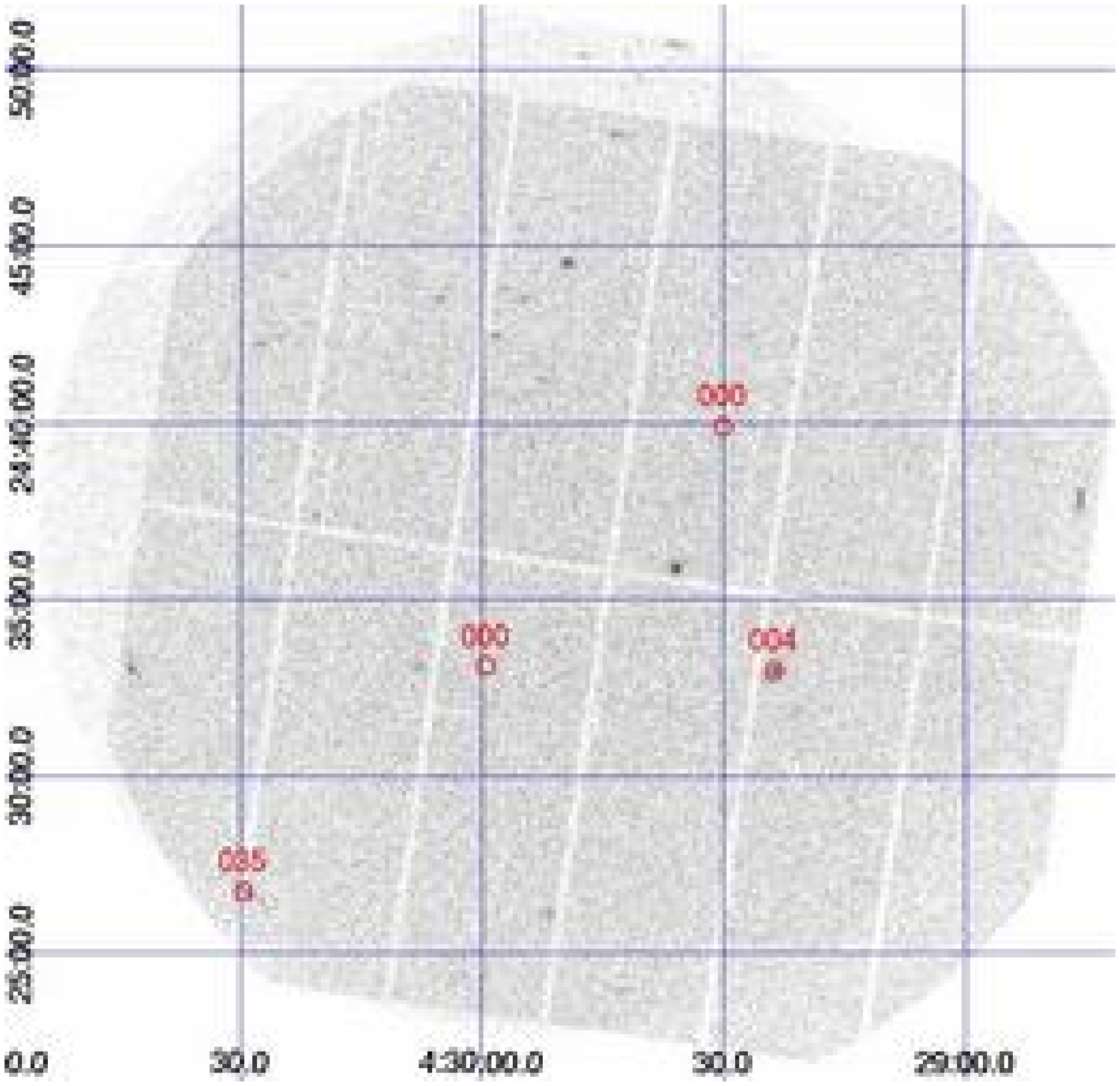}}}
}
\hbox{
{\resizebox{0.45\hsize}{!}{\includegraphics{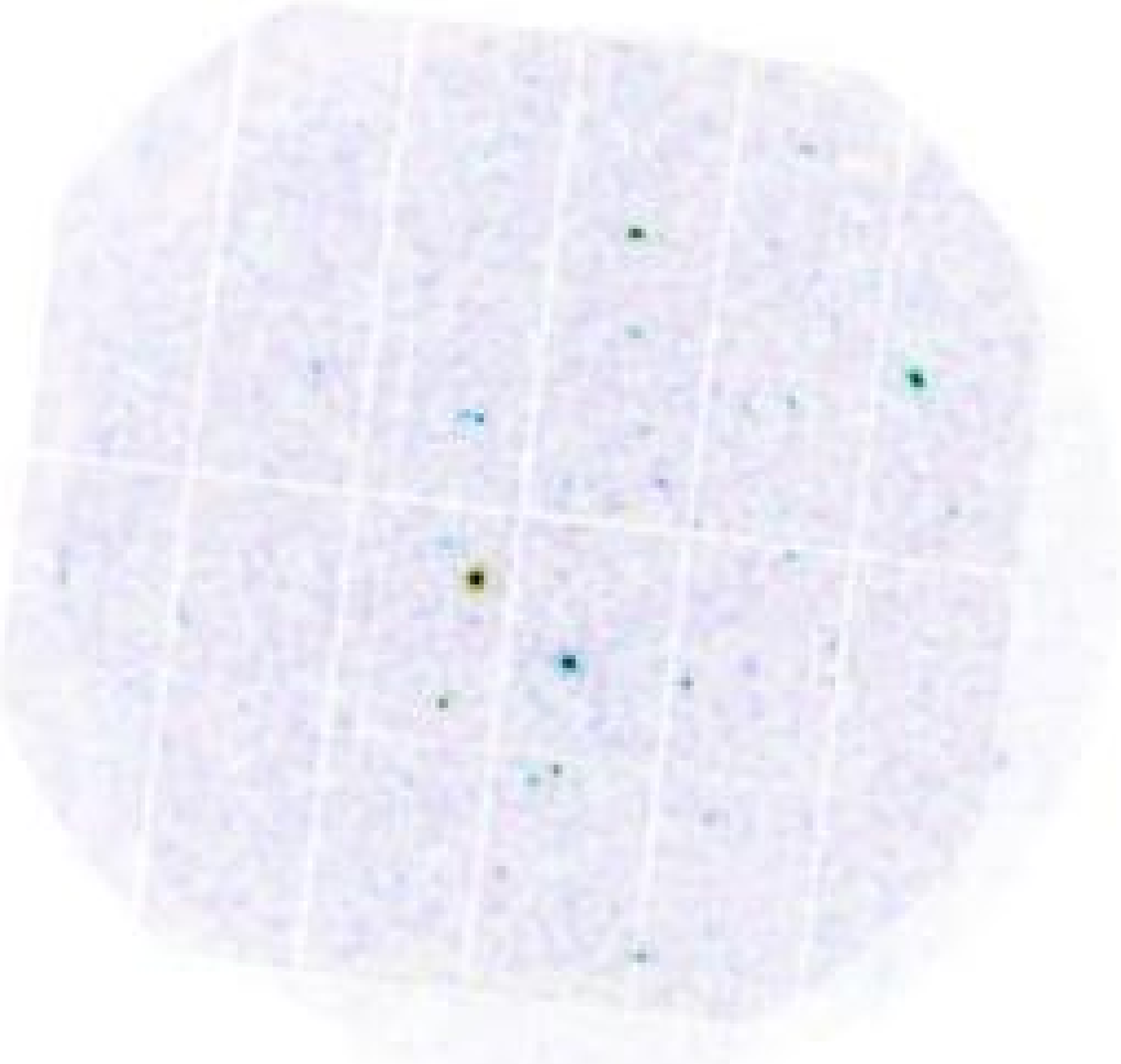}}}
{\resizebox{0.45\hsize}{!}{\includegraphics{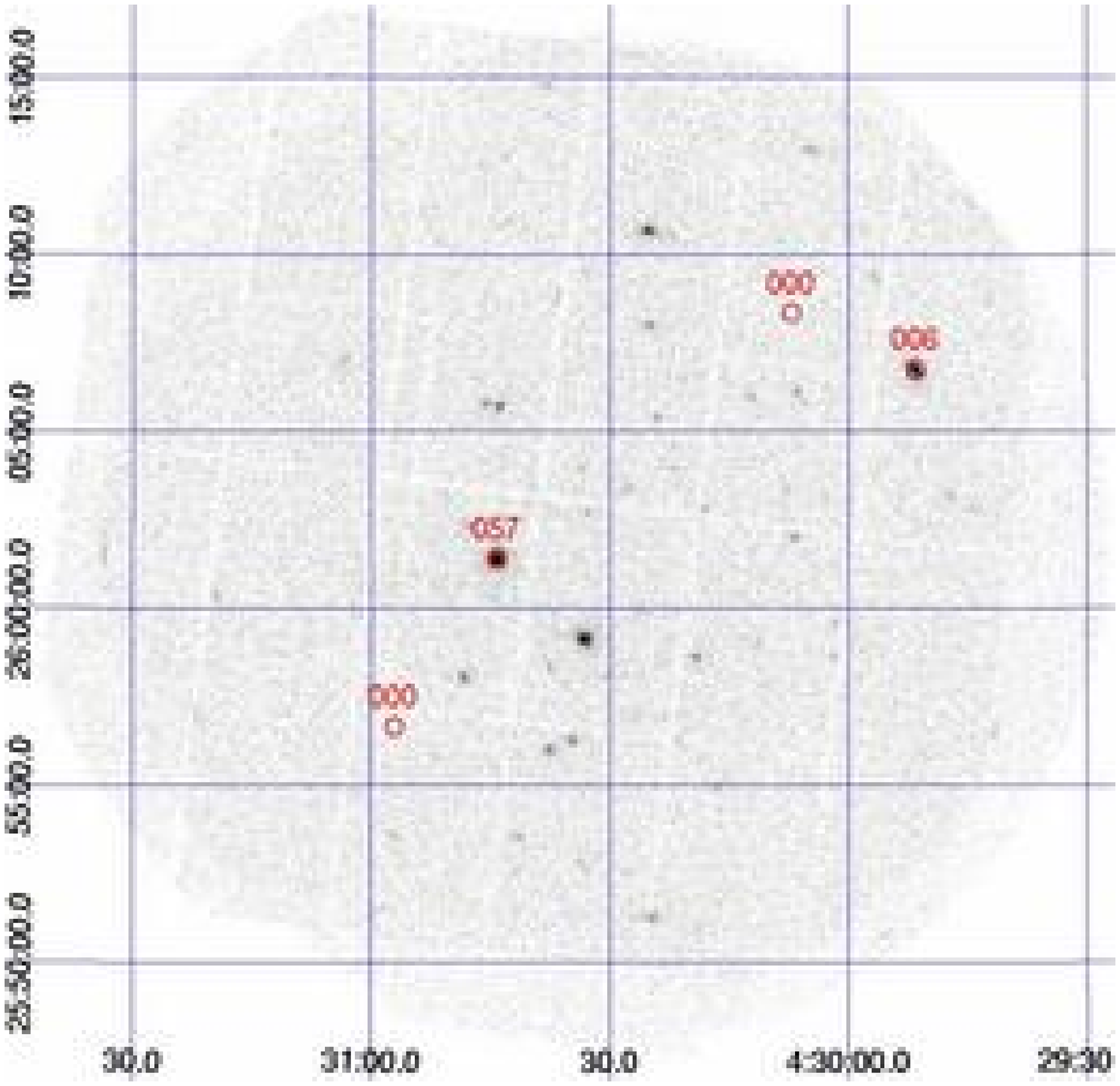}}}
}
\hbox{
{\resizebox{0.45\hsize}{!}{\includegraphics{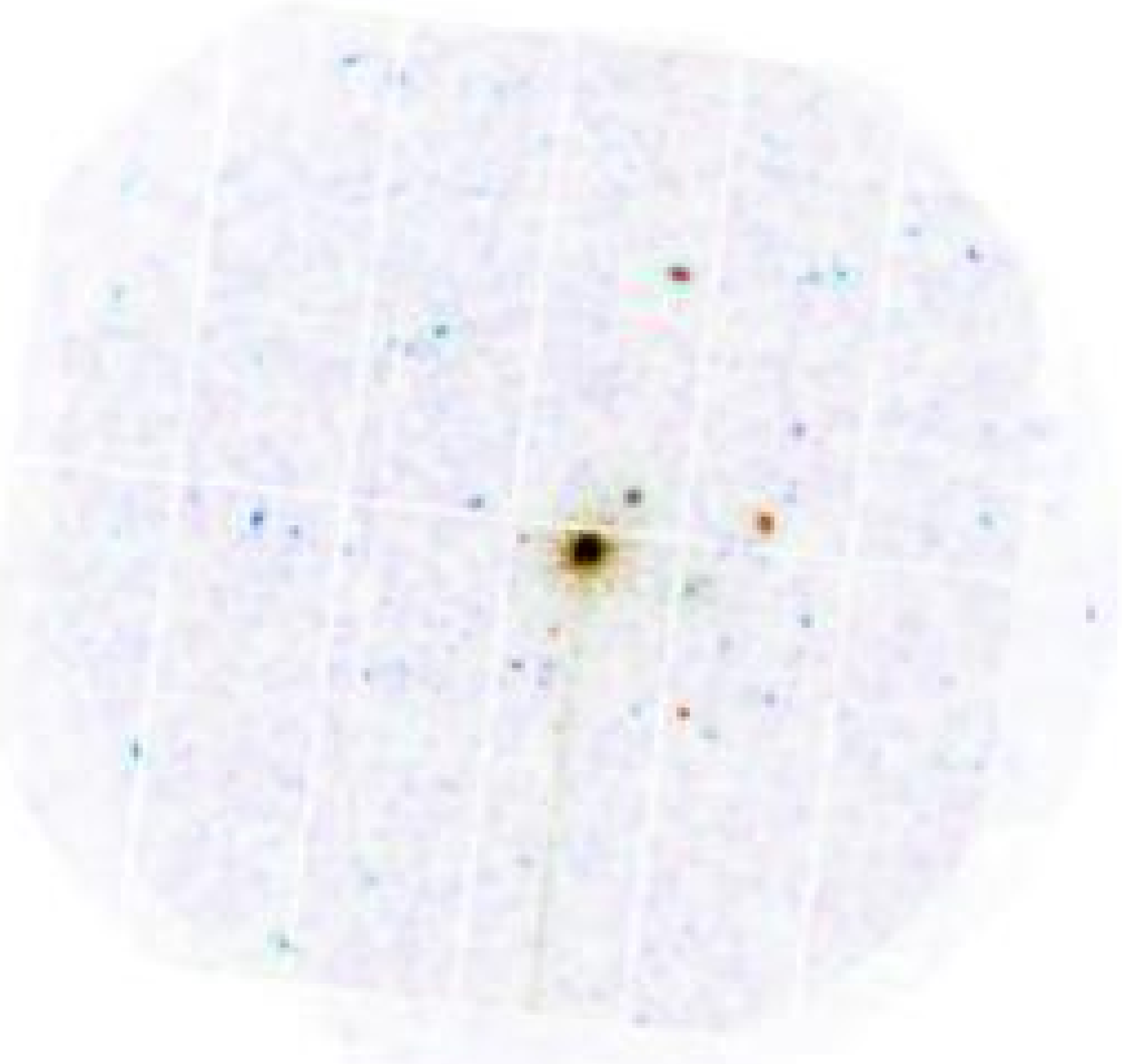}}}
{\resizebox{0.45\hsize}{!}{\includegraphics{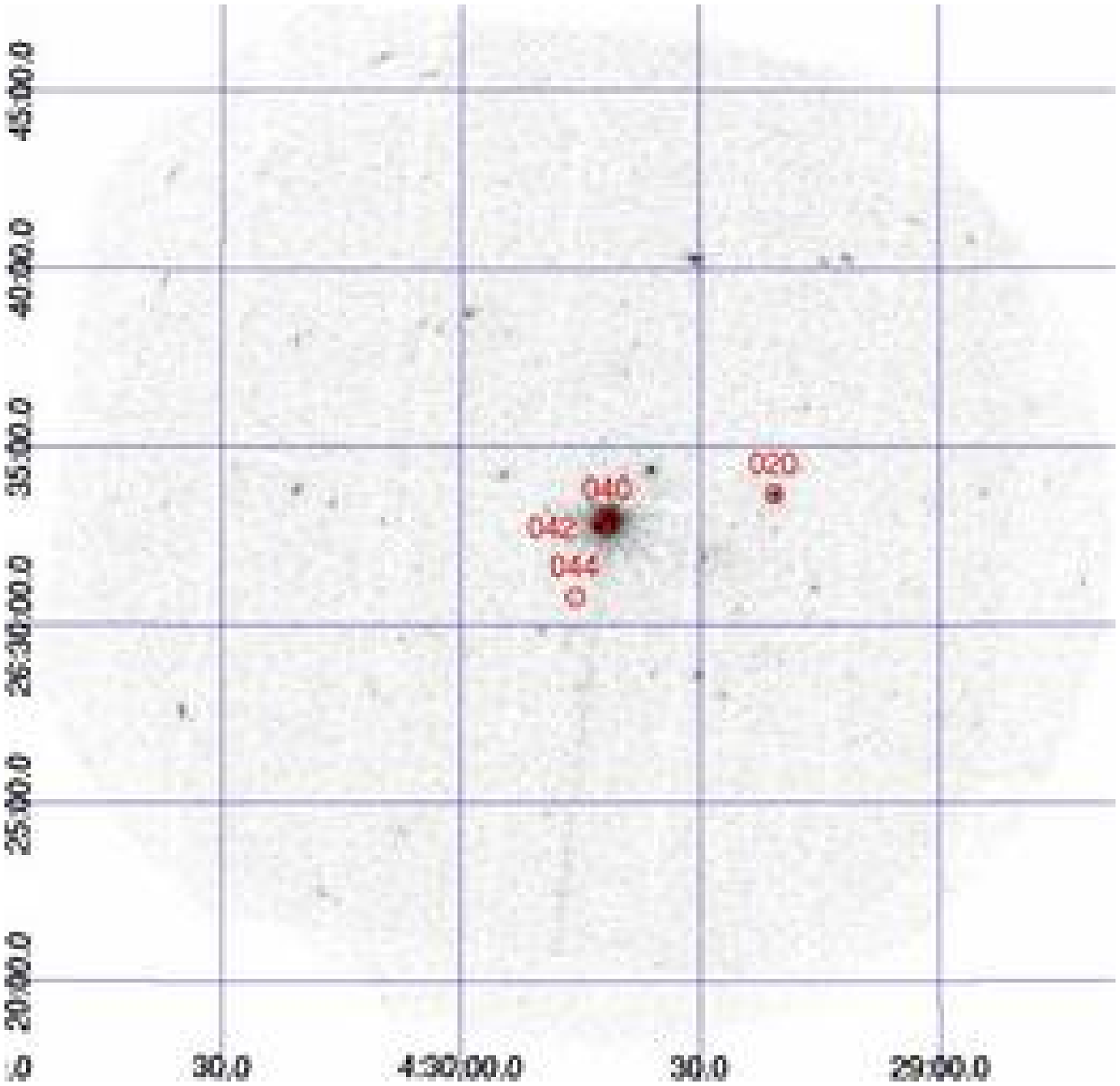}}}
}
\caption{Co-added EPIC images of field XEST-13, XEST-14, and XEST-15 (from top to bottom). Left: Smoothed images, color coded for hardness; right: 
      coordinate grid and TMC identifications included.\label{atlas5}} 
\end{figure*}

\begin{figure*}
\hbox{
{\resizebox{0.45\hsize}{!}{\includegraphics{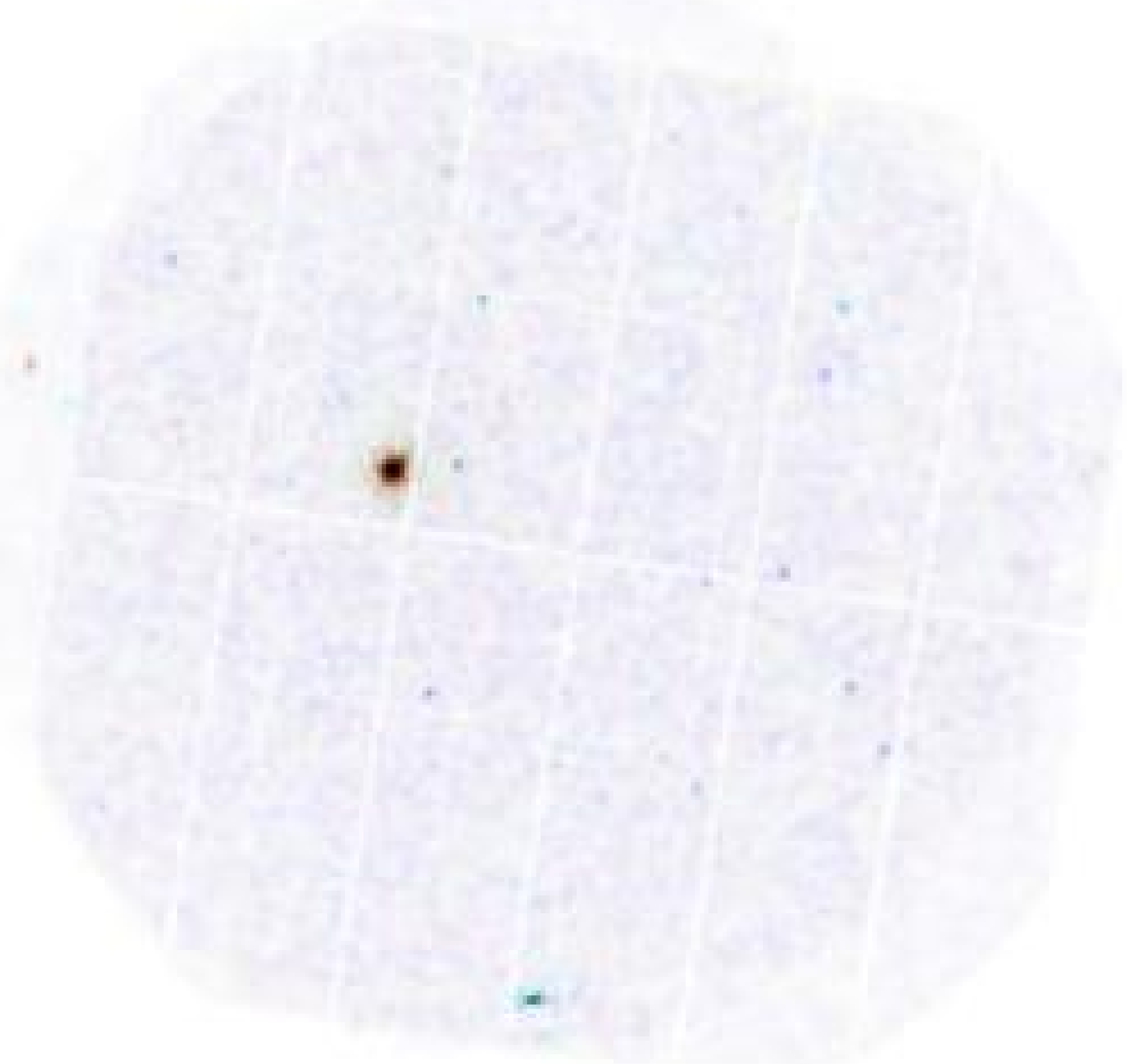}}}
{\resizebox{0.45\hsize}{!}{\includegraphics{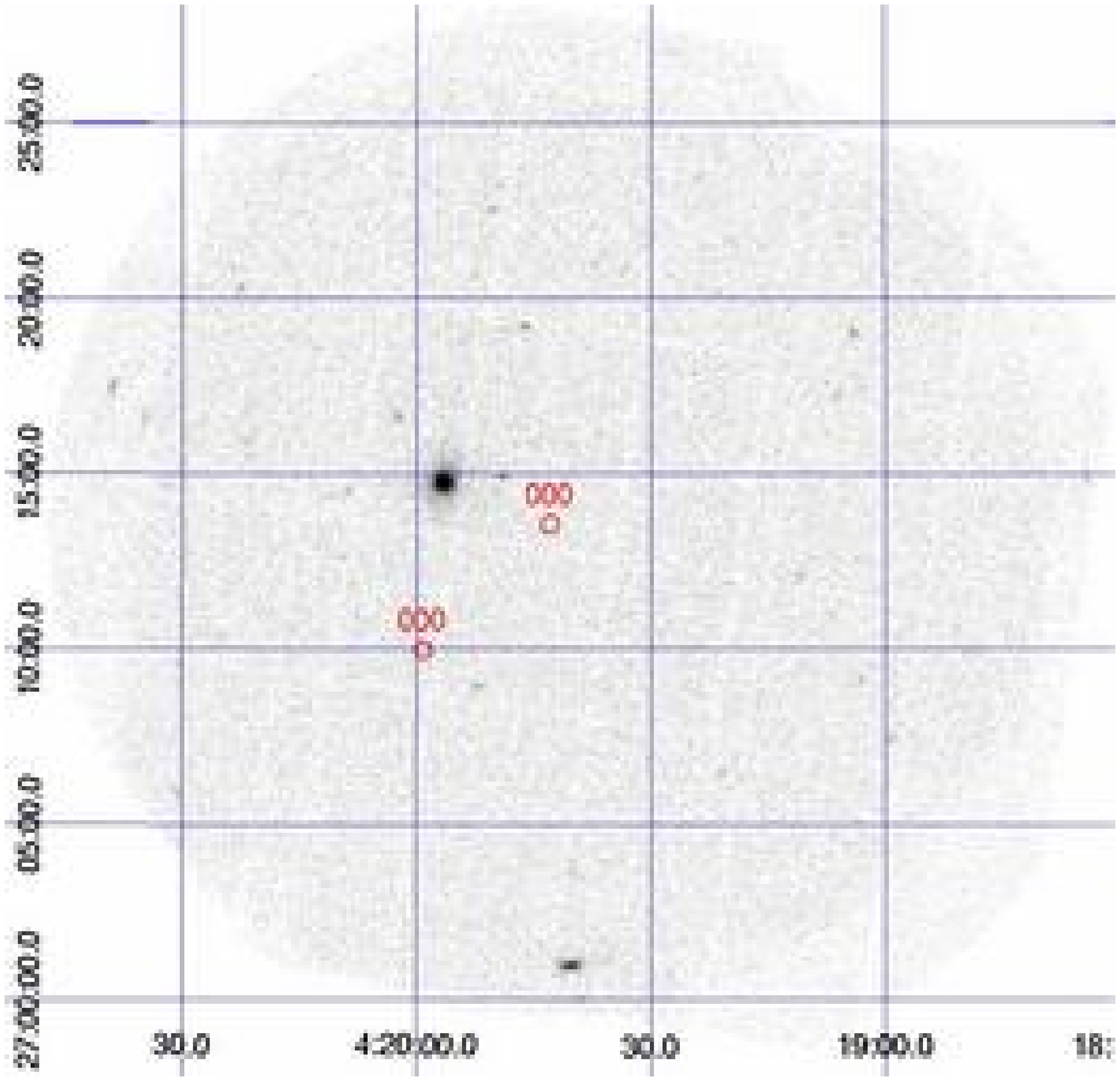}}}
}
\hbox{
{\resizebox{0.45\hsize}{!}{\includegraphics{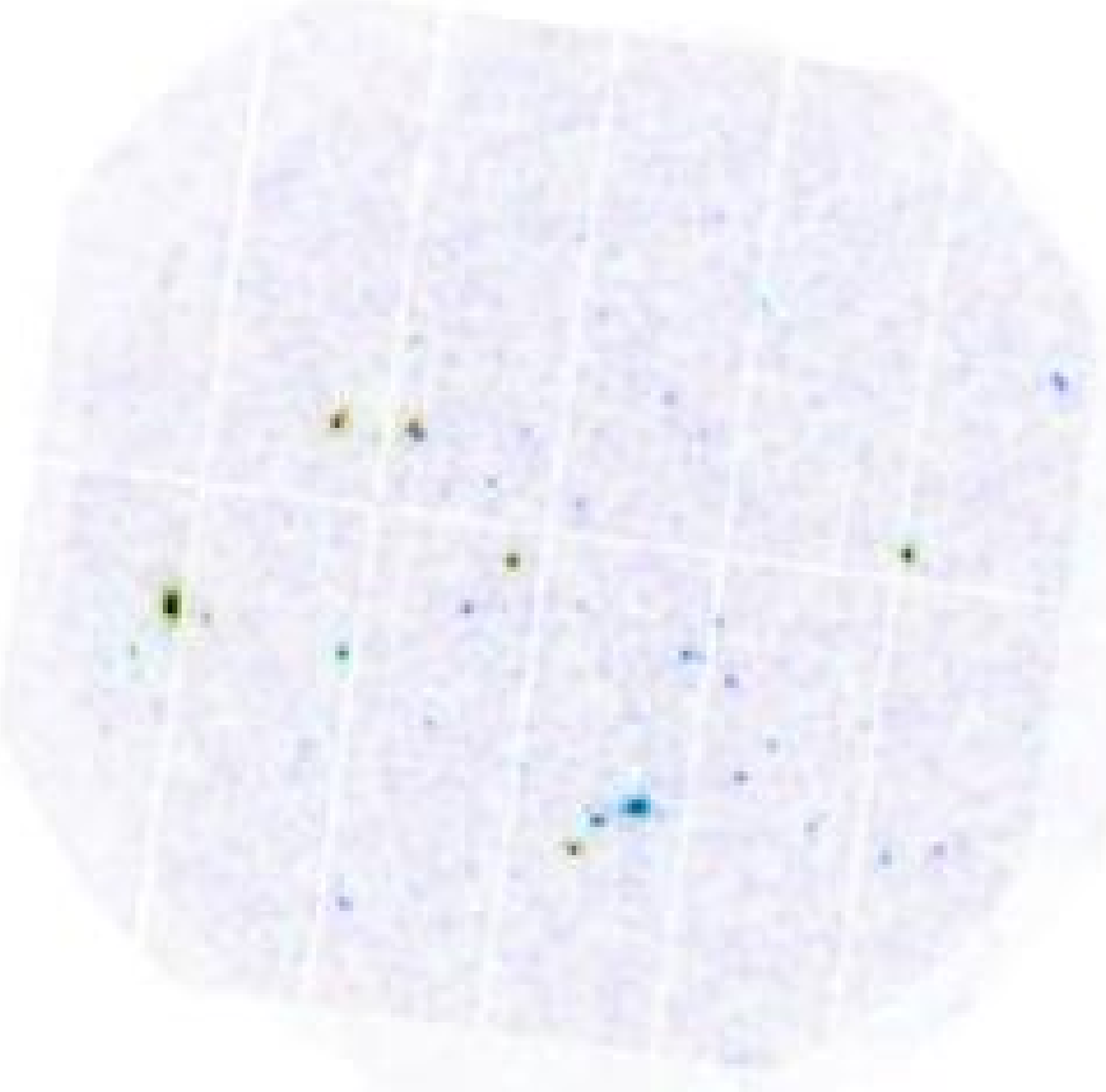}}}
{\resizebox{0.45\hsize}{!}{\includegraphics{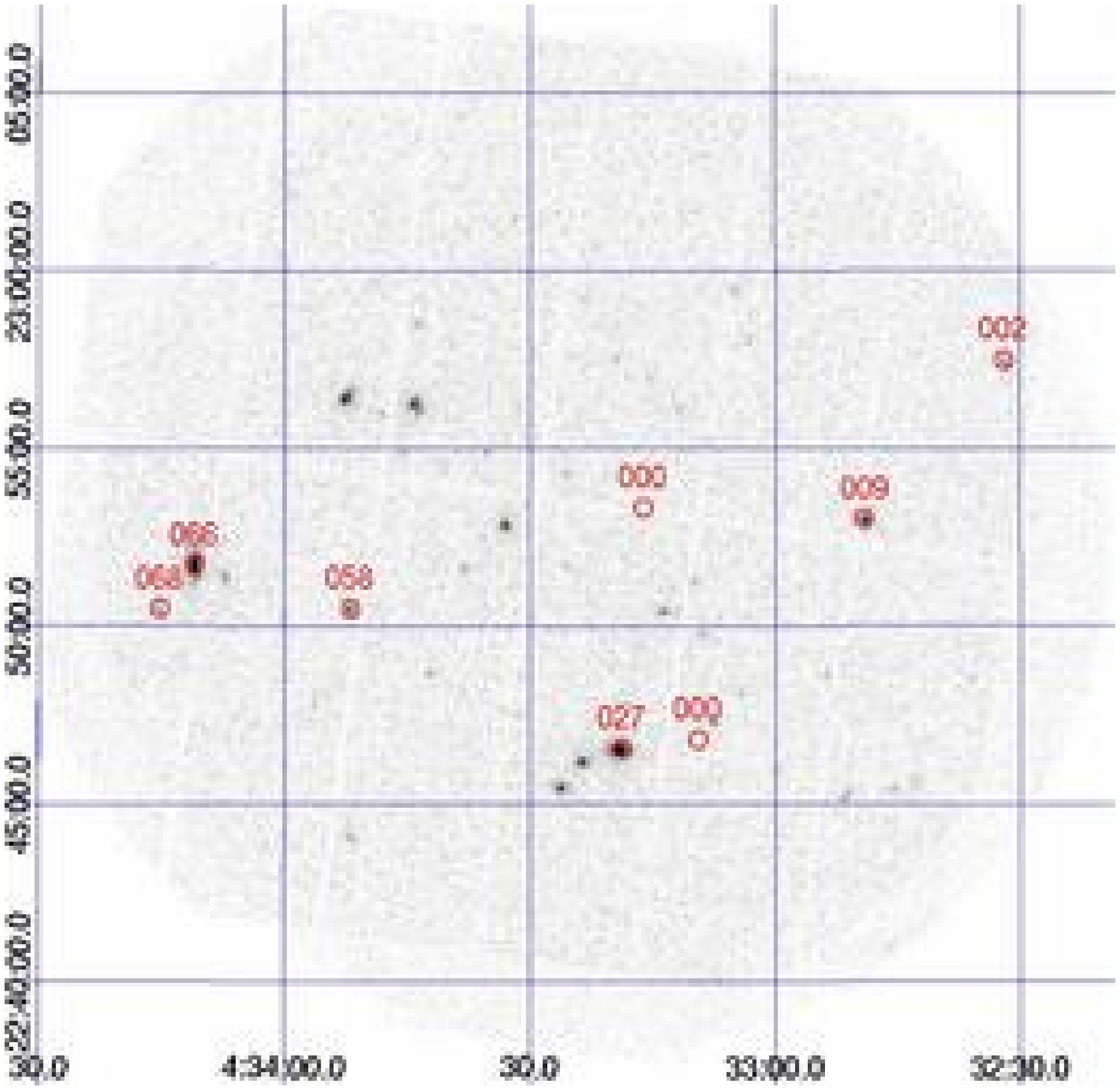}}}
}
\hbox{
{\resizebox{0.45\hsize}{!}{\includegraphics{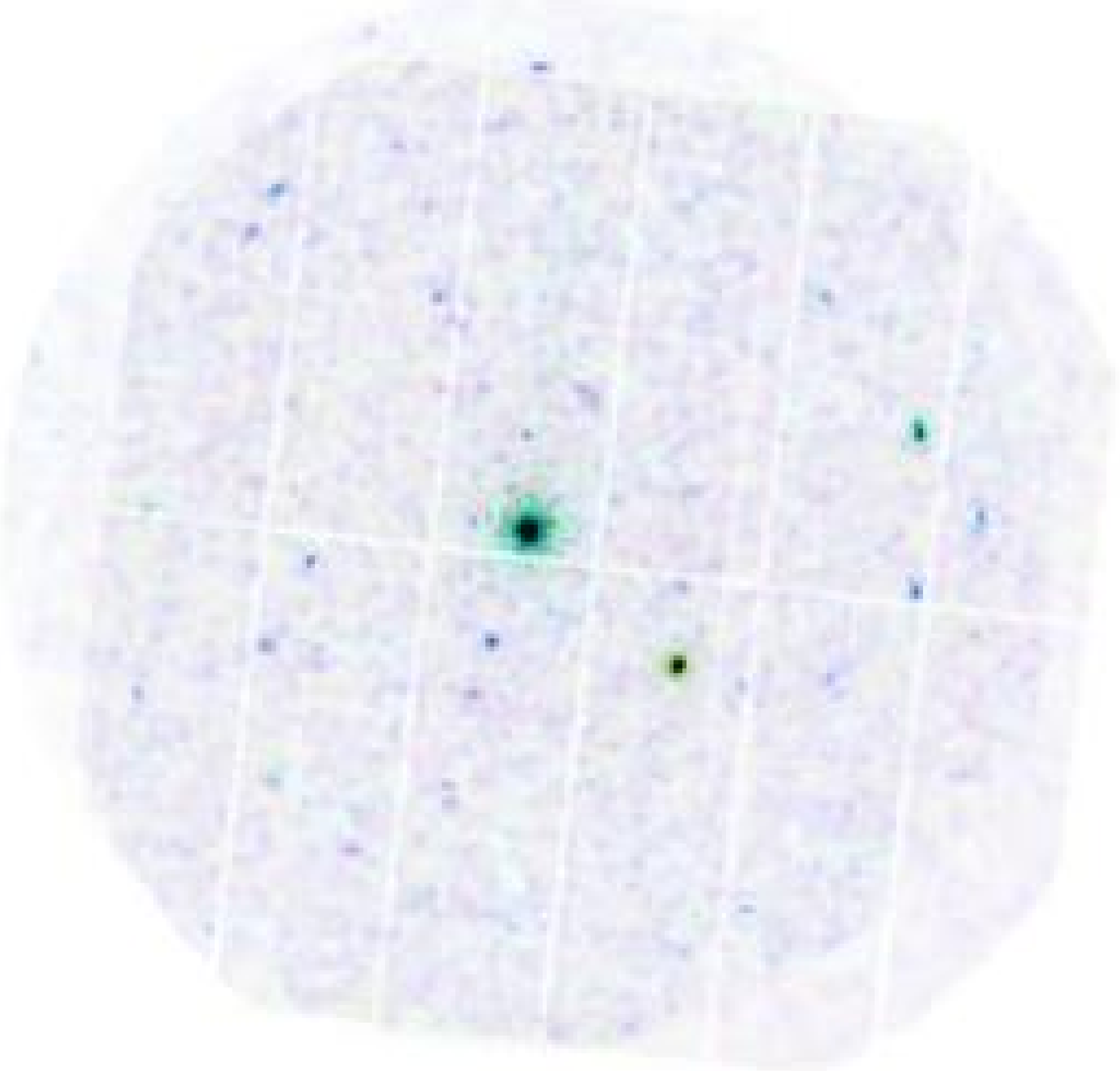}}}
{\resizebox{0.45\hsize}{!}{\includegraphics{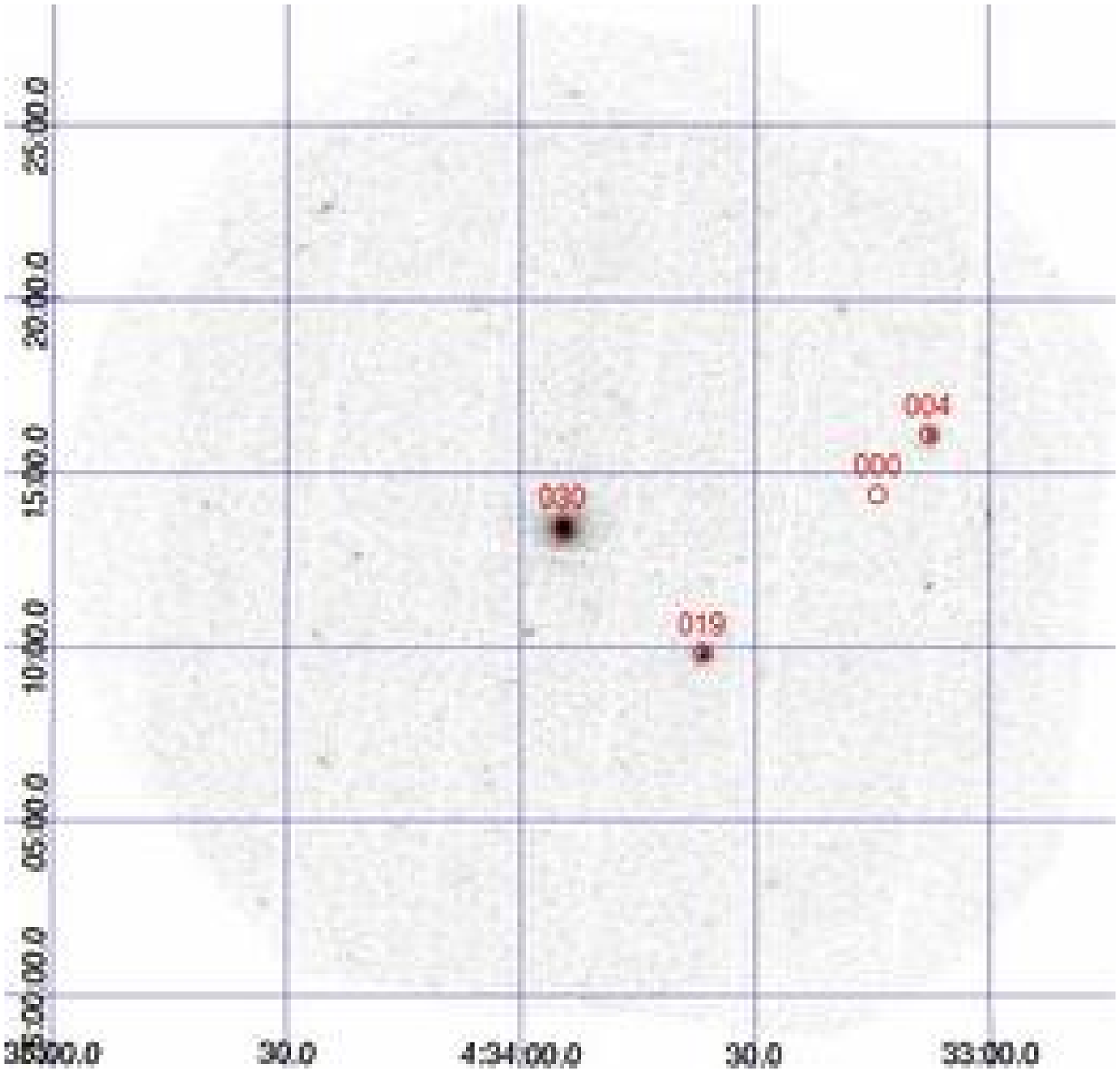}}}
}
\caption{Co-added EPIC images of field XEST-16, XEST-17, and XEST-18 (from top to bottom). Left: Smoothed images, color coded for hardness; right: 
      coordinate grid and TMC identifications included.\label{atlas6}} 
\end{figure*}

\begin{figure*}
\hbox{
{\resizebox{0.45\hsize}{!}{\includegraphics{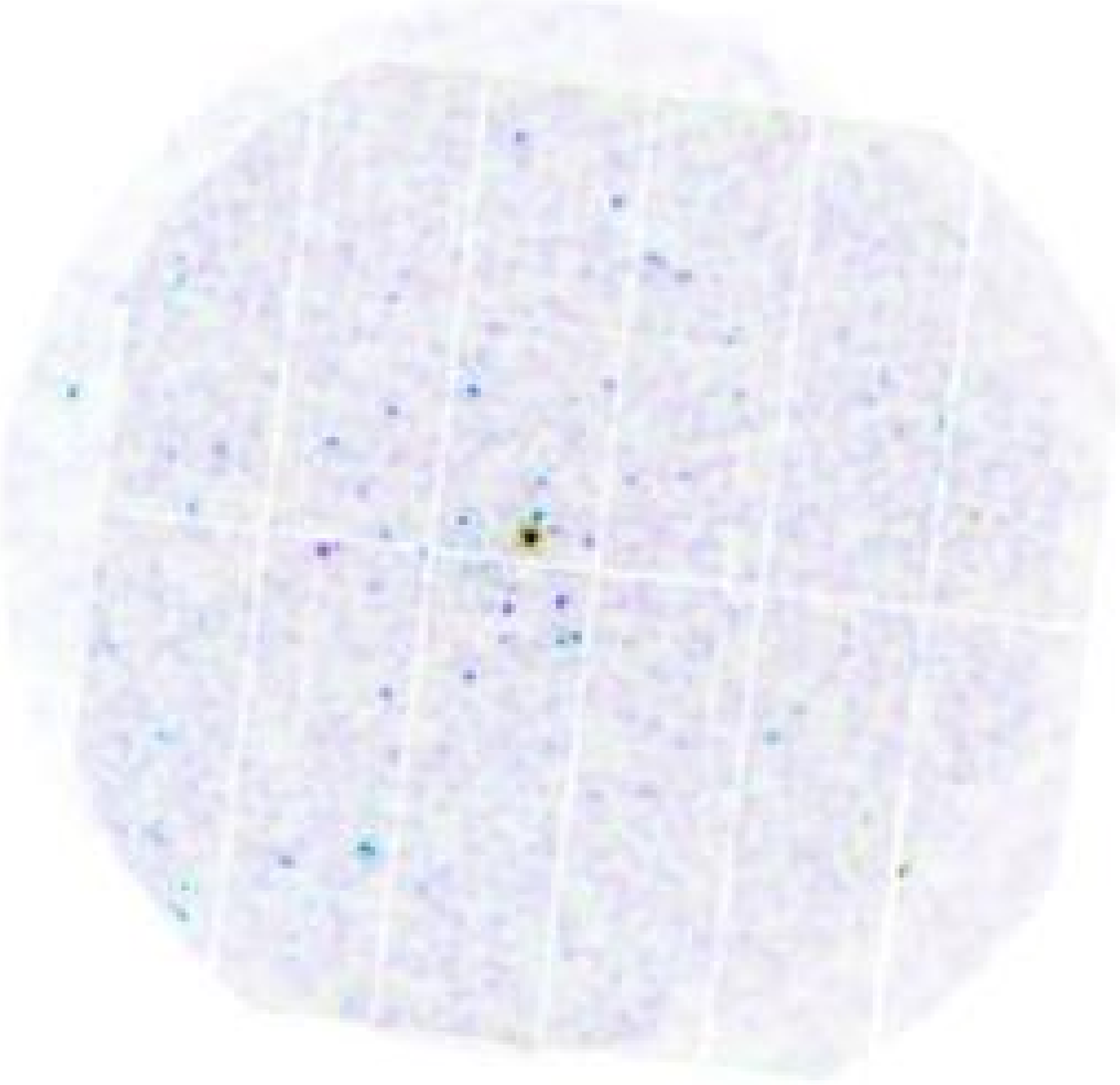}}}
{\resizebox{0.45\hsize}{!}{\includegraphics{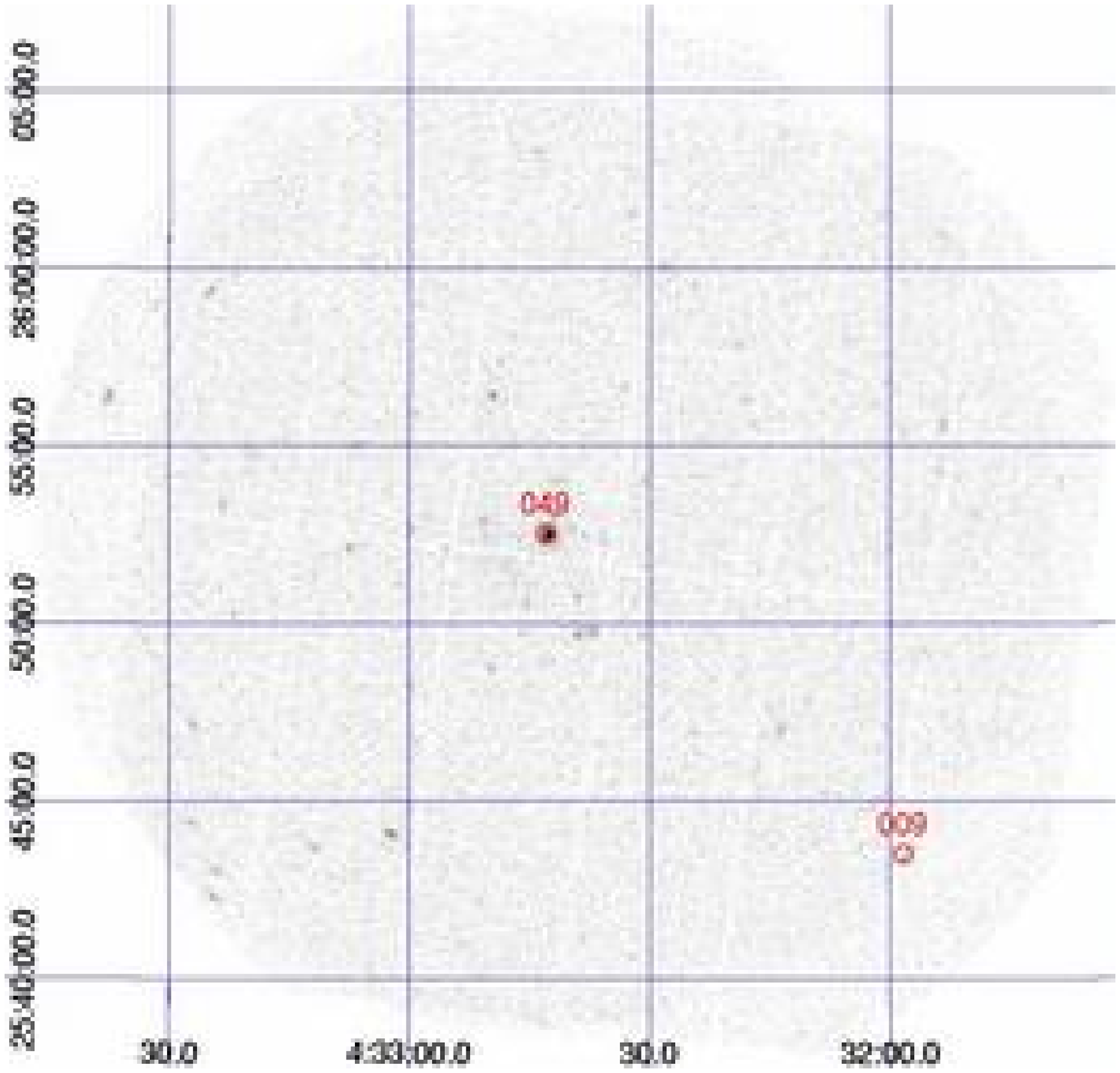}}}
}
\hbox{
{\resizebox{0.45\hsize}{!}{\includegraphics{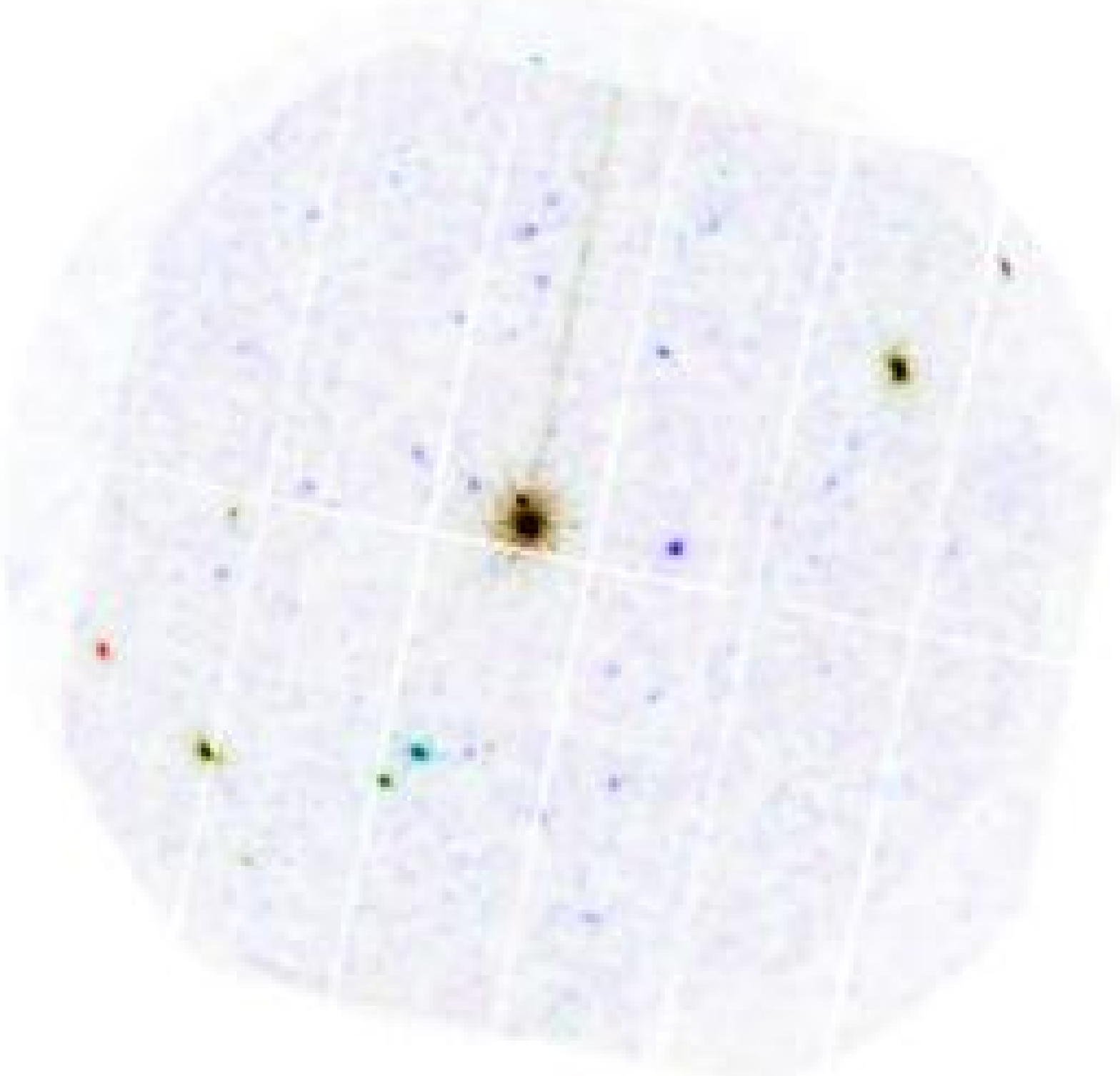}}}
{\resizebox{0.45\hsize}{!}{\includegraphics{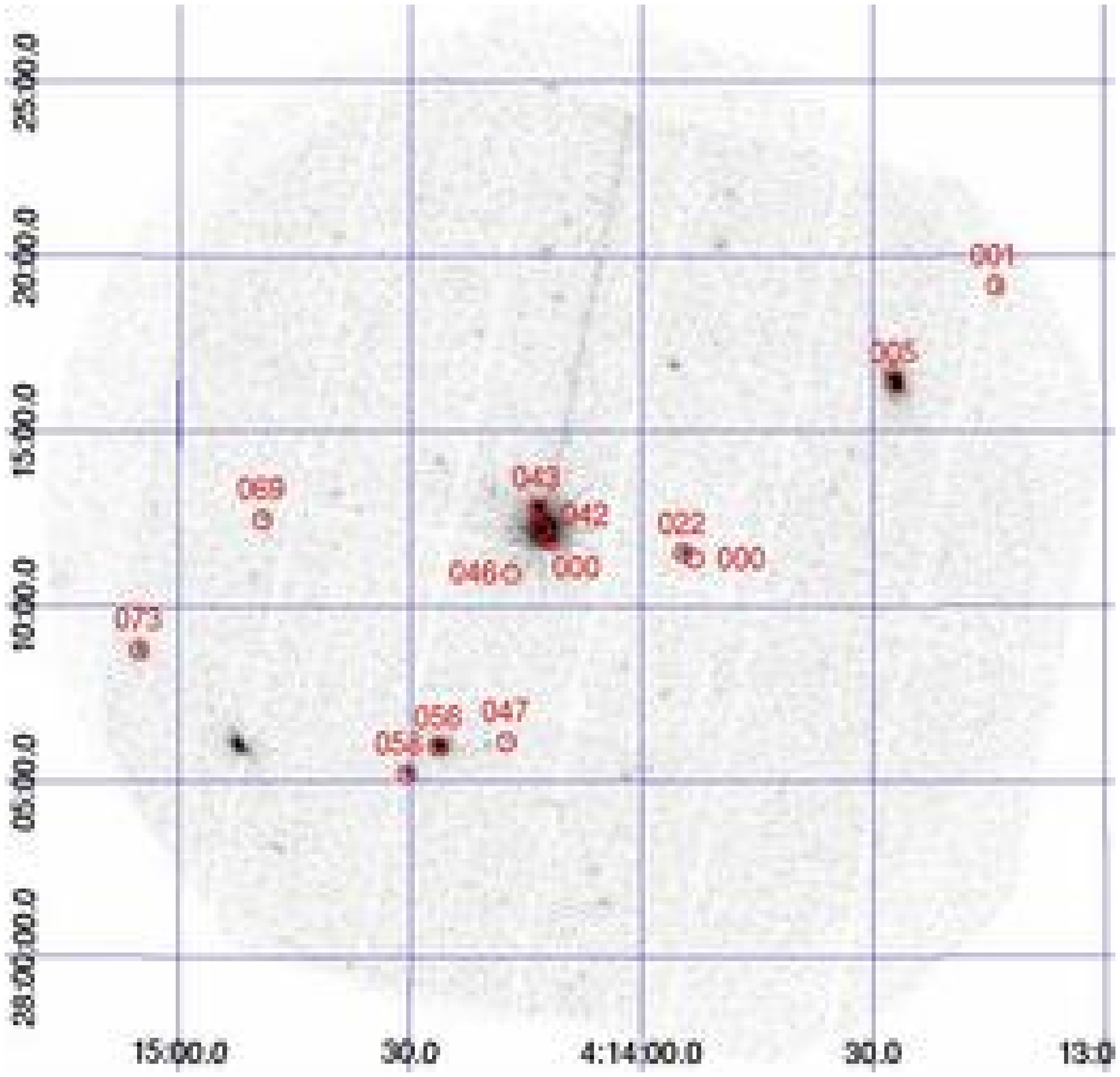}}}
}
\hbox{
{\resizebox{0.45\hsize}{!}{\includegraphics{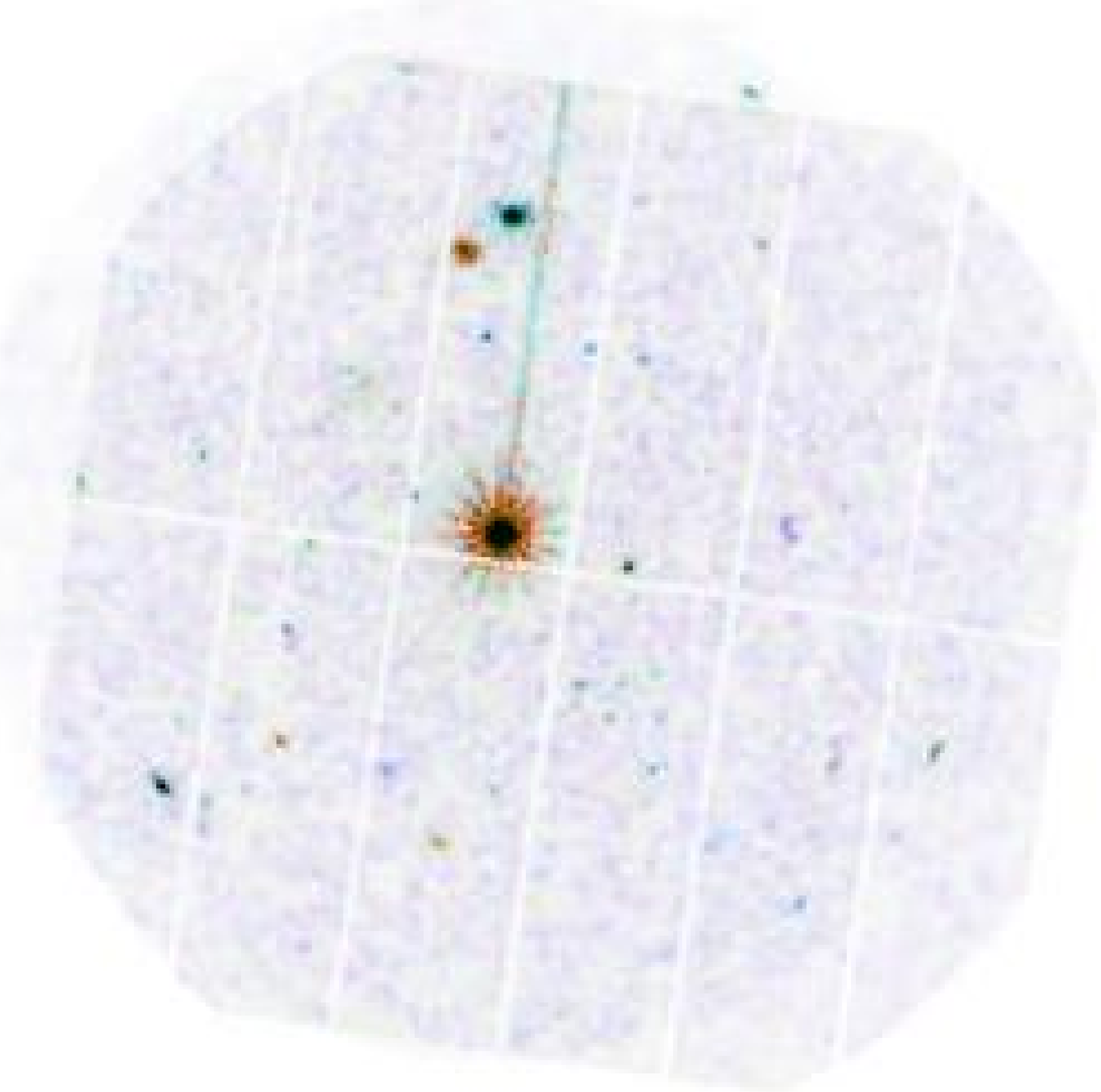}}}
{\resizebox{0.45\hsize}{!}{\includegraphics{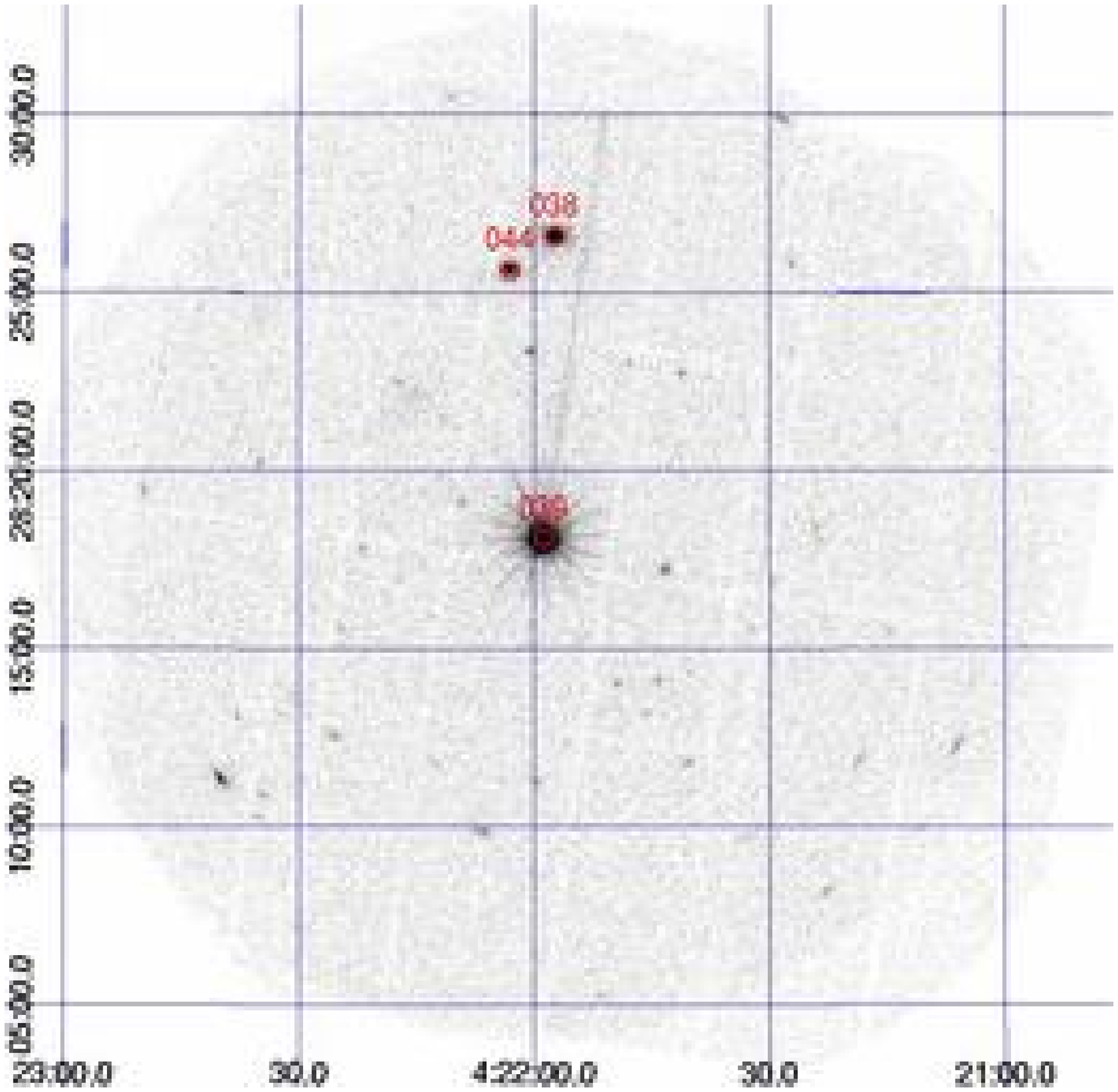}}}
}
\caption{Co-added EPIC images of field XEST-19, XEST-20, and XEST-21 (from top to bottom). Left: Smoothed images, color coded for hardness; right: 
      coordinate grid and TMC identifications included.\label{atlas7}} 
\end{figure*}

\begin{figure*}
\hbox{
{\resizebox{0.45\hsize}{!}{\includegraphics{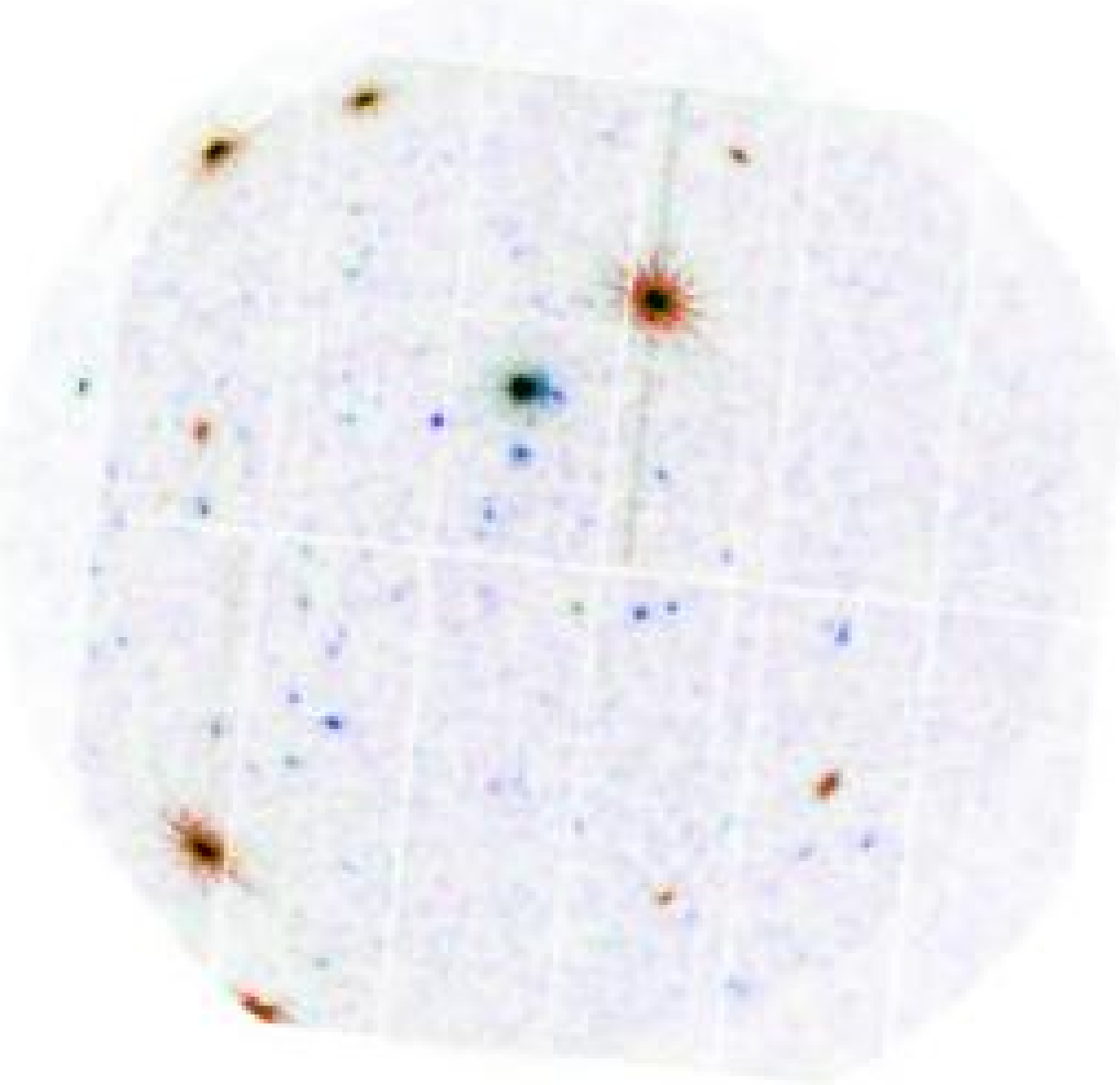}}}
{\resizebox{0.45\hsize}{!}{\includegraphics{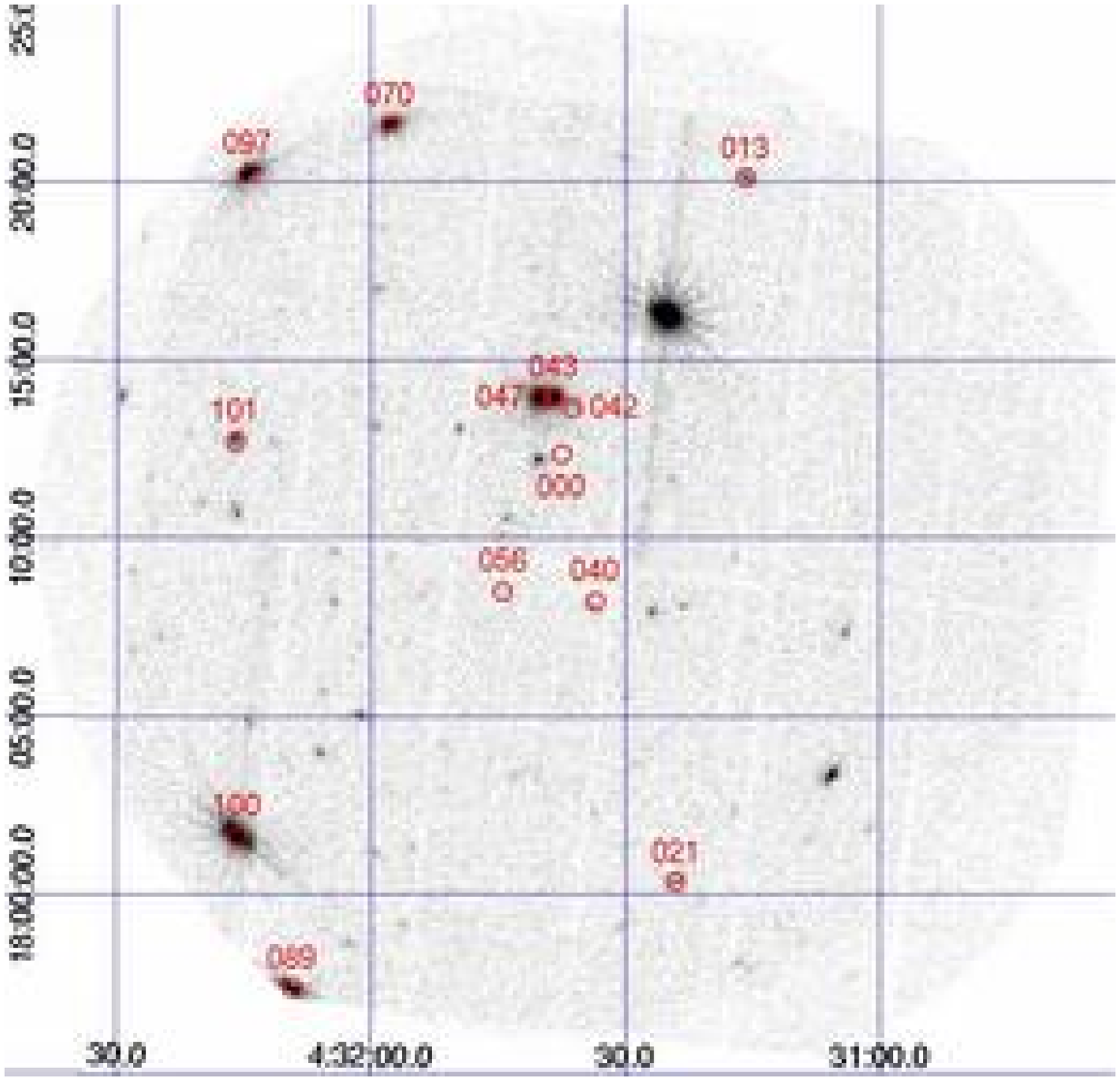}}}
}
\hbox{
{\resizebox{0.45\hsize}{!}{\includegraphics{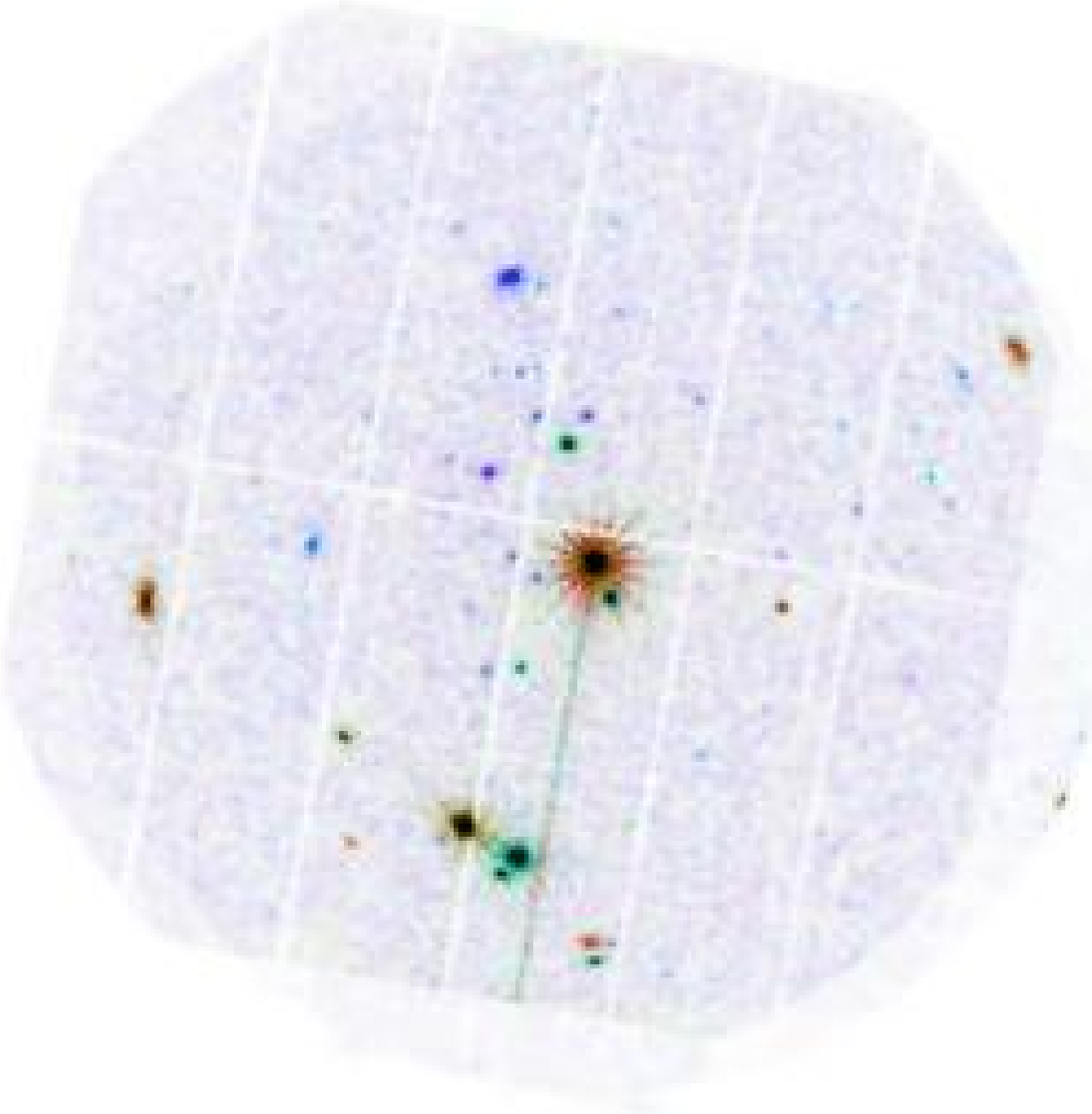}}}
{\resizebox{0.45\hsize}{!}{\includegraphics{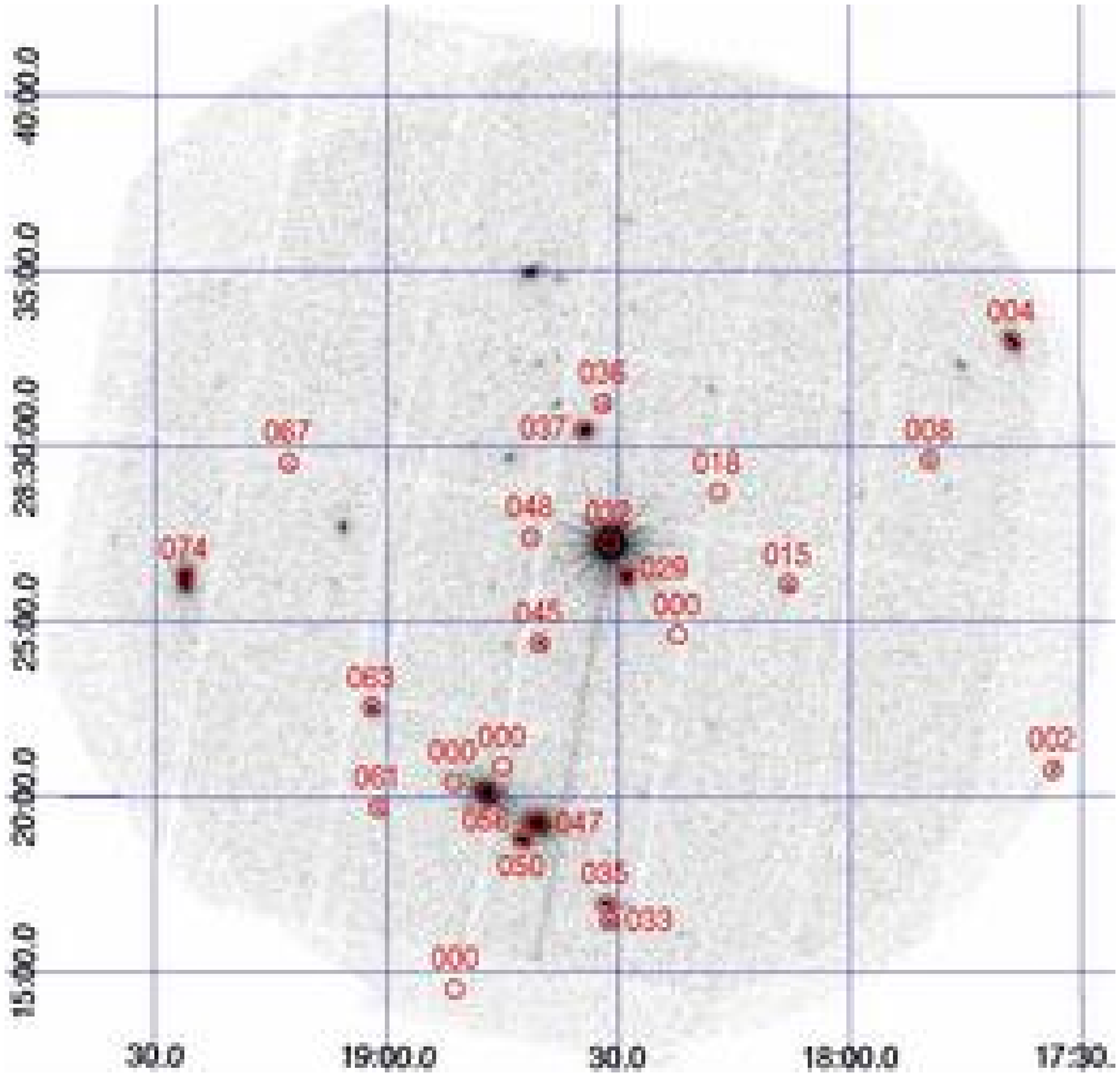}}}
}
\hbox{
{\resizebox{0.45\hsize}{!}{\includegraphics{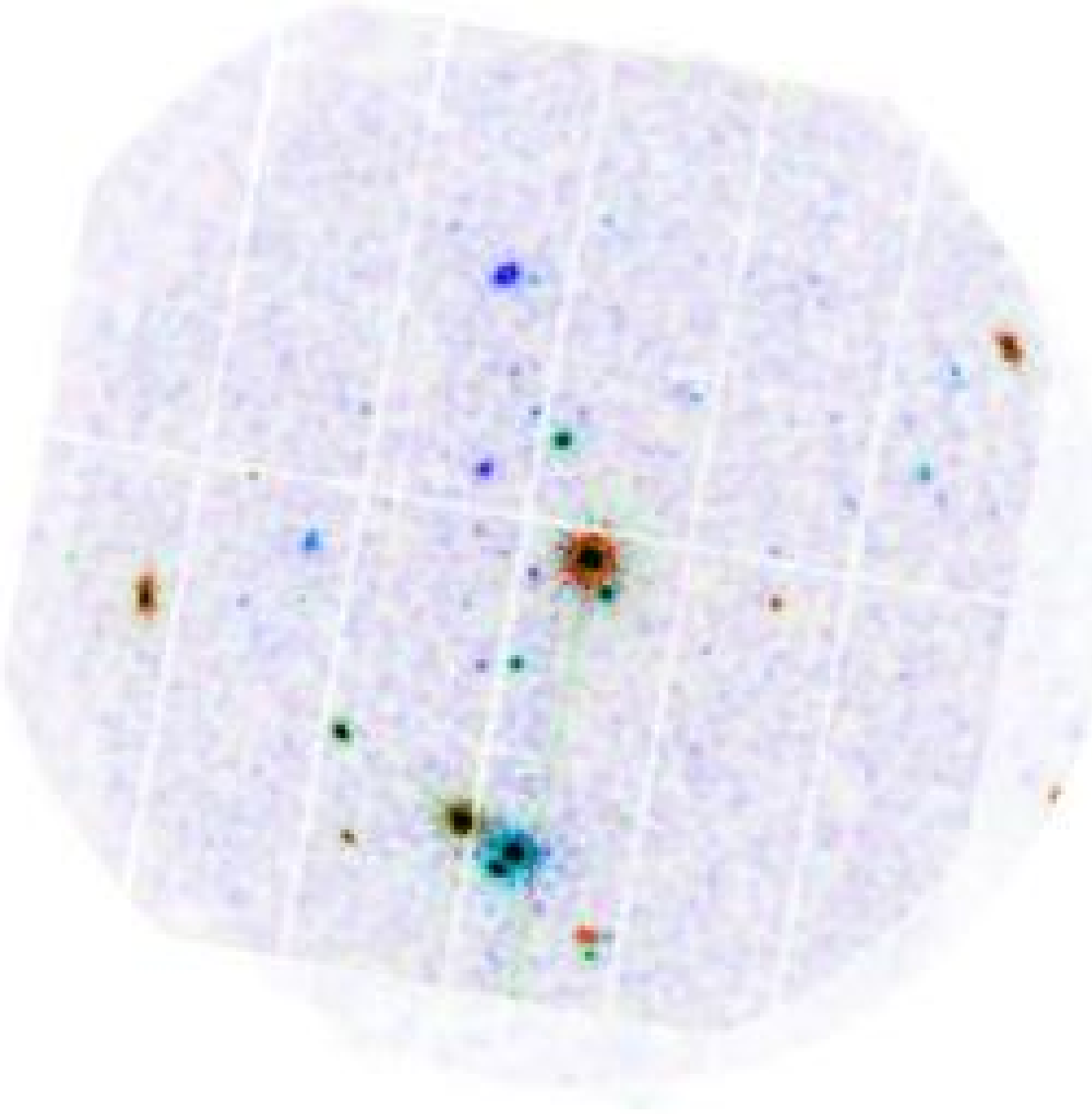}}}
{\resizebox{0.45\hsize}{!}{\includegraphics{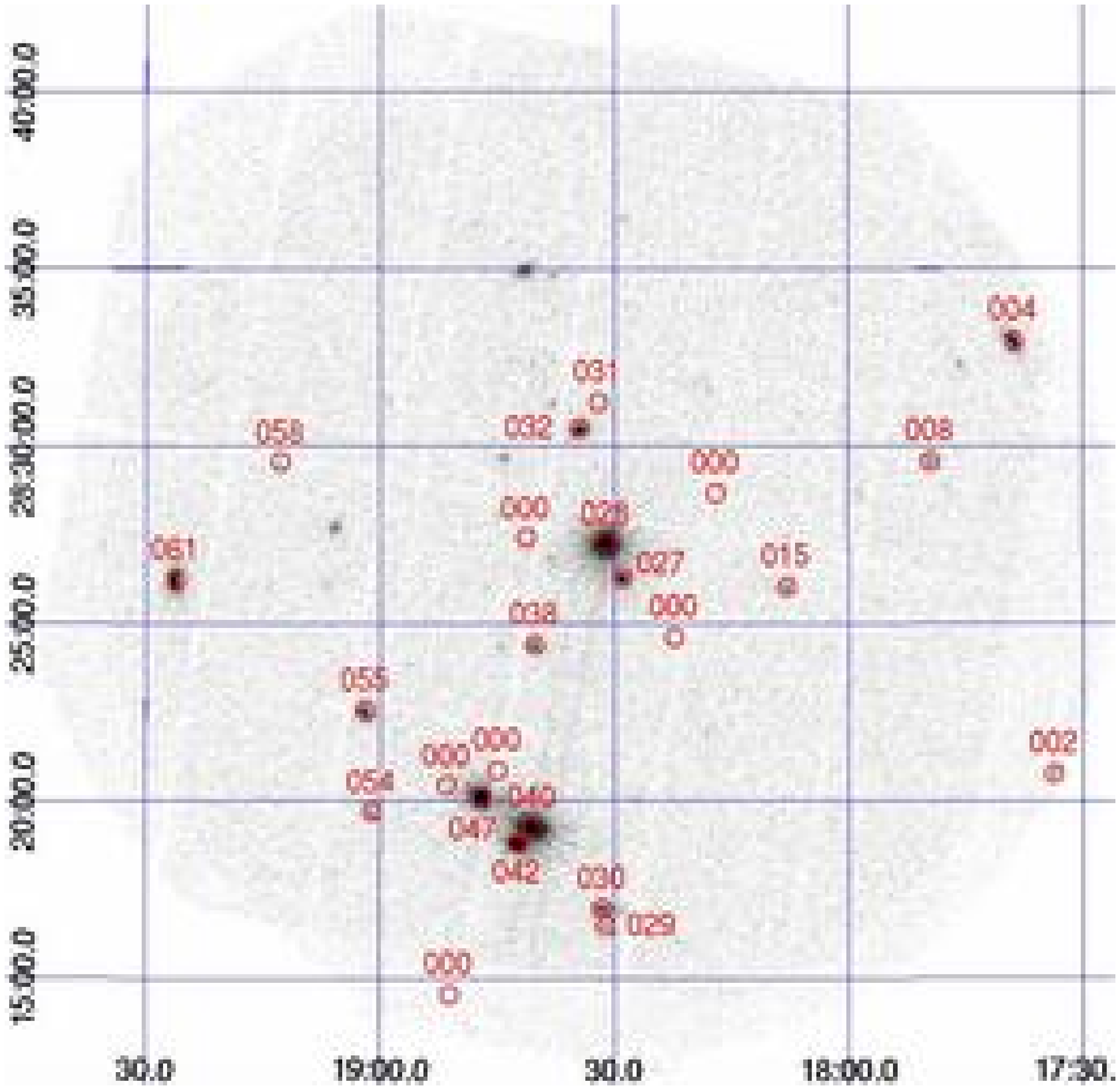}}}
}
\caption{Co-added EPIC images of field XEST-22, XEST-23, and XEST-24 (from top to bottom). Left: Smoothed images, color coded for hardness; right: 
      coordinate grid and TMC identifications included.\label{atlas8}} 
\end{figure*}

\begin{figure*}
\hbox{
{\resizebox{0.45\hsize}{!}{\includegraphics{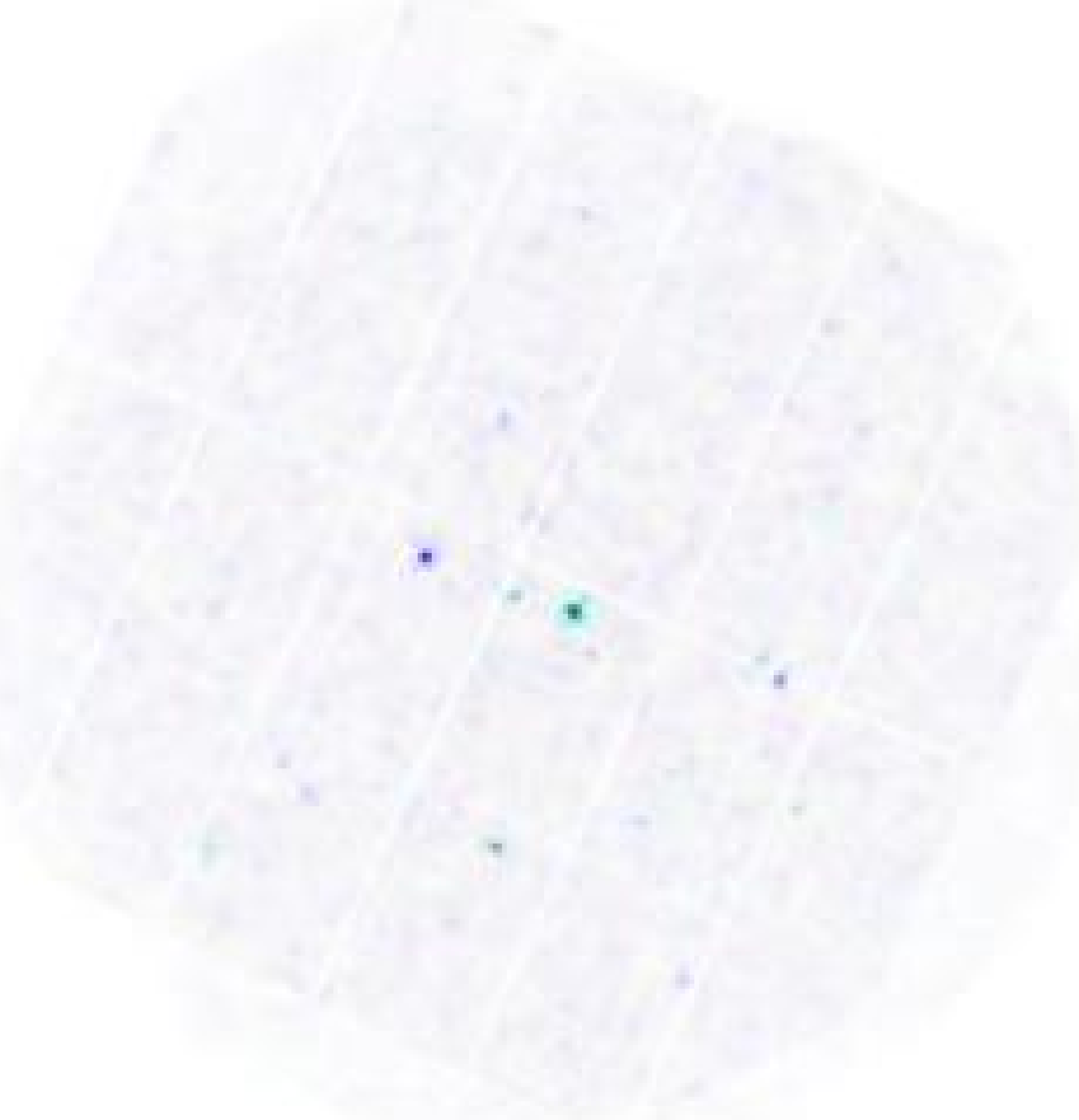}}}
{\resizebox{0.45\hsize}{!}{\includegraphics{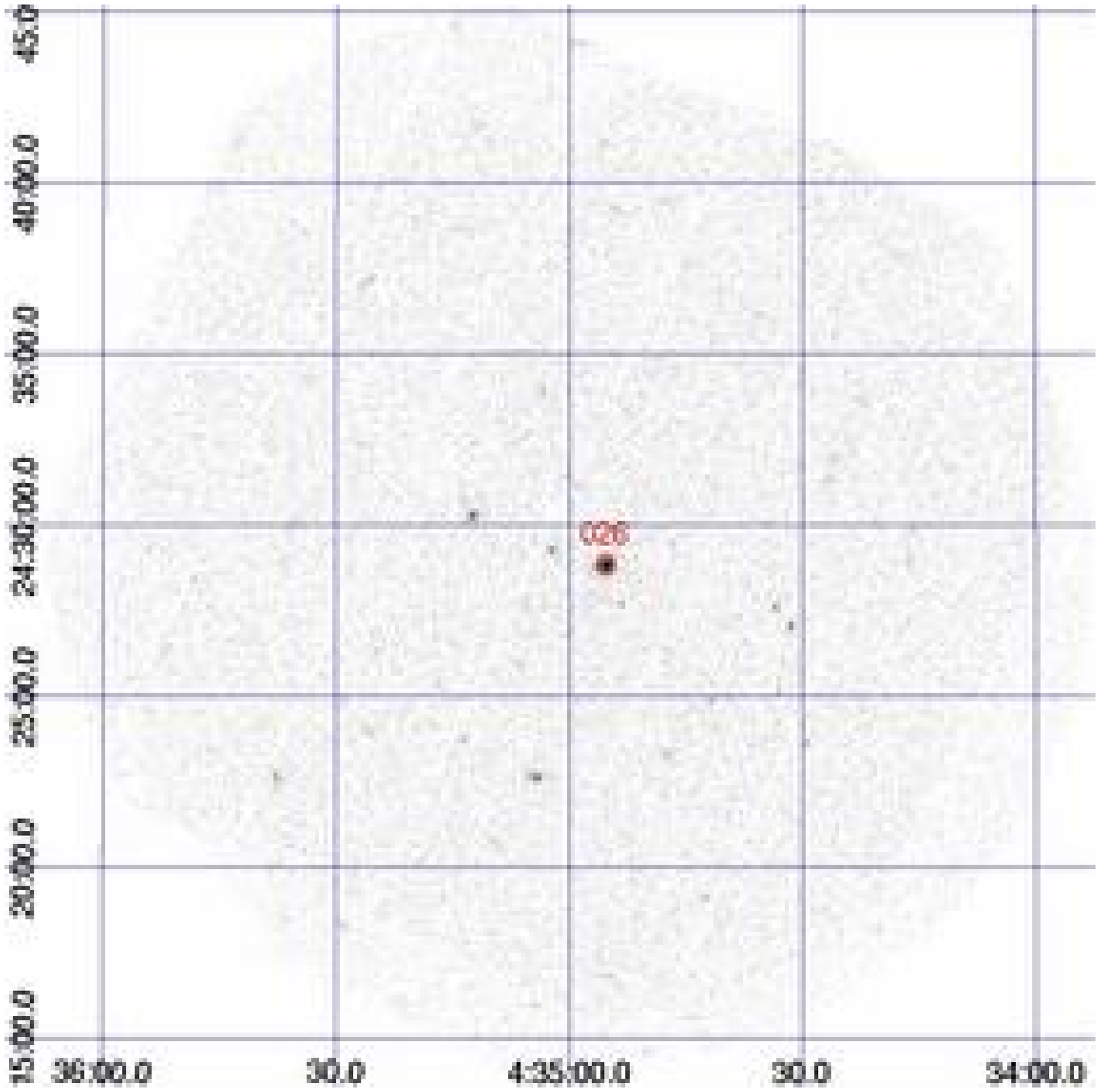}}}
}
\hbox{
{\resizebox{0.45\hsize}{!}{\includegraphics{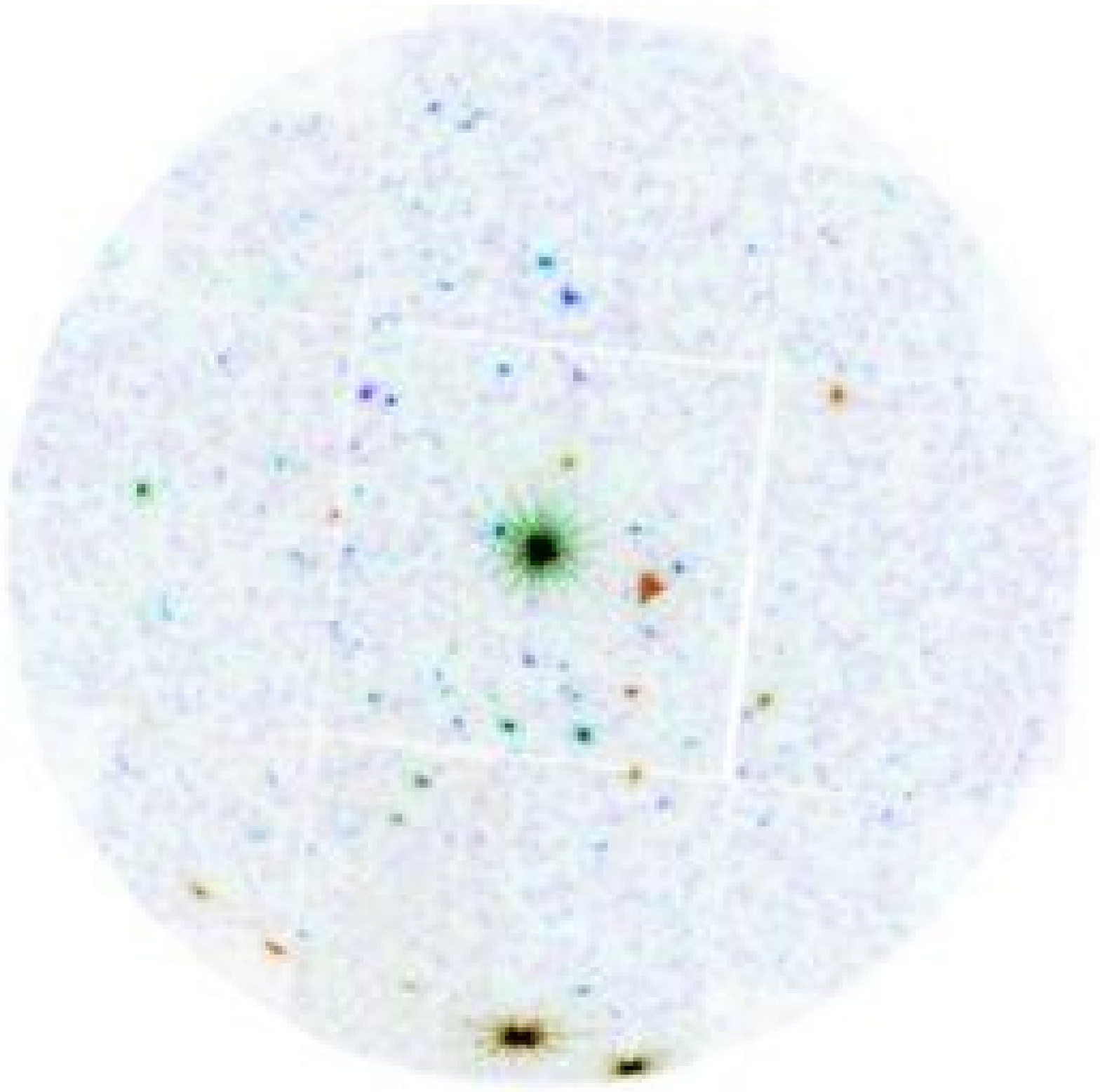}}}
{\resizebox{0.45\hsize}{!}{\includegraphics{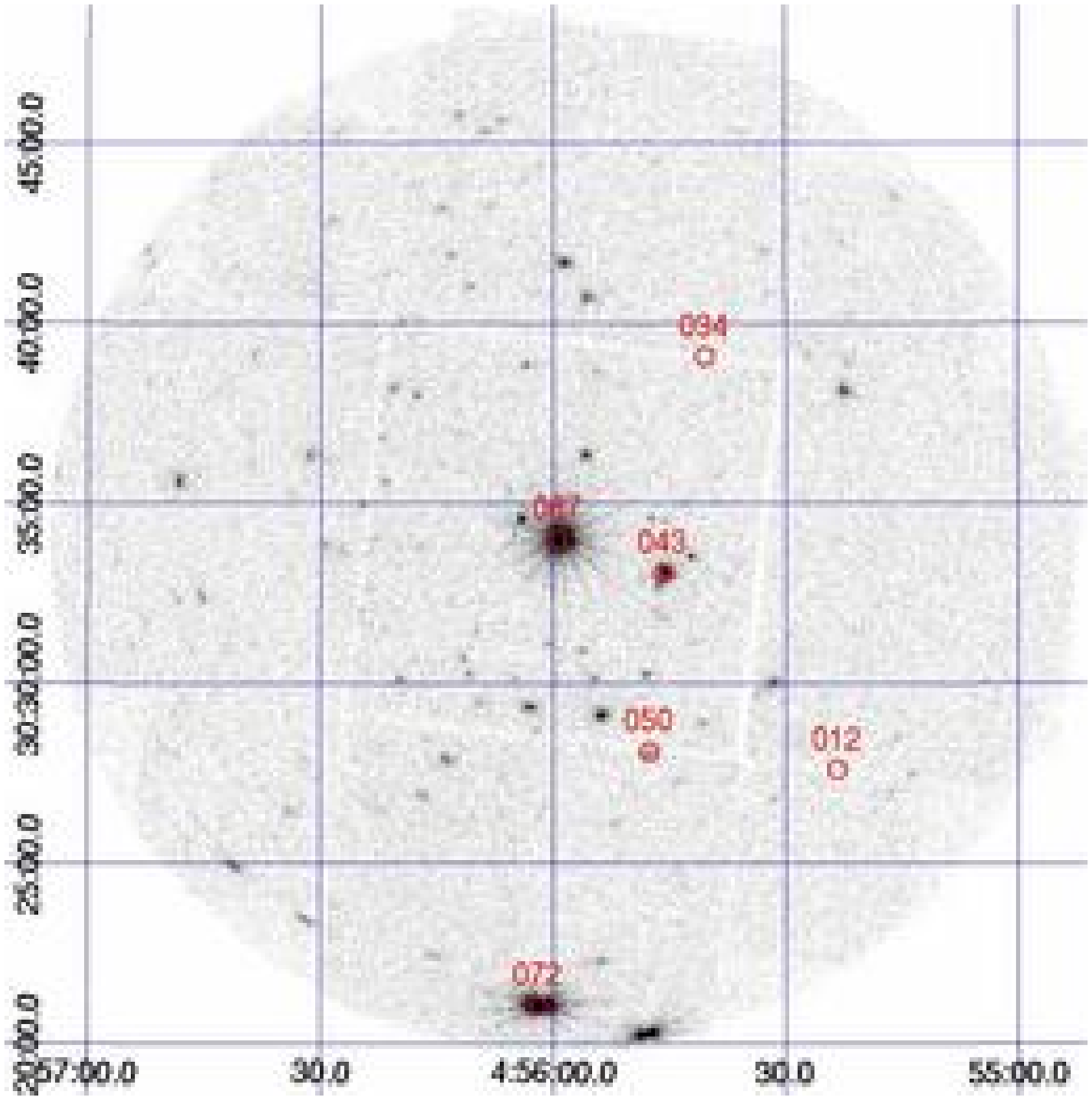}}}
}
\hbox{
{\resizebox{0.45\hsize}{!}{\includegraphics{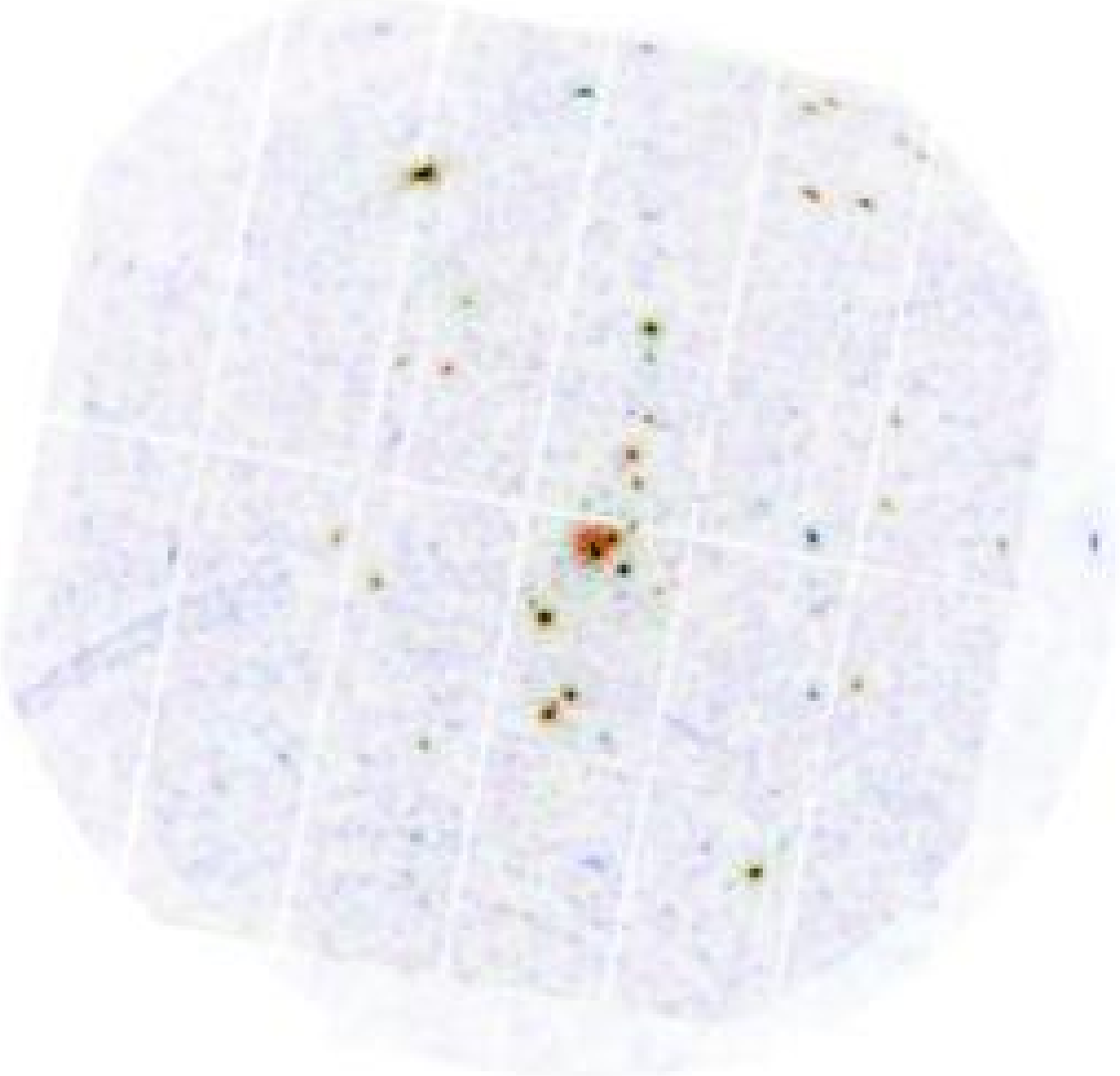}}}
{\resizebox{0.45\hsize}{!}{\includegraphics{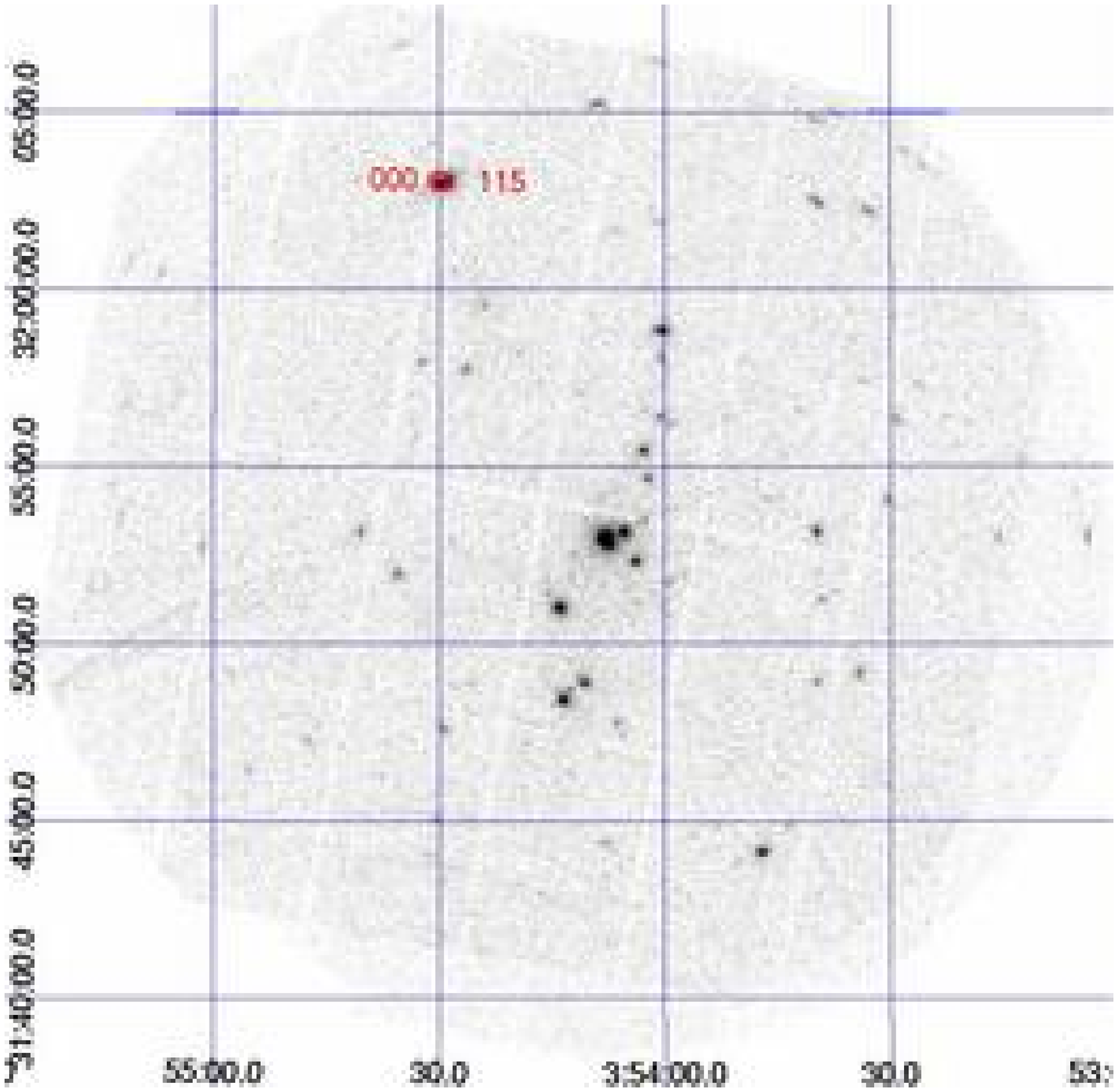}}}
}
\caption{Co-added EPIC images of field XEST-25, XEST-26, and XEST-27 (from top to bottom). Left: Smoothed images, color coded for hardness; right: 
      coordinate grid and TMC identifications included.\label{atlas9}} 
\end{figure*}

\begin{figure*}
\hbox{
{\resizebox{0.45\hsize}{!}{\includegraphics{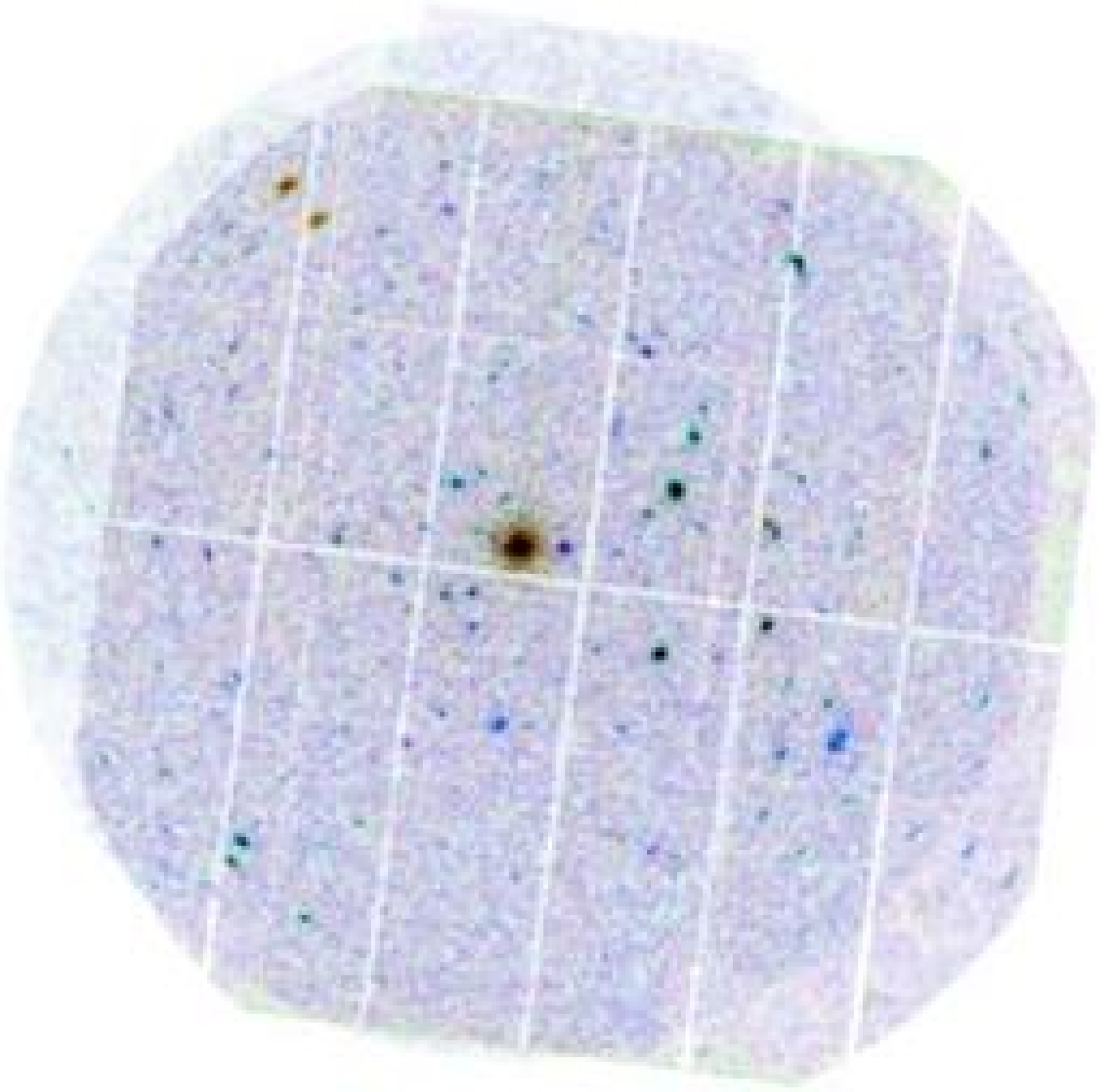}}}
{\resizebox{0.45\hsize}{!}{\includegraphics{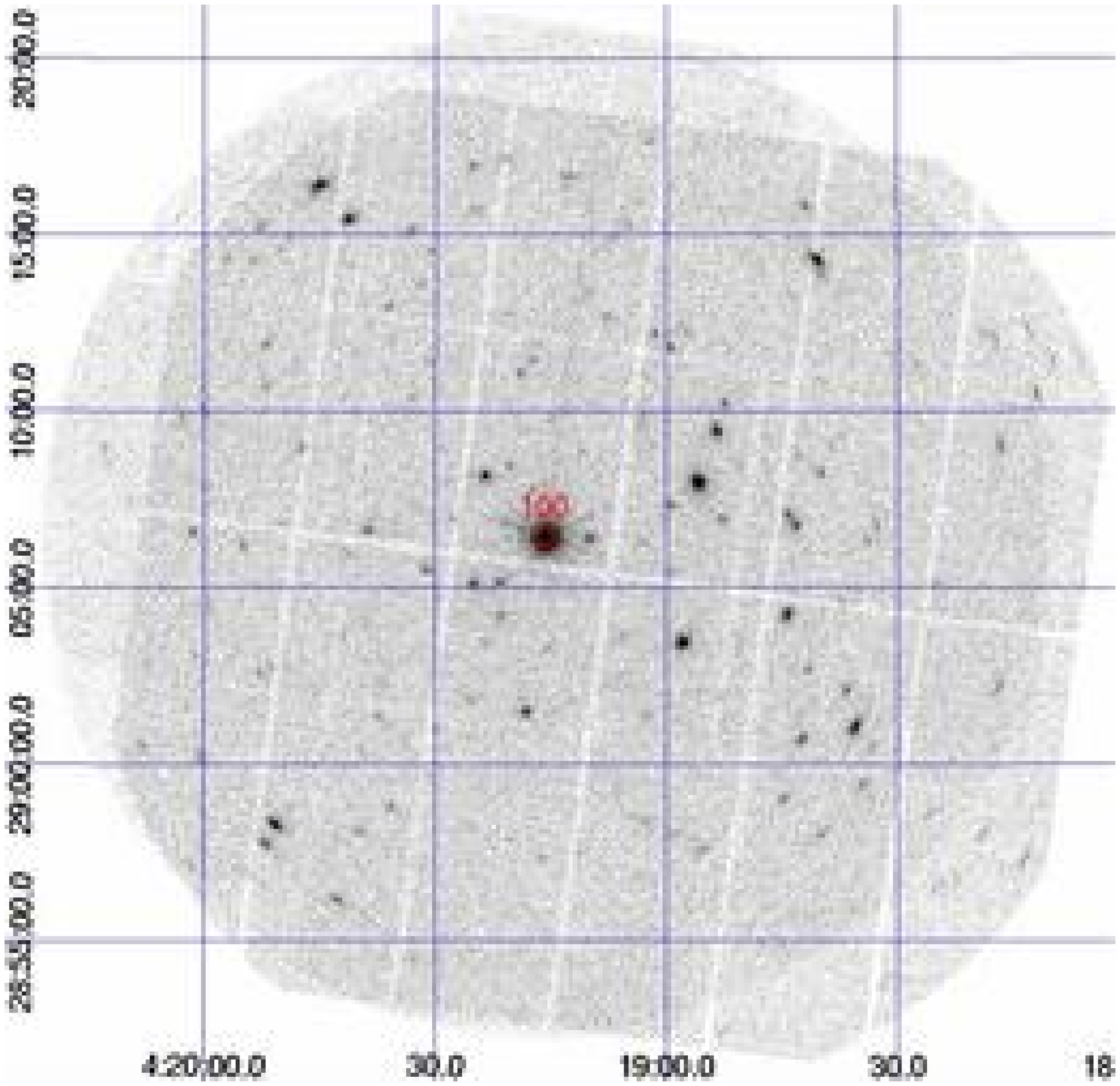}}}
}
\caption{Co-added EPIC images of field XEST-28. Left: Smoothed image, color coded for hardness; right: 
      coordinate grid and TMC identifications included.\label{atlas10}} 
\end{figure*}

Fig.~\ref{atlas1} - \ref{atlas10} show all 28 XEST EPIC exposures. Each field of view is
presented in two versions. The left panel shows a co-added EPIC image, logarithmically compressed 
in intensity and slightly smoothed. The colors code for hardness, where hardness increases from 
red to yellow to green to blue. Saturated (bright) stellar images are black. The  right panel
shows an unsmoothed, co-added EPIC image with an RA(J2000.0) - dec(J2000.0) coordinate grid.
Also plotted are the locations of known TMC members (red circles), labeled with their XEST source IDs.
The latter are located directly above the source circles except in crowded areas where
labels may have been shifted.

\begin{acknowledgements}
The XEST project was made possible by the co-operation of many individuals. In particular,
we thank the {\it XMM-Newton} SOC team in Vilspa for its excellent support before, during, and
after the observations. We warmly thank the International Space Science Institute (ISSI) in Bern 
for their financial support of the project and their hospitality during several XEST team
meetings at ISSI. Thanks go in particular to Vittorio Manno, Brigitte Fasler, and Saliba F. Saliba 
for their  efforts to make these  events enjoyable. In the course of the scientific analysis,
we obtained advice from numerous colleagues, of which we mention in particular Claude Catala,
Laurence DeWarf, Ed Fitzpatrick, Sylvain Guieu, Antonio Maggio, and Karl Stapelfeldt. Francesco Damiani
is thanked for giving us access to  his private version of PWXDETECT for {\it XMM-Newton} data.
We are grateful to the referee, Marc Gagn\'e, for his competent review of this paper and his 
constructive comments that helped improve the paper.
This research is based on observations obtained with {\it XMM-Newton}, an ESA science mission
with instruments and contributions directly funded by ESA member states and the USA (NASA).   
This publication makes use of data products from the
Two Micron All Sky Survey (2MASS), which is a joint project of the University of Massachusetts
and the Infrared Processing and Analysis Center/California Institute of Technology,
funded by the National Aeronautics and Space Administration and the National Science  
Foundation. Further, our research has made use of the SIMBAD database,
operated at CDS, Strasbourg, France. We have  made use of the ASURV
statistical software package maintained by Penn State. X-ray astronomy research at PSI has been supported by the 
Swiss National Science Foundation (grants 20-66875.01 and 20-109255/1). MA acknowledges support 
by NASA grants NNG05GF92G. The Palermo group acknowledges financial contributions from contract 
ASI-INAF I/023/05/0.   
\end{acknowledgements}

\clearpage

\clearpage

\Online

\noindent\hskip 1.5truecm {\bf References cited in the tables:}

\noindent\hskip 1.5truecm 1     = \citet{andrews05}   

\noindent\hskip 1.5truecm 2     =  \citet{bouvier06}   

\noindent\hskip 1.5truecm 3	     = \citet{briceno93}   

\noindent\hskip 1.5truecm 4          = \citet{briceno98}   

\noindent\hskip 1.5truecm 5          = \citet{briceno02}   

\noindent\hskip 1.5truecm 6      =    \citet{boehm93}     

\noindent\hskip 1.5truecm 7	     = \citet{cohen79}     

\noindent\hskip 1.5truecm 8        =    \citet{dewarf03}  

\noindent\hskip 1.5truecm 9         =   \citet{dewarf06}  

\noindent\hskip 1.5truecm 10          = \citet{doppmann05}  

\noindent\hskip 1.5truecm 11   = \citet{duchene99b}   

\noindent\hskip 1.5truecm 12	     =  \citet{duchene02}

\noindent\hskip 1.5truecm 13          = \citet{duchene04}   

\noindent\hskip 1.5truecm 14	     = \citet{fernandez95} 

\noindent\hskip 1.5truecm 15	 =  \citet{finkenzeller84}

\noindent\hskip 1.5truecm 16	     = \citet{ghez93}      

\noindent\hskip 1.5truecm 17	     = \citet{ghez97}      

\noindent\hskip 1.5truecm 18           = \citet{guieu06}     

\noindent\hskip 1.5truecm 19           = \citet{guieu06b}     

\noindent\hskip 1.5truecm 20          = \citet{hartigan03}  

\noindent\hskip 1.5truecm 21	     = \citet{hartigan94}  

\noindent\hskip 1.5truecm 22           = \citet{hartmann02}  

\noindent\hskip 1.5truecm 23     = \citet{hartmann05}  

\noindent\hskip 1.5truecm 24	     = \citet{herbig86}    

\noindent\hskip 1.5truecm 25	     = \citet{itoh05}      

\noindent\hskip 1.5truecm 26	     = \citet{jensen03}  
  
\noindent\hskip 1.5truecm 27  = \citet{kenyon95}    

\noindent\hskip 1.5truecm 28          = \citet{kenyon93}

\noindent\hskip 1.5truecm 29           = \citet{kenyon98}    

\noindent\hskip 1.5truecm 30	     = \citet{leinert89}   

\noindent\hskip 1.5truecm 31	     = \citet{leinert93}   

\noindent\hskip 1.5truecm 32       =   \citet{leinert97}    

\noindent\hskip 1.5truecm 33          = \citet{luhman04}    

\noindent\hskip 1.5truecm 34          = \citet{luhman03}    

\noindent\hskip 1.5truecm 35           = \citet{martin00}    

\noindent\hskip 1.5truecm 36	     = \citet{martin99b}    

\noindent\hskip 1.5truecm 37	     = \citet{martin94}    

\noindent\hskip 1.5truecm 38	     = \citet{martin99a}

\noindent\hskip 1.5truecm 39          = \citet{martin01}    

\noindent\hskip 1.5truecm 40   =    \citet{mccabe06}                                

\noindent\hskip 1.5truecm 41	      = \citet{mohanty05}   

\noindent\hskip 1.5truecm 42	     = \citet{moneti91}    

\noindent\hskip 1.5truecm 43	     = \citet{monin98}     

\noindent\hskip 1.5truecm 44   =       \citet{muzerolle98}  

\noindent\hskip 1.5truecm 45	     = \citet{muzerolle03} 

\noindent\hskip 1.5truecm 46      = \citet{muzerolle05} 

\noindent\hskip 1.5truecm 47	     = \citet{prato97}     

\noindent\hskip 1.5truecm 48      =      \citet{rebull04} and references therein  

\noindent\hskip 1.5truecm 49	     = \citet{reipurth93}  

\noindent\hskip 1.5truecm 50	     = \citet{richichi94}  

\noindent\hskip 1.5truecm 51	     = \citet{rodriguez98}

\noindent\hskip 1.5truecm 52	     = \citet{simon92}  and references therein    

\noindent\hskip 1.5truecm 53           = \citet{simon95}     

\noindent\hskip 1.5truecm 54	     = \citet{smith05}     

\noindent\hskip 1.5truecm 55    =      \citet{strom94}   

\noindent\hskip 1.5truecm 56	     = \citet{walter88}    

\noindent\hskip 1.5truecm 57	     = \citet{white03}     

\noindent\hskip 1.5truecm 58          = \citet{white01}     

\noindent\hskip 1.5truecm 59     = \citet{white04}     

\noindent\hskip 1.5truecm 60     = \citet{mundt83}

\noindent\hskip 1.5truecm 61     = \citet{welty95}
   
\noindent\hskip 1.5truecm 62     = \citet{woitas03}

\noindent\hskip 1.5truecm 63     = \citet{calvet04}

\noindent\hskip 1.5truecm 64     = \citet{reipurth02}

\noindent\hskip 1.5truecm 65     = \citet{walter03} and references therein

\noindent\hskip 1.5truecm 66     = \citet{luhman06}  

\noindent\hskip 1.5truecm 67     = \citet{simon96}  

\setcounter{table}{3}
\begin{table*}[h!]
\caption{X-ray parameters of targets in XEST (1): Positions and count rates}
\begin{tabular}{llrrrrrrrrr}
\hline
\hline
XEST & Name & RA$_{\rm X}$  & Dec$_{\rm X}$               &  Poserr    & Offset     & ML$_{\rm det}^a$ & Scts   & $T_{\rm exp}$  &  Rate  & Var$^b$           \\
     &      & h\ \ m \ \ s  & $\deg\ \ \arcmin\ \ \arcsec$   & ($\arcsec$)&($\arcsec$) &                  &        &  (s)           &  (ct~s$^{-1}$)  &   \\
\hline
27-115 &  HBC 352           &  3 54 29.54 & 32 03 02.2 & 1.04 & 0.89 &   9395 &      3832 &  15780 &    0.2429 & 0 \\
27-000 &  HBC 353           &  3 54 30.17 & 32 03 04.3 & 0.00 & 0.00 &      0 & $<$   183 &  14469 & $<$0.0127 & 0 \\
06-005 &  HBC 358 AB        &  4 03 49.27 & 26 10 53.1 & 1.48 & 1.22 &   1073 &       879 &   7371 &    0.1194 & 0 \\
06-007 &  HBC 359           &  4 03 50.82 & 26 10 53.0 & 1.46 & 0.34 &   2921 &      1605 &   7668 &    0.2094 & 0 \\
06-059 &  L1489 IRS         &  4 04 43.07 & 26 18 56.3 & 1.46 & 0.10 &   7632 &      3336 &  46889 &    0.0712 & 0 \\
20-001 &  LkCa 1            &  4 13 14.03 & 28 19 09.9 & 1.58 & 1.71 &    455 &       336 &   7820 &    0.0430 & 0 \\
20-005 &  Anon 1            &  4 13 27.28 & 28 16 23.3 & 1.53 & 1.64 &  24658 &      8299 &  21632 &    0.3836 & 0 \\
20-000 &  IRAS 04108+2803 A &  4 13 53.29 & 28 11 23.4 & 0.00 & 0.00 &      0 & $<$    26 &  29957 & $<$0.0009 & 0 \\
20-022 &  IRAS 04108+2803 B &  4 13 54.72 & 28 11 32.2 & 1.55 & 0.70 &   1145 &       696 &  30670 &    0.0227 & 1 \\
20-000 &  2M J04141188+28   &  4 14 11.88 & 28 11 53.5 & 0.00 & 0.00 &      0 & $<$    54 &  48645 & $<$0.0011 & 0 \\
20-042 &  V773 Tau ABC      &  4 14 12.92 & 28 12 11.8 & 1.53 & 0.60 & 213164 &     59321 &  48974 &    1.2113 & 0 \\
20-043 &  FM Tau            &  4 14 13.57 & 28 12 48.6 & 1.54 & 0.61 &   2627 &      2988 &  48901 &    0.0611 & 0 \\
20-046 &  CW Tau            &  4 14 16.99 & 28 11 00.6 & 1.84 & 2.80 &     11 &        52 &  35943 &    0.0015 & 0 \\
20-047 &  CIDA 1            &  4 14 17.72 & 28 06 09.8 & 1.69 & 1.46 &     63 &       105 &  33985 &    0.0031 & 0 \\
20-056 &  MHO 2/1           &  4 14 26.41 & 28 06 00.9 & 1.54 & 1.21 &   6260 &      2735 &  32359 &    0.0845 & 0 \\
20-058 &  MHO 3             &  4 14 30.58 & 28 05 14.7 & 1.55 & 0.40 &   1234 &       766 &  26856 &    0.0285 & 0 \\
20-069 &  FO Tau AB         &  4 14 49.23 & 28 12 31.3 & 1.60 & 1.06 &    238 &       240 &  33352 &    0.0072 & 0 \\
20-073 &  CIDA 2            &  4 15 05.16 & 28 08 46.5 & 1.56 & 0.30 &    814 &       584 &  22342 &    0.0261 & 0 \\
23-002 &  CY Tau            &  4 17 33.61 & 28 20 47.7 & 1.08 & 1.77 &    683 &       457 &  10256 &    0.0446 & 1 \\
24-002 &  CY Tau            &  4 17 33.73 & 28 20 47.9 & 1.01 & 1.00 &    407 &       284 &   8366 &    0.0340 & 2 \\
23-004 &  LkCa 5            &  4 17 38.91 & 28 32 59.6 & 1.04 & 0.98 &   5286 &      2855 &  32431 &    0.0880 & 0 \\
24-004 &  LkCa 5            &  4 17 38.94 & 28 33 00.2 & 0.94 & 0.30 &   4767 &      2221 &  18626 &    0.1192 & 2 \\
23-008 &  CIDA 3            &  4 17 49.67 & 28 29 35.9 & 1.10 & 0.48 &    292 &       358 &  45960 &    0.0078 & 0 \\
24-008 &  CIDA 3            &  4 17 49.59 & 28 29 35.3 & 0.99 & 1.28 &    338 &       332 &  26423 &    0.0126 & 0 \\
23-015 &  V410 X3           &  4 18 07.98 & 28 26 02.8 & 1.09 & 0.94 &    576 &       561 &  64938 &    0.0086 & 0 \\
24-015 &  V410 X3           &  4 18 07.97 & 28 26 02.1 & 0.99 & 1.61 &    388 &       341 &  37758 &    0.0090 & 0 \\
23-018 &  V410 A13          &  4 18 17.12 & 28 28 40.5 & 1.45 & 1.41 &     20 &        72 &  66746 &    0.0011 & 0 \\
24-000 &  V410 A13          &  4 18 17.11 & 28 28 41.9 & 0.00 & 0.00 &      0 & $<$    24 &  34966 & $<$0.0007 & 0 \\
23-000 &  V410 A24          &  4 18 22.39 & 28 24 37.6 & 0.00 & 0.00 &      0 & $<$    26 &  68500 & $<$0.0004 & 0 \\
24-000 &  V410 A24          &  4 18 22.39 & 28 24 37.6 & 0.00 & 0.00 &      0 & $<$    30 &  42210 & $<$0.0007 & 0 \\
23-029 &  V410 A25          &  4 18 29.15 & 28 26 18.9 & 1.04 & 0.69 &   4542 &      3586 &  76359 &    0.0470 & 0 \\
24-027 &  V410 A25          &  4 18 29.12 & 28 26 19.1 & 0.94 & 0.26 &   2447 &      1812 &  43216 &    0.0419 & 0 \\
23-032 &  V410 Tau ABC      &  4 18 31.08 & 28 27 16.7 & 1.03 & 0.57 & 365212 &    101132 &  78334 &    1.2910 & 1 \\
24-028 &  V410 Tau ABC      &  4 18 31.09 & 28 27 16.8 & 0.93 & 0.61 & 137848 &     38215 &  44779 &    0.8534 & 3 \\
23-033 &  DD Tau AB         &  4 18 31.11 & 28 16 28.7 & 1.08 & 0.40 &   1006 &       885 &  39614 &    0.0223 & 1 \\
24-029 &  DD Tau AB         &  4 18 31.10 & 28 16 29.3 & 1.04 & 0.50 &    295 &       320 &  22841 &    0.0140 & 0 \\
23-035 &  CZ Tau AB         &  4 18 31.60 & 28 16 59.2 & 1.05 & 0.71 &   1756 &      1271 &  41620 &    0.0305 & 0 \\
24-030 &  CZ Tau AB         &  4 18 31.57 & 28 16 59.2 & 0.97 & 0.75 &   1509 &       995 &  23898 &    0.0417 & 0 \\
23-036 &  IRAS 04154+2823   &  4 18 31.94 & 28 31 15.6 & 1.11 & 1.20 &    211 &       308 &  64773 &    0.0048 & 1 \\
24-031 &  IRAS 04154+2823   &  4 18 31.88 & 28 31 14.9 & 1.59 & 2.04 &     18 &        61 &  36863 &    0.0017 & 0 \\
23-037 &  V410 X2           &  4 18 34.40 & 28 30 30.1 & 1.04 & 0.67 &   5641 &      3007 &  65004 &    0.0463 & 0 \\
24-032 &  V410 X2           &  4 18 34.38 & 28 30 30.6 & 0.94 & 1.01 &   3959 &      2015 &  36402 &    0.0554 & 1 \\
23-045 &  V410 X4           &  4 18 40.24 & 28 24 25.1 & 1.06 & 0.61 &    859 &       760 &  69092 &    0.0110 & 0 \\
24-038 &  V410 X4           &  4 18 40.26 & 28 24 25.2 & 0.95 & 0.80 &   1153 &       778 &  39074 &    0.0199 & 2 \\
23-047 &  V892 Tau          &  4 18 40.64 & 28 19 15.9 & 1.03 & 0.48 &  58194 &     20108 &  48862 &    0.4115 & 0 \\
24-040 &  V892 Tau          &  4 18 40.62 & 28 19 15.6 & 0.93 & 0.10 & 111462 &     31125 &  28058 &    1.1093 & 1 \\
23-048 &  LR 1              &  4 18 41.29 & 28 27 26.9 & 1.22 & 1.97 &     45 &       134 &  69240 &    0.0019 & 1 \\
24-000 &  LR 1              &  4 18 41.33 & 28 27 25.0 & 0.00 & 0.00 &      0 & $<$    67 &  41278 & $<$0.0016 & 0 \\
23-050 &  V410 X7           &  4 18 42.53 & 28 18 49.9 & 1.04 & 0.41 &   2808 &      2319 &  46487 &    0.0499 & 1 \\
24-042 &  V410 X7           &  4 18 42.49 & 28 18 50.0 & 0.93 & 0.24 &  10724 &      6309 &  26634 &    0.2369 & 1 \\
23-000 &  V410 A20          &  4 18 45.06 & 28 20 52.8 & 0.00 & 0.00 &      0 & $<$    23 &  32774 & $<$0.0007 & 0 \\
24-000 &  V410 A20          &  4 18 45.06 & 28 20 52.8 & 0.00 & 0.00 &      0 & $<$    15 &  17241 & $<$0.0009 & 0 \\
23-056 &  Hubble 4          &  4 18 47.03 & 28 20 07.7 & 1.03 & 0.42 &  95259 &     27648 &  45560 &    0.6069 & 0 \\
24-047 &  Hubble 4          &  4 18 47.02 & 28 20 07.7 & 0.93 & 0.48 &  43119 &     14030 &  26926 &    0.5211 & 0 \\
23-000 &  KPNO-Tau 2        &  4 18 51.16 & 28 14 33.2 & 0.00 & 0.00 &      0 & $<$    28 &  15428 & $<$0.0018 & 0 \\
24-000 &  KPNO-Tau 2        &  4 18 51.16 & 28 14 33.2 & 0.00 & 0.00 &      0 & $<$    15 &   8642 & $<$0.0018 & 0 \\
23-000 &  CoKu Tau 1        &  4 18 51.48 & 28 20 26.5 & 0.00 & 0.00 &      0 & $<$   128 &  48084 & $<$0.0027 & 0 \\
24-000 &  CoKu Tau 1        &  4 18 51.48 & 28 20 26.5 & 0.00 & 0.00 &      0 & $<$    57 &  27535 & $<$0.0021 & 0 \\
23-061 &  V410 X6           &  4 19 01.07 & 28 19 42.1 & 1.10 & 0.54 &    298 &       369 &  38775 &    0.0095 & 0 \\
\hline
\end{tabular}
\normalsize
\end{table*}
 
 \setcounter{table}{3}
\begin{table*}[h!]
\caption{(Continued)}
\begin{tabular}{llrrrrrrrrr}
\hline
\hline
XEST & Name & RA$_{\rm X}$  & Dec$_{\rm X}$               &  Poserr    & Offset     & ML$_{\rm det}^a$ & Scts   & $T_{\rm exp}$  &  Rate   & Var$^b$	  \\
     &      & h\ \ m \ \ s  & $\deg\ \ \arcmin\ \ \arcsec$   & ($\arcsec$)&($\arcsec$) &                &        &  (s)           &  (ct~s$^{-1}$) &   \\
\hline
24-054 &  V410 X6           &  4 19 01.08 & 28 19 42.2 & 0.98 & 0.44 &    514 &       442 &  22047 &    0.0201 & 2 \\
23-063 &  V410 X5           &  4 19 01.94 & 28 22 34.4 & 1.05 & 1.31 &   1688 &      1179 &  43645 &    0.0270 & 0 \\
24-055 &  V410 X5           &  4 19 01.94 & 28 22 33.8 & 0.95 & 0.80 &   2422 &      1339 &  24869 &    0.0539 & 2 \\
23-067 &  FQ Tau AB         &  4 19 12.77 & 28 29 33.3 & 1.21 & 0.56 &    100 &       178 &  34347 &    0.0052 & 0 \\
24-058 &  FQ Tau AB         &  4 19 12.77 & 28 29 34.3 & 1.20 & 1.31 &     64 &       101 &  18919 &    0.0054 & 0 \\
28-100 &  BP Tau            &  4 19 15.90 & 29 06 26.6 & 2.24 & 0.84 & 142902 &     46278 & 197539 &    0.2343 & 1 \\
23-074 &  V819 Tau AB       &  4 19 26.26 & 28 26 14.5 & 1.04 & 0.20 &  19437 &      7785 &  28888 &    0.2695 & 0 \\
24-061 &  V819 Tau AB       &  4 19 26.26 & 28 26 14.7 & 0.94 & 0.40 &   9399 &      3967 &  16434 &    0.2414 & 0 \\
16-000 &  IRAS 04166+2706   &  4 19 43.00 & 27 13 33.7 & 0.00 & 0.00 &      0 & $<$    30 &  30078 & $<$0.0010 & 0 \\
16-000 &  IRAS 04169+2702   &  4 19 58.45 & 27 09 57.1 & 0.00 & 0.00 &      0 & $<$    38 &  23883 & $<$0.0016 & 0 \\
11-000 &  CFHT-Tau 19       &  4 21 07.95 & 27 02 20.4 & 0.00 & 0.00 &      0 & $<$    18 &  22031 & $<$0.0008 & 0 \\
11-000 &  IRAS 04181+2655   &  4 21 10.90 & 27 02 06.0 & 0.00 & 0.00 &      0 & $<$    18 &  22755 & $<$0.0008 & 0 \\
11-000 &  IRAS 04181+2654AB &  4 21 11.47 & 27 01 09.4 & 0.00 & 0.00 &      0 & $<$    34 &  24662 & $<$0.0014 & 0 \\
11-023 &  2M J04213459      &  4 21 34.51 & 27 01 38.7 & 1.88 & 1.22 &    140 &       171 &  38717 &    0.0044 & 0 \\
01-028 &  IRAS 04187+1927   &  4 21 43.23 & 19 34 12.5 & 1.55 & 0.81 &   2842 &      1460 &  30378 &    0.0481 & 0 \\
11-037 &  CFHT-Tau 10       &  4 21 46.39 & 26 59 28.4 & 2.25 & 1.61 &      9 &        35 &  48347 &    0.0007 & 0 \\
11-000 &  2M J04215450+2652 &  4 21 54.51 & 26 52 31.5 & 0.00 & 0.00 &      0 & $<$    17 &  37210 & $<$0.0004 & 0 \\
21-038 &  RY Tau            &  4 21 57.40 & 28 26 35.4 & 1.36 & 0.10 &  17357 &      6495 &  23482 &    0.2766 & 0 \\
21-039 &  HD 283572         &  4 21 58.87 & 28 18 06.1 & 1.36 & 0.64 & 282825 &     76762 &  37198 &    2.0636 & 0 \\
01-045 &  T Tau N(+Sab)     &  4 21 59.44 & 19 32 05.8 & 1.54 & 0.62 & 211562 &     59050 &  72108 &    0.8189 & 0 \\
11-054 &  Haro 6-5 B        &  4 22 00.72 & 26 57 32.4 & 1.88 & 0.29 &     28 &       188 &  51812 &    0.0036 & 0 \\
11-057 &  FS Tau AC         &  4 22 02.19 & 26 57 30.9 & 1.78 & 0.42 &  20767 &      7398 &  51050 &    0.1449 & 3 \\
21-044 &  LkCa 21           &  4 22 03.12 & 28 25 39.1 & 1.36 & 0.28 &   6256 &      2932 &  25588 &    0.1146 & 0 \\
01-054 &  RX J0422.1+1934   &  4 22 04.91 & 19 34 48.8 & 1.54 & 0.87 &  54106 &     17801 &  52504 &    0.3390 & 0 \\
01-062 &  2M J04221332+1934 &  4 22 13.24 & 19 34 40.2 & 1.72 & 1.51 &     69 &       117 &  33836 &    0.0035 & 0 \\
11-079 &  CFHT-Tau 21       &  4 22 16.80 & 26 54 58.2 & 1.82 & 1.22 &    286 &       274 &  39533 &    0.0070 & 0 \\
02-013 &  FV Tau AB         &  4 26 53.48 & 26 06 54.9 & 1.60 & 0.84 &    781 &       662 &  31329 &    0.0212 & 0 \\
02-000 &  FV Tau/c AB       &  4 26 54.41 & 26 06 51.0 & 0.00 & 0.00 &      0 & $<$    94 &  35140 & $<$0.0027 & 0 \\
02-016 &  KPNO-Tau 13       &  4 26 57.34 & 26 06 27.2 & 1.63 & 1.21 &    217 &       288 &  22911 &    0.0126 & 0 \\
02-000 &  DG Tau B          &  4 27 02.66 & 26 05 30.5 & 0.00 & 0.00 &      0 & $<$    27 &  38087 & $<$0.0007 & 0 \\
02-022 &  DG Tau A          &  4 27 04.70 & 26 06 15.5 & 1.59 & 0.80 &   1669 &      1185 &  40865 &    0.0290 & 2 \\
02-000 &  KPNO-Tau 4        &  4 27 28.00 & 26 12 05.3 & 0.00 & 0.00 &      0 & $<$    41 &  50463 & $<$0.0008 & 0 \\
02-000 &  IRAS 04248+2612AB &  4 27 57.31 & 26 19 18.3 & 0.00 & 0.00 &      0 & $<$    25 &  17943 & $<$0.0014 & 0 \\
15-020 &  JH 507            &  4 29 20.67 & 26 33 40.2 & 1.56 & 0.73 &   2000 &      1102 &  28851 &    0.0382 & 0 \\
13-004 &  GV Tau AB         &  4 29 23.69 & 24 33 00.2 & 1.63 & 0.55 &    518 &       615 &  20976 &    0.0293 & 1 \\
13-000 &  IRAS 04264+2433   &  4 29 30.08 & 24 39 55.1 & 0.00 & 0.00 &      0 & $<$    68 &  26868 & $<$0.0025 & 0 \\
15-040 &  DH Tau AB         &  4 29 41.56 & 26 32 58.5 & 1.55 & 0.20 & 130666 &     36137 &  33680 &    1.0729 & 3 \\
15-042 &  DI Tau AB         &  4 29 42.51 & 26 32 49.3 & 1.56 & 0.40 &    926 &      3305 &  33773 &    0.0979 & 0 \\
15-044 &  KPNO-Tau 5        &  4 29 45.81 & 26 30 47.9 & 1.82 & 2.06 &     40 &        86 &  32576 &    0.0026 & 0 \\
14-006 &  IQ Tau A          &  4 29 51.56 & 26 06 44.9 & 2.02 & 0.00 &   3811 &      1972 &  22748 &    0.0867 & 1 \\
13-000 &  CFHT-Tau 20       &  4 29 59.51 & 24 33 07.9 & 0.00 & 0.00 &      0 & $<$    22 &  30935 & $<$0.0007 & 0 \\
14-000 &  KPNO-Tau 6        &  4 30 07.24 & 26 08 20.8 & 0.00 & 0.00 &      0 & $<$    27 &  26206 & $<$0.0010 & 0 \\
13-035 &  FX Tau AB         &  4 30 29.59 & 24 26 47.2 & 1.65 & 2.22 &    313 &       378 &   7123 &    0.0532 & 0 \\
14-057 &  DK Tau AB         &  4 30 44.25 & 26 01 25.5 & 2.01 & 1.00 &   8459 &      3697 &  37386 &    0.0989 & 0 \\
14-000 &  KPNO-Tau 7        &  4 30 57.19 & 25 56 39.5 & 0.00 & 0.00 &      0 & $<$    31 &  25489 & $<$0.0012 & 0 \\
22-013 &  MHO 9             &  4 31 15.83 & 18 20 05.9 & 1.62 & 1.48 &    777 &       611 &  37684 &    0.0162 & 0 \\
22-021 &  MHO 4             &  4 31 24.18 & 18 00 21.6 & 1.63 & 1.71 &    700 &       569 &  37887 &    0.0150 & 0 \\
22-040 &  L1551 IRS5        &  4 31 34.02 & 18 08 04.2 & 1.68 & 1.11 &    139 &       191 &  74201 &    0.0026 & 0 \\
22-042 &  LkHa 358          &  4 31 36.13 & 18 13 43.3 & 1.66 & 0.00 &    115 &       273 &  76104 &    0.0036 & 0 \\
22-000 &  HH 30             &  4 31 37.47 & 18 12 24.5 & 0.00 & 0.00 &      0 & $<$    21 &  81198 & $<$0.0003 & 0 \\
22-043 &  HL Tau            &  4 31 38.38 & 18 13 57.1 & 1.60 & 1.04 &   6052 &      5618 &  75594 &    0.0743 & 0 \\
22-047 &  XZ Tau AB         &  4 31 40.05 & 18 13 56.6 & 1.60 & 0.66 &  60086 &     18391 &  75388 &    0.2440 & 1 \\
22-056 &  L1551 NE          &  4 31 44.53 & 18 08 32.3 & 1.98 & 1.39 &     13 &        47 &  80720 &    0.0006 & 0 \\
03-005 &  HK Tau AB         &  4 31 50.57 & 24 24 16.1 & 1.77 & 2.00 &     60 &        86 &  19411 &    0.0045 & 0 \\
22-070 &  V710 Tau BA       &  4 31 57.72 & 18 21 37.2 & 1.60 & 1.34 &  11966 &      4639 &  35159 &    0.1320 & 0 \\
19-009 &  JH 665            &  4 31 58.50 & 25 43 30.6 & 2.27 & 1.07 &    108 &       123 &   8791 &    0.0141 & 0 \\
22-089 &  L1551 51          &  4 32 09.45 & 17 57 22.9 & 1.60 & 2.57 &  11539 &      4287 &  13801 &    0.3107 & 2 \\
22-097 &  V827 Tau          &  4 32 14.50 & 18 20 14.5 & 1.60 & 1.02 &  36001 &     10530 &  18303 &    0.5753 & 0 \\
03-016 &  Haro 6-13         &  4 32 15.40 & 24 28 59.9 & 1.59 & 0.24 &    338 &       288 &  20435 &    0.0141 & 0 \\
\hline
\end{tabular}
\normalsize
\end{table*}
 
 \setcounter{table}{3}
\begin{table*}[h!]
\caption{(Continued)}
\begin{tabular}{llrrrrrrrrr}
\hline
\hline
XEST & Name & RA$_{\rm X}$  & Dec$_{\rm X}$               &  Poserr    & Offset     & ML$_{\rm det}^a$ & Scts   & $T_{\rm exp}$  &  Rate   & Var$^b$	  \\
     &      & h\ \ m \ \ s  & $\deg\ \ \arcmin\ \ \arcsec$   & ($\arcsec$)&($\arcsec$) &                &        &  (s)           &  (ct~s$^{-1}$) &   \\
\hline
22-100 &  V826 Tau          &  4 32 15.91 & 18 01 39.1 & 1.60 & 1.08 &  94853 &     27095 &  36121 &    0.7501 & 0 \\
22-101 &  MHO 5             &  4 32 16.07 & 18 12 46.5 & 1.62 & 0.10 &   1457 &       963 &  51887 &    0.0186 & 0 \\
03-017 &  CFHT-Tau 7        &  4 32 17.83 & 24 22 13.1 & 1.71 & 1.94 &     29 &       121 &  30227 &    0.0040 & 0 \\
03-019 &  V928 Tau AB       &  4 32 18.90 & 24 22 26.1 & 1.54 & 1.14 &   4972 &      2243 &  30636 &    0.0732 & 0 \\
03-022 &  FY Tau            &  4 32 30.68 & 24 19 58.1 & 1.55 & 1.58 &   4931 &      2110 &  25663 &    0.0822 & 0 \\
03-023 &  FZ Tau            &  4 32 31.83 & 24 20 04.1 & 1.59 & 1.46 &    213 &       440 &  19003 &    0.0232 & 0 \\
17-002 &  IRAS 04295+2251   &  4 32 32.07 & 22 57 26.4 & 1.36 & 0.41 &    322 &       303 &  17292 &    0.0176 & 2 \\
19-049 &  UZ Tau E+W(AB)    &  4 32 42.89 & 25 52 32.6 & 2.19 & 2.52 &   4110 &      2085 &  44868 &    0.0465 & 0 \\
17-009 &  JH 112            &  4 32 49.09 & 22 53 01.9 & 1.33 & 0.94 &   1517 &       867 &  24276 &    0.0357 & 0 \\
03-031 &  CFHT-Tau 5        &  4 32 50.28 & 24 22 11.4 & 1.73 & 0.24 &     54 &        82 &  15890 &    0.0052 & 0 \\
04-003 &  CFHT-Tau 5        &  4 32 50.31 & 24 22 11.1 & 1.73 & 0.74 &    367 &       312 &  26369 &    0.0119 & 0 \\
03-035 &  MHO 8             &  4 33 01.97 & 24 21 02.7 & 1.66 & 2.70 &    105 &       130 &  17386 &    0.0075 & 0 \\
04-009 &  MHO 8             &  4 33 02.02 & 24 20 59.0 & 1.77 & 1.14 &    181 &       172 &  28001 &    0.0062 & 0 \\
04-010 &  GH Tau AB         &  4 33 06.33 & 24 09 32.5 & 1.78 & 2.13 &     96 &       169 &  11180 &    0.0151 & 0 \\
04-012 &  V807 Tau  SNab    &  4 33 06.75 & 24 09 54.6 & 1.68 & 1.56 &  10380 &      4068 &  20068 &    0.2027 & 0 \\
18-004 &  KPNO-Tau 14       &  4 33 07.79 & 26 16 05.7 & 2.13 & 0.94 &   1424 &       840 &  19063 &    0.0441 & 0 \\
17-000 &  CFHT-Tau 12       &  4 33 09.46 & 22 46 48.7 & 0.00 & 0.00 &      0 & $<$    10 &  30963 & $<$0.0003 & 0 \\
04-016 &  V830 Tau          &  4 33 09.98 & 24 33 43.4 & 1.68 & 0.68 &  44530 &     13508 &  15863 &    0.8516 & 3 \\
18-000 &  IRAS S04301+261   &  4 33 14.36 & 26 14 23.5 & 0.00 & 0.00 &      0 & $<$    17 &  22545 & $<$0.0008 & 0 \\
17-000 &  IRAS 04302+2247   &  4 33 16.50 & 22 53 20.4 & 0.00 & 0.00 &      0 & $<$    28 &  29406 & $<$0.0010 & 0 \\
17-027 &  IRAS 04303+2240   &  4 33 19.10 & 22 46 34.1 & 1.31 & 0.43 &   9238 &      3723 &  30569 &    0.1218 & 0 \\
04-034 &  GI Tau            &  4 33 34.06 & 24 21 17.9 & 1.69 & 0.90 &    792 &      1707 &  46957 &    0.0364 & 0 \\
04-035 &  GK Tau AB         &  4 33 34.52 & 24 21 08.0 & 1.68 & 2.17 &  17201 &      6635 &  47269 &    0.1404 & 0 \\
18-019 &  IS Tau AB         &  4 33 36.83 & 26 09 48.3 & 2.13 & 1.05 &   3999 &      1835 &  31294 &    0.0586 & 0 \\
17-058 &  CI Tau            &  4 33 52.02 & 22 50 30.8 & 1.34 & 0.66 &    647 &       485 &  23775 &    0.0204 & 1 \\
18-030 &  IT Tau AB         &  4 33 54.72 & 26 13 26.7 & 2.12 & 0.84 &  58042 &     17849 &  44176 &    0.4041 & 0 \\
17-066 &  JH 108            &  4 34 11.01 & 22 51 44.7 & 1.32 & 0.34 &   9235 &      3875 &  14379 &    0.2695 & 1 \\
17-068 &  CFHT-BD Tau 1     &  4 34 15.31 & 22 50 33.1 & 1.45 & 2.17 &    163 &       204 &  14010 &    0.0146 & 2 \\
25-026 &  AA Tau            &  4 34 55.39 & 24 28 53.6 & 1.82 & 0.57 &   3334 &      1471 &  24628 &    0.0597 & 0 \\
09-010 &  HO Tau AB         &  4 35 20.31 & 22 32 11.1 & 1.58 & 3.82 &     82 &       122 &  18311 &    0.0067 & 0 \\
08-019 &  FF Tau AB         &  4 35 20.90 & 22 54 24.3 & 1.60 & 0.10 &   3812 &      2076 &  30501 &    0.0681 & 0 \\
12-040 &  DN Tau            &  4 35 27.34 & 24 14 58.9 & 1.57 & 0.41 &  28146 &      9246 &  47293 &    0.1955 & 0 \\
12-000 &  IRAS 04325+2402AB &  4 35 35.39 & 24 08 19.4 & 0.00 & 0.00 &      0 & $<$    44 &  34141 & $<$0.0013 & 0 \\
12-059 &  CoKu Tau 3 AB     &  4 35 40.95 & 24 11 09.0 & 1.57 & 0.24 &  69838 &     20015 &  38317 &    0.5224 & 0 \\
09-022 &  KPNO-Tau 8        &  4 35 41.91 & 22 34 10.9 & 1.44 & 1.20 &   3919 &      1858 &  27822 &    0.0668 & 0 \\
08-037 &  HQ Tau AB         &  4 35 47.39 & 22 50 22.0 & 1.59 & 0.75 &  25465 &      9258 &  30355 &    0.3050 & 0 \\
09-026 &  HQ Tau AB         &  4 35 47.33 & 22 50 20.9 & 1.43 & 0.81 &  20604 &      7171 &  14030 &    0.5111 & 3 \\
08-043 &  KPNO-Tau 15       &  4 35 51.13 & 22 52 40.1 & 1.59 & 0.41 &  25254 &     10329 &  46282 &    0.2232 & 0 \\
09-031 &  KPNO-Tau 15       &  4 35 51.01 & 22 52 40.3 & 1.49 & 1.26 &    633 &       529 &  11974 &    0.0443 & 0 \\
08-000 &  KPNO-Tau 9        &  4 35 51.43 & 22 49 11.9 & 0.00 & 0.00 &      0 & $<$    28 &  37232 & $<$0.0008 & 0 \\
09-000 &  KPNO-Tau 9        &  4 35 51.43 & 22 49 11.9 & 0.00 & 0.00 &      0 & $<$    26 &  17141 & $<$0.0016 & 0 \\
08-048 &  HP Tau AB         &  4 35 52.78 & 22 54 22.9 & 1.59 & 0.20 &   3573 &      5706 &  49366 &    0.1156 & 0 \\
08-051a&  HP Tau/G3 AB      &  4 35 53.50 & 22 54 09.0 & 0.00 & 0.00 & 134269 &      3006 &  48905 &    0.0648 & 0 \\
08-051 &  HP Tau/G2         &  4 35 54.14 & 22 54 12.8 & 1.59 & 0.71 & 134269 &     39787 &  48905 &    0.8136 & 0 \\
08-058 &  Haro 6-28 AB      &  4 35 56.80 & 22 54 37.1 & 1.63 & 1.23 &    290 &       614 &  49680 &    0.0124 & 0 \\
08-000 &  CFHT-BD Tau 2     &  4 36 10.39 & 22 59 56.0 & 0.00 & 0.00 &      0 & $<$    29 &  36650 & $<$0.0008 & 0 \\
08-080 &  CFHT-BD Tau 3     &  4 36 38.91 & 22 58 13.2 & 1.86 & 1.36 &     20 &        64 &  25098 &    0.0026 & 0 \\
05-005 &  CFHT-Tau 6        &  4 39 04.10 & 25 44 26.4 & 1.90 & 1.89 &     48 &        84 &  21111 &    0.0040 & 0 \\
05-000 &  IRAS 04361+2547   &  4 39 13.89 & 25 53 20.9 & 0.00 & 0.00 &      0 & $<$    35 &  11917 & $<$0.0030 & 0 \\
05-013 &  GN Tau AB         &  4 39 20.86 & 25 45 00.9 & 1.77 & 1.38 &    301 &       248 &  22952 &    0.0108 & 0 \\
05-017 &  IRAS 04365+2535   &  4 39 35.25 & 25 41 45.4 & 1.93 & 1.07 &     34 &        75 &  31546 &    0.0024 & 0 \\
05-024 &  IRAS 04369+2539   &  4 39 55.70 & 25 45 01.7 & 1.75 & 0.74 &    435 &       335 &  20227 &    0.0166 & 0 \\
07-011 &  JH 223            &  4 40 49.54 & 25 51 20.0 & 1.87 & 0.90 &    374 &       318 &  24335 &    0.0131 & 0 \\
07-022 &  Haro 6-32         &  4 41 04.32 & 25 57 55.5 & 1.89 & 1.23 &    351 &       311 &  17222 &    0.0181 & 0 \\
07-000 &  ITG 33 A          &  4 41 08.26 & 25 56 07.5 & 0.00 & 0.00 &      0 & $<$    21 &  22254 & $<$0.0010 & 0 \\
07-000 &  CFHT-Tau 8        &  4 41 10.78 & 25 55 11.7 & 0.00 & 0.00 &      0 & $<$    16 &  23759 & $<$0.0007 & 0 \\
07-000 &  IRAS 04381+2540   &  4 41 12.68 & 25 46 35.4 & 0.00 & 0.00 &      0 & $<$    32 &  44259 & $<$0.0007 & 0 \\
07-041 &  IRAS 04385+2550AB &  4 41 38.80 & 25 56 27.4 & 1.86 & 0.66 &    503 &       397 &  16994 &    0.0234 & 2 \\
10-017 &  CoKuLk332/G2 AB   &  4 42 05.47 & 25 22 56.2 & 1.68 & 0.29 &  17061 &      6651 &  36423 &    0.1826 & 0 \\
\hline
\end{tabular}
\normalsize
\end{table*}
 
 \setcounter{table}{3}
\begin{table*}[h!]
\caption{(Continued)}
\begin{tabular}{llrrrrrrrrr}
\hline
\hline
XEST & Name & RA$_{\rm X}$  & Dec$_{\rm X}$               &  Poserr    & Offset     & ML$_{\rm det}^a$ & Scts   & $T_{\rm exp}$  &  Rate   & Var$^b$	  \\
     &      & h\ \ m \ \ s  & $\deg\ \ \arcmin\ \ \arcsec$   & ($\arcsec$)&($\arcsec$) &                &        &  (s)           &  (ct~s$^{-1}$) &   \\
\hline
10-018 &  CoKuLk332/G1 AB   &  4 42 07.32 & 25 23 03.2 & 1.69 & 0.14 &    243 &      1448 &  30835 &    0.0470 & 0 \\
10-020 &  V955 Tau AB       &  4 42 07.76 & 25 23 11.6 & 1.70 & 0.24 &    308 &      1360 &  23616 &    0.0576 & 0 \\
10-034 &  CIDA 7            &  4 42 21.02 & 25 20 35.5 & 1.83 & 1.10 &     71 &       119 &  43153 &    0.0028 & 2 \\
10-045 &  DP Tau            &  4 42 37.72 & 25 15 36.8 & 1.83 & 0.75 &     64 &        93 &  30465 &    0.0031 & 2 \\
10-060 &  GO Tau            &  4 43 03.12 & 25 20 19.8 & 1.72 & 1.08 &    611 &       479 &  21286 &    0.0225 & 0 \\
26-012 &  2M J04552333+30   &  4 55 23.12 & 30 27 38.2 & 1.98 & 3.15 &     17 &        54 &  28154 &    0.0019 & 0 \\
26-034 &  2M J04554046+30   &  4 55 40.34 & 30 39 07.1 & 2.19 & 2.09 &      4 &        45 &  35843 &    0.0013 & 0 \\
26-043 &  AB Aur            &  4 55 45.83 & 30 33 03.3 & 1.73 & 1.10 &   7917 &      3472 &  69171 &    0.0502 & 0 \\
26-050 &  2MJ04554757/801   &  4 55 47.83 & 30 28 05.3 & 1.76 & 4.13 &    426 &       418 &  43514 &    0.0096 & 2 \\
26-067 &  SU Aur            &  4 55 59.34 & 30 34 00.9 & 1.73 & 0.87 & 245679 &     59141 &  78106 &    0.7572 & 1 \\
26-072 &  HBC 427           &  4 56 02.05 & 30 21 03.9 & 1.73 & 0.40 &  91805 &     25437 &  28045 &    0.9070 & 1 \\
\hline
\multicolumn{6}{l}{Additional sources from Chandra} & Sig.$^a$ & & & & \\
\hline
C1-0   &  KPNO-Tau 10       &  4 17 49.55 & 28 13 31.9 & 0.00 & 0.00 &    0.0 &         0 &  17734 &    0.0000 & 0 \\
C1-1   &  IRAS 04158+2805   &  4 18 58.15 & 28 12 23.3 & 0.93 & 0.24 &    6.0 &       100 &  17734 &    0.0056 & 0 \\
C2-1   &  Haro 6-5 B        &  4 22 00.71 & 26 57 32.2 & 0.07 & 0.33 &   15.4 &        32 &  29674 &    0.0011 & 0 \\
C2-2   &  FS Tau AC         &  4 22 02.20 & 26 57 30.4 & 0.04 & 0.29 &  101.0 &       254 &  29674 &    0.0086 & 0 \\
C3-1   &  FV Tau/c AB       &  4 26 54.34 & 26 06 51.3 & 0.21 & 0.99 &    6.2 &        13 &  29717 &    0.0005 & 0 \\
C3-2   &  DG Tau B          &  4 27 02.58 & 26 05 30.8 & 0.22 & 1.12 &    4.3 &         9 &  29717 &    0.0003 & 0 \\
C4-1   &  GV Tau AB         &  4 29 23.74 & 24 33 00.5 & 0.05 & 0.24 &   26.7 &        57 &  24650 &    0.0023 & 0 \\
C5-2   &  HN Tau AB         &  4 33 39.34 & 17 51 51.3 & 1.40 & 1.11 &    3.6 &        22 &   4679 &    0.0047 & 0 \\
C5-1   &  L1551 55          &  4 32 43.70 & 18 02 54.4 & 1.17 & 1.95 &    6.7 &        74 &   4679 &    0.0160 & 0 \\
C5-4   &  HD 28867          &  4 33 33.06 & 18 01 00.1 & 0.01 & 1.18 &   39.2 &      1582 &   4679 &    0.3390 & 0 \\
C5-3   &  DM Tau            &  4 33 48.63 & 18 10 11.5 & 1.28 & 1.97 &    5.1 &        33 &   4679 &    0.0071 & 0 \\
C6-1   &  CFHT-BD Tau 4     &  4 39 47.50 & 26 01 40.8 & 0.14 & 0.27 &   16.8 &        30 &  19317 &    0.0016 & 0 \\
C6-0   &  L1527 IRS         &  4 39 53.59 & 26 03 05.5 & 0.00 & 0.00 &    0.0 &         0 &  19317 &    0.0000 & 0 \\
C6-0   &  CFHT-Tau 17       &  4 40 01.75 & 25 56 29.2 & 0.00 & 0.00 &    0.0 &         0 &  19317 &    0.0000 & 0 \\
C6-2   &  IRAS 04370+2559   &  4 40 08.02 & 26 05 25.5 & 0.09 & 0.29 &   74.6 &       229 &  19317 &    0.0119 & 0 \\
\hline
\end{tabular}
\begin{minipage}{0.87\textwidth}
\footnotetext{
\hskip -0.5truecm $^a$ Maximum likelihood for detection for {\it XMM-Newton} data, CIAO WAVDETECT 'Significance' for {\it Chandra} data\\
$^b$ Variability flag: 0 = no or only low-level variability; 1 = clear flaring, flare intervals removed in spectral fit;  2 = clear flaring observed but flare intervals not removed; 3 = slow decay of flare throughout observation, all data used\\
NOTES on individual objects:
\begin{itemize}
\item CoKu Tau 1 = XEST-23-000: {\it Chandra} observation C1 may contain a marginal off-axis detection  ($\approx$6 counts)
\item V892 Tau=  XEST-23-047 = XEST-24-040: Companion at 4\arcsec\ is visible as a faint source in {\it Chandra} observation C1
\item HP Tau/G2 and G3 = XEST-08-051 (separation: 10\arcsec) were treated as one source. The ratios of the counts and rates were derived from PSF fitting in the image
\end{itemize}
}
\end{minipage}
\label{tab4}
\normalsize
\end{table*}

\clearpage

\setcounter{table}{4}
\begin{table*}[h!]
\scriptsize
\caption{X-ray parameters of targets in XEST (2): Plasma parameters from the DEM fits}
\begin{tabular}{lllllrlrrrr}
\hline
\hline
XEST & Name & $N_{\rm H}$ \hfill{\scriptsize (1$\sigma$ range)}                 &  $T_0$ \hfill{\scriptsize (1$\sigma$ range)}  &  $\beta$ \hfill{\scriptsize (1$\sigma$ range)} & EM$_t^a$                & $L_{\rm X}^b$ \hfill{\scriptsize (range)}                  & log         & $T_{\rm av}$ & $\chi^2_{\rm red}$ & dof       \\
   &      & ($10^{22}$~cm$^{-2}$) &  (MK)   &          & ($10^{52})$             & ($10^{30}$~erg~s$^{-1}$)               & $L_{\rm X}/L_*$   & (MK)   &              &           \\
\hline
27-115 &  HBC 352           & 0.22\hfill{~\scriptsize(0.19,0.25)} &  6.5\hfill{~\scriptsize( 5.0, 8.2)} & -0.60\hfill{~\scriptsize(-0.79,-0.47)} &  25.21 &     2.657\hfill{~\scriptsize(2.43, 2.87)} &    -3.03 & 12.00 &  0.85 & 129  \\
27-000 &  HBC 353           & 0.17\hfill                          & 10.0\hfill                          & -1.00\hfill                            &    ... & $<$ 0.176                                 & $<$-4.04 &   ... &   ... &  ... \\
06-005 &  HBC 358 AB        & 0.01\hfill{~\scriptsize(0.00,0.04)} &  4.7\hfill{~\scriptsize( 3.0, 7.5)} & -0.60\hfill{~\scriptsize(-1.24,-0.34)} &   3.79 &     0.383\hfill{~\scriptsize(0.37, 0.44)} &    -3.45 &  9.42 &  0.74 & 21   \\
06-007 &  HBC 359           & 0.01\hfill{~\scriptsize(0.00,0.02)} &  7.2\hfill{~\scriptsize( 5.8, 8.9)} & -1.25\hfill{~\scriptsize(-1.84,-0.91)} &   6.65 &     0.663\hfill{~\scriptsize(0.64, 0.69)} &    -3.18 &  9.20 &  1.05 & 57   \\
06-059 &  L1489 IRS         & 6.63\hfill{~\scriptsize(6.18,7.09)} & 50.9\hfill{~\scriptsize(30.9,60.5)} & -3.00\hfill{~\scriptsize(-3.00, 0.02)} &  30.39 &     4.471\hfill{~\scriptsize(4.07, 4.91)} &    -3.63 & 40.42 &  0.94 & 99   \\
20-001 &  LkCa 1            & 0.07\hfill{~\scriptsize(0.04,0.15)} &  4.9\hfill{~\scriptsize( 3.7, 5.7)} & -3.00\hfill{~\scriptsize(-3.00,-1.94)} &   2.66 &     0.232\hfill{~\scriptsize(0.20, 0.37)} &    -3.80 &  4.35 &  0.57 & 18   \\
20-005 &  Anon 1            & 0.33\hfill{~\scriptsize(0.30,0.35)} &  7.6\hfill{~\scriptsize( 6.4, 9.1)} & -1.01\hfill{~\scriptsize(-1.26,-0.88)} &  40.27 &     4.139\hfill{~\scriptsize(3.84, 4.42)} &    -3.38 & 10.72 &  0.92 & 228  \\
20-000 &  IRAS 04108+2803 A &                                 ... &                                 ... &                                    ... &    ... &       ...                                 &      ... &   ... &   ... &  ... \\
20-022 &  IRAS 04108+2803 B & 5.68\hfill{~\scriptsize(2.99,8.19)} &  4.5\hfill{~\scriptsize( 2.0,31.6)} & -1.00\hfill                            &   4.39 &     0.417\hfill{~\scriptsize(0.07, 0.57)} &    -3.57 &  6.93 &  0.68 & 8    \\
20-000 &  2M J04141188+28   & 0.18\hfill                          &  7.9\hfill                          & -3.00\hfill                            &    ... & $<$ 0.028                                 & $<$-3.31 &   ... &   ... &  ... \\
20-042 &  V773 Tau ABC      & 0.17\hfill{~\scriptsize(0.17,0.17)} &  8.7\hfill{~\scriptsize( 8.4, 9.1)} & -0.87\hfill{~\scriptsize(-0.93,-0.82)} &  89.77 &     9.488\hfill{~\scriptsize(9.39, 9.54)} &    -3.36 & 12.96 &  1.06 & 526  \\
20-043 &  FM Tau            & 0.15\hfill{~\scriptsize(0.13,0.17)} &  6.1\hfill{~\scriptsize( 3.9, 9.6)} & -0.02\hfill{~\scriptsize(-0.29, 0.11)} &   4.59 &     0.532\hfill{~\scriptsize(0.51, 0.56)} &    -3.52 & 18.47 &  1.12 & 78   \\
20-046 &  CW Tau            & 6.49\hfill{~\scriptsize(4.02,8.12)} &  4.5\hfill{~\scriptsize( 2.0,12.4)} & -3.00\hfill{~\scriptsize(-3.00,-2.21)} &  33.14 &     2.844\hfill{~\scriptsize(0.28, 4.00)} &    -3.17 &  4.07 &  1.76 & 2    \\
20-047 &  CIDA 1            & 0.22\hfill{~\scriptsize(0.03,0.47)} &  5.0\hfill{~\scriptsize( 2.5,10.7)} & -1.00\hfill                            &   0.26 &     0.025\hfill{~\scriptsize(0.01, 0.07)} &      ... &  7.57 &  0.68 & 3    \\
20-056 &  MHO 2/1           & 0.92\hfill{~\scriptsize(0.83,1.15)} & 17.7\hfill{~\scriptsize(10.4,28.9)} & -1.20\hfill{~\scriptsize(-3.00,-0.35)} &  14.44 &     1.636\hfill{~\scriptsize(1.48, 2.39)} &    -3.12 & 20.53 &  0.98 & 85   \\
20-058 &  MHO 3             & 0.61\hfill{~\scriptsize(0.48,1.02)} & 10.0\hfill{~\scriptsize( 2.0,15.4)} & -1.40\hfill{~\scriptsize(-3.00,-0.74)} &   4.46 &     0.459\hfill{~\scriptsize(0.36, 2.18)} &    -3.88 & 11.76 &  0.86 & 20   \\
20-069 &  FO Tau AB         & 0.25\hfill{~\scriptsize(0.17,0.77)} & 12.5\hfill{~\scriptsize( 2.0,18.7)} & -2.01\hfill{~\scriptsize(-3.00,-0.37)} &   0.64 &     0.065\hfill{~\scriptsize(0.05, 0.52)} &    -4.66 & 12.33 &  1.29 & 4    \\
20-073 &  CIDA 2            & 0.13\hfill{~\scriptsize(0.06,0.27)} &  6.0\hfill{~\scriptsize( 3.3, 7.8)} & -2.87\hfill{~\scriptsize(-3.00,-1.49)} &   1.96 &     0.178\hfill{~\scriptsize(0.14, 0.36)} &    -3.84 &  5.34 &  0.80 & 18   \\
23-002 &  CY Tau            & 0.00\hfill{~\scriptsize(0.00,0.17)} &  6.7\hfill{~\scriptsize( 2.2,10.3)} & -1.05\hfill{~\scriptsize(-3.00,-0.36)} &   1.32 &     0.133\hfill{~\scriptsize(0.13, 0.29)} &    -4.16 &  9.42 &  1.71 & 10   \\
24-002 &  CY Tau            & 0.02\hfill                          &  9.5\hfill{~\scriptsize( 6.6,13.7)} & -0.79\hfill{~\scriptsize(-2.00,-0.29)} &   1.79 &     0.194\hfill{~\scriptsize(0.19, 0.30)} &    -4.00 & 14.46 &  0.80 & 12   \\
23-004 &  LkCa 5            & 0.02\hfill{~\scriptsize(0.01,0.03)} &  7.9\hfill{~\scriptsize( 6.4, 9.1)} & -1.33\hfill{~\scriptsize(-1.86,-0.91)} &   4.32 &     0.432\hfill{~\scriptsize(0.41, 0.45)} &    -3.52 &  9.67 &  0.59 & 85   \\
24-004 &  LkCa 5            & 0.04\hfill{~\scriptsize(0.02,0.06)} &  5.6\hfill{~\scriptsize( 3.7, 8.3)} & -0.50\hfill{~\scriptsize(-0.96,-0.28)} &   6.46 &     0.681\hfill{~\scriptsize(0.64, 0.73)} &    -3.32 & 11.62 &  0.99 & 72   \\
23-008 &  CIDA 3            & 1.38\hfill{~\scriptsize(0.78,2.09)} &  5.7\hfill{~\scriptsize( 2.0,31.6)} & -0.67\hfill{~\scriptsize(-1.67, 0.45)} &   2.67 &     0.277\hfill{~\scriptsize(0.12, 1.62)} &    -3.29 & 10.45 &  1.15 & 11   \\
24-008 &  CIDA 3            & 0.85\hfill{~\scriptsize(0.67,1.09)} & 31.6\hfill{~\scriptsize(22.3,31.6)} & -1.00\hfill                            &   1.95 &     0.254\hfill{~\scriptsize(0.22, 0.31)} &    -3.33 & 33.60 &  0.87 & 7    \\
23-015 &  V410 X3           & 0.13\hfill{~\scriptsize(0.07,0.25)} &  4.9\hfill{~\scriptsize( 3.1, 8.7)} & -1.22\hfill{~\scriptsize(-2.75,-0.71)} &   0.63 &     0.059\hfill{~\scriptsize(0.05, 0.10)} &    -3.74 &  6.67 &  0.76 & 13   \\
24-015 &  V410 X3           & 0.24\hfill{~\scriptsize(0.00,0.52)} &  3.5\hfill{~\scriptsize( 2.0, 9.9)} & -1.98\hfill{~\scriptsize(-3.00,-1.04)} &   1.10 &     0.092\hfill{~\scriptsize(0.03, 0.47)} &    -3.55 &  3.81 &  0.79 & 5    \\
23-018 &  V410 A13          & 0.42\hfill{~\scriptsize(0.00,0.78)} &  4.0\hfill{~\scriptsize( 2.0,20.6)} & -1.00\hfill                            &   0.10 &     0.010\hfill{~\scriptsize(0.00, 0.03)} &    -4.18 &  6.21 &  1.11 & 2    \\
24-000 &  V410 A13          & 0.50\hfill                          & 10.0\hfill                          & -1.00\hfill                            &    ... & $<$ 0.011                                 & $<$-4.14 &   ... &   ... &  ... \\
23-000 &  V410 A24          & 4.37\hfill                          & 10.0\hfill                          & -1.00\hfill                            &    ... & $<$ 0.042                                 & $<$-5.41 &   ... &   ... &  ... \\
24-000 &  V410 A24          & 4.37\hfill                          & 10.0\hfill                          & -1.00\hfill                            &    ... & $<$ 0.078                                 & $<$-5.14 &   ... &   ... &  ... \\
23-029 &  V410 A25          & 0.59\hfill{~\scriptsize(0.53,0.65)} & 25.9\hfill{~\scriptsize(23.1,29.1)} & -3.00\hfill{~\scriptsize(-3.00,-2.17)} &   5.98 &     0.676\hfill{~\scriptsize(0.63, 0.72)} &    -3.96 & 21.75 &  1.27 & 90   \\
24-027 &  V410 A25          & 0.57\hfill{~\scriptsize(0.49,0.66)} & 33.6\hfill{~\scriptsize(26.4,40.0)} & -3.00\hfill{~\scriptsize(-3.00,-1.10)} &   4.79 &     0.587\hfill{~\scriptsize(0.55, 0.64)} &    -4.02 & 27.79 &  1.11 & 47   \\
23-032 &  V410 Tau ABC      & 0.01\hfill{~\scriptsize(0.01,0.02)} &  9.0\hfill{~\scriptsize( 7.7,10.1)} & -1.14\hfill{~\scriptsize(-1.43,-0.85)} &  36.33 &     3.762\hfill{~\scriptsize(3.68, 3.84)} &    -3.35 & 11.74 &  0.95 & 115  \\
24-028 &  V410 Tau ABC      & 0.02\hfill{~\scriptsize(0.02,0.03)} & 10.4\hfill{~\scriptsize( 9.4,10.8)} & -1.17\hfill{~\scriptsize(-1.28,-0.96)} &  44.47 &     4.663\hfill{~\scriptsize(4.58, 4.76)} &    -3.26 & 13.20 &  1.10 & 414  \\
23-033 &  DD Tau AB         & 0.23\hfill{~\scriptsize(0.16,0.37)} &  3.8\hfill{~\scriptsize( 2.0,31.6)} &  0.93\hfill{~\scriptsize( 0.19, 1.00)} &   0.69 &     0.093\hfill{~\scriptsize(0.08, 0.11)} &    -4.15 & 37.84 &  1.37 & 6    \\
24-029 &  DD Tau AB         & 0.29\hfill{~\scriptsize(0.21,0.45)} & 31.6\hfill{~\scriptsize( 2.0,31.6)} &  0.13\hfill{~\scriptsize(-3.00, 1.00)} &   1.00 &     0.139\hfill{~\scriptsize(0.12, 0.16)} &    -3.97 & 41.71 &  1.37 & 5    \\
23-035 &  CZ Tau AB         & 0.34\hfill{~\scriptsize(0.24,0.43)} &  5.0\hfill{~\scriptsize( 3.9, 6.5)} & -3.00\hfill{~\scriptsize(-3.00,-2.57)} &   4.86 &     0.423\hfill{~\scriptsize(0.28, 0.64)} &    -3.39 &  4.41 &  1.09 & 28   \\
24-030 &  CZ Tau AB         & 0.33\hfill{~\scriptsize(0.24,0.37)} &  5.0\hfill{~\scriptsize( 4.6, 6.4)} & -3.00\hfill{~\scriptsize(-3.00,-2.73)} &   6.22 &     0.547\hfill{~\scriptsize(0.37, 0.64)} &    -3.28 &  4.45 &  1.86 & 22   \\
23-036 &  IRAS 04154+2823   & 6.80\hfill{~\scriptsize(4.36,13.1)} & 31.6\hfill{~\scriptsize( 2.0,31.6)} & -1.00\hfill                            &   1.44 &     0.188\hfill{~\scriptsize(0.13, 0.88)} &    -3.43 & 33.60 &  0.89 & 8    \\
24-031 &  IRAS 04154+2823   & 5.43\hfill{~\scriptsize(1.89,10.6)} &  9.1\hfill{~\scriptsize( 2.0,31.6)} & -1.00\hfill                            &   1.56 &     0.164\hfill{~\scriptsize(0.03, 0.72)} &    -3.48 & 12.58 &  1.16 & 1    \\
23-037 &  V410 X2           & 1.47\hfill{~\scriptsize(1.32,1.59)} &  2.9\hfill{~\scriptsize( 2.0, 5.2)} & -1.65\hfill{~\scriptsize(-1.82,-1.48)} &  76.83 &     6.192\hfill{~\scriptsize(3.06,11.49)} &    -3.27 &  3.48 &  0.81 & 84   \\
24-032 &  V410 X2           & 1.46\hfill{~\scriptsize(0.95,1.60)} &  2.0\hfill{~\scriptsize( 2.0,14.1)} & -1.47\hfill{~\scriptsize(-1.75,-1.15)} & 109.71 &     7.853\hfill{~\scriptsize(1.18,11.25)} &    -3.17 &  2.79 &  0.58 & 24   \\
23-045 &  V410 X4           & 0.94\hfill{~\scriptsize(0.79,1.16)} & 15.0\hfill{~\scriptsize( 9.9,19.1)} & -3.00\hfill{~\scriptsize(-3.00,-1.34)} &   2.10 &     0.211\hfill{~\scriptsize(0.17, 0.30)} &    -4.38 & 12.73 &  0.74 & 19   \\
24-038 &  V410 X4           & 0.74\hfill{~\scriptsize(0.32,1.43)} & 17.6\hfill{~\scriptsize( 2.0,31.6)} & -2.84\hfill{~\scriptsize(-3.00, 0.42)} &   2.06 &     0.213\hfill{~\scriptsize(0.13, 1.69)} &    -4.37 & 15.15 &  0.12 & 4    \\
23-047 &  V892 Tau          & 0.94\hfill{~\scriptsize(0.89,0.98)} & 12.0\hfill{~\scriptsize(10.5,14.5)} & -1.18\hfill{~\scriptsize(-1.50,-1.04)} &  86.14 &     9.207\hfill{~\scriptsize(8.51, 9.90)} &    -4.51 & 14.86 &  1.17 & 377  \\
24-040 &  V892 Tau          & 1.04\hfill{~\scriptsize(0.90,1.31)} &  9.0\hfill{~\scriptsize( 3.3,13.7)} & -1.39\hfill{~\scriptsize(-1.97,-1.04)} &  78.05 &     7.944\hfill{~\scriptsize(6.02,18.12)} &    -4.57 & 10.74 &  0.90 & 86   \\
23-048 &  LR 1              & 2.77\hfill{~\scriptsize(2.11,3.66)} &  2.0\hfill{~\scriptsize( 2.0,31.6)} & -1.00\hfill                            &   5.33 &     0.417\hfill{~\scriptsize(0.29, 0.55)} &    -3.64 &  3.63 &  0.59 & 5    \\
24-000 &  LR 1              & 4.16\hfill                          & 10.0\hfill                          & -1.00\hfill                            &    ... & $<$ 0.169                                 & $<$-4.03 &   ... &   ... &  ... \\
23-050 &  V410 X7           & 0.80\hfill{~\scriptsize(0.70,0.92)} & 12.6\hfill{~\scriptsize( 8.5,17.9)} & -1.34\hfill{~\scriptsize(-2.48,-0.88)} &   8.60 &     0.913\hfill{~\scriptsize(0.78, 1.16)} &    -3.32 & 14.68 &  0.98 & 44   \\
24-042 &  V410 X7           & 0.72\hfill{~\scriptsize(0.67,0.78)} & 29.4\hfill{~\scriptsize(19.4,32.5)} & -2.99\hfill{~\scriptsize(-3.00,-1.06)} &  25.28 &     2.975\hfill{~\scriptsize(2.82, 3.14)} &    -2.81 & 24.57 &  0.93 & 82   \\
23-000 &  V410 A20          & 4.27\hfill                          & 10.0\hfill                          & -1.00\hfill                            &    ... & $<$ 0.075                                 & $<$-4.42 &   ... &   ... &  ... \\
24-000 &  V410 A20          & 4.27\hfill                          & 10.0\hfill                          & -1.00\hfill                            &    ... & $<$ 0.094                                 & $<$-4.32 &   ... &   ... &  ... \\
23-056 &  Hubble 4          & 0.25\hfill{~\scriptsize(0.24,0.25)} &  9.1\hfill{~\scriptsize( 8.6, 9.6)} & -1.08\hfill{~\scriptsize(-1.17,-0.99)} &  51.25 &     5.342\hfill{~\scriptsize(5.24, 5.38)} &    -3.29 & 12.14 &  1.02 & 383  \\
24-047 &  Hubble 4          & 0.24\hfill{~\scriptsize(0.23,0.25)} &  9.2\hfill{~\scriptsize( 8.6,10.0)} & -1.16\hfill{~\scriptsize(-1.29,-1.03)} &  44.83 &     4.668\hfill{~\scriptsize(4.52, 4.76)} &    -3.35 & 11.96 &  0.98 & 280  \\
23-000 &  KPNO-Tau 2        & 0.07\hfill                          &  7.9\hfill                          & -3.00\hfill                            &    ... & $<$ 0.012                                 & $<$-3.38 &   ... &   ... &  ... \\
24-000 &  KPNO-Tau 2        & 0.07\hfill                          &  7.9\hfill                          & -3.00\hfill                            &    ... & $<$ 0.012                                 & $<$-3.38 &   ... &   ... &  ... \\
23-000 &  CoKu Tau 1        &                                 ... &                                 ... &                                    ... &    ... &       ...                                 &      ... &   ... &   ... &  ... \\
24-000 &  CoKu Tau 1        &                                 ... &                                 ... &                                    ... &    ... &       ...                                 &      ... &   ... &   ... &  ... \\
23-061 &  V410 X6           & 0.36\hfill{~\scriptsize(0.26,0.43)} &  3.2\hfill{~\scriptsize( 2.1, 4.6)} & -1.00\hfill                            &   1.39 &     0.124\hfill{~\scriptsize(0.08, 0.17)} &    -3.79 &  5.15 &  1.26 & 13   \\
\hline
\end{tabular}
\normalsize
\end{table*}
 
\setcounter{table}{4}
\begin{table*}[h!]
\scriptsize
\caption{(Continued)}
\begin{tabular}{lllllrlrrrr}
\hline
\hline
XEST & Name & $N_{\rm H}$ {\scriptsize \hfill(1$\sigma$ range)}         &  $T_0$ \hfill{\scriptsize (1$\sigma$ range)} &  $\beta$ \hfill{\scriptsize (1$\sigma$ range)} & EM$_t^a$                & $L_{\rm X}^b$ \hfill{\scriptsize (range)}                    & log           & $T_{\rm av}$   & $\chi^2_{\rm red}$   & dof    \\
   &      & ($10^{22}$~cm$^{-2}$) &  (MK)   &          & ($10^{52})$         & ($10^{30}$~erg~s$^{-1}$)                 &  $L_{\rm X}/L_*$    & (MK)           &                      &  \\
\hline
24-054 &  V410 X6           & 0.45\hfill{~\scriptsize(0.32,0.63)} &  5.0\hfill{~\scriptsize( 2.4, 9.1)} & -1.19\hfill{~\scriptsize(-2.55,-0.71)} &   2.93 &     0.278\hfill{~\scriptsize(0.18, 0.57)} &    -3.44 &  6.86 &  1.21 & 9    \\
23-063 &  V410 X5           & 0.45\hfill{~\scriptsize(0.37,0.56)} &  7.0\hfill{~\scriptsize( 4.6,11.6)} & -1.16\hfill{~\scriptsize(-1.90,-0.77)} &   3.87 &     0.387\hfill{~\scriptsize(0.31, 0.52)} &    -2.92 &  9.30 &  1.17 & 32   \\
24-055 &  V410 X5           & 0.61\hfill{~\scriptsize(0.52,0.72)} &  4.0\hfill{~\scriptsize( 2.2, 7.0)} & -0.50\hfill{~\scriptsize(-0.76,-0.30)} &  11.07 &     1.115\hfill{~\scriptsize(0.87, 1.59)} &    -2.46 &  9.16 &  0.94 & 41   \\
23-067 &  FQ Tau AB         & 0.50\hfill{~\scriptsize(0.22,0.93)} &  5.0\hfill{~\scriptsize( 2.0,10.1)} & -3.00\hfill{~\scriptsize(-3.00,-1.51)} &   1.36 &     0.120\hfill{~\scriptsize(0.05, 0.83)} &    -3.83 &  4.46 &  1.79 & 6    \\
24-058 &  FQ Tau AB         & 0.35\hfill                          &  5.9\hfill{~\scriptsize( 2.0,10.0)} & -3.00\hfill                            &   5.59 &     0.049\hfill{~\scriptsize(0.04, 0.16)} &    -4.22 &  5.17 &  0.87 & 1    \\
28-100 &  BP Tau            & 0.06\hfill{~\scriptsize(0.06,0.07)} &  7.1\hfill{~\scriptsize( 6.7, 7.6)} & -0.67\hfill{~\scriptsize(-0.73,-0.60)} &  12.84 &     1.365\hfill{~\scriptsize(1.35, 1.40)} &    -3.43 & 12.34 &  1.10 & 430  \\
23-074 &  V819 Tau AB       & 0.21\hfill{~\scriptsize(0.20,0.23)} &  4.9\hfill{~\scriptsize( 4.3, 5.4)} & -0.99\hfill{~\scriptsize(-1.10,-0.90)} &  25.41 &     2.445\hfill{~\scriptsize(2.33, 2.61)} &    -3.16 &  7.45 &  1.04 & 199  \\
24-061 &  V819 Tau AB       & 0.22\hfill{~\scriptsize(0.18,0.26)} &  4.5\hfill{~\scriptsize( 3.5, 5.7)} & -1.00\hfill{~\scriptsize(-1.13,-0.87)} &  23.24 &     2.205\hfill{~\scriptsize(1.91, 2.66)} &    -3.20 &  6.87 &  0.90 & 110  \\
16-000 &  IRAS 04166+2706   &                                 ... &                                 ... &                                    ... &    ... &       ...                                 &      ... &   ... &   ... &  ... \\
16-000 &  IRAS 04169+2702   &                                 ... &                                 ... &                                    ... &    ... &       ...                                 &      ... &   ... &   ... &  ... \\
11-000 &  CFHT-Tau 19       & 1.31\hfill                          &  7.9\hfill                          & -3.00\hfill                            &    ... & $<$ 0.047                                 & $<$-3.77 &   ... &   ... &  ... \\
11-000 &  IRAS 04181+2655   &                                 ... &                                 ... &                                    ... &    ... &       ...                                 &      ... &   ... &   ... &  ... \\
11-000 &  IRAS 04181+2654AB &                                 ... &                                 ... &                                    ... &    ... &       ...                                 &      ... &   ... &   ... &  ... \\
11-023 &  2M J04213459      & 0.31\hfill                          &  5.4\hfill{~\scriptsize( 3.8,10.0)} & -1.37\hfill{~\scriptsize(-3.00,-0.91)} &   0.49 &     0.043\hfill{~\scriptsize(0.04, 0.05)} &    -3.76 &  6.79 &  0.90 & 4    \\
01-028 &  IRAS 04187+1927   & 0.62\hfill{~\scriptsize(0.55,0.70)} &  6.3\hfill{~\scriptsize( 4.7,12.1)} & -1.02\hfill{~\scriptsize(-1.28,-0.84)} &   9.09 &     0.909\hfill{~\scriptsize(0.75, 1.15)} &      ... &  9.15 &  0.94 & 79   \\
11-037 &  CFHT-Tau 10       & 0.65\hfill                          & 10.0\hfill                          & -3.00\hfill                            &   0.16 &     0.015                                 &    -3.73 &  8.59 &   ... &  ... \\
11-000 &  2M J04215450+2652 & 0.54\hfill                          &  7.9\hfill                          & -3.00\hfill                            &    ... & $<$ 0.008                                 & $<$-3.17 &   ... &   ... &  ... \\
21-038 &  RY Tau            & 0.77\hfill{~\scriptsize(0.70,0.86)} &  2.0\hfill{~\scriptsize( 2.0, 4.2)} &  0.20\hfill{~\scriptsize(-0.03, 0.41)} &  49.07 &     5.520\hfill{~\scriptsize(4.82, 6.38)} &    -3.72 & 15.70 &  1.18 & 158  \\
21-039 &  HD 283572         & 0.08\hfill{~\scriptsize(0.07,0.08)} & 10.4\hfill{~\scriptsize(10.0,10.7)} & -0.94\hfill{~\scriptsize(-1.02,-0.87)} & 114.33 &    13.003\hfill{~\scriptsize(2.68,13.05)} &    -3.28 & 14.43 &  1.12 & 522  \\
01-045 &  T Tau N(+Sab)     & 0.27\hfill{~\scriptsize(0.27,0.28)} & 11.3\hfill{~\scriptsize(10.8,11.8)} & -0.65\hfill{~\scriptsize(-0.72,-0.58)} &  71.84 &     8.048\hfill{~\scriptsize(7.99, 8.15)} &    -3.63 & 17.62 &  1.21 & 627  \\
11-054 &  Haro 6-5 B        & 19.3\hfill{~\scriptsize(12.9,23.8)} &  6.6\hfill{~\scriptsize( 2.0,16.7)} & -3.00\hfill{~\scriptsize(-3.00,-2.30)} &  18.95 &    17.551\hfill{~\scriptsize(1.49,22.38)} &    -1.01 &  5.74 &  1.26 & 7    \\
11-057 &  FS Tau AC         & 1.42\hfill{~\scriptsize(1.34,1.48)} & 39.8\hfill{~\scriptsize(34.6,44.8)} & -2.90\hfill{~\scriptsize(-3.00,-1.53)} &  24.99 &     3.224\hfill{~\scriptsize(3.09, 3.36)} &    -2.58 & 32.76 &  1.13 & 155  \\
21-044 &  LkCa 21           & 0.08\hfill{~\scriptsize(0.06,0.10)} &  8.5\hfill{~\scriptsize( 6.0, 9.7)} & -2.21\hfill{~\scriptsize(-3.00,-1.31)} &   6.65 &     0.646\hfill{~\scriptsize(0.60, 0.71)} &    -3.57 &  8.18 &  1.07 & 82   \\
01-054 &  RX J0422.1+1934   & 0.28\hfill{~\scriptsize(0.27,0.28)} & 13.6\hfill{~\scriptsize(12.8,14.5)} & -1.64\hfill{~\scriptsize(-1.93,-1.22)} &  29.59 &     3.110\hfill{~\scriptsize(3.05, 3.12)} &      ... & 14.46 &  1.05 & 411  \\
01-062 &  2M J04221332+1934 & 0.37\hfill{~\scriptsize(0.19,0.60)} &  4.4\hfill{~\scriptsize( 2.5, 7.7)} & -3.00\hfill{~\scriptsize(-3.00,-1.63)} &   0.67 &     0.057\hfill{~\scriptsize(0.03, 0.22)} &    -3.07 &  3.92 &  1.06 & 6    \\
11-079 &  CFHT-Tau 21       & 1.19\hfill                          & 31.6\hfill{~\scriptsize(23.4,31.6)} & -1.00\hfill                            &   1.16 &     0.151\hfill{~\scriptsize(0.15, 0.15)} &    -3.99 & 33.60 &  0.73 & 8    \\
02-013 &  FV Tau AB         & 1.20\hfill{~\scriptsize(0.87,1.50)} &  2.0\hfill{~\scriptsize( 2.0,31.6)} &  0.32\hfill{~\scriptsize(-0.41, 1.00)} &   4.51 &     0.532\hfill{~\scriptsize(0.37, 0.72)} &    -3.94 & 18.76 &  1.06 & 13   \\
02-000 &  FV Tau/c AB       & 0.59\hfill                          & 10.0\hfill                          & -1.00\hfill                            &    ... & $<$ 0.047                                 & $<$-4.24 &   ... &   ... &  ... \\
02-016 &  KPNO-Tau 13       & 0.32\hfill{~\scriptsize(0.23,0.45)} &  5.0\hfill{~\scriptsize( 2.2, 8.1)} & -0.97\hfill{~\scriptsize(-1.93,-0.53)} &   1.42 &     0.138\hfill{~\scriptsize(0.10, 0.24)} &    -3.62 &  7.71 &  0.79 & 11   \\
02-000 &  DG Tau B          &                                 ... &                                 ... &                                    ... &    ... &       ...                                 &      ... &   ... &   ... &  ... \\
02-022 &  DG Tau A          &                                 ... &                                 ... &                                    ... &    ... &       ...                                 &      ... &   ... &   ... &  ... \\
02-000 &  KPNO-Tau 4        & 0.44\hfill                          &  7.9\hfill                          & -3.00\hfill                            &    ... & $<$ 0.013                                 & $<$-3.09 &   ... &   ... &  ... \\
02-000 &  IRAS 04248+2612AB &                                 ... &                                 ... &                                    ... &    ... &       ...                                 &      ... &   ... &   ... &  ... \\
15-020 &  JH 507            & 0.26\hfill{~\scriptsize(0.19,0.36)} &  3.8\hfill{~\scriptsize( 2.7, 5.2)} & -1.56\hfill{~\scriptsize(-1.97,-1.25)} &   5.20 &     0.455\hfill{~\scriptsize(0.34, 0.75)} &    -3.55 &  4.62 &  0.47 & 31   \\
13-004 &  GV Tau AB         &                                 ... &                                 ... &                                    ... &    ... &       ...                                 &      ... &   ... &   ... &  ... \\
13-000 &  IRAS 04264+2433   &                                 ... &                                 ... &                                    ... &    ... &       ...                                 &      ... &   ... &   ... &  ... \\
15-040 &  DH Tau AB         & 0.20\hfill{~\scriptsize(0.19,0.21)} & 11.5\hfill{~\scriptsize(11.0,12.1)} & -1.38\hfill{~\scriptsize(-1.49,-1.26)} &  80.60 &     8.458\hfill{~\scriptsize(8.23, 8.64)} &    -2.41 & 13.42 &  1.06 & 381  \\
15-042 &  DI Tau AB         & 0.14\hfill{~\scriptsize(0.13,0.16)} &  9.5\hfill{~\scriptsize( 8.4,11.3)} & -1.43\hfill{~\scriptsize(-2.01,-1.18)} &  15.39 &     1.568\hfill{~\scriptsize(1.51, 1.65)} &    -3.39 & 11.07 &  0.92 & 111  \\
15-044 &  KPNO-Tau 5        & 0.01\hfill                          &  7.9\hfill{~\scriptsize( 4.6,12.6)} & -3.00\hfill{~\scriptsize(-3.00,-0.66)} &   0.10 &     0.010\hfill{~\scriptsize(0.01, 0.01)} &    -3.95 &  6.88 &  1.11 & 2    \\
14-006 &  IQ Tau A          & 0.49\hfill{~\scriptsize(0.36,0.80)} & 10.1\hfill{~\scriptsize( 2.0,19.6)} & -0.95\hfill{~\scriptsize(-3.00,-0.06)} &   3.89 &     0.416\hfill{~\scriptsize(0.33, 1.17)} &    -3.91 & 14.06 &  1.57 & 9    \\
13-000 &  CFHT-Tau 20       & 0.65\hfill                          &  7.9\hfill                          & -3.00\hfill                            &    ... & $<$ 0.017                                 & $<$-4.49 &   ... &   ... &  ... \\
14-000 &  KPNO-Tau 6        & 0.16\hfill                          &  7.9\hfill                          & -3.00\hfill                            &    ... & $<$ 0.008                                 & $<$-3.17 &   ... &   ... &  ... \\
13-035 &  FX Tau AB         & 0.26\hfill{~\scriptsize(0.18,0.55)} &  7.9\hfill{~\scriptsize( 2.0,10.1)} & -3.00\hfill{~\scriptsize(-3.00,-0.94)} &   5.26 &     0.502\hfill{~\scriptsize(0.39, 2.36)} &    -3.89 &  6.86 &  0.98 & 5    \\
14-057 &  DK Tau AB         & 0.25\hfill{~\scriptsize(0.23,0.27)} &  9.2\hfill{~\scriptsize( 6.7,13.3)} & -0.81\hfill{~\scriptsize(-1.57,-0.52)} &   8.54 &     0.916\hfill{~\scriptsize(0.87, 0.96)} &    -3.74 & 13.90 &  0.86 & 106  \\
14-000 &  KPNO-Tau 7        & 0.01\hfill                          &  7.9\hfill                          & -3.00\hfill                            &    ... & $<$ 0.006                                 & $<$-3.33 &   ... &   ... &  ... \\
22-013 &  MHO 9             & 0.05\hfill{~\scriptsize(0.02,0.08)} &  6.8\hfill{~\scriptsize( 5.6, 7.9)} & -3.00\hfill{~\scriptsize(-3.00,-1.92)} &   0.86 &     0.080\hfill{~\scriptsize(0.07, 0.09)} &    -4.02 &  5.93 &  1.05 & 20   \\
22-021 &  MHO 4             & 0.16\hfill{~\scriptsize(0.11,0.22)} &  6.1\hfill{~\scriptsize( 3.8,11.2)} & -1.15\hfill{~\scriptsize(-3.00,-0.60)} &   1.25 &     0.123\hfill{~\scriptsize(0.10, 0.16)} &    -3.18 &  8.26 &  0.73 & 16   \\
22-040 &  L1551 IRS5        & 0.17\hfill{~\scriptsize(0.11,0.29)} & 31.6\hfill{~\scriptsize(21.4,31.6)} & -1.00\hfill                            &   0.17 &     0.021\hfill{~\scriptsize(0.02, 0.02)} &    -5.68 & 33.60 &  1.40 & 6    \\
22-042 &  LkHa 358          & 3.01\hfill{~\scriptsize(2.26,4.05)} &  3.0\hfill{~\scriptsize( 2.0,30.9)} & -1.00\hfill                            &   4.58 &     0.404\hfill{~\scriptsize(0.29, 0.53)} &    -3.75 &  4.99 &  0.40 & 11   \\
22-000 &  HH 30             &                                 ... &                                 ... &                                    ... &    ... &       ...                                 &      ... &   ... &   ... &  ... \\
22-043 &  HL Tau            & 2.81\hfill{~\scriptsize(2.59,3.04)} & 20.1\hfill{~\scriptsize(13.6,29.9)} & -1.40\hfill{~\scriptsize(-3.00,-0.80)} &  26.84 &     3.838\hfill{~\scriptsize(3.22, 4.73)} &    -3.19 & 21.80 &  1.14 & 87   \\
22-047 &  XZ Tau AB         & 0.28\hfill{~\scriptsize(0.23,0.33)} &  8.5\hfill{~\scriptsize( 6.0,11.5)} & -0.90\hfill{~\scriptsize(-1.45,-0.62)} &   9.14 &     0.962\hfill{~\scriptsize(0.86, 1.12)} &    -3.12 & 12.49 &  1.07 & 38   \\
22-056 &  L1551 NE          & 7.66\hfill{~\scriptsize(3.96,13.8)} & 31.6\hfill                          & -1.00\hfill                            &   0.46 &     0.059\hfill{~\scriptsize(0.03, 0.09)} &      ... & 33.60 &  0.10 & 1    \\
03-005 &  HK Tau AB         & 0.92\hfill{~\scriptsize(0.49,1.89)} & 63.1\hfill{~\scriptsize(14.1,63.1)} & -1.00\hfill                            &   0.53 &     0.079\hfill{~\scriptsize(0.06, 0.12)} &    -4.44 & 51.83 &  3.19 & 1    \\
22-070 &  V710 Tau BA       & 0.28\hfill{~\scriptsize(0.26,0.30)} &  7.9\hfill{~\scriptsize( 6.7, 8.8)} & -1.44\hfill{~\scriptsize(-1.71,-1.22)} &  13.81 &     1.378\hfill{~\scriptsize(1.32, 1.49)} &    -3.49 &  9.33 &  0.94 & 158  \\
19-009 &  JH 665            & 0.14\hfill{~\scriptsize(0.05,0.37)} &  5.0\hfill{~\scriptsize( 2.0,11.8)} & -0.94\hfill{~\scriptsize(-3.00,-0.26)} &   0.81 &     0.079\hfill{~\scriptsize(0.06, 0.23)} &    -4.10 &  7.78 &  0.58 & 3    \\
22-089 &  L1551 51          & 0.11\hfill{~\scriptsize(0.10,0.12)} &  7.7\hfill{~\scriptsize( 7.0, 8.4)} & -1.63\hfill{~\scriptsize(-1.88,-1.42)} &  18.67 &     1.841\hfill{~\scriptsize(1.78, 1.91)} &    -2.99 &  8.61 &  1.13 & 204  \\
22-097 &  V827 Tau          & 0.06\hfill{~\scriptsize(0.05,0.06)} &  9.4\hfill{~\scriptsize( 8.7,10.7)} & -0.74\hfill{~\scriptsize(-0.96,-0.65)} &  37.02 &     4.010\hfill{~\scriptsize(3.88, 4.01)} &    -3.02 & 14.64 &  1.11 & 273  \\
03-016 &  Haro 6-13         & 1.26\hfill{~\scriptsize(0.44,1.41)} &  2.5\hfill{~\scriptsize( 2.0,31.6)} & -1.00\hfill                            &   9.49 &     0.799\hfill{~\scriptsize(0.14, 0.91)} &    -4.01 &  4.29 &  1.04 & 6    \\
\hline
\end{tabular}
\normalsize
\end{table*}
 
\setcounter{table}{4}
\begin{table*}[h!]
\scriptsize
\caption{(Continued)}
\begin{tabular}{lllllrlrrrr}
\hline
\hline
XEST & Name & $N_{\rm H}$ {\scriptsize \hfill(1$\sigma$ range)}         &  $T_0$ \hfill{\scriptsize (1$\sigma$ range)} &  $\beta$ \hfill{\scriptsize (1$\sigma$ range)} & EM$_t^a$                & $L_{\rm X}^b$ \hfill{\scriptsize (range)}                    & log           & $T_{\rm av}$   & $\chi^2_{\rm red}$   & dof    \\
   &      & ($10^{22}$~cm$^{-2}$) &  (MK)   &          & ($10^{52})$         & ($10^{30}$~erg~s$^{-1}$)                 &  $L_{\rm X}/L_*$    & (MK)           &                      &  \\
\hline
22-100 &  V826 Tau          & 0.07\hfill{~\scriptsize(0.06,0.07)} &  9.1\hfill{~\scriptsize( 8.6, 9.4)} & -1.26\hfill{~\scriptsize(-1.36,-1.16)} &  44.06 &     4.523\hfill{~\scriptsize(4.42, 4.57)} &    -2.90 & 11.27 &  1.12 & 359  \\
22-101 &  MHO 5             & 0.09\hfill{~\scriptsize(0.06,0.14)} &  5.8\hfill{~\scriptsize( 4.2, 8.6)} & -1.62\hfill{~\scriptsize(-3.00,-1.09)} &   1.18 &     0.111\hfill{~\scriptsize(0.10, 0.14)} &    -3.58 &  6.63 &  0.95 & 31   \\
03-017 &  CFHT-Tau 7        & 0.01\hfill                          &  8.1\hfill{~\scriptsize( 5.9,12.0)} & -3.00\hfill{~\scriptsize(-3.00,-0.90)} &   0.21 &     0.020\hfill{~\scriptsize(0.02, 0.02)} &    -4.07 &  7.03 &  1.82 & 3    \\
03-019 &  V928 Tau AB       & 0.38\hfill{~\scriptsize(0.31,0.44)} &  5.0\hfill{~\scriptsize( 3.9, 7.0)} & -1.40\hfill{~\scriptsize(-1.78,-1.23)} &  11.21 &     1.046\hfill{~\scriptsize(0.81, 1.33)} &    -3.71 &  6.27 &  0.93 & 46   \\
03-022 &  FY Tau            & 0.31\hfill{~\scriptsize(0.28,0.36)} & 13.7\hfill{~\scriptsize( 8.7,19.6)} & -0.99\hfill{~\scriptsize(-2.31,-0.40)} &   7.30 &     0.807\hfill{~\scriptsize(0.76, 0.90)} &    -3.68 & 17.73 &  0.54 & 34   \\
03-023 &  FZ Tau            & 0.53\hfill{~\scriptsize(0.33,0.82)} &  4.8\hfill{~\scriptsize( 2.0,12.0)} & -1.55\hfill{~\scriptsize(-2.44,-1.05)} &   7.05 &     0.644\hfill{~\scriptsize(0.35, 2.91)} &    -3.77 &  5.65 &  0.70 & 16   \\
17-002 &  IRAS 04295+2251   & 3.75\hfill{~\scriptsize(2.74,5.26)} & 16.8\hfill{~\scriptsize( 5.2,27.0)} & -3.00\hfill{~\scriptsize(-3.00,-1.21)} &  14.56 &     1.489\hfill{~\scriptsize(0.82, 4.80)} &    -3.49 & 14.28 &  0.46 & 7    \\
19-049 &  UZ Tau E+W(AB)    & 0.44\hfill{~\scriptsize(0.29,0.53)} &  2.7\hfill{~\scriptsize( 2.0, 7.4)} & -0.75\hfill{~\scriptsize(-0.95,-0.58)} &   9.98 &     0.890\hfill{~\scriptsize(0.51, 1.35)} &    -3.59 &  5.44 &  0.74 & 61   \\
17-009 &  JH 112            & 0.72\hfill{~\scriptsize(0.57,0.86)} &  3.9\hfill{~\scriptsize( 2.0, 7.8)} & -0.90\hfill{~\scriptsize(-1.20,-0.61)} &   8.71 &     0.820\hfill{~\scriptsize(0.52, 1.48)} &    -3.54 &  6.58 &  0.64 & 22   \\
03-031 &  CFHT-Tau 5        & 0.84\hfill                          &  8.1\hfill{~\scriptsize( 3.5,11.5)} & -3.00\hfill{~\scriptsize(-3.00,-1.27)} &   1.91 &     0.181\hfill{~\scriptsize(0.14, 0.26)} &    -3.20 &  7.03 &  1.26 & 4    \\
04-003 &  CFHT-Tau 5        & 0.84\hfill{~\scriptsize(0.51,1.48)} & 11.9\hfill{~\scriptsize( 2.0,17.8)} & -3.00\hfill{~\scriptsize(-3.00,-1.07)} &   2.51 &     0.247\hfill{~\scriptsize(0.14, 2.28)} &    -3.07 & 10.14 &  0.63 & 7    \\
03-035 &  MHO 8             & 0.21\hfill{~\scriptsize(0.13,0.34)} &  4.0\hfill{~\scriptsize( 2.1, 6.1)} & -1.00\hfill                            &   0.70 &     0.065\hfill{~\scriptsize(0.05, 0.12)} &    -4.00 &  6.21 &  2.00 & 3    \\
04-009 &  MHO 8             & 0.68\hfill{~\scriptsize(0.46,0.82)} &  2.5\hfill{~\scriptsize( 2.0, 4.4)} & -3.00\hfill{~\scriptsize(-3.00,-2.51)} &   6.18 &     0.444\hfill{~\scriptsize(0.13, 0.93)} &    -3.17 &  2.43 &  0.97 & 5    \\
04-010 &  GH Tau AB         & 0.09\hfill                          & 12.6\hfill{~\scriptsize( 9.2,18.8)} & -1.00\hfill                            &   1.00 &     0.109\hfill{~\scriptsize(0.10, 0.12)} &    -4.46 & 16.59 &  0.71 & 7    \\
04-012 &  V807 Tau SNab     & 0.10\hfill{~\scriptsize(0.08,0.12)} &  5.4\hfill{~\scriptsize( 4.4, 6.4)} & -1.18\hfill{~\scriptsize(-1.40,-0.99)} &  10.92 &     1.049\hfill{~\scriptsize(0.97, 1.15)} &    -3.89 &  7.35 &  1.06 & 84   \\
18-004 &  KPNO-Tau 14       & 0.85\hfill{~\scriptsize(0.69,1.31)} & 12.6\hfill{~\scriptsize( 2.0,21.3)} & -1.00\hfill{~\scriptsize(-2.30,-0.40)} &   8.44 &     0.923\hfill{~\scriptsize(0.76, 3.08)} &    -2.66 & 16.50 &  1.05 & 27   \\
17-000 &  CFHT-Tau 12       & 0.61\hfill                          &  7.9\hfill                          & -3.00\hfill                            &    ... & $<$ 0.007                                 & $<$-4.30 &   ... &   ... &  ... \\
04-016 &  V830 Tau          & 0.05\hfill{~\scriptsize(0.04,0.05)} &  9.2\hfill{~\scriptsize( 8.4, 9.6)} & -0.53\hfill{~\scriptsize(-0.64,-0.42)} &  46.62 &     5.181\hfill{~\scriptsize(5.07, 5.23)} &    -2.76 & 16.23 &  1.22 & 309  \\
18-000 &  IRAS S04301+261   & 1.13\hfill                          & 10.0\hfill                          & -1.00\hfill                            &    ... & $<$ 0.023                                 & $<$-3.62 &   ... &   ... &  ... \\
17-000 &  IRAS 04302+2247   &                                 ... &                                 ... &                                    ... &    ... &       ...                                 &      ... &   ... &   ... &  ... \\
17-027 &  IRAS 04303+2240   & 1.66\hfill{~\scriptsize(1.44,1.79)} &  2.0\hfill{~\scriptsize( 2.0, 6.8)} &  0.07\hfill{~\scriptsize(-0.10, 0.48)} &  41.37 &     5.006\hfill{~\scriptsize(3.50, 5.97)} &    -3.23 & 12.95 &  0.95 & 114  \\
04-034 &  GI Tau            & 0.38\hfill{~\scriptsize(0.33,0.45)} &  4.0\hfill{~\scriptsize( 2.0, 6.8)} & -0.19\hfill{~\scriptsize(-0.33,-0.05)} &   7.76 &     0.833\hfill{~\scriptsize(0.73, 1.06)} &    -3.66 & 12.52 &  0.76 & 49   \\
04-035 &  GK Tau AB         & 0.45\hfill{~\scriptsize(0.41,0.71)} &  7.8\hfill{~\scriptsize( 2.0, 9.5)} & -1.03\hfill{~\scriptsize(-1.31,-0.72)} &  14.24 &     1.471\hfill{~\scriptsize(1.33, 4.41)} &    -3.56 & 10.92 &  1.02 & 80   \\
18-019 &  IS Tau AB         & 0.40\hfill{~\scriptsize(0.35,0.50)} & 14.2\hfill{~\scriptsize( 8.4,16.6)} & -2.49\hfill{~\scriptsize(-3.00,-1.17)} &   6.48 &     0.658\hfill{~\scriptsize(0.60, 0.85)} &    -3.59 & 12.83 &  1.07 & 56   \\
17-058 &  CI Tau            & 0.55\hfill{~\scriptsize(0.38,1.18)} & 17.7\hfill{~\scriptsize( 2.0,31.6)} & -1.00\hfill                            &   1.69 &     0.195\hfill{~\scriptsize(0.16, 0.89)} &    -4.23 & 21.78 &  0.68 & 9    \\
18-030 &  IT Tau AB         & 0.69\hfill{~\scriptsize(0.66,0.72)} & 22.6\hfill{~\scriptsize(17.1,34.5)} & -0.86\hfill{~\scriptsize(-3.00,-0.38)} &  53.13 &     6.492\hfill{~\scriptsize(6.28, 6.73)} &    -3.15 & 27.22 &  0.92 & 427  \\
17-066 &  JH 108            & 0.20\hfill{~\scriptsize(0.17,0.24)} &  9.4\hfill{~\scriptsize( 6.5,12.5)} & -0.85\hfill{~\scriptsize(-1.49,-0.44)} &  11.48 &     1.233\hfill{~\scriptsize(1.14, 1.36)} &    -2.97 & 13.93 &  1.07 & 38   \\
17-068 &  CFHT-BD Tau 1     & 0.56\hfill                          & 17.0\hfill{~\scriptsize(12.6,25.7)} & -1.00\hfill                            &   1.48 &     0.169\hfill{~\scriptsize(0.17, 0.17)} &    -2.59 & 21.04 &  1.10 & 4    \\
25-026 &  AA Tau            & 1.09\hfill{~\scriptsize(0.96,1.20)} & 25.7\hfill{~\scriptsize(20.6,31.0)} & -3.00\hfill{~\scriptsize(-3.00,-1.79)} &  11.00 &     1.241\hfill{~\scriptsize(1.11, 1.36)} &    -3.39 & 21.55 &  0.91 & 42   \\
09-010 &  HO Tau AB         & 0.20\hfill                          &  5.3\hfill{~\scriptsize( 3.7, 8.0)} & -1.00\hfill                            &   0.48 &     0.047\hfill{~\scriptsize(0.05, 0.05)} &    -4.14 &  8.01 &  1.06 & 5    \\
08-019 &  FF Tau AB         & 0.31\hfill{~\scriptsize(0.26,0.41)} &  8.0\hfill{~\scriptsize( 4.7,10.6)} & -1.50\hfill{~\scriptsize(-2.47,-1.00)} &   7.99 &     0.796\hfill{~\scriptsize(0.69, 1.12)} &    -3.52 &  9.21 &  1.07 & 59   \\
12-040 &  DN Tau            & 0.07\hfill{~\scriptsize(0.07,0.08)} & 10.5\hfill{~\scriptsize( 9.5,11.5)} & -1.25\hfill{~\scriptsize(-1.51,-1.02)} &  11.03 &     1.155\hfill{~\scriptsize(1.14, 1.17)} &    -3.52 & 12.93 &  1.12 & 227  \\
12-000 &  IRAS 04325+2402AB &                                 ... &                                 ... &                                    ... &    ... &       ...                                 &      ... &   ... &   ... &  ... \\
12-059 &  CoKu Tau 3 AB     & 0.38\hfill{~\scriptsize(0.37,0.39)} & 14.2\hfill{~\scriptsize(13.3,15.0)} & -2.07\hfill{~\scriptsize(-2.36,-1.80)} &  56.63 &     5.851\hfill{~\scriptsize(5.71, 5.94)} &    -2.81 & 13.80 &  1.06 & 388  \\
09-022 &  KPNO-Tau 8        & 0.11\hfill{~\scriptsize(0.08,0.16)} &  3.4\hfill{~\scriptsize( 2.0, 5.8)} & -0.11\hfill{~\scriptsize(-0.23, 0.01)} &   4.66 &     0.504\hfill{~\scriptsize(0.45, 0.60)} &    -2.21 & 12.62 &  0.75 & 59   \\
08-037 &  HQ Tau AB         & 0.40\hfill{~\scriptsize(0.37,0.45)} &  8.0\hfill{~\scriptsize( 6.4, 9.6)} & -1.16\hfill{~\scriptsize(-1.43,-0.99)} &  24.25 &     2.476\hfill{~\scriptsize(2.29, 2.84)} &      ... & 10.57 &  1.07 & 135  \\
09-026 &  HQ Tau AB         & 0.56\hfill{~\scriptsize(0.52,0.61)} &  6.3\hfill{~\scriptsize( 4.7, 7.5)} & -0.84\hfill{~\scriptsize(-0.95,-0.71)} &  79.94 &     8.161\hfill{~\scriptsize(7.43, 9.56)} &      ... & 10.05 &  0.84 & 166  \\
08-043 &  KPNO-Tau 15       & 0.38\hfill{~\scriptsize(0.36,0.43)} & 13.7\hfill{~\scriptsize( 7.1,15.0)} & -0.33\hfill{~\scriptsize(-0.45, 0.10)} &  21.46 &     2.624\hfill{~\scriptsize(2.49, 2.78)} &    -2.31 & 23.46 &  0.91 & 248  \\
09-031 &  KPNO-Tau 15       & 0.37\hfill{~\scriptsize(0.26,0.52)} &  6.3\hfill{~\scriptsize( 3.6,10.1)} & -1.07\hfill{~\scriptsize(-2.05,-0.59)} &   4.85 &     0.483\hfill{~\scriptsize(0.37, 0.77)} &    -3.05 &  8.92 &  1.22 & 11   \\
08-000 &  KPNO-Tau 9        & 0.01\hfill                          &  7.9\hfill                          & -3.00\hfill                            &    ... & $<$ 0.004                                 & $<$-3.13 &   ... &   ... &  ... \\
09-000 &  KPNO-Tau 9        & 0.01\hfill                          &  7.9\hfill                          & -3.00\hfill                            &    ... & $<$ 0.007                                 & $<$-2.89 &   ... &   ... &  ... \\
08-048 &  HP Tau AB         & 0.53\hfill{~\scriptsize(0.48,0.60)} &  8.8\hfill{~\scriptsize( 6.6,11.4)} & -1.09\hfill{~\scriptsize(-1.44,-0.89)} &  24.55 &     2.550\hfill{~\scriptsize(2.28, 3.03)} &    -3.33 & 11.81 &  1.07 & 129  \\
08-051a&  HP Tau/G3 AB      & 0.41\hfill{~\scriptsize(0.39,0.42)} &  9.2\hfill{~\scriptsize( 8.6, 9.9)} & -1.37\hfill{~\scriptsize(-1.46,-1.28)} &  12.85 &     1.293\hfill{~\scriptsize(1.09, 1.39)} &    -3.33 & 11.03 &  1.04 & 382  \\
08-051 &  HP Tau/G2         & 0.41\hfill{~\scriptsize(0.39,0.42)} &  9.2\hfill{~\scriptsize( 8.6, 9.9)} & -1.37\hfill{~\scriptsize(-1.46,-1.28)} &  94.61 &     9.653\hfill{~\scriptsize(9.15, 9.90)} &    -3.41 & 11.03 &  1.04 & 382  \\
08-058 &  Haro 6-28 AB      & 0.38\hfill{~\scriptsize(0.33,0.47)} & 10.0\hfill{~\scriptsize( 6.5,12.2)} & -2.10\hfill{~\scriptsize(-3.00,-1.16)} &   2.49 &     0.248\hfill{~\scriptsize(0.22, 0.34)} &    -3.27 &  9.80 &  0.85 & 32   \\
08-000 &  CFHT-BD Tau 2     & 0.01\hfill                          &  7.9\hfill                          & -3.00\hfill                            &    ... & $<$ 0.004                                 & $<$-4.02 &   ... &   ... &  ... \\
08-080 &  CFHT-BD Tau 3     & 0.18\hfill                          &  7.9\hfill{~\scriptsize( 4.0,17.8)} & -3.00\hfill                            &   0.13 &     0.012\hfill{~\scriptsize(0.01, 0.01)} &    -3.36 &  6.88 &  0.16 & 1    \\
05-005 &  CFHT-Tau 6        & 0.47\hfill{~\scriptsize(0.33,0.64)} &  7.9\hfill                          & -3.00\hfill                            &   0.75 &     0.071\hfill{~\scriptsize(0.04, 0.09)} &    -3.11 &  6.88 &  1.53 & 3    \\
05-000 &  IRAS 04361+2547   &                                 ... &                                 ... &                                    ... &    ... &       ...                                 &      ... &   ... &   ... &  ... \\
05-013 &  GN Tau AB         & 2.34\hfill{~\scriptsize(1.48,2.80)} &  4.2\hfill{~\scriptsize( 2.0,17.7)} & -1.00\hfill                            &   8.35 &     0.785\hfill{~\scriptsize(0.21, 1.00)} &    -3.55 &  6.52 &  1.33 & 6    \\
05-017 &  IRAS 04365+2535   & 30.7\hfill{~\scriptsize(20.0,51.8)} & 10.0\hfill                          & -1.00\hfill                            &  18.70 &     1.986\hfill{~\scriptsize(1.23, 3.52)} &    -3.63 & 13.65 &  0.28 & 2    \\
05-024 &  IRAS 04369+2539   & 7.69\hfill{~\scriptsize(5.40,8.71)} &  6.4\hfill{~\scriptsize( 2.0,31.6)} & -1.00\hfill                            &  46.50 &     4.671\hfill{~\scriptsize(1.18, 5.55)} &    -4.23 &  9.32 &  1.12 & 7    \\
07-011 &  JH 223            & 0.09\hfill                          &  9.3\hfill{~\scriptsize( 7.6,10.6)} & -3.00\hfill{~\scriptsize(-3.00,-1.74)} &   0.66 &     0.064\hfill{~\scriptsize(0.06, 0.07)} &    -4.03 &  8.00 &  1.22 & 10   \\
07-022 &  Haro 6-32         & 0.08\hfill{~\scriptsize(0.03,0.14)} & 12.0\hfill{~\scriptsize( 5.8,14.7)} & -3.00\hfill{~\scriptsize(-3.00,-0.76)} &   1.02 &     0.101\hfill{~\scriptsize(0.09, 0.12)} &    -3.66 & 10.28 &  0.50 & 5    \\
07-000 &  ITG 33 A          &                                 ... &                                 ... &                                    ... &    ... &       ...                                 &      ... &   ... &   ... &  ... \\
07-000 &  CFHT-Tau 8        & 0.32\hfill                          &  7.9\hfill                          & -3.00\hfill                            &    ... & $<$ 0.008                                 & $<$-4.06 &   ... &   ... &  ... \\
07-000 &  IRAS 04381+2540   &                                 ... &                                 ... &                                    ... &    ... &       ...                                 &      ... &   ... &   ... &  ... \\
07-041 &  IRAS 04385+2550AB & 0.80\hfill{~\scriptsize(0.69,1.02)} & 31.6\hfill{~\scriptsize(17.9,31.6)} & -1.00\hfill                            &   3.08 &     0.401\hfill{~\scriptsize(0.37, 0.50)} &    -3.24 & 33.60 &  0.74 & 10   \\
10-017 &  CoKuLk332/G2 AB   & 0.53\hfill{~\scriptsize(0.49,0.59)} &  6.3\hfill{~\scriptsize( 4.8, 7.3)} & -1.14\hfill{~\scriptsize(-1.26,-1.02)} &  32.92 &     3.257\hfill{~\scriptsize(2.91, 3.94)} &    -3.11 &  8.59 &  0.83 & 163  \\
\hline
\end{tabular}
\normalsize
\end{table*}
 
\setcounter{table}{4}
\begin{table*}[h!]
\scriptsize
\caption{(Continued)}
\begin{tabular}{lllllrlrrrr}
\hline
\hline
XEST & Name & $N_{\rm H}$ {\scriptsize \hfill(1$\sigma$ range)}         &  $T_0$ \hfill{\scriptsize (1$\sigma$ range)} &  $\beta$ \hfill{\scriptsize (1$\sigma$ range)} & EM$_t^a$                & $L_{\rm X}^b$ \hfill{\scriptsize (range)}                    & log           & $T_{\rm av}$   & $\chi^2_{\rm red}$   & dof    \\
   &      & ($10^{22}$~cm$^{-2}$) &  (MK)   &          & ($10^{52})$         & ($10^{30}$~erg~s$^{-1}$)                 &  $L_{\rm X}/L_*$    & (MK)           &                      &  \\
\hline
10-018 &  CoKuLk332/G1 AB   & 0.33\hfill{~\scriptsize(0.24,0.58)} & 10.0\hfill{~\scriptsize( 2.0,15.5)} & -0.70\hfill{~\scriptsize(-1.73,-0.26)} &   4.50 &     0.493\hfill{~\scriptsize(0.42, 1.11)} &    -4.12 & 15.78 &  0.64 & 21   \\
10-020 &  V955 Tau AB       & 0.89\hfill{~\scriptsize(0.70,1.30)} &  9.5\hfill{~\scriptsize( 2.0,21.1)} & -0.83\hfill{~\scriptsize(-1.90,-0.41)} &  15.07 &     1.622\hfill{~\scriptsize(1.23, 4.55)} &    -3.38 & 14.25 &  0.91 & 36   \\
10-034 &  CIDA 7            & 0.32\hfill                          &  4.3\hfill{~\scriptsize( 2.4, 7.7)} & -1.88\hfill{~\scriptsize(-3.00,-0.90)} &   0.46 &     0.040\hfill{~\scriptsize(0.04, 0.05)} &    -3.68 &  4.70 &  0.76 & 6    \\
10-045 &  DP Tau            &                                 ... &                                 ... &                                    ... &    ... &       ...                                 &      ... &   ... &   ... &  ... \\
10-060 &  GO Tau            & 0.31\hfill{~\scriptsize(0.25,0.44)} &  6.2\hfill{~\scriptsize( 2.0,10.2)} & -0.23\hfill{~\scriptsize(-0.71, 0.13)} &   2.24 &     0.249\hfill{~\scriptsize(0.22, 0.36)} &    -3.76 & 15.54 &  1.56 & 13   \\
26-012 &  2M J04552333+30   & 0.01\hfill                          & 12.6\hfill{~\scriptsize( 8.3,18.6)} & -3.00\hfill                            &   0.13 &     0.013\hfill{~\scriptsize(0.01, 0.01)} &    -3.65 & 10.75 &  1.55 & 2    \\
26-034 &  2M J04554046+30   & 0.05\hfill                          & 10.2\hfill{~\scriptsize( 6.5,17.0)} & -3.00\hfill                            &   0.11 &     0.011\hfill{~\scriptsize(0.01, 0.01)} &    -3.87 &  8.79 &  0.52 & 1    \\
26-043 &  AB Aur            & 0.05\hfill{~\scriptsize(0.03,0.07)} &  4.8\hfill{~\scriptsize( 4.3, 5.5)} & -1.55\hfill{~\scriptsize(-1.78,-1.41)} &   3.80 &     0.349\hfill{~\scriptsize(0.32, 0.38)} &    -5.73 &  5.70 &  1.25 & 125  \\
26-050 &  2MJ04554757/801   & 0.10\hfill{~\scriptsize(0.06,0.15)} & 10.8\hfill{~\scriptsize( 8.3,12.3)} & -3.00\hfill{~\scriptsize(-3.00,-1.76)} &   0.71 &     0.070\hfill{~\scriptsize(0.06, 0.08)} &    -3.78 &  9.29 &  1.01 & 21   \\
26-067 &  SU Aur            & 0.47\hfill{~\scriptsize(0.43,0.48)} &  6.4\hfill{~\scriptsize( 6.2, 7.6)} & -1.11\hfill{~\scriptsize(-1.21,-1.06)} &  95.36 &     9.464\hfill{~\scriptsize(8.42, 9.70)} &    -3.61 &  8.89 &  1.56 & 397  \\
26-072 &  HBC 427           & 0.04\hfill{~\scriptsize(0.03,0.05)} &  9.3\hfill{~\scriptsize( 7.8,10.8)} & -0.81\hfill{~\scriptsize(-1.08,-0.57)} &  33.01 &     3.558\hfill{~\scriptsize(3.39, 3.57)} &    -3.08 & 14.11 &  0.90 & 107  \\
\hline
\multicolumn{11}{l}{Additional sources from Chandra}\\
\hline
C1-0   &  KPNO-Tau 10       &                                 ... &                                 ... &                                    ... &    ... &       ...                                 &      ... &   ... &   ... &  ... \\
C1-1   &  IRAS 04158+2805   & 4.36\hfill{~\scriptsize(2.58,6.12)} &  2.3\hfill{~\scriptsize( 2.0,31.6)} &  0.04\hfill{~\scriptsize(-3.00, 1.00)} &   8.10 &     0.882                                 &    -2.34 & 13.10 &  0.88 & 6    \\
C2-1   &  Haro 6-5 B        &                                 ... &                                 ... &                                    ... &    ... &       ...                                 &      ... &   ... &   ... &  ... \\
C2-2   &  FS Tau AC         & 1.10\hfill{~\scriptsize(0.95,1.38)} &  3.0\hfill{~\scriptsize( 2.0,31.6)} &  0.99\hfill{~\scriptsize( 0.26, 1.00)} &   2.84 &     0.391                                 &    -3.50 & 39.18 &  1.13 & 11   \\
C3-1   &  FV Tau/c AB       &                                 ... &                                 ... &                                    ... &    ... &       ...                                 &      ... &   ... &   ... &  ... \\
C3-2   &  DG Tau B          &                                 ... &                                 ... &                                    ... &    ... &       ...                                 &      ... &   ... &   ... &  ... \\
C4-1   &  GV Tau AB         &                                 ... &                                 ... &                                    ... &    ... &       ...                                 &      ... &   ... &   ... &  ... \\
C5-2   &  HN Tau AB         &                                 ... &                                 ... &                                    ... &    ... &       ...                                 &      ... &   ... &   ... &  ... \\
C5-1   &  L1551 55          &                                 ... &                                 ... &                                    ... &    ... &       ...                                 &      ... &   ... &   ... &  ... \\
C5-4   &  HD 28867          &                                 ... &                                 ... &                                    ... &    ... &       ...                                 &      ... &   ... &   ... &  ... \\
C5-3   &  DM Tau            &                                 ... &                                 ... &                                    ... &    ... &       ...                                 &      ... &   ... &   ... &  ... \\
C6-1   &  CFHT-BD Tau 4     &                                 ... &                                 ... &                                    ... &    ... &       ...                                 &      ... &   ... &   ... &  ... \\
C6-0   &  L1527 IRS         &                                 ... &                                 ... &                                    ... &    ... &       ...                                 &      ... &   ... &   ... &  ... \\
C6-0   &  CFHT-Tau 17       &                                 ... &                                 ... &                                    ... &    ... &       ...                                 &      ... &   ... &   ... &  ... \\
C6-2   &  IRAS 04370+2559   & 1.31\hfill{~\scriptsize(1.19,1.51)} & 31.6\hfill{~\scriptsize(23.4,31.6)} & -1.00\hfill                            &   5.94 &     0.773                                 &    -3.02 & 33.60 &  1.03 & 12   \\
\hline
\end{tabular}
\begin{minipage}{0.95\textwidth}
\footnotetext{
\hskip -0.5truecm $^a$ EM$_t$ is sum of EM over all DEM bins from log$T$ = 6.0 to log$T$ = 7.9; given in units of $10^{52}$~cm$^{-3}$\\
$^b$ $L_{\rm X}$ for [0.3,10]~keV, in units of $10^{30}$~erg~s$^{-1}$\\
NOTES on individual objects:
\begin{itemize}
\item For several detected,  faint {\it Chandra} sources, only 1-$T$ fits were derived; see Table~\ref{tab6}  
\item Fit performed with MOS spectra for: HBC 352 (XEST-27-115), CY Tau (23-002), CY Tau (24-002),  
FQ Tau (23-067),  2M-J04213459 (11-023), FX Tau (13-035), V827 Tau (22-097), FZ Tau (03-023), CFHT-Tau 5 (03-031), GH Tau (04-010), V955 Tau (10-020),
2M~J04552333+30, 2M~J04554046+30, 2M~J04554757/801, AB Aur, SU Aur, 
2M~J04552333+30 (26-012), 2M~J04554046+30 (26-034), 2M~J04554757/801 (26-050), AB Aur (26-043), SU Aur (26-067)
\item HP Tau/G2 and G3 = XEST-08-051 (separation: 10\arcsec) were fitted as one source. The ratio of the normalizations was derived from PSF fitting in the image
\item CW Tau = XEST-20-046: fit is of low quality. CW Tau may be a two-absorber X-ray source \citep{guedel06b} 
\item DG Tau A (XEST-02-022), GV Tau (XEST-13-004), and DP Tau  (XEST-10-045) require 2-$T$ fits with two absorbers; see Table~\ref{tab6}  
\item L1489 IRS = XEST-06-049 required  Fe = 0.46 (AG89)  for acceptable fit 
\item HD~283572 = XEST-21-039 required  Fe = 0.27 (AG89)  for acceptable fit 
\item IRAS 04303+2240 = XEST-17-027 required  Fe = 0.43 (AG89)  for acceptable fit 
\item HL Tau = XEST-22-043 required Fe = 0.77 (AG89) for acceptable fit    
\item L1551 IRS 5 = XEST-22-040: the lightly absorbed source cannot originate from heavily absorbed protostar
\item CFHT-Tau 10 = XEST-11-037: Too few counts in spectrum for reliable spectral fit. Adopted model parameters are characteristic for low-mass stars,
 and $N_{\rm H}$ has been derived from $A_{\rm V}$. $L_{\rm X}$ has been determined from count rate-to-flux conversion using spectral model in XSPEC
\end{itemize}
}
\end{minipage}
\label{tab5}
\normalsize
\end{table*}

\clearpage
 
\setcounter{table}{5}
\begin{table*}[h!]
\caption{X-ray parameters of targets in XEST (3): Plasma parameters from the 1-$T$ and 2-$T$ fits}
\begin{tabular}{lllrrrrrrrrr}
\hline
\hline
XEST & Name & $N_{\rm H}$  {\scriptsize \hfill(1$\sigma$ range)}              &  $T_1^a$  &  $T_2$    & EM$_1^b$               & EM$_2^b$             & $L_{\rm X}^c$                     & log      & $T_{\rm av}$ & $\chi^{2\ d}_{\rm red}$ & dof     \\
   &      & ($10^{22}$~cm$^{-2}$) &  (MK)   &  (MK)     & ($10^{52})$ & ($10^{52})$       & ($10^{30})$                & $L_{\rm X}/L_*$ & (MK)         &                      &          \\
\hline
27-115 &  HBC 352           & 0.19\hfill{\scriptsize~(0.17,0.21)} &   7.54 &   23.77 &  10.35 &  11.52 &  2.307 & -3.09 &  13.81 &  0.87  &  128 \\
27-000 &  HBC 353           &                                 ... &    ... &     ... &    ... &    ... &    ... &   ... &    ... & ...    &  ... \\
06-005 &  HBC 358 AB        & 0.00\hfill{\scriptsize~(0.00,0.03)} &   4.29 &   14.26 &   1.41 &   2.12 &  0.346 & -3.49 &   8.82 &  0.73  &   20 \\
06-007 &  HBC 359           & 0.00\hfill{\scriptsize~(0.00,0.02)} &   5.33 &   15.19 &   3.06 &   3.53 &  0.628 & -3.20 &   9.34 &  1.09  &   56 \\
06-059 &  L1489 IRS         & 6.54\hfill{\scriptsize~(6.08,6.88)} &    ... &   50.20 &    ... &  26.57 &  4.003 & -3.67 &  50.20 &  1.02  &  100 \\
20-001 &  LkCa 1            & 0.07\hfill{\scriptsize~(0.00,0.21)} &   1.39 &    8.35 &   1.76 &   0.82 &  0.205 & -3.85 &   2.46 &  1.27  &   15 \\
20-005 &  Anon 1            & 0.28\hfill{\scriptsize~(0.27,0.29)} &   8.58 &   24.12 &  19.05 &  14.11 &  3.473 & -3.46 &  13.32 &  1.02  &  228 \\
20-000 &  IRAS 04108+2803 A &                                 ... &    ... &     ... &    ... &    ... &    ... &   ... &    ... & ...    &  ... \\
20-022 &  IRAS 04108+2803 B & 7.86\hfill{\scriptsize~(3.40,28.0)} &    ... &   11.59 &    ... &   5.95 &  0.588 & -3.42 &  11.59 &  0.61  &    8 \\
20-000 &  2M J04141188+28   &                                 ... &    ... &     ... &    ... &    ... &    ... &   ... &    ... & ...    &  ... \\
20-042 &  V773 Tau ABC      & 0.14\hfill{\scriptsize~(0.14,0.15)} &   8.81 &   28.17 &  41.39 &  37.63 &  8.588 & -3.40 &  15.32 &  1.28  &  525 \\
20-043 &  FM Tau            & 0.14\hfill{\scriptsize~(0.12,0.15)} &   7.19 &   29.45 &   1.41 &   3.06 &  0.494 & -3.55 &  18.87 &  1.17  &   77 \\
20-046 &  CW Tau            &                                 ... &    ... &     ... &    ... &    ... &    ... &   ... &    ... & ...    &  ... \\
20-047 &  CIDA 1            & 0.34\hfill{\scriptsize~(0.04,1.10)} &   6.38 &     ... &   0.24 &    ... &  0.031 &   ... &   6.38 &  0.02  &    2 \\
20-056 &  MHO 2/1           & 1.01\hfill{\scriptsize~(0.84,1.15)} &   8.00 &   28.87 &   6.35 &  10.35 &  1.853 & -3.07 &  17.72 &  0.99  &   84 \\
20-058 &  MHO 3             & 0.50\hfill{\scriptsize~(0.40,0.69)} &  11.01 &   35.94 &   2.12 &   1.18 &  0.357 & -3.99 &  16.80 &  0.88  &   19 \\
20-069 &  FO Tau AB         & 0.18\hfill{\scriptsize~(0.10,0.50)} &    ... &   13.80 &    ... &   0.47 &  0.052 & -4.76 &  13.80 &  1.41  &    4 \\
20-073 &  CIDA 2            & 0.03\hfill{\scriptsize~(0.00,0.06)} &   7.77 &     ... &   1.18 &    ... &  0.115 & -4.03 &   7.77 &  0.79  &   19 \\
23-002 &  CY Tau            & 0.01\hfill{\scriptsize~(0.00,0.16)} &   7.77 &     ... &   1.18 &    ... &  0.108 & -4.25 &   7.77 &  1.85  &    9 \\
24-002 &  CY Tau            & 0.12\hfill{\scriptsize~(0.03,0.22)} &   8.81 &     ... &   2.12 &    ... &  0.202 & -3.98 &   8.81 &  1.19  &   12 \\
23-004 &  LkCa 5            & 0.03\hfill{\scriptsize~(0.01,0.04)} &   4.75 &   17.16 &   2.35 &   2.35 &  0.440 & -3.51 &   9.03 &  0.75  &   84 \\
24-004 &  LkCa 5            & 0.01\hfill{\scriptsize~(0.00,0.02)} &   8.35 &   28.29 &   3.06 &   2.35 &  0.588 & -3.38 &  14.19 &  1.10  &   71 \\
23-008 &  CIDA 3            & 0.71\hfill{\scriptsize~(0.52,0.84)} &    ... &   38.38 &    ... &   0.94 &  0.115 & -3.67 &  38.38 &  0.89  &    9 \\
24-008 &  CIDA 3            & 0.43\hfill{\scriptsize~(0.19,0.68)} &    ... &  126.38 &    ... &   1.18 &  0.226 & -3.38 & 126.38 &  0.62  &    7 \\
23-015 &  V410 X3           & 0.14\hfill{\scriptsize~(0.09,0.24)} &   4.41 &   63.07 &   0.49 &   0.16 &  0.071 & -3.66 &   8.58 &  0.65  &   12 \\
24-015 &  V410 X3           & 0.00\hfill{\scriptsize~(0.00,0.07)} &   7.88 &     ... &   0.31 &    ... &  0.031 & -4.02 &   7.88 &  0.78  &    6 \\
23-018 &  V410 A13          & 0.29\hfill{\scriptsize~(0.02,0.62)} & =10.00 &     ... &   0.05 &    ... &  0.006 & -4.40 &  10.00 &  0.50  &    3 \\
24-000 &  V410 A13          &                                 ... &    ... &     ... &    ... &    ... &    ... &   ... &    ... & ...    &  ... \\
23-000 &  V410 A24          &                                 ... &    ... &     ... &    ... &    ... &    ... &   ... &    ... & ...    &  ... \\
24-000 &  V410 A24          &                                 ... &    ... &     ... &    ... &    ... &    ... &   ... &    ... & ...    &  ... \\
23-029 &  V410 A25          & 0.71\hfill{\scriptsize~(0.60,0.80)} &   1.28 &   22.14 & 103.48 &   6.35 &  3.807 & -3.21 &   1.51 &  1.11  &   89 \\
24-027 &  V410 A25          & 0.65\hfill{\scriptsize~(0.51,0.73)} &   1.62 &   27.83 &  11.05 &   4.94 &  1.174 & -3.72 &   3.90 &  1.04  &   46 \\
23-032 &  V410 Tau ABC      & 0.00\hfill{\scriptsize~(0.00,0.01)} &   8.81 &   24.81 &  19.52 &  14.11 &  3.523 & -3.38 &  13.60 &  1.03  &  115 \\
24-028 &  V410 Tau ABC      & 0.02\hfill{\scriptsize~(0.01,0.02)} &   8.70 &   22.03 &  19.28 &  23.05 &  4.402 & -3.28 &  14.43 &  1.30  &  413 \\
23-033 &  DD Tau AB         & 0.28\hfill{\scriptsize~(0.22,0.54)} &   5.10 &  139.48 &   0.26 &   0.52 &  0.122 & -4.03 &  46.29 &  1.41  &    5 \\
24-029 &  DD Tau AB         & 0.27\hfill{\scriptsize~(0.18,0.40)} &    ... &   38.72 &    ... &   0.94 &  0.132 & -4.00 &  38.72 &  1.09  &    6 \\
23-035 &  CZ Tau AB         & 0.17\hfill{\scriptsize~(0.14,0.21)} &   7.42 &     ... &   2.12 &    ... &  0.205 & -3.71 &   7.42 &  0.86  &   29 \\
24-030 &  CZ Tau AB         & 0.19\hfill{\scriptsize~(0.15,0.23)} &   7.07 &     ... &   3.06 &    ... &  0.292 & -3.55 &   7.07 &  1.43  &   23 \\
23-036 &  IRAS 04154+2823   & 4.10\hfill{\scriptsize~(2.49,6.38)} &    ... &   67.83 &    ... &   1.18 &  0.176 & -3.45 &  67.83 &  0.98  &    7 \\
24-031 &  IRAS 04154+2823   & 7.67\hfill{\scriptsize~(1.15,25.8)} &    ... &   14.14 &    ... &   2.89 &  0.282 & -3.25 &  14.14 &  0.95  &    1 \\
23-037 &  V410 X2           & 1.23\hfill{\scriptsize~(1.14,1.31)} &   7.54 &   32.81 &  16.93 &   2.12 &  1.973 & -3.77 &   8.88 &  0.81  &   83 \\
24-032 &  V410 X2           & 1.43\hfill{\scriptsize~(1.12,1.67)} &   3.48 &   15.19 &  46.33 &   9.88 &  4.953 & -3.37 &   4.51 &  0.57  &   23 \\
23-045 &  V410 X4           & 0.86\hfill{\scriptsize~(0.71,1.05)} &    ... &   14.72 &    ... &   1.76 &  0.172 & -4.46 &  14.72 &  0.59  &   16 \\
24-038 &  V410 X4           & 0.56\hfill{\scriptsize~(0.28,1.12)} &    ... &   19.13 &    ... &   1.58 &  0.160 & -4.50 &  19.13 &  0.11  &    5 \\
23-047 &  V892 Tau          & 0.86\hfill{\scriptsize~(0.83,0.89)} &  10.55 &   28.87 &  32.69 &  39.04 &  7.923 & -4.57 &  18.25 &  1.15  &  372 \\
24-040 &  V892 Tau          & 1.25\hfill{\scriptsize~(1.12,1.40)} &   4.87 &   22.49 & 108.65 &  35.98 & 13.779 & -4.33 &   7.13 &  0.87  &   85 \\
23-048 &  LR 1              & 0.63\hfill{\scriptsize~(0.31,1.60)} &    ... &  277.10 &    ... &   0.24 &  0.045 & -4.60 & 277.10 &  0.30  &    3 \\
24-000 &  LR 1              &                                 ... &    ... &     ... &    ... &    ... &    ... &   ... &    ... & ...    &  ... \\
23-050 &  V410 X7           & 0.77\hfill{\scriptsize~(0.64,0.91)} &   9.97 &   18.09 &   3.06 &   3.29 &  0.626 & -3.49 &  13.58 &  0.73  &   28 \\
24-042 &  V410 X7           & 0.86\hfill{\scriptsize~(0.73,1.01)} &   7.88 &   30.72 &  10.58 &  20.70 &  3.584 & -2.73 &  19.39 &  0.90  &   81 \\
23-000 &  V410 A20          &                                 ... &    ... &     ... &    ... &    ... &    ... &   ... &    ... & ...    &  ... \\
24-000 &  V410 A20          &                                 ... &    ... &     ... &    ... &    ... &    ... &   ... &    ... & ...    &  ... \\
23-056 &  Hubble 4          & 0.31\hfill{\scriptsize~(0.30,0.33)} &   4.29 &   18.55 &  33.39 &  35.04 &  6.491 & -3.20 &   9.08 &  1.10  &  382 \\
24-047 &  Hubble 4          & 0.21\hfill{\scriptsize~(0.20,0.22)} &   8.70 &   22.96 &  20.46 &  19.52 &  4.167 & -3.40 &  13.97 &  1.04  &  279 \\
23-000 &  KPNO-Tau 2        &                                 ... &    ... &     ... &    ... &    ... &    ... &   ... &    ... & ...    &  ... \\
24-000 &  KPNO-Tau 2        &                                 ... &    ... &     ... &    ... &    ... &    ... &   ... &    ... & ...    &  ... \\
23-000 &  CoKu Tau 1        &                                 ... &    ... &     ... &    ... &    ... &    ... &   ... &    ... & ...    &  ... \\
24-000 &  CoKu Tau 1        &                                 ... &    ... &     ... &    ... &    ... &    ... &   ... &    ... & ...    &  ... \\
23-061 &  V410 X6           & 0.73\hfill{\scriptsize~(0.51,0.99)} &   3.25 &     ... &   6.35 &    ... &  0.541 & -3.15 &   3.25 &  0.64  &    7 \\
\hline
\end{tabular}
\normalsize
\end{table*}
 
 \setcounter{table}{5}
\begin{table*}[h!]
\caption{(Continued)}
\begin{tabular}{lllrrrrrrrrr}
\hline
\hline
XEST & Name & $N_{\rm H}$ \hfill{\scriptsize (1$\sigma$ range)}                 &  $T_1^a$  &  $T_2$    & EM$_1^b$               & EM$_2^b$                   & $L_{\rm X}^c$                     & log      & $T_{\rm av}$ & $\chi^{2\ d}_{\rm red}$ & dof    \\
     &      & ($10^{22}$~cm$^{-2}$) &  (MK)   &  (MK)     & ($10^{52}) $ & ($10^{52})$ & ($10^{30})$  & $L_{\rm X}/L_*$& (MK)         &                         &         \\
\hline
24-054 &  V410 X6           & 0.37\hfill{\scriptsize~(0.28,0.64)} &   7.19 &   50.20 &   1.65 &   0.47 &  0.214 & -3.56 &  11.07 &  1.21  &    8 \\
23-063 &  V410 X5           & 0.40\hfill{\scriptsize~(0.34,0.48)} &   6.96 &   18.20 &   1.65 &   1.65 &  0.322 & -3.00 &  11.25 &  1.08  &   31 \\
24-055 &  V410 X5           & 0.81\hfill{\scriptsize~(0.75,0.85)} &   4.64 &   55.19 &  15.05 &   3.06 &  1.823 & -2.24 &   7.05 &  0.88  &   40 \\
23-067 &  FQ Tau AB         & 0.45\hfill{\scriptsize~(0.26,0.80)} &   6.26 &     ... &   0.86 &    ... &  0.085 & -3.98 &   6.26 &  1.48  &    7 \\
24-058 &  FQ Tau AB         & 0.35\hfill                          &   6.84 &     ... &   0.41 &    ... &  0.040 & -4.31 &   6.84 &  0.88  &    1 \\
28-100 &  BP Tau            & 0.09\hfill{\scriptsize~(0.08,0.09)} &   4.75 &   22.26 &   9.64 &   7.53 &  1.482 & -3.39 &   9.35 &  1.10  &  427 \\
23-074 &  V819 Tau AB       & 0.21\hfill{\scriptsize~(0.20,0.23)} &   4.64 &   17.51 &  15.52 &   9.64 &  2.356 & -3.17 &   7.72 &  1.05  &  198 \\
24-061 &  V819 Tau AB       & 0.20\hfill{\scriptsize~(0.17,0.22)} &   4.17 &   14.72 &  11.29 &   9.64 &  1.945 & -3.26 &   7.46 &  0.85  &  109 \\
16-000 &  IRAS 04166+2706   &                                 ... &    ... &     ... &    ... &    ... &    ... &   ... &    ... & ...    &  ... \\
16-000 &  IRAS 04169+2702   &                                 ... &    ... &     ... &    ... &    ... &    ... &   ... &    ... & ...    &  ... \\
11-000 &  CFHT-Tau 19       &                                 ... &    ... &     ... &    ... &    ... &    ... &   ... &    ... & ...    &  ... \\
11-000 &  IRAS 04181+2655   &                                 ... &    ... &     ... &    ... &    ... &    ... &   ... &    ... & ...    &  ... \\
11-000 &  IRAS 04181+2654AB &                                 ... &    ... &     ... &    ... &    ... &    ... &   ... &    ... & ...    &  ... \\
11-023 &  2M J04213459      & 0.12\hfill{\scriptsize~(0.01,1.10)} &   9.97 &     ... &   0.24 &    ... &  0.024 & -4.02 &   9.97 &  0.63  &    3 \\
01-028 &  IRAS 04187+1927   & 0.68\hfill{\scriptsize~(0.64,0.80)} &   5.10 &   21.68 &   7.48 &   3.53 &  1.072 &   ... &   8.11 &  0.87  &   78 \\
11-037 &  CFHT-Tau 10       & 0.65\hfill                          & =10.00 &     ... &   0.07 &    ... &  0.007 & -4.06 &  10.00 &  0.36  &    2 \\
11-000 &  2M J04215450+2652 &                                 ... &    ... &     ... &    ... &    ... &    ... &   ... &    ... & ...    &  ... \\
21-038 &  RY Tau            & 0.76\hfill{\scriptsize~(0.66,0.85)} &   5.91 &   37.57 &  15.29 &  27.28 &  5.242 & -3.75 &  19.34 &  1.15  &  157 \\
21-039 &  HD 283572         & 0.08\hfill{\scriptsize~(0.08,0.08)} &   8.46 &   23.30 &  59.50 &  61.61 & 12.687 & -3.30 &  14.16 &  1.16  &  522 \\
01-045 &  T Tau N(+Sab)     & 0.34\hfill{\scriptsize~(0.33,0.35)} &   4.52 &   23.65 &  33.86 &  57.85 &  9.395 & -3.56 &  12.84 &  1.36  &  626 \\
11-054 &  Haro 6-5 B        & 22.0\hfill{\scriptsize~(14.4,33.6)} &    ... &   12.17 &    ... &  69.61 &  6.848 & -1.42 &  12.17 &  0.90  &    8 \\
11-057 &  FS Tau AC         & 1.70\hfill{\scriptsize~(1.53,1.87)} &   3.83 &   36.17 &  36.92 &  24.69 &  6.434 & -2.28 &   9.42 &  1.06  &  154 \\
21-044 &  LkCa 21           & 0.10\hfill{\scriptsize~(0.08,0.12)} &   4.29 &   13.22 &   4.00 &   3.53 &  0.698 & -3.53 &   7.27 &  1.13  &   81 \\
01-054 &  RX J0422.1+1934   & 0.27\hfill{\scriptsize~(0.27,0.28)} &   8.46 &   21.91 &  10.82 &  18.11 &  3.010 &   ... &  15.35 &  1.04  &  410 \\
01-062 &  2M J04221332+1934 & 0.40\hfill{\scriptsize~(0.17,0.65)} &   4.64 &    9.97 &   0.63 &   0.05 &  0.059 & -3.06 &   4.89 &  1.11  &    5 \\
11-079 &  CFHT-Tau 21       & 1.48\hfill{\scriptsize~(1.09,1.96)} &    ... &   30.26 &    ... &   1.41 &  0.169 & -3.94 &  30.26 &  0.69  &    7 \\
02-013 &  FV Tau AB         & 1.51\hfill{\scriptsize~(0.97,2.21)} &   3.94 &   39.65 &   8.94 &   2.82 &  1.176 & -3.59 &   6.86 &  1.04  &   12 \\
02-000 &  FV Tau/c AB       &                                 ... &    ... &     ... &    ... &    ... &    ... &   ... &    ... & ...    &  ... \\
02-016 &  KPNO-Tau 13       & 0.30\hfill{\scriptsize~(0.22,0.42)} &   6.03 &   21.33 &   0.82 &   0.42 &  0.122 & -3.68 &   9.26 &  0.71  &   10 \\
02-000 &  DG Tau B          &                                 ... &    ... &     ... &    ... &    ... &    ... &   ... &    ... & ...    &  ... \\
02-022 &  DG Tau A          & 0.11\hfill{\scriptsize~(0.08,0.14)} &   3.71 &   22.96 &   0.80 &   1.69 &  0.252 & -4.41 &  12.80 &  0.74  &   33 \\
02-000 &  KPNO-Tau 4        &                                 ... &    ... &     ... &    ... &    ... &    ... &   ... &    ... & ...    &  ... \\
02-000 &  IRAS 04248+2612AB &                                 ... &    ... &     ... &    ... &    ... &    ... &   ... &    ... & ...    &  ... \\
15-020 &  JH 507            & 0.27\hfill{\scriptsize~(0.21,0.34)} &   3.94 &   20.75 &   4.23 &   0.94 &  0.456 & -3.55 &   5.33 &  0.46  &   30 \\
13-004 &  GV Tau AB         & 0.12\hfill{\scriptsize~(0.01,0.79)} &   5.80 &   47.77 &   0.45 &   5.48 &  0.647 & -4.03 &  40.75 &  1.08  &    5 \\
13-000 &  IRAS 04264+2433   &                                 ... &    ... &     ... &    ... &    ... &    ... &   ... &    ... & ...    &  ... \\
15-040 &  DH Tau AB         & 0.18\hfill{\scriptsize~(0.18,0.19)} &   8.58 &   23.54 &  33.16 &  42.80 &  8.003 & -2.43 &  15.15 &  0.94  &  380 \\
15-042 &  DI Tau AB         & 0.12\hfill{\scriptsize~(0.11,0.14)} &   7.65 &   18.20 &   6.58 &   7.76 &  1.437 & -3.42 &  12.23 &  0.96  &  110 \\
15-044 &  KPNO-Tau 5        & 0.01\hfill                          &   6.61 &     ... &   0.09 &    ... &  0.009 & -3.99 &   6.61 &  0.44  &    3 \\
14-006 &  IQ Tau A          & 0.41\hfill{\scriptsize~(0.33,0.52)} &    ... &   14.26 &    ... &   3.29 &  0.313 & -4.03 &  14.26 &  1.57  &   10 \\
13-000 &  CFHT-Tau 20       &                                 ... &    ... &     ... &    ... &    ... &    ... &   ... &    ... & ...    &  ... \\
14-000 &  KPNO-Tau 6        &                                 ... &    ... &     ... &    ... &    ... &    ... &   ... &    ... & ...    &  ... \\
13-035 &  FX Tau AB         & 0.20\hfill{\scriptsize~(0.14,0.29)} &   8.46 &     ... &   0.39 &    ... &  0.393 & -4.00 &   8.46 &  0.68  &    6 \\
14-057 &  DK Tau AB         & 0.31\hfill{\scriptsize~(0.28,0.34)} &   4.17 &   18.67 &   4.70 &   6.58 &  1.061 & -3.67 &  10.00 &  0.83  &  105 \\
14-000 &  KPNO-Tau 7        &                                 ... &    ... &     ... &    ... &    ... &    ... &   ... &    ... & ...    &  ... \\
22-013 &  MHO 9             & 0.20\hfill{\scriptsize~(0.12,0.28)} &   1.04 &    5.45 &   6.11 &   1.18 &  0.235 & -3.56 &   1.36 &  0.77  &   19 \\
22-021 &  MHO 4             & 0.13\hfill{\scriptsize~(0.09,0.16)} &   5.10 &   13.45 &   0.52 &   0.54 &  0.101 & -3.26 &   8.37 &  0.75  &   14 \\
22-040 &  L1551 IRS5        & 0.18\hfill{\scriptsize~(0.10,0.27)} &    ... &   27.83 &    ... &   0.16 &  0.019 & -5.72 &  27.83 &  1.29  &    6 \\
22-042 &  LkHa 358          & 1.78\hfill{\scriptsize~(0.85,3.24)} &    ... &   36.64 &    ... &   0.49 &  0.066 & -4.54 &  36.64 &  0.44  &   11 \\
22-000 &  HH 30             &                                 ... &    ... &     ... &    ... &    ... &    ... &   ... &    ... & ...    &  ... \\
22-043 &  HL Tau            & 2.79\hfill{\scriptsize~(2.55,3.21)} &  22.38 &  159.42 &  21.64 &   2.12 &  3.285 & -3.25 &  26.66 &  1.14  &   85 \\
22-047 &  XZ Tau AB         & 0.24\hfill{\scriptsize~(0.21,0.27)} &   8.70 &   26.43 &   4.00 &   3.76 &  0.849 & -3.18 &  14.91 &  1.07  &   37 \\
22-056 &  L1551 NE          & 5.51\hfill{\scriptsize~(1.64,13.5)} &    ... &   43.13 &    ... &   0.33 &  0.045 &   ... &  43.13 &  0.03  &    2 \\
03-005 &  HK Tau AB         & 0.89\hfill{\scriptsize~(0.40,1.90)} &    ... &   79.54 &    ... &   0.47 &  0.080 & -4.43 &  79.54 &  2.46  &    2 \\
22-070 &  V710 Tau BA       & 0.23\hfill{\scriptsize~(0.22,0.25)} &   7.77 &   18.32 &   6.82 &   4.94 &  1.166 & -3.56 &  11.14 &  0.94  &  157 \\
19-009 &  JH 665            & 0.13\hfill{\scriptsize~(0.02,0.27)} &   6.84 &     ... &   0.71 &    ... &  0.059 & -4.23 &   6.84 &  0.99  &    4 \\
22-089 &  L1551 51          & 0.08\hfill{\scriptsize~(0.07,0.09)} &   7.30 &   16.81 &   9.88 &   6.35 &  1.611 & -3.05 &  10.12 &  1.14  &  203 \\
22-097 &  V827 Tau          & 0.05\hfill{\scriptsize~(0.04,0.05)} &   8.35 &   22.38 &  13.64 &  21.64 &  3.702 & -3.06 &  15.29 &  1.16  &  272 \\
03-016 &  Haro 6-13         & 0.54\hfill{\scriptsize~(0.41,0.72)} &    ... &   27.25 &    ... &   1.29 &  0.151 & -4.73 &  27.25 &  0.78  &    8 \\
\hline
\end{tabular}
\normalsize
\end{table*}
 
 \setcounter{table}{5}
\begin{table*}[h!]
\caption{(Continued)}
\begin{tabular}{lllrrrrrrrrr}
\hline
\hline
XEST & Name & $N_{\rm H}$ \hfill{\scriptsize (1$\sigma$ range)}                 &  $T_1^a$  &  $T_2$    & EM$_1^b$               & EM$_2^b$                   & $L_{\rm X}^c$                     & log      & $T_{\rm av}$ & $\chi^{2\ d}_{\rm red}$ & dof    \\
     &      & ($10^{22}$~cm$^{-2}$) &  (MK)   &  (MK)     & ($10^{52}) $ & ($10^{52})$ & ($10^{30})$  & $L_{\rm X}/L_*$& (MK)         &                         &         \\
\hline
22-100 &  V826 Tau          & 0.06\hfill{\scriptsize~(0.06,0.06)} &   7.54 &   17.97 &  19.52 &  22.34 &  4.210 & -2.93 &  11.99 &  1.37  &  373 \\
22-101 &  MHO 5             & 0.09\hfill{\scriptsize~(0.06,0.13)} &   2.55 &    8.46 &   0.28 &   0.80 &  0.108 & -3.59 &   6.19 &  1.06  &   30 \\
03-017 &  CFHT-Tau 7        & 0.47\hfill{\scriptsize~(0.24,1.46)} &   3.13 &     ... &   2.12 &    ... &  0.176 & -3.13 &   3.13 &  0.80  &    3 \\
03-019 &  V928 Tau AB       & 0.40\hfill{\scriptsize~(0.33,0.46)} &   4.64 &   20.99 &   9.17 &   2.59 &  1.096 & -3.69 &   6.47 &  0.92  &   45 \\
03-022 &  FY Tau            & 0.30\hfill{\scriptsize~(0.27,0.34)} &   9.28 &   27.94 &   2.35 &   4.47 &  0.769 & -3.70 &  19.11 &  0.99  &   33 \\
03-023 &  FZ Tau            & 0.28\hfill{\scriptsize~(0.22,0.47)} &    ... &   11.48 &    ... &   2.70 &  0.268 & -4.15 &  11.48 &  0.83  &   17 \\
17-002 &  IRAS 04295+2251   & 3.43\hfill{\scriptsize~(2.67,4.59)} &    ... &   20.17 &    ... &   9.88 &  1.025 & -3.65 &  20.17 &  0.33  &    8 \\
19-049 &  UZ Tau E+W(AB)    & 0.41\hfill{\scriptsize~(0.33,0.44)} &   3.83 &   19.01 &   4.94 &   3.06 &  0.736 & -3.67 &   7.07 &  0.77  &   60 \\
17-009 &  JH 112            & 0.90\hfill{\scriptsize~(0.68,1.02)} &   4.06 &   30.26 &  13.17 &   2.12 &  1.399 & -3.31 &   5.36 &  0.55  &   21 \\
03-031 &  CFHT-Tau 5        & 1.51\hfill{\scriptsize~(0.88,2.29)} &   4.41 &     ... &  10.82 &    ... &  0.962 & -2.48 &   4.41 &  0.88  &    4 \\
04-003 &  CFHT-Tau 5        & 0.33\hfill{\scriptsize~(0.21,0.55)} &    ... &   19.83 &    ... &   1.18 &  0.115 & -3.40 &  19.83 &  0.51  &    6 \\
03-035 &  MHO 8             & 0.29\hfill{\scriptsize~(0.18,0.58)} &   4.99 &     ... &   0.87 &    ... &  0.080 & -3.91 &   4.99 &  0.54  &    3 \\
04-009 &  MHO 8             & 0.72\hfill{\scriptsize~(0.50,0.97)} &   3.25 &     ... &   4.42 &    ... &  0.376 & -3.24 &   3.25 &  0.50  &    6 \\
04-010 &  GH Tau AB         & 0.11\hfill{\scriptsize~(0.06,0.19)} &    ... &   12.29 &    ... &   1.20 &  0.120 & -4.41 &  12.29 &  0.78  &    6 \\
04-012 &  V807 Tau  SNab    & 0.10\hfill{\scriptsize~(0.09,0.12)} &   4.52 &   17.04 &   6.82 &   4.23 &  1.030 & -3.89 &   7.51 &  1.22  &   83 \\
18-004 &  KPNO-Tau 14       & 0.63\hfill{\scriptsize~(0.59,0.70)} &    ... &   22.84 &    ... &   5.88 &  0.647 & -2.82 &  22.84 &  1.15  &   28 \\
17-000 &  CFHT-Tau 12       &                                 ... &    ... &     ... &    ... &    ... &    ... &   ... &    ... & ...    &  ... \\
04-016 &  V830 Tau          & 0.04\hfill{\scriptsize~(0.04,0.04)} &   8.12 &   22.49 &  15.05 &  30.57 &  4.807 & -2.80 &  16.07 &  1.24  &  308 \\
18-000 &  IRAS S04301+261   &                                 ... &    ... &     ... &    ... &    ... &    ... &   ... &    ... & ...    &  ... \\
17-000 &  IRAS 04302+2247   &                                 ... &    ... &     ... &    ... &    ... &    ... &   ... &    ... & ...    &  ... \\
17-027 &  IRAS 04303+2240   & 1.59\hfill{\scriptsize~(1.41,1.78)} &   4.06 &   39.19 &  32.22 &  19.99 &  5.548 & -3.18 &   9.67 &  0.89  &  114 \\
04-034 &  GI Tau            & 0.41\hfill{\scriptsize~(0.36,0.47)} &   4.87 &   32.58 &   4.23 &   3.76 &  0.872 & -3.64 &  11.91 &  0.75  &   48 \\
04-035 &  GK Tau AB         & 0.40\hfill{\scriptsize~(0.37,0.55)} &   8.81 &   28.87 &   7.29 &   4.47 &  1.244 & -3.64 &  13.83 &  1.04  &   79 \\
18-019 &  IS Tau AB         & 0.65\hfill{\scriptsize~(0.50,0.80)} &   4.41 &   17.86 &   9.41 &   4.70 &  1.310 & -3.29 &   7.03 &  1.08  &   55 \\
17-058 &  CI Tau            & 0.62\hfill{\scriptsize~(0.31,1.10)} &   7.88 &   43.71 &   0.94 &   1.65 &  0.329 & -4.01 &  23.44 &  0.52  &   10 \\
18-030 &  IT Tau AB         & 0.68\hfill{\scriptsize~(0.65,0.71)} &  11.71 &   37.33 &  11.76 &  38.57 &  6.307 & -3.17 &  28.47 &  0.90  &  426 \\
17-066 &  JH 108            & 0.17\hfill{\scriptsize~(0.15,0.21)} &   8.81 &   24.93 &   4.70 &   5.64 &  1.115 & -3.02 &  15.54 &  1.07  &   36 \\
17-068 &  CFHT-BD Tau 1     & 0.65\hfill{\scriptsize~(0.44,0.80)} &    ... &   15.77 &    ... &   1.88 &  0.174 & -2.58 &  15.77 &  1.38  &    2 \\
25-026 &  AA Tau            & 0.94\hfill{\scriptsize~(0.84,1.03)} &    ... &   27.13 &    ... &   8.94 &  1.039 & -3.47 &  27.13 &  0.82  &   43 \\
09-010 &  HO Tau AB         & 0.64\hfill{\scriptsize~(0.02,1.48)} &   3.25 &     ... &   0.47 &    ... &  0.289 & -3.36 &   3.25 &  0.36  &    3 \\
08-019 &  FF Tau AB         & 0.27\hfill{\scriptsize~(0.21,0.28)} &   7.42 &   15.07 &   3.53 &   3.29 &  0.658 & -3.61 &  10.45 &  1.13  &   58 \\
12-040 &  DN Tau            & 0.06\hfill{\scriptsize~(0.05,0.06)} &   8.81 &   23.42 &   5.17 &   5.17 &  1.072 & -3.56 &  14.36 &  1.14  &  226 \\
12-000 &  IRAS 04325+2402AB &                                 ... &    ... &     ... &    ... &    ... &    ... &   ... &    ... & ...    &  ... \\
12-059 &  CoKu Tau 3 AB     & 0.50\hfill{\scriptsize~(0.48,0.52)} &   4.41 &   18.67 &  41.86 &  43.98 &  8.207 & -2.66 &   9.24 &  1.01  &  387 \\
09-022 &  KPNO-Tau 8        & 0.11\hfill{\scriptsize~(0.09,0.13)} &   4.41 &   20.99 &   1.65 &   3.06 &  0.459 & -2.25 &  12.16 &  0.80  &   58 \\
08-037 &  HQ Tau AB         & 0.35\hfill{\scriptsize~(0.32,0.37)} &   8.70 &   24.58 &  11.99 &   8.00 &  2.074 &   ... &  13.18 &  1.11  &  134 \\
09-026 &  HQ Tau AB         & 0.66\hfill{\scriptsize~(0.62,0.70)} &   4.75 &   24.23 &  75.72 &  34.81 & 10.724 &   ... &   7.93 &  0.83  &  165 \\
08-043 &  KPNO-Tau 15       & 0.38\hfill{\scriptsize~(0.37,0.41)} &   8.81 &   33.51 &   5.64 &  15.52 &  2.526 & -2.33 &  23.47 &  0.92  &  260 \\
09-031 &  KPNO-Tau 15       & 0.39\hfill{\scriptsize~(0.25,0.55)} &   5.68 &   24.58 &   3.29 &   2.12 &  0.553 & -2.99 &  10.08 &  1.22  &   10 \\
08-000 &  KPNO-Tau 9        &                                 ... &    ... &     ... &    ... &    ... &    ... &   ... &    ... & ...    &  ... \\
09-000 &  KPNO-Tau 9        &                                 ... &    ... &     ... &    ... &    ... &    ... &   ... &    ... & ...    &  ... \\
08-048 &  HP Tau AB         & 0.48\hfill{\scriptsize~(0.45,0.51)} &   8.58 &   23.88 &  10.11 &  10.58 &  2.187 & -3.39 &  14.48 &  1.08  &  128 \\
08-051a&  HP Tau/G3 AB      & 0.53\hfill{\scriptsize~(0.52,0.56)} &   4.64 &   20.64 &  12.46 &   6.82 &  1.839 & -3.17 &   7.87 &  1.08  &  381 \\
08-051 &  HP Tau/G2         & 0.53\hfill{\scriptsize~(0.52,0.56)} &   4.64 &   20.64 &  93.83 &  51.27 & 13.830 & -3.26 &   7.86 &  1.08  &  381 \\
08-058 &  Haro 6-28 AB      & 0.34\hfill{\scriptsize~(0.29,0.45)} &    ... &   10.55 &    ... &   2.12 &  0.202 & -3.36 &  10.55 &  0.87  &   33 \\
08-000 &  CFHT-BD Tau 2     &                                 ... &    ... &     ... &    ... &    ... &    ... &   ... &    ... & ...    &  ... \\
08-080 &  CFHT-BD Tau 3     & 0.11\hfill{\scriptsize~(0.00,0.41)} & =10.00 &     ... &   0.09 &    ... &  0.009 & -3.49 &  10.00 &  0.12  &    1 \\
05-005 &  CFHT-Tau 6        & 0.32\hfill{\scriptsize~(0.18,0.46)} &   9.97 &     ... &   0.40 &    ... &  0.040 & -3.36 &   9.97 &  1.59  &    2 \\
05-000 &  IRAS 04361+2547   &                                 ... &    ... &     ... &    ... &    ... &    ... &   ... &    ... & ...    &  ... \\
05-013 &  GN Tau AB         & 2.42\hfill{\scriptsize~(1.28,3.33)} &    ... &   14.38 &    ... &   4.23 &  0.423 & -3.82 &  14.38 &  1.60  &    5 \\
05-017 &  IRAS 04365+2535   & 24.3\hfill{\scriptsize~(9.76,63.0)} &    ... &   49.16 &    ... &   3.29 &  0.499 & -4.23 &  49.16 &  0.48  &    1 \\
05-024 &  IRAS 04369+2539   & 6.15\hfill{\scriptsize~(4.93,7.81)} &    ... &   32.70 &    ... &  10.35 &  1.315 & -4.78 &  32.70 &  1.17  &    7 \\
07-011 &  JH 223            & 0.09\hfill{\scriptsize~(0.05,0.14)} &   8.70 &     ... &   0.71 &    ... &  0.061 & -4.06 &   8.70 &  1.04  &   10 \\
07-022 &  Haro 6-32         & 0.04\hfill{\scriptsize~(0.00,0.08)} &    ... &   11.25 &    ... &   0.94 &  0.085 & -3.74 &  11.25 &  0.47  &    6 \\
07-000 &  ITG 33 A          &                                 ... &    ... &     ... &    ... &    ... &    ... &   ... &    ... & ...    &  ... \\
07-000 &  CFHT-Tau 8        &                                 ... &    ... &     ... &    ... &    ... &    ... &   ... &    ... & ...    &  ... \\
07-000 &  IRAS 04381+2540   &                                 ... &    ... &     ... &    ... &    ... &    ... &   ... &    ... & ...    &  ... \\
07-041 &  IRAS 04385+2550AB & 0.76\hfill{\scriptsize~(0.60,0.97)} &    ... &   32.12 &    ... &   2.82 &  0.367 & -3.28 &  32.12 &  0.70  &    9 \\
10-017 &  CoKuLk332/G2 AB   & 0.63\hfill{\scriptsize~(0.59,0.67)} &   4.75 &   22.96 &  33.63 &  11.52 &  4.313 & -2.99 &   7.10 &  0.82  &  162 \\
\hline
\end{tabular}
\normalsize
\end{table*}
 
 \setcounter{table}{5}
\begin{table*}[h!]
\caption{(Continued)}
\begin{tabular}{lllrrrrrrrrr}
\hline
\hline
XEST & Name & $N_{\rm H}$ \hfill{\scriptsize (1$\sigma$ range)}                 &  $T_1^a$  &  $T_2$    & EM$_1^b$               & EM$_2^b$                   & $L_{\rm X}^c$                     & log      & $T_{\rm av}$ & $\chi^{2\ d}_{\rm red}$ & dof    \\
     &      & ($10^{22}$~cm$^{-2}$) &  (MK)   &  (MK)     & ($10^{52}) $ & ($10^{52})$ & ($10^{30})$  & $L_{\rm X}/L_*$& (MK)         &                         &         \\
\hline
10-018 &  CoKuLk332/G1 AB   & 0.23\hfill{\scriptsize~(0.15,0.31)} &  11.48 &   43.01 &   1.88 &   1.65 &  0.414 & -4.20 &  21.26 &  0.64  &   20 \\
10-020 &  V955 Tau AB       & 1.09\hfill{\scriptsize~(0.83,1.34)} &   4.17 &   25.86 &  27.99 &  10.11 &  3.617 & -3.03 &   6.77 &  1.00  &   33 \\
10-034 &  CIDA 7            & 0.88\hfill{\scriptsize~(0.58,1.11)} &   2.20 &     ... &  11.52 &    ... &  0.884 & -2.34 &   2.20 &  0.58  &    6 \\
10-045 &  DP Tau            & 0.00\hfill{\scriptsize~(0.00,1.05)} &   3.25 &   43.83 &   0.05 &   0.61 &  0.101 & -3.88 &  36.40 &  1.27  &    7 \\
10-060 &  GO Tau            & 0.35\hfill{\scriptsize~(0.26,0.55)} &   6.26 &   36.29 &   1.18 &   1.18 &  0.308 & -3.67 &  15.07 &  1.53  &   12 \\
26-012 &  2M J04552333+30   & 0.07\hfill{\scriptsize~(0.00,0.24)} & =10.00 &     ... &   0.15 &    ... &  0.015 & -3.59 &  10.00 &  1.07  &    2 \\
26-034 &  2M J04554046+30   & 0.04\hfill{\scriptsize~(0.00,0.18)} & =10.00 &     ... &   0.10 &    ... &  0.010 & -3.91 &  10.00 &  0.17  &    1 \\
26-043 &  AB Aur            & 0.06\hfill{\scriptsize~(0.04,0.07)} &   2.43 &    7.54 &   1.18 &   2.59 &  0.353 & -5.73 &   5.29 &  1.36  &  124 \\
26-050 &  2MJ04554757/801   & 0.06\hfill{\scriptsize~(0.02,0.12)} &   8.58 &   12.17 &   0.24 &   0.24 &  0.056 & -3.88 &  10.22 &  1.03  &   20 \\
26-067 &  SU Aur            & 0.56\hfill{\scriptsize~(0.54,0.60)} &   5.22 &   23.30 &  87.01 &  32.45 & 11.641 & -3.52 &   7.84 &  1.68  &  396 \\
26-072 &  HBC 427           & 0.02\hfill{\scriptsize~(0.01,0.03)} &   9.04 &   28.06 &  14.98 &  15.07 &  3.276 & -3.11 &  15.96 &  0.92  &  106 \\
\hline
\multicolumn{12}{l}{Additional sources from Chandra}\\
\hline
C1-0   &  KPNO-Tau 10       &                                 ... &    ... &     ... &    ... &    ... &    ... &   ... &    ... & ...    &  ... \\
C1-1   &  IRAS 04158+2805   & 3.24\hfill{\scriptsize~(1.95,5.47)} &    ... &   70.72 &    ... &   2.59 &  0.414 & -2.67 &  70.72 &  0.70  &    7 \\
C2-1   &  Haro 6-5 B        & 26.4\hfill{\scriptsize~(10.4,37.2)} &    ... &   15.54 &    ... &  43.27 &  3.774 & -1.68 &  15.54 & 142.3C &  512 \\
C2-2   &  FS Tau AC         & 0.82\hfill{\scriptsize~(0.58,1.05)} &    ... &   70.96 &    ... &   1.95 &  0.329 & -3.57 &  70.96 &  1.02  &   12 \\
C3-1   &  FV Tau/c AB       & 10.8\hfill{\scriptsize~(5.70,16.4)} &    ... &   16.46 &    ... &   4.94 &  0.430 & -3.27 &  16.46 &  96.3C &  512 \\
C3-2   &  DG Tau B          & 51.3\hfill{\scriptsize~(11.0,63.0)} &    ... &   11.59 &    ... & 111.94 &  9.689 & -3.34 &  11.59 & 70.20C &  512 \\
C4-1   &  GV Tau AB         &                                 ... &    ... &     ... &    ... &    ... &    ... &   ... &    ... & ...    &  ... \\
C5-2   &  HN Tau AB         & 0.20\hfill                          & =10.00 &     ... &    ... &    ... &  0.153 & -3.80 &  10.00 & ...    &  ... \\
C5-1   &  L1551 55          & 0.14\hfill                          & =10.00 &     ... &    ... &    ... &  0.597 & -3.38 &  10.00 & ...    &  ... \\
C5-4   &  HD 28867          & 0.05\hfill                          & =10.00 &     ... &    ... &    ... &  6.867 & -4.85 &  10.00 & ...    &  ... \\
C5-3   &  DM Tau            & 0.12\hfill                          & =10.00 &     ... &    ... &    ... &  0.181 & -3.80 &  10.00 & ...    &  ... \\
C6-1   &  CFHT-BD Tau 4     & 0.52\hfill                          &    ... &   14.26 &    ... &   0.94 &  0.096 & -3.40 &  14.26 & 0.087  &    3 \\
C6-0   &  L1527 IRS         &                                 ... &    ... &     ... &    ... &    ... &    ... &   ... &    ... & ...    &  ... \\
C6-0   &  CFHT-Tau 17       &                                 ... &    ... &     ... &    ... &    ... &    ... &   ... &    ... & ...    &  ... \\
C6-2   &  IRAS 04370+2559   & 1.09\hfill{\scriptsize~(0.90,1.33)} &    ... &   50.55 &    ... &   4.47 &  0.666 & -3.08 &  50.55 &  0.97  &   12 \\
\hline
\end{tabular}
\begin{minipage}{0.93\textwidth}
\footnotetext{
\hskip -0.5truecm $^a$ '=' sign before number indicates that parameter was fixed\\
$^b$ EM in units of $10^{52}$~cm$^{-3}$\\
$^c$ $L_{\rm X}$ for [0.3,10]~keV, in units of $10^{30}$~erg~s$^{-1}$\\
$^d$ Numbers followed by 'C' denote C statistic (for low-background {\it Chandra} data)\\
NOTES on individual objects: 
\begin{itemize}
\item Fit performed for MOS1 + MOS2 spectra for: HBC 352 (XEST-27-115), CY Tau (23-002), CY Tau (24-002), FQ Tau (23-067),	
      2M-J04213459 (11-023), FX Tau (13-035), V827 Tau (22-097), FZ Tau (03-023), CFHT-Tau 5 (03-031), GH Tau (04-010), 
      V955 Tau (10-020), 2M~J04552333+30 (26-012), 2M~J04554046+30 (26-034), 2M~J04554757/801 (26-050), AB Aur (26-043),  
      and SU Aur (26-067)  
\item HN Tau A (C5-2), L1551 55 (C5-1), DM Tau (C5-3), and HD~28867 (C5-4):  $T$ = 10~MK was adopted for {\it Chandra} HRC observations, and $N_{\rm H}$ was derived from $A_{\rm V}$ assuming
     standard gas-to-dust ratios
\item DG Tau A (XEST-02-022), GV Tau (XEST-13-004), and DP Tau (XEST-10-045) require different $N_{\rm H}$ for the two components (also CW Tau; see \citealt{guedel06b}). 
      $N_{\rm H}$ for cooler component is given. For hotter component: $N_{\rm H} = 1.8, 4.1,$ and $3.8$, respectively	
\item CW Tau = XEST-20-046:  spectrum cannot be reliably fitted. CW Tau may be two-absorber X-ray source \citep{guedel06b} 
\item HP Tau/G2 and G3 = XEST-08-051 (separation: 10\arcsec) were fitted as one source. The ratio of the normalizations was derived from PSF fitting in the image
\item BP Tau = XEST-28-100: for an acceptable fit, two abundances were adjusted: O = 0.267, Fe = 0.10 (AG89)  
\item LkCa 1 = XEST-20-001: for an acceptable fit, Fe was adjusted: Fe = 0.57 (AG89)    
\item HL Tau = XEST-22-043: for an acceptable fit, Fe was adjusted: Fe = 1.06 (AG89)    
\item L1551 IRS 5 = XEST-22-040: the lightly absorbed source cannot originate from heavily absorbed protostar
\end{itemize}
}
\end{minipage}
\label{tab6}
\normalsize
\end{table*}

\clearpage

\setcounter{table}{6}
\begin{table*}[h!]
\caption{Fundamental parameters of targets in XEST (1): Names and coordinates}
\begin{tabular}{llcrlll}
\hline
\hline
XEST & Name & 2MASS$^a$ & IRAS$^b$ & Alternative names  &  RA(J2000.0)$^c$   & Dec(J2000.0)$^c$                 \\
     &      &           &          &      & h\ \ m \ \ s  & $\deg\ \ \arcmin\ \ \arcsec$ \\
\hline
27-115 &  HBC 352           &  03542950+3203013  & ...          & NTTS 035120+3154SW                    & 3 54 29.51 & 32 03 01.4     \\
27-000 &  HBC 353           &  03543017+3203043  & ...          & NTTS 035120+3154NE                    & 3 54 30.17 & 32 03 04.3     \\
06-005 &  HBC 358 AB        &  04034930+2610520  & ...          & NTTS 040047+2603W                     & 4 03 49.31 & 26 10 52.0     \\
06-007 &  HBC 359           &  04035084+2610531  & ...          & TTS 040047+2603                       & 4 03 50.84 & 26 10 53.2     \\
06-059 &  L1489 IRS         &  04044307+2618563  & 04016+2610   & ...                                   & 4 04 43.07 & 26 18 56.4     \\
20-001 &  LkCa 1            &  04131414+2819108  & ...          & HBC 365, V1095 Tau, JH 141            & 4 13 14.14 & 28 19 10.8     \\
20-005 &  Anon 1            &  04132722+2816247  & ...          & HBC 366, V1096 Tau                    & 4 13 27.23 & 28 16 24.8     \\
20-000 &  IRAS 04108+2803 A &  04135328+2811233  & 04108+2803A  & L1495N IRS                            & 4 13 53.29 & 28 11 23.4     \\
20-022 &  IRAS 04108+2803 B &  04135471+2811328  & 04108+2803   & ...                                   & 4 13 54.72 & 28 11 32.9     \\
20-000 &  2M J04141188+28   &  04141188+2811535  & ...          & ...                                   & 4 14 11.88 & 28 11 53.5     \\
20-042 &  V773 Tau ABC      &  04141291+2812124  & 04111+2804   & HBC 367, HD 283447                    & 4 14 12.92 & 28 12 12.4     \\
20-043 &  FM Tau            &  04141358+2812492  & ...          & HBC 23, Haro 6-1                      & 4 14 13.58 & 28 12 49.2     \\
20-046 &  CW Tau            &  04141700+2810578  & 04112+2803   & HBC 25                                & 4 14 17.00 & 28 10 57.8     \\
20-047 &  CIDA 1            &  04141760+2806096  & ...          & C4101                                 & 4 14 17.61 & 28 06 09.7     \\
20-056 &  MHO 2/1           &  04142639+2805597  & 04113+2758   & ...                                   & 4 14 26.40 & 28 05 59.7     \\
20-058 &  MHO 3             &  04143054+2805147  & 04114+2757   & ...                                   & 4 14 30.55 & 28 05 14.7     \\
20-069 &  FO Tau AB         &  04144928+2812305  & 04117+2804   & HBC 369                               & 4 14 49.29 & 28 12 30.6     \\
20-073 &  CIDA 2            &  04150515+2808462  & ...          & C3601                                 & 4 15 05.16 & 28 08 46.2     \\
23-002 &  CY Tau            &  04173372+2820468  & C04144+2813  & HBC 28                                & 4 17 33.73 & 28 20 46.9     \\
24-002 &  CY Tau            &  04173372+2820468  & C04144+2813  & HBC 28                                & 4 17 33.73 & 28 20 46.9     \\
23-004 &  LkCa 5            &  04173893+2833005  & ...          & HBC 371, JH 153                       & 4 17 38.94 & 28 33 00.5     \\
24-004 &  LkCa 5            &  04173893+2833005  & ...          & HBC 371, JH 153                       & 4 17 38.94 & 28 33 00.5     \\
23-008 &  CIDA 3            &  04174965+2829362  & ...          & C2701, V410 X1, Kim3-76               & 4 17 49.65 & 28 29 36.3     \\
24-008 &  CIDA 3            &  04174965+2829362  & ...          & C2701, V410 X1, Kim3-76               & 4 17 49.65 & 28 29 36.3     \\
23-015 &  V410 X3           &  04180796+2826036  & ...          & ...                                   & 4 18 07.96 & 28 26 03.7     \\
24-015 &  V410 X3           &  04180796+2826036  & ...          & ...                                   & 4 18 07.96 & 28 26 03.7     \\
23-018 &  V410 A13          &  04181710+2828419  & ...          & MHO 10                                & 4 18 17.11 & 28 28 41.9     \\
24-000 &  V410 A13          &  04181710+2828419  & ...          & MHO 10                                & 4 18 17.11 & 28 28 41.9     \\
23-000 &  V410 A24          &  04182239+2824375  & ...          & ...                                   & 4 18 22.39 & 28 24 37.6     \\
24-000 &  V410 A24          &  04182239+2824375  & ...          & ...                                   & 4 18 22.39 & 28 24 37.6     \\
23-029 &  V410 A25          &  04182909+2826191  & ...          & ...                                   & 4 18 29.10 & 28 26 19.1     \\
24-027 &  V410 A25          &  04182909+2826191  & ...          & ...                                   & 4 18 29.10 & 28 26 19.1     \\
23-032 &  V410 Tau ABC      &  04183110+2827162  & C04152+2820  & HBC 29, HD 283518                     & 4 18 31.10 & 28 27 16.2     \\
24-028 &  V410 Tau ABC      &  04183110+2827162  & C04152+2820  & HBC 29, HD 283518                     & 4 18 31.10 & 28 27 16.2     \\
23-033 &  DD Tau AB         &  04183112+2816290  & 04154+2809   & HBC 30                                & 4 18 31.13 & 28 16 29.0     \\
24-029 &  DD Tau AB         &  04183112+2816290  & 04154+2809   & HBC 30                                & 4 18 31.13 & 28 16 29.0     \\
23-035 &  CZ Tau AB         &  04183158+2816585  & ...          & HBC 31                                & 4 18 31.59 & 28 16 58.5     \\
24-030 &  CZ Tau AB         &  04183158+2816585  & ...          & HBC 31                                & 4 18 31.59 & 28 16 58.5     \\
23-036 &  IRAS 04154+2823   &  04183203+2831153  & 04154+2823   & ...                                   & 4 18 32.03 & 28 31 15.4     \\
24-031 &  IRAS 04154+2823   &  04183203+2831153  & 04154+2823   & ...                                   & 4 18 32.03 & 28 31 15.4     \\
23-037 &  V410 X2           &  04183444+2830302  & ...          & ...                                   & 4 18 34.45 & 28 30 30.2     \\
24-032 &  V410 X2           &  04183444+2830302  & ...          & ...                                   & 4 18 34.45 & 28 30 30.2     \\
23-045 &  V410 X4           &  04184023+2824245  & ...          & ...                                   & 4 18 40.23 & 28 24 24.5     \\
24-038 &  V410 X4           &  04184023+2824245  & ...          & ...                                   & 4 18 40.23 & 28 24 24.5     \\
23-047 &  V892 Tau          &  04184061+2819155  & 04155+2812   & HBC 373, Elias 1                      & 4 18 40.62 & 28 19 15.5     \\
24-040 &  V892 Tau          &  04184061+2819155  & 04155+2812   & HBC 373, Elias 1                      & 4 18 40.62 & 28 19 15.5     \\
23-048 &  LR 1              &  04184133+2827250  & ...          & ...                                   & 4 18 41.33 & 28 27 25.0     \\
24-000 &  LR 1              &  04184133+2827250  & ...          & ...                                   & 4 18 41.33 & 28 27 25.0     \\
23-050 &  V410 X7           &  04184250+2818498  & ...          & MHO-11                                & 4 18 42.50 & 28 18 49.8     \\
24-042 &  V410 X7           &  04184250+2818498  & ...          & MHO-11                                & 4 18 42.50 & 28 18 49.8     \\
23-000 &  V410 A20          &  04184505+2820528  & ...          & ...                                   & 4 18 45.06 & 28 20 52.8     \\
24-000 &  V410 A20          &  04184505+2820528  & ...          & ...                                   & 4 18 45.06 & 28 20 52.8     \\
23-056 &  Hubble 4          &  04184703+2820073  & 04157+2813   & V1023 Tau, HBC 374                    & 4 18 47.04 & 28 20 07.3     \\
24-047 &  Hubble 4          &  04184703+2820073  & 04157+2813   & V1023 Tau, HBC 374                    & 4 18 47.04 & 28 20 07.3     \\
23-000 &  KPNO-Tau 2        &  04185115+2814332  & ...          & ...                                   & 4 18 51.16 & 28 14 33.2     \\
24-000 &  KPNO-Tau 2        &  04185115+2814332  & ...          & ...                                   & 4 18 51.16 & 28 14 33.2     \\
23-000 &  CoKu Tau 1        &  04185147+2820264  & ...          & HBC 375                               & 4 18 51.48 & 28 20 26.5     \\
24-000 &  CoKu Tau 1        &  04185147+2820264  & ...          & HBC 375                               & 4 18 51.48 & 28 20 26.5     \\
23-061 &  V410 X6           &  04190110+2819420  & ...          & Kim3-89                               & 4 19 01.11 & 28 19 42.0     \\
\hline
\end{tabular}
\normalsize
\end{table*}
 
\setcounter{table}{6}
\begin{table*}[h!]
\caption{(Continued)}
\begin{tabular}{llcrlll}
\hline
\hline
XEST & Name & 2MASS$^a$ & IRAS$^b$ & Alternative names  & RA(J2000.0)$^c$   & Dec(J2000.0)$^c$                 \\
     &      &           &          &      & h\ \ m \ \ s  & $\deg\ \ \arcmin\ \ \arcsec$ \\
\hline
24-054 &  V410 X6           &  04190110+2819420  & ...          & Kim3-89                               & 4 19 01.11 & 28 19 42.0     \\
23-063 &  V410 X5           &  04190197+2822332  & ...          & MHO-12                                & 4 19 01.98 & 28 22 33.2     \\
24-055 &  V410 X5           &  04190197+2822332  & ...          & MHO-12                                & 4 19 01.98 & 28 22 33.2     \\
23-067 &  FQ Tau AB         &  04191281+2829330  & C04161+2822  & HBC 377, Haro 6-3                     & 4 19 12.81 & 28 29 33.1     \\
24-058 &  FQ Tau AB         &  04191281+2829330  & C04161+2822  & HBC 377, Haro 6-3                     & 4 19 12.81 & 28 29 33.1     \\
28-100 &  BP Tau            &  04191583+2906269  & 04161+2859   & HD 281934, HBC 32                     & 4 19 15.84 & 29 06 26.9     \\
23-074 &  V819 Tau AB       &  04192625+2826142  & C04162+2819  & HBC 378, TAP 27                       & 4 19 26.26 & 28 26 14.3     \\
24-061 &  V819 Tau AB       &  04192625+2826142  & C04162+2819  & HBC 378, TAP 27                       & 4 19 26.26 & 28 26 14.3     \\
16-000 &  IRAS 04166+2706   &  ...               & 04166+2706   & ...                                   & 4 19 43.00 & 27 13 33.7 (S) \\
16-000 &  IRAS 04169+2702   &  04195844+2709570  & 04169+2702   & ...                                   & 4 19 58.45 & 27 09 57.1     \\
11-000 &  CFHT-Tau 19       &  04210795+2702204  & ...          & ...                                   & 4 21 07.95 & 27 02 20.4     \\
11-000 &  IRAS 04181+2655   &  ...               & 04181+2655   & ...                                   & 4 21 10.90 & 27 02 06.0 (S) \\
11-000 &  IRAS 04181+2654AB &  04211146+2701094  & 04181+2654   & ...                                   & 4 21 11.47 & 27 01 09.4     \\
11-023 &  2M J04213459      &  04213459+2701388  & ...          & ...                                   & 4 21 34.60 & 27 01 38.9     \\
01-028 &  IRAS 04187+1927   &  04214323+1934133  & 04187+1927   & ...                                   & 4 21 43.24 & 19 34 13.3     \\
11-037 &  CFHT-Tau 10       &  04214631+2659296  & ...          & ...                                   & 4 21 46.31 & 26 59 29.6     \\
11-000 &  2M J04215450+2652 &  04215450+2652315  &              & ...                                   & 4 21 54.51 & 26 52 31.5     \\
21-038 &  RY Tau            &  04215740+2826355  & 04188+2819   & HBC 34, HD 283571                     & 4 21 57.40 & 28 26 35.5     \\
21-039 &  HD 283572         &  04215884+2818066  & C04185+2811  & V987 Tau, HBC 380                     & 4 21 58.84 & 28 18 06.6     \\
01-045 &  T Tau N(+Sab)     &  04215943+1932063  & 04190+1924   & HD 284419, HBC 35                     & 4 21 59.43 & 19 32 06.4     \\
11-054 &  Haro 6-5 B        &  04220069+2657324  & ...          & Haro 381, FS Tau B                    & 4 22 00.70 & 26 57 32.5     \\
11-057 &  FS Tau AC         &  04220217+2657304  & 04189+2650   & HBC 383, Haro 6-5 A                   & 4 22 02.18 & 26 57 30.5     \\
21-044 &  LkCa 21           &  04220313+2825389  & C04185+2818  & V1071 Tau, HBC 382                    & 4 22 03.14 & 28 25 39.0     \\
01-054 &  RX J0422.1+1934   &  04220496+1934483  & ...          & ...                                   & 4 22 04.96 & 19 34 48.3     \\
01-062 &  2M J04221332+1934 &  04221332+1934392  &              & ...                                   & 4 22 13.32 & 19 34 39.2     \\
11-079 &  CFHT-Tau 21       &  04221675+2654570  & ...          & ...                                   & 4 22 16.76 & 26 54 57.1     \\
02-013 &  FV Tau AB         &  04265352+2606543  & C04238+2600  & HBC 386, Haro 6-8                     & 4 26 53.53 & 26 06 54.4     \\
02-000 &  FV Tau/c AB       &  04265440+2606510  & ...          & HBC 387                               & 4 26 54.41 & 26 06 51.0     \\
02-016 &  KPNO-Tau 13       &  04265732+2606284  & ...          & ...                                   & 4 26 57.33 & 26 06 28.4     \\
02-000 &  DG Tau B          &  04270266+2605304  & ...          & ...                                   & 4 27 02.66 & 26 05 30.5     \\
02-022 &  DG Tau A          &  04270469+2606163  & 04240+2559   & HBC 37                                & 4 27 04.70 & 26 06 16.3     \\
02-000 &  KPNO-Tau 4        &  04272799+2612052  & ...          & ...                                   & 4 27 28.00 & 26 12 05.3     \\
02-000 &  IRAS 04248+2612AB &  04275730+2619183  & 04248+2612   & HH 31 IRS 2                           & 4 27 57.31 & 26 19 18.3     \\
15-020 &  JH 507            &  04292071+2633406  & ...          & J1-507                                & 4 29 20.71 & 26 33 40.7     \\
13-004 &  GV Tau AB         &  04292373+2433002  & 04263+2426   & HBC 389, Haro 6-10 AB                 & 4 29 23.73 & 24 33 00.3     \\
13-000 &  IRAS 04264+2433   &  04293008+2439550  & 04264+2433   & ...                                   & 4 29 30.08 & 24 39 55.1     \\
15-040 &  DH Tau AB         &  04294155+2632582  & 04267+2626   & HBC 38                                & 4 29 41.56 & 26 32 58.3     \\
15-042 &  DI Tau AB         &  04294247+2632493  & ...          & HBC 39                                & 4 29 42.48 & 26 32 49.3     \\
15-044 &  KPNO-Tau 5        &  04294568+2630468  & ...          & ...                                   & 4 29 45.68 & 26 30 46.8     \\
14-006 &  IQ Tau A          &  04295156+2606448  & 04267+2600   & HBC 41, LkHa 265                      & 4 29 51.56 & 26 06 44.9     \\
13-000 &  CFHT-Tau 20       &  04295950+2433078  & ...          & ...                                   & 4 29 59.51 & 24 33 07.9     \\
14-000 &  KPNO-Tau 6        &  04300724+2608207  & ...          & ...                                   & 4 30 07.24 & 26 08 20.8     \\
13-035 &  FX Tau AB         &  04302961+2426450  & 04267+2420   & HBC 44, Haro 6-11                     & 4 30 29.61 & 24 26 45.0     \\
14-057 &  DK Tau AB         &  04304425+2601244  & 04276+2554   & HBC 45                                & 4 30 44.25 & 26 01 24.5     \\
14-000 &  KPNO-Tau 7        &  04305718+2556394  & ...          & ...                                   & 4 30 57.19 & 25 56 39.5     \\
22-013 &  MHO 9             &  04311578+1820072  & ...          & ...                                   & 4 31 15.78 & 18 20 07.2     \\
22-021 &  MHO 4             &  04312405+1800215  & ...          & RXJ0431.4+1800                        & 4 31 24.06 & 18 00 21.5     \\
22-040 &  L1551 IRS5        &  04313407+1808049  & 04287+1801   & ...                                   & 4 31 34.08 & 18 08 04.9     \\
22-042 &  LkHa 358          &  04313613+1813432  &              & CoKu Tau 2, HBC 394                   & 4 31 36.13 & 18 13 43.3     \\
22-000 &  HH 30             &  04313747+1812244  & 04287+1806   & V1213 Tau                             & 4 31 37.47 & 18 12 24.5     \\
22-043 &  HL Tau            &  04313843+1813576  & 04287+1807   & HBC 49, Haro 6-14                     & 4 31 38.44 & 18 13 57.7     \\
22-047 &  XZ Tau AB         &  04314007+1813571  & ...          & HBC 50, Haro 6-15                     & 4 31 40.07 & 18 13 57.2     \\
22-056 &  L1551 NE          &  04314444+1808315  & ...          & ...                                   & 4 31 44.45 & 18 08 31.5     \\
03-005 &  HK Tau AB         &  04315056+2424180  & 04288+2417   & HBC 48, Haro 6-12                     & 4 31 50.57 & 24 24 18.1     \\
22-070 &  V710 Tau BA       &  04315779+1821380  & 04290+1815   & HBC 395+51, LkHa266S                  & 4 31 57.79 & 18 21 38.1     \\
19-009 &  JH 665            &  04315844+2543299  & ...          & J1-665                                & 4 31 58.44 & 25 43 29.9     \\
22-089 &  L1551 51          &  04320926+1757227  & 042916+1751  & V1075 Tau, HBC 397                    & 4 32 09.27 & 17 57 22.8     \\
22-097 &  V827 Tau          &  04321456+1820147  & ...          & HBC 399, TAP 42                       & 4 32 14.57 & 18 20 14.7     \\
03-016 &  Haro 6-13         &  04321540+2428597  & 04292+2422   & V806 Tau, HBC 396                     & 4 32 15.41 & 24 28 59.7     \\
\hline
\end{tabular}
\normalsize
\end{table*}
 
\setcounter{table}{6}
\begin{table*}[h!]
\caption{(Continued)}
\begin{tabular}{llcrlll}
\hline
\hline
XEST & Name & 2MASS$^a$ & IRAS$^b$ & Alternative names  & RA(J2000.0)$^c$   & Dec(J2000.0)$^c$                 \\
     &      &           &          &      & h\ \ m \ \ s  & $\deg\ \ \arcmin\ \ \arcsec$ \\
\hline
22-100 &  V826 Tau          &  04321583+1801387  & ...          & HBC 400, TAP 43                       & 4 32 15.84 & 18 01 38.7     \\
22-101 &  MHO 5             &  04321606+1812464  & ...          & ...                                   & 4 32 16.07 & 18 12 46.4     \\
03-017 &  CFHT-Tau 7        &  04321786+2422149  & ...          & ...                                   & 4 32 17.86 & 24 22 15.0     \\
03-019 &  V928 Tau AB       &  04321885+2422271  & ...          & HK Tau/G2, JH 91, HBC 398             & 4 32 18.86 & 24 22 27.1     \\
03-022 &  FY Tau            &  04323058+2419572  & 04294+2413   & HBC 401, Haro 6-17                    & 4 32 30.58 & 24 19 57.3     \\
03-023 &  FZ Tau            &  04323176+2420029  & ...          & HBC 402, Haro 6-18                    & 4 32 31.76 & 24 20 03.0     \\
17-002 &  IRAS 04295+2251   &  04323205+2257266  & 04295+2251   & L1536 IRS                             & 4 32 32.05 & 22 57 26.7     \\
19-049 &  UZ Tau E+W(AB)    &  04324303+2552311  & 04296+2546   & HBC 52                                & 4 32 43.04 & 25 52 31.1     \\
17-009 &  JH 112            &  04324911+2253027  & 04298+2246   & ...                                   & 4 32 49.11 & 22 53 02.8     \\
03-031 &  CFHT-Tau 5        &  04325026+2422115  & ...          & ...                                   & 4 32 50.27 & 24 22 11.6     \\
04-003 &  CFHT-Tau 5        &  04325026+2422115  & ...          & ...                                   & 4 32 50.27 & 24 22 11.6     \\
03-035 &  MHO 8             &  04330197+2421000  & ...          & ...                                   & 4 33 01.98 & 24 21 00.0     \\
04-009 &  MHO 8             &  04330197+2421000  & ...          & ...                                   & 4 33 01.98 & 24 21 00.0     \\
04-010 &  GH Tau AB         &  04330622+2409339  & 04300+2403   & HBC 55, Haro 6-20                     & 4 33 06.22 & 24 09 34.0     \\
04-012 &  V807 Tau  SNab    &  04330664+2409549  & ...          & HBC 404, Elias 12                     & 4 33 06.64 & 24 09 55.0     \\
18-004 &  KPNO-Tau 14       &  04330781+2616066  & ...          & ...                                   & 4 33 07.81 & 26 16 06.6     \\
17-000 &  CFHT-Tau 12       &  04330945+2246487  & ...          & ...                                   & 4 33 09.46 & 22 46 48.7     \\
04-016 &  V830 Tau          &  04331003+2433433  & C04301+2427  & HBC 405, TAP 46                       & 4 33 10.03 & 24 33 43.4     \\
18-000 &  IRAS S04301+261   &  04331435+2614235  & ...          & ...                                   & 4 33 14.36 & 26 14 23.5     \\
17-000 &  IRAS 04302+2247   &  04331650+2253204  & 04302+2247   & Butterfly Star                        & 4 33 16.50 & 22 53 20.4     \\
17-027 &  IRAS 04303+2240   &  04331907+2246342  & 04303+2240   & L1536S                                & 4 33 19.07 & 22 46 34.2     \\
04-034 &  GI Tau            &  04333405+2421170  & 04305+2414   & HBC 56, Haro 6-21                     & 4 33 34.06 & 24 21 17.0     \\
04-035 &  GK Tau AB         &  04333456+2421058  & ...          & HBC 57, Haro 6-22                     & 4 33 34.56 & 24 21 05.9     \\
18-019 &  IS Tau AB         &  04333678+2609492  & 04308+2607   & HBC 59, Haro 6-23                     & 4 33 36.79 & 26 09 49.2     \\
17-058 &  CI Tau            &  04335200+2250301  & 04308+2244   & HBC 61, Haro 6-25                     & 4 33 52.00 & 22 50 30.2     \\
18-030 &  IT Tau AB         &  04335470+2613275  & ...          & Haro 6-26                             & 4 33 54.70 & 26 13 27.5     \\
17-066 &  JH 108            &  04341099+2251445  & ...          & ...                                   & 4 34 10.99 & 22 51 44.5     \\
17-068 &  CFHT-BD Tau 1     &  04341527+2250309  & ...          & ...                                   & 4 34 15.27 & 22 50 31.0     \\
25-026 &  AA Tau            &  04345542+2428531  & 04318+2422   & HBC 63                                & 4 34 55.42 & 24 28 53.2     \\
09-010 &  HO Tau AB         &  04352020+2232146  & C04323+2226  & HBC 64, Haro 6-27                     & 4 35 20.20 & 22 32 14.6     \\
08-019 &  FF Tau AB         &  04352089+2254242  & ...          & HBC 409                               & 4 35 20.90 & 22 54 24.2     \\
12-040 &  DN Tau            &  04352737+2414589  & 04324+2408   & HBC 65                                & 4 35 27.37 & 24 14 58.9     \\
12-000 &  IRAS 04325+2402AB &  04353539+2408194  & 04325+2402   & L1535 IRS                             & 4 35 35.39 & 24 08 19.4     \\
12-059 &  CoKu Tau 3 AB     &  04354093+2411087  & ...          & HBC 411                               & 4 35 40.94 & 24 11 08.8     \\
09-022 &  KPNO-Tau 8        &  04354183+2234115  & ...          & ...                                   & 4 35 41.84 & 22 34 11.6     \\
08-037 &  HQ Tau AB         &  04354733+2250216  & 04327+2244   & ...                                   & 4 35 47.34 & 22 50 21.7     \\
09-026 &  HQ Tau AB         &  04354733+2250216  & 04327+2244   & ...                                   & 4 35 47.34 & 22 50 21.7     \\
08-043 &  KPNO-Tau 15       &  04355109+2252401  & ...          & ...                                   & 4 35 51.10 & 22 52 40.1     \\
09-031 &  KPNO-Tau 15       &  04355109+2252401  & ...          & ...                                   & 4 35 51.10 & 22 52 40.1     \\
08-000 &  KPNO-Tau 9        &  04355143+2249119  & ...          & ...                                   & 4 35 51.43 & 22 49 11.9     \\
09-000 &  KPNO-Tau 9        &  04355143+2249119  & ...          & ...                                   & 4 35 51.43 & 22 49 11.9     \\
08-048 &  HP Tau AB         &  04355277+2254231  & 04328+2248   & HBC 66, LkHa 258                      & 4 35 52.78 & 22 54 23.1     \\
08-051a&  HP Tau/G3 AB      &  04355349+2254089  & ...          & HBC 414                               & 4 35 53.50 & 22 54 09.0     \\
08-051 &  HP Tau/G2         &  04355415+2254134  & ...          & V1025 Tau, HBC 415                    & 4 35 54.15 & 22 54 13.5     \\
08-058 &  Haro 6-28 AB      &  04355684+2254360  & ...          & V1026 Tau, HBC 416                    & 4 35 56.84 & 22 54 36.0     \\
08-000 &  CFHT-BD Tau 2     &  04361038+2259560  & ...          & ...                                   & 4 36 10.39 & 22 59 56.0     \\
08-080 &  CFHT-BD Tau 3     &  04363893+2258119  & ...          & ...                                   & 4 36 38.94 & 22 58 11.9     \\
05-005 &  CFHT-Tau 6        &  04390396+2544264  & ...          & ...                                   & 4 39 03.96 & 25 44 26.4     \\
05-000 &  IRAS 04361+2547   &  04391389+2553208  & 04361+2547   & TMR 1, ITG 12                         & 4 39 13.89 & 25 53 20.9     \\
05-013 &  GN Tau AB         &  04392090+2545021  & 04362+2539   & Haro 6-31                             & 4 39 20.91 & 25 45 02.1     \\
05-017 &  IRAS 04365+2535   &  04393519+2541447  & 04365+2535   & TMC-1A                                & 4 39 35.19 & 25 41 44.7     \\
05-024 &  IRAS 04369+2539   &  04395574+2545020  & 04369+2539   & Elias 18, IC 2087/IR                  & 4 39 55.75 & 25 45 02.0     \\
07-011 &  JH 223            &  04404950+2551191  & ...          & ...                                   & 4 40 49.51 & 25 51 19.2     \\
07-022 &  Haro 6-32         &  04410424+2557561  & ...          & ...                                   & 4 41 04.24 & 25 57 56.1     \\
07-000 &  ITG 33 A          &  04410826+2556074  & ...          & ...                                   & 4 41 08.26 & 25 56 07.5     \\
07-000 &  CFHT-Tau 8        &  04411078+2555116  & ...          & ...                                   & 4 41 10.78 & 25 55 11.7     \\
07-000 &  IRAS 04381+2540   &  04411267+2546354  & 04381+2540   & TMC-1, ITG 35                         & 4 41 12.68 & 25 46 35.4     \\
07-041 &  IRAS 04385+2550AB &  04413882+2556267  & 04385+2550   & Haro 6-33                             & 4 41 38.82 & 25 56 26.8     \\
10-017 &  CoKuLk332/G2 AB   &  04420548+2522562  & ...          & V999 Tau, HBC 422                     & 4 42 05.49 & 25 22 56.3     \\
\hline
\end{tabular}
\normalsize
\end{table*}
 
\setcounter{table}{6}
\begin{table*}[h!]
\caption{(Continued)}
\begin{tabular}{llcrlll}
\hline
\hline
XEST & Name & 2MASS$^a$ & IRAS$^b$ & Alternative names  & RA(J2000.0)$^c$   & Dec(J2000.0)$^c$                 \\
     &      &           &          &      & h\ \ m \ \ s  & $\deg\ \ \arcmin\ \ \arcsec$ \\
\hline
10-018 &  CoKuLk332/G1 AB   &  04420732+2523032  & ...          & HBC 423                               & 4 42 07.33 & 25 23 03.2     \\
10-020 &  V955 Tau AB       &  04420777+2523118  & 04390+2517   & HBC 69, JH 226, LkHa 332              & 4 42 07.77 & 25 23 11.8     \\
10-034 &  CIDA 7            &  04422101+2520343  & ...          & C31                                   & 4 42 21.02 & 25 20 34.4     \\
10-045 &  DP Tau            &  04423769+2515374  & 04395+2509   & HBC 70                                & 4 42 37.70 & 25 15 37.5     \\
10-060 &  GO Tau            &  04430309+2520187  & C04400+2514  & HBC 71                                & 4 43 03.09 & 25 20 18.8     \\
26-012 &  2M J04552333+30   &  04552333+3027366  & ...          & ...                                   & 4 55 23.33 & 30 27 36.6     \\
26-034 &  2M J04554046+30   &  04554046+3039057  & ...          & ...                                   & 4 55 40.46 & 30 39 05.7     \\
26-043 &  AB Aur            &  04554582+3033043  & 04525+3028   & HD 31293, HBC 78                      & 4 55 45.83 & 30 33 04.4     \\
26-050 &  2MJ04554757/801   &  04554757+3028077  & ...          & ...                                   & 4 55 47.57 & 30 28 07.7     \\
26-067 &  SU Aur            &  04555938+3034015  & 04528+3029   & HBC 79, HD 282624                     & 4 55 59.38 & 30 34 01.6     \\
26-072 &  HBC 427           &  04560201+3021037  & ...          & V397 Aur                              & 4 56 02.02 & 30 21 03.8     \\
\hline
\multicolumn{7}{l}{Additional sources from Chandra}\\
\hline
C1-0   &  KPNO-Tau 10       &  04174955+2813318  & ...          & ...                                   & 4 17 49.55 & 28 13 31.9     \\
C1-1   &  IRAS 04158+2805   &  04185813+2812234  & 04158+2805   & ...                                   & 4 18 58.14 & 28 12 23.5     \\
C2-1   &  Haro 6-5 B        &  04220069+2657324  & ...          & Haro 381, FS Tau B                    & 4 22 00.70 & 26 57 32.5     \\
C2-2   &  FS Tau AC         &  04220217+2657304  & 04189+2650   & HBC 383, Haro 6-5 A                   & 4 22 02.18 & 26 57 30.5     \\
C3-1   &  FV Tau/c AB       &  04265440+2606510  & ...          & HBC 387                               & 4 26 54.41 & 26 06 51.0     \\
C3-2   &  DG Tau B          &  04270266+2605304  & ...          & ...                                   & 4 27 02.66 & 26 05 30.5     \\
C4-1   &  GV Tau AB         &  04292373+2433002  & 04263+2426   & HBC 389, Haro 6-10 AB                 & 4 29 23.73 & 24 33 00.3     \\
C5-2   &  HN Tau AB         &  04333935+1751523  & ...          & ...                                   & 4 33 39.35 & 17 51 52.4     \\
C5-1   &  L1551 55          &  04324373+1802563  & 042950+1757  & ...                                   & 4 32 43.73 & 18 02 56.3     \\
C5-4   &  HD 28867          &  04333297+1801004  & 04306+1754   & HR 1442, L1551 53                     & 4 33 32.98 & 18 01 00.4     \\
C5-3   &  DM Tau            &  04334871+1810099  & 04309+1803   & HBC 62                                & 4 33 48.72 & 18 10 10.0     \\
C6-1   &  CFHT-BD Tau 4     &  04394748+2601407  & 04368+2557N3 & ...                                   & 4 39 47.48 & 26 01 40.8     \\
C6-0   &  L1527 IRS         &  ...               & 04368+2557   & ...                                   & 4 39 53.59 & 26 03 05.5 (S) \\
C6-0   &  CFHT-Tau 17       &  04400174+2556292  & ...          & ...                                   & 4 40 01.75 & 25 56 29.2     \\
C6-2   &  IRAS 04370+2559   &  04400800+2605253  & 04370+2559   & ...                                   & 4 40 08.00 & 26 05 25.4     \\
\hline
\end{tabular}
\begin{minipage}{0.96\textwidth}
\footnotetext{
\hskip -0.5truecm $^a$ Nearest 2MASS entry within 5\arcsec\ to coordinates given by references 5, 18, 33, or SIMBAD. Unlikely identifications in parentheses \\
$^b$ Nearest IRAS catalog entry, within 10\arcsec\\
$^c$ 2MASS coordinates. For unlikely identifications, SIMBAD (S) or reference 5 (B)\\
}
\end{minipage}
\label{tab7}
\normalsize
\end{table*}

\clearpage

\setcounter{table}{7}
\begin{table*}[h!]
\caption{Fundamental parameters of targets in XEST (2): Multiplicity}
\begin{tabular}{llrrr}
\hline
\hline
XEST & Name & Comp & Separations	  & Refs\\
     &      &      & (\arcsec) &	 \\
\hline
27-115 &  HBC 352           & 1 & ...             & ...                                  \\
27-000 &  HBC 353           & 1 & ...             & ...                                  \\
06-005 &  HBC 358 AB        & 3 & 0.15, 1.55      &  11, 20, 31, 40                      \\
06-007 &  HBC 359           & 1 & ...             & ...                                  \\
06-059 &  L1489 IRS         & 1 & ...             & ...                                  \\
20-001 &  LkCa 1            & 1 & ...             & ...                                  \\
20-005 &  Anon 1            & 1 & ...             & ...                                  \\
20-000 &  IRAS 04108+2803 A & 1 & ...             & ...                                  \\
20-022 &  IRAS 04108+2803 B & 1 & ...             & ...                                  \\
20-000 &  2M J04141188+28   & 1 & ...             & ...                                  \\
20-042 &  V773 Tau ABC      & 4 & 0.002,0.11,0.2  &  16, 17, 31, 58, 61, 62              \\
20-043 &  FM Tau            & 1 & ...             & ...                                  \\
20-046 &  CW Tau            & 1 & ...             & ...                                  \\
20-047 &  CIDA 1            & 1 & ...             & ...                                  \\
20-056 &  MHO 2/1           & 2 & 4.0             &  13                                  \\
20-058 &  MHO 3             & 1 & ...             & ...                                  \\
20-069 &  FO Tau AB         & 2 & 0.16            &  16, 20, 31, 58                      \\
20-073 &  CIDA 2            & 1 & ...             & ...                                  \\
23-002 &  CY Tau            & 1 & ...             & ...                                  \\
24-002 &  CY Tau            & 1 & ...             & ...                                  \\
23-004 &  LkCa 5            & 1 & ...             & ...                                  \\
24-004 &  LkCa 5            & 1 & ...             & ...                                  \\
23-008 &  CIDA 3            & 1 & ...             & ...                                  \\
24-008 &  CIDA 3            & 1 & ...             & ...                                  \\
23-015 &  V410 X3           & 1 & ...             & ...                                  \\
24-015 &  V410 X3           & 1 & ...             & ...                                  \\
23-018 &  V410 A13          & 1 & ...             & ...                                  \\
24-000 &  V410 A13          & 1 & ...             & ...                                  \\
23-000 &  V410 A24          & 1 & ...             & ...                                  \\
24-000 &  V410 A24          & 1 & ...             & ...                                  \\
23-029 &  V410 A25          & 1 & ...             & ...                                  \\
24-027 &  V410 A25          & 1 & ...             & ...                                  \\
23-032 &  V410 Tau ABC      & 3 & 0.12, 0.29      &  16, 17, 58                          \\
24-028 &  V410 Tau ABC      & 3 & 0.12, 0.29      &  16, 17, 58                          \\
23-033 &  DD Tau AB         & 2 & 0.54            &  16, 20, 31, 40, 47, 58              \\
24-029 &  DD Tau AB         & 2 & 0.54            &  16, 20, 31, 40, 47, 58              \\
23-035 &  CZ Tau AB         & 2 & 0.33            &  31, 40, 58                          \\
24-030 &  CZ Tau AB         & 2 & 0.33            &  31, 40, 58                          \\
23-036 &  IRAS 04154+2823   & 1 & ...             & ...                                  \\
24-031 &  IRAS 04154+2823   & 1 & ...             & ...                                  \\
23-037 &  V410 X2           & 1 & ...             & ...                                  \\
24-032 &  V410 X2           & 1 & ...             & ...                                  \\
23-045 &  V410 X4           & 1 & ...             & ...                                  \\
24-038 &  V410 X4           & 1 & ...             & ...                                  \\
23-047 &  V892 Tau          & 3 & 0.05, 4         &  32, 54                              \\
24-040 &  V892 Tau          & 3 & 0.05, 4         &  32, 54                              \\
23-048 &  LR 1              & 1 & ...             & ...                                  \\
24-000 &  LR 1              & 1 & ...             & ...                                  \\
23-050 &  V410 X7           & 1 & ...             & ...                                  \\
24-042 &  V410 X7           & 1 & ...             & ...                                  \\
23-000 &  V410 A20          & 1 & ...             & ...                                  \\
24-000 &  V410 A20          & 1 & ...             & ...                                  \\
23-056 &  Hubble 4          & 1 & ...             & ...                                  \\
24-047 &  Hubble 4          & 1 & ...             & ...                                  \\
23-000 &  KPNO-Tau 2        & 1 & ...             & ...                                  \\
24-000 &  KPNO-Tau 2        & 1 & ...             & ...                                  \\
23-000 &  CoKu Tau 1        & 1 & ...             & ...                                  \\
24-000 &  CoKu Tau 1        & 1 & ...             & ...                                  \\
23-061 &  V410 X6           & 1 & ...             & ...                                  \\
\hline
\end{tabular}
\normalsize
\end{table*}
 
\setcounter{table}{7}
\begin{table*}[h!]
\caption{(Continued)}
\begin{tabular}{llrrr}
\hline
\hline
XEST & Name & Comp & Separations       & Refs\\
     &	   &	  & (\arcsec) &      \\
\hline
24-054 &  V410 X6           & 1 & ...             & ...                                  \\
23-063 &  V410 X5           & 1 & ...             & ...                                  \\
24-055 &  V410 X5           & 1 & ...             & ...                                  \\
23-067 &  FQ Tau AB         & 2 & 0.73            &  20, 31, 40, 58                      \\
24-058 &  FQ Tau AB         & 2 & 0.73            &  20, 31, 40, 58                      \\
28-100 &  BP Tau            & 1 & ...             & ...                                  \\
23-074 &  V819 Tau AB       & 2 & 10.5            &  31$^a$                              \\
24-061 &  V819 Tau AB       & 2 & 10.5            &  31$^a$                              \\
16-000 &  IRAS 04166+2706   & 1 & ...             & ...                                  \\
16-000 &  IRAS 04169+2702   & 1 & ...             & ...                                  \\
11-000 &  CFHT-Tau 19       & 1 & ...             & ...                                  \\
11-000 &  IRAS 04181+2655   & 1 & ...             & ...                                  \\
11-000 &  IRAS 04181+2654AB & 2 & ...             & ...                                  \\
11-023 &  2M J04213459      & 1 & ...             & ...                                  \\
01-028 &  IRAS 04187+1927   & 1 & ...             & ...                                  \\
11-037 &  CFHT-Tau 10       & 1 & ...             & ...                                  \\
11-000 &  2M J04215450+2652 & 1 & ...             & ...                                  \\
21-038 &  RY Tau            & 1 & ...             & ...                                  \\
21-039 &  HD 283572         & 1 & ...             & ...                                  \\
01-045 &  T Tau N(+Sab)     & 3 & 0.67, 0.13      &  12, 47, 58                          \\
11-054 &  Haro 6-5 B        & 1 & ...             & ...                                  \\
11-057 &  FS Tau AC         & 2 & 0.30            &  20, 31, 40, 52, 53, 58              \\
21-044 &  LkCa 21           & 1 & ...             & ...                                  \\
01-054 &  RX J0422.1+1934   & 1 & ...             & ...                                  \\
01-062 &  2M J04221332+1934 & 1 & ...             & ...                                  \\
11-079 &  CFHT-Tau 21       & 1 & ...             & ...                                  \\
02-013 &  FV Tau AB         & 2 & 0.71            &  16, 31, 40, 47, 53, 58              \\
02-000 &  FV Tau/c AB       & 2 & 0.67            &  31, 40, 52, 53, 58                  \\
02-016 &  KPNO-Tau 13       & 1 & ...             & ...                                  \\
02-000 &  DG Tau B          & 1 & ...             & ...                                  \\
02-022 &  DG Tau A          & 1 & ...             & ...                                  \\
02-000 &  KPNO-Tau 4        & 1 & ...             & ...                                  \\
02-000 &  IRAS 04248+2612AB & 2 & 4.55            &  13                                  \\
15-020 &  JH 507            & 1 & ...             & ...                                  \\
13-004 &  GV Tau AB         & 2 & 1.2             &  30, 31, 47, 53, 58                  \\
13-000 &  IRAS 04264+2433   & 1 & ...             & ...                                  \\
15-040 &  DH Tau AB         & 2 & 2.3             &  25                                  \\
15-042 &  DI Tau AB         & 2 & 0.12            &  16, 53                              \\
15-044 &  KPNO-Tau 5        & 1 & ...             & ...                                  \\
14-006 &  IQ Tau A          & 1 & ...             &  $^b$                                \\
13-000 &  CFHT-Tau 20       & 1 & ...             & ...                                  \\
14-000 &  KPNO-Tau 6        & 1 & ...             & ...                                  \\
13-035 &  FX Tau AB         & 2 & 0.85            &  11, 16, 31, 40, 53, 58              \\
14-057 &  DK Tau AB         & 2 & 2.27            &  11, 31, 40, 43, 47, 52, 53, 58      \\
14-000 &  KPNO-Tau 7        & 1 & ...             & ...                                  \\
22-013 &  MHO 9             & 1 & ...             & ...                                  \\
22-021 &  MHO 4             & 1 & ...             & ...                                  \\
22-040 &  L1551 IRS5        & 2 & 0.3             &  51                                  \\
22-042 &  LkHa 358          & 1 & ...             & ...                                  \\
22-000 &  HH 30             & 1 & ...             & ...                                  \\
22-043 &  HL Tau            & 1 & ...             & ...                                  \\
22-047 &  XZ Tau AB         & 2 & 0.3             &  16, 20, 31,  47,  58                \\
22-056 &  L1551 NE          & 2 & 0.5             &  64                                  \\
03-005 &  HK Tau AB         & 2 & 2.4             &  11, 31, 40, 42, 43, 53, 58          \\
22-070 &  V710 Tau BA       & 2 & 3.1             &  21, 31, 40, 43, 58                  \\
19-009 &  JH 665            & 1 & ...             & ...                                  \\
22-089 &  L1551 51          & 1 & ...             & ...                                  \\
22-097 &  V827 Tau          & 1 & ...             & ...                                  \\
03-016 &  Haro 6-13         & 1 & ...             & ...                                  \\
\hline
\end{tabular}
\normalsize
\end{table*}
 
\setcounter{table}{7}
\begin{table*}[h!]
\caption{(Continued)}
\begin{tabular}{llrrr}
\hline
\hline
XEST & Name & Comp & Separations       & Refs\\
     &	   &	  & (\arcsec) &      \\
\hline
22-100 &  V826 Tau          & 1 & ...             & ...                                  \\
22-101 &  MHO 5             & 1 & ...             & ...                                  \\
03-017 &  CFHT-Tau 7        & 1 & ...             & ...                                  \\
03-019 &  V928 Tau AB       & 2 & 0.17            &  16, 31, 53, 58                      \\
03-022 &  FY Tau            & 1 & ...             & $^c$                                 \\
03-023 &  FZ Tau            & 1 & ...             & ...                                  \\
17-002 &  IRAS 04295+2251   & 1 & ...             & ...                                  \\
19-049 &  UZ Tau E+W(AB)    & 3 & 3.5, 0.36       &  16, 17, 20, 21, 31, 40, 52, 53, 58  \\
17-009 &  JH 112            & 1 & ...             & ...                                  \\
03-031 &  CFHT-Tau 5        & 1 & ...             & ...                                  \\
04-003 &  CFHT-Tau 5        & 1 & ...             & ...                                  \\
03-035 &  MHO 8             & 1 & ...             & ...                                  \\
04-009 &  MHO 8             & 1 & ...             & ...                                  \\
04-010 &  GH Tau AB         & 2 & 0.31            &  16, 31, 40, 58                      \\
04-012 &  V807 Tau  SNab    & 3 & 0.3, 0.023      &  16, 20, 31, 53, 58                  \\
18-004 &  KPNO-Tau 14       & 1 & ...             & ...                                  \\
17-000 &  CFHT-Tau 12       & 1 & ...             & ...                                  \\
04-016 &  V830 Tau          & 1 & ...             & ...                                  \\
18-000 &  IRAS S04301+261   & 1 & ...             & ...                                  \\
17-000 &  IRAS 04302+2247   & 1 & ...             & ...                                  \\
17-027 &  IRAS 04303+2240   & 1 & ...             & ...                                  \\
04-034 &  GI Tau            & 1 & ...             & ...                                  \\
04-035 &  GK Tau AB         & 2 & 2.5             &  21, 49                              \\
18-019 &  IS Tau AB         & 2 & 0.22            &  16, 20, 53, 58                      \\
17-058 &  CI Tau            & 1 & ...             & ...                                  \\
18-030 &  IT Tau AB         & 2 & 2.4             &  11, 40, 52, 53, 58                  \\
17-066 &  JH 108            & 1 & ...             & ...                                  \\
17-068 &  CFHT-BD Tau 1     & 1 & ...             & ...                                  \\
25-026 &  AA Tau            & 1 & ...             & ...                                  \\
09-010 &  HO Tau AB         & 2 & 6.9             &  21$^d$                              \\
08-019 &  FF Tau AB         & 2 & 0.03            &  52, 53                              \\
12-040 &  DN Tau            & 1 & ...             & ...                                  \\
12-000 &  IRAS 04325+2402AB & 1 & 8.15            &  13                                  \\
12-059 &  CoKu Tau 3 AB     & 2 & 2.06            &  31, 53, 58                          \\
09-022 &  KPNO-Tau 8        & 1 & ...             & ...                                  \\
08-037 &  HQ Tau AB         & 2 & 0.009           &  52$^e$                              \\
09-026 &  HQ Tau AB         & 2 & 0.009           &  52$^e$                              \\
08-043 &  KPNO-Tau 15       & 1 & ...             & ...                                  \\
09-031 &  KPNO-Tau 15       & 1 & ...             & ...                                  \\
08-000 &  KPNO-Tau 9        & 1 & ...             & ...                                  \\
09-000 &  KPNO-Tau 9        & 1 & ...             & ...                                  \\
08-048 &  HP Tau AB         & 2 & 0.017           &  50, 53                              \\
08-051a&  HP Tau/G3 AB      & 2 & 0.022           &  50, 53                              \\
08-051 &  HP Tau/G2         & 1 & ...             & ...                                  \\
08-058 &  Haro 6-28 AB      & 2 & 0.65            &  20, 31, 40, 53, 58                  \\
08-000 &  CFHT-BD Tau 2     & 1 & ...             & ...                                  \\
08-080 &  CFHT-BD Tau 3     & 1 & ...             & ...                                  \\
05-005 &  CFHT-Tau 6        & 1 & ...             & ...                                  \\
05-000 &  IRAS 04361+2547   & 1 & ...             & ...                                  \\
05-013 &  GN Tau AB         & 2 & 0.36            & 52, 53, 58, 67$^f$                   \\
05-017 &  IRAS 04365+2535   & 1 & ...             & ...                                  \\
05-024 &  IRAS 04369+2539   & 1 & ...             & ...                                  \\
07-011 &  JH 223            & 1 & ...             & ...                                  \\
07-022 &  Haro 6-32         & 1 & ...             & ...                                  \\
07-000 &  ITG 33 A          & 1 & ...             &  $^g$                                \\
07-000 &  CFHT-Tau 8        & 1 & ...             & ...                                  \\
07-000 &  IRAS 04381+2540   & 1 & ...             & ...                                  \\
07-041 &  IRAS 04385+2550AB & 2 & 18.9            &  13                                  \\
10-017 &  CoKuLk332/G2 AB   & 2 & 0.27            &  20, 31, 58                          \\
\hline
\end{tabular}
\normalsize
\end{table*}
 
\setcounter{table}{7}
\begin{table*}[h!]
\caption{(Continued)}
\begin{tabular}{llrrr}
\hline
\hline
XEST & Name & Comp & Separations       & Refs\\
     &	   &	  & (\arcsec) &      \\
\hline
10-018 &  CoKuLk332/G1 AB   & 2 & 0.23            &  16, 31, 58                          \\
10-020 &  V955 Tau AB       & 2 & 0.34            &  20, 31, 40, 58                      \\
10-034 &  CIDA 7            & 1 & ...             & ...                                  \\
10-045 &  DP Tau            & 1 & ...             & ...                                  \\
10-060 &  GO Tau            & 1 & ...             & ...                                  \\
26-012 &  2M J04552333+30   & 1 & ...             & ...                                  \\
26-034 &  2M J04554046+30   & 1 & ...             & ...                                  \\
26-043 &  AB Aur            & 1 & ...             & ...                                  \\
26-050 &  2MJ04554757/801   & 1 & ...             & ...                                  \\
26-067 &  SU Aur            & 1 & ...             & ...                                  \\
26-072 &  HBC 427           & 1 & ...             & ...                                  \\
\hline
\multicolumn{5}{l}{Additional sources from Chandra}\\
\hline
C1-0   &  KPNO-Tau 10       & 1 & ...             & ...                                  \\
C1-1   &  IRAS 04158+2805   & 1 & ...             & ...                                  \\
C2-1   &  Haro 6-5 B        & 1 & ...             & ...                                  \\
C2-2   &  FS Tau AC         & 2 & 0.25            &  20, 31, 52, 53, 58                  \\
C3-1   &  FV Tau/c AB       & 2 & 0.67            &  31, 40, 52, 53, 58                  \\
C3-2   &  DG Tau B          & 1 & ...             & ...                                  \\
C4-1   &  GV Tau AB         & 2 & 1.2             &  30, 31, 47, 53, 58                  \\
C5-2   &  HN Tau AB         & 2 & 3.1             &  11, 21, 31, 42, 43, 58              \\
C5-1   &  L1551 55          & 1 & ...             & ...                                  \\
C5-4   &  HD 28867          & 3 & 3.08            &  65                                  \\
C5-3   &  DM Tau            & 1 & ...             & ...                                  \\
C6-1   &  CFHT-BD Tau 4     & 1 & ...             & ...                                  \\
C6-0   &  L1527 IRS         & 1 & ...             & ...                                  \\
C6-0   &  CFHT-Tau 17       & 1 & ...             & ...                                  \\
C6-2   &  IRAS 04370+2559   & 1 & ...             & ...                                  \\
\hline
\end{tabular}
\par\begin{minipage}{0.56\textwidth}
\footnotetext{
\hskip -0.5truecm $^a$  V819 Tau B suspected by \citet{woitas01} to be a background star\\ 
$^b$ IQ Tau~B, 10.2\arcsec\ from IQ Tau~A = XEST-14-006,  has been interpreted as an extincted   
background G star by \citet{hartigan94}\\
$^c$ FY Tau is a suspected binary with separation of 0.153\arcsec\ \citep{richichi94}\\ 
$^d$ HO Tau B suspected by \citet{hartigan94} to be a background star\\ 
$^e$ HQ Tau B not confirmed by \citet{simon95}\\
$^f$ GN Tau AB: A component detected from lunar occulation by \citet{simon95} at a projected separation of 0.041\arcsec\ may be identical to GN Tau B at $\approx 0.36$\arcsec\ \citep{simon96}\\ 
$^g$ ITG 33~B,  5.2\arcsec\ from ITG 33~A, has been interpreted as a reddened background object by \citet{martin00} \\
}
\end{minipage}
\label{tab8}
\normalsize
\end{table*}

\clearpage
 
\setcounter{table}{8}
\begin{table*}[h!]
\caption{Fundamental parameters of targets in XEST (3): Photometry and spectroscopy}
\begin{tabular}{llrrrrrrr}
\hline
\hline
XEST & Name & Spec$^a$  & Refs & $A_{\rm V}^a$ & $A_{\rm J}$  & $T_{\rm eff}^a$ &  $L_*^b$        & Refs	 \\
     &      &		  &      & (mag)	& (mag)   &  (K)	    & ($L_{\odot})$	&	 \\
\hline
27-115 &  HBC 352           & G0        &   27      &        0.87 &  0.25 &        6030 &               0.740 &   27           \\
27-000 &  HBC 353           & G5        &   27      &        0.97 &  0.28 &        5770 &               0.500 &   27           \\
06-005 &  HBC 358 AB        & M2        &   27      &        0.21 &  0.06 &        3580 &               0.280 &   27           \\
06-007 &  HBC 359           & M2        &   27      &        0.49 &  0.14 &        3580 &               0.260 &   27           \\
06-059 &  L1489 IRS         & K4        &   59      &       10.20 &   ... &        4500 &               4.900 &   10, 59       \\
20-001 &  LkCa 1            & M4        &    5      &        0.00 &  0.00 &        3270 &               0.380 &    5, 27       \\
20-005 &  Anon 1            & M0        &    5      &        1.32 &  1.03 &        3850 &               2.600 &    5, 27       \\
20-000 &  IRAS 04108+2803 A & ...       &  ...      &         ... &   ... &         ... &                 ... &   ...          \\
20-022 &  IRAS 04108+2803 B & ...       &  ...      &         ... &   ... &        3500 &               0.400 &   10           \\
20-000 &  2M J04141188+28   & M6.25     &   33      &         ... &  0.28 &        2962 &               0.015 &   33           \\
20-042 &  V773 Tau ABC      & K2/M0     &   58      &        1.39 &  0.31 &   4898/3873 &    1.89/ 1.17/ 5.60 &    5, 58       \\
20-043 &  FM Tau            & M0        &    5      &        0.69 &  0.59 &        3850 &               0.460 &    5, 27       \\
20-046 &  CW Tau            & K3        &    5      &        2.29 &  0.55 &        4730 &               1.100 &    5, 27       \\
20-047 &  CIDA 1            & M5.5      &   57      &         ... &   ... &         ... &                 ... &    ...         \\
20-056 &  MHO 2/1           & M2.5/2.5  &    5      &        4.58 &  1.85 &   3488/3488 &    0.42/ 0.14/ 0.56 &    5,  4       \\
20-058 &  MHO 3             & K7        &    5      &        6.01 &  2.15 &        4060 &               0.910 &    5,  4       \\
20-069 &  FO Tau AB         & M2        &    5      &        1.87 &  0.63 &   3556/3556 &    0.48/ 0.45/ 0.77 &    5, 58, 27   \\
20-073 &  CIDA 2            & M5.5      &    5      &        0.83 &  0.28 &        3058 &               0.320 &    5, 27       \\
23-002 &  CY Tau            & M1.5      &    5      &        0.10 &  0.23 &        3632 &               0.500 &    5, 27       \\
24-002 &  CY Tau            & M1.5      &    5      &        0.10 &  0.23 &        3632 &               0.500 &    5, 27       \\
23-004 &  LkCa 5            & M2        &   27      &        0.10 &  0.11 &        3560 &               0.370 &    5, 27       \\
24-004 &  LkCa 5            & M2        &   27      &        0.10 &  0.11 &        3560 &               0.370 &    5, 27       \\
23-008 &  CIDA 3            & M4        &    5      &         ... &  0.26 &        3270 &               0.140 &    5           \\
24-008 &  CIDA 3            & M4        &    5      &         ... &  0.26 &        3270 &               0.140 &    5           \\
23-015 &  V410 X3           & M6        &    5      &        0.54 &  0.17 &        2990 &               0.084 &    5, 58       \\
24-015 &  V410 X3           & M6        &    5      &        0.54 &  0.17 &        2990 &               0.084 &    5, 58       \\
23-018 &  V410 A13          & M5.75     &    5      &        2.78 &  0.79 &        3024 &               0.039 &    5, 4        \\
24-000 &  V410 A13          & M5.75     &    5      &        2.78 &  0.79 &        3024 &               0.039 &    5, 4        \\
23-000 &  V410 A24          & G1        &    5      &         ... &  6.73 &        5945 &               2.800 &    5           \\
24-000 &  V410 A24          & G1        &    5      &         ... &  6.73 &        5945 &               2.800 &    5           \\
23-029 &  V410 A25          & M1        &    5      &         ... &  6.55 &        3705 &               1.600 &    5           \\
24-027 &  V410 A25          & M1        &    5      &         ... &  6.55 &        3705 &               1.600 &    5           \\
23-032 &  V410 Tau ABC      & K4        &   58      &        0.67 &  0.00 &   4602/3076 &    2.15/ 0.05/ 2.20 &    58          \\
24-028 &  V410 Tau ABC      & K4        &   58      &        0.67 &  0.00 &   4602/3076 &    2.15/ 0.05/ 2.20 &    58          \\
23-033 &  DD Tau AB         & M3        &    5      &  0.39/ 0.39 &  0.05 &   3412/3412 &    0.13/ 0.13/ 0.34 &    5, 58       \\
24-029 &  DD Tau AB         & M3        &    5      &  0.39/ 0.39 &  0.05 &   3412/3412 &    0.13/ 0.13/ 0.34 &    5, 58       \\
23-035 &  CZ Tau AB         & M3        &    5      &        1.32 &  0.48 &        3415 &               0.270 &    5, 27       \\
24-030 &  CZ Tau AB         & M3        &    5      &        1.32 &  0.48 &        3415 &               0.270 &    5, 27       \\
23-036 &  IRAS 04154+2823   & M2.5      &    5      &         ... &  4.26 &        3488 &               0.130 &    5           \\
24-031 &  IRAS 04154+2823   & M2.5      &    5      &         ... &  4.26 &        3488 &               0.130 &    5           \\
23-037 &  V410 X2           & M0        &    5      &         ... &  6.03 &        3850 &               3.000 &    5           \\
24-032 &  V410 X2           & M0        &    5      &         ... &  6.03 &        3850 &               3.000 &    5           \\
23-045 &  V410 X4           & M4        &    5      &         ... &  5.29 &        3270 &               1.300 &    5           \\
24-038 &  V410 X4           & M4        &    5      &         ... &  5.29 &        3270 &               1.300 &    5           \\
23-047 &  V892 Tau          & B9        &    5      &        5.93 &  2.39 &       10500 &              77.000 &    5, 27       \\
24-040 &  V892 Tau          & B9        &    5      &        5.93 &  2.39 &       10500 &              77.000 &    5, 27       \\
23-048 &  LR 1              & K4.5      &    5      &         ... &  6.40 &        4470 &               0.470 &    5           \\
24-000 &  LR 1              & K4.5      &    5      &         ... &  6.40 &        4470 &               0.470 &    5           \\
23-050 &  V410 X7           & M0.75     &    5      &        8.18 &  2.34 &        3741 &               0.500 &    5, 58       \\
24-042 &  V410 X7           & M0.75     &    5      &        8.18 &  2.34 &        3741 &               0.500 &    5, 58       \\
23-000 &  V410 A20          & K3        &    5      &         ... &  6.57 &        4730 &               0.510 &    5           \\
24-000 &  V410 A20          & K3        &    5      &         ... &  6.57 &        4730 &               0.510 &    5           \\
23-056 &  Hubble 4          & K7        &    5      &        0.76 &  0.68 &        4060 &               2.700 &    5, 27       \\
24-047 &  Hubble 4          & K7        &    5      &        0.76 &  0.68 &        4060 &               2.700 &    5, 27       \\
23-000 &  KPNO-Tau 2        & M6.75     &   18      &        0.37 &  0.00 &        2889 &               0.007 &   18, 5        \\
24-000 &  KPNO-Tau 2        & M6.75     &   18      &        0.37 &  0.00 &        2889 &               0.007 &   18, 5        \\
23-000 &  CoKu Tau 1        & K7        &   59      &        6.80 &   ... &        4000 &               0.150 &   59           \\
24-000 &  CoKu Tau 1        & K7        &   59      &        6.80 &   ... &        4000 &               0.150 &   59           \\
23-061 &  V410 X6           & M5.5      &    5      &         ... &  0.17 &        3058 &               0.200 &    5           \\
\hline
\end{tabular}
\normalsize
\end{table*}
 
\setcounter{table}{8}
\begin{table*}[h!]
\caption{(Continued)}
\begin{tabular}{llrrrrrrr}
\hline
\hline
XEST & Name & Spec$^a$  & Refs & $A_{\rm V}^a$ & $A_{\rm J}$   & $T_{\rm eff}^a$ &  $L_*^b$        & Refs    \\
   &      &           &      & (mag)   & (mag)   &  (K)            & ($L_{\odot})$     &        \\
\hline
24-054 &  V410 X6           & M5.5      &    5      &         ... &  0.17 &        3058 &               0.200 &    5           \\
23-063 &  V410 X5           & M5.5      &    5      &        3.36 &  0.72 &        3058 &               0.083 &    5, 58       \\
24-055 &  V410 X5           & M5.5      &    5      &        3.36 &  0.72 &        3058 &               0.083 &    5, 58       \\
23-067 &  FQ Tau AB         & M3/M3.5   &   20      &  1.95/ 1.80 &  0.16 &   3416/3345 &    0.20/ 0.28/ 0.21 &    5, 20       \\
24-058 &  FQ Tau AB         & M3/M3.5   &   20      &  1.95/ 1.80 &  0.16 &   3416/3345 &    0.20/ 0.28/ 0.21 &    5, 20       \\
28-100 &  BP Tau            & K7        &   27      &        0.49 &  0.14 &        4060 &               0.950 &   27           \\
23-074 &  V819 Tau AB       & K7        &    5      &        1.35 &  0.48 &        4060 &               0.910 &    5, 27       \\
24-061 &  V819 Tau AB       & K7        &    5      &        1.35 &  0.48 &        4060 &               0.910 &    5, 27       \\
16-000 &  IRAS 04166+2706   & $<$M0     &   66      &         ... &   ... &         ... &                 ... &   ...          \\
16-000 &  IRAS 04169+2702   & ...       &   ...     &         ... &   ... &         ... &               0.800 &   13           \\
11-000 &  CFHT-Tau 19       & M5.25     &   18      &        7.30 &   ... &        3100 &               0.072 &   18           \\
11-000 &  IRAS 04181+2655   & ...       &   ...     &         ... &   ... &        4000 &               1.800 &   10           \\
11-000 &  IRAS 04181+2654AB & ...       &   ...     &         ... &   ... &         ... &    0.25/ 0.25/ 0.50 &   13           \\
11-023 &  2M J04213459      & M5.5      &   33      &         ... &  0.49 &        3058 &               0.065 &   33           \\
01-028 &  IRAS 04187+1927   & M0        &   27      &         ... &   ... &        3850 &                 ... &   27           \\
11-037 &  CFHT-Tau 10       & M5.75     &   66      &        3.59 &   ... &        3030 &               0.021 &   18, 19       \\
11-000 &  2M J04215450+2652 & M8.5      &   66      &        2.97 &   ... &        2642 &               0.003 &   19           \\
21-038 &  RY Tau            & K1        &   27      &        1.84 &  0.53 &        5080 &               7.600 &   27           \\
21-039 &  HD 283572         & G5        &   27      &        0.38 &  0.11 &        5770 &               6.500 &   27           \\
01-045 &  T Tau N(+Sab)     & K0        &   27      &        1.39 &  0.40 &        5250 &               8.910 &   27           \\
11-054 &  Haro 6-5 B        & K5        &   59      &        9.96 &   ... &        4395 &               0.047 &   59           \\
11-057 &  FS Tau AC         & M0/M3.5   &   20      &  4.95/ 5.15 &  0.53 &   3876/3345 &    0.15/ 0.17/ 0.32 &   27, 20       \\
21-044 &  LkCa 21           & M3        &   27      &        0.73 &  0.21 &        3470 &               0.620 &   27           \\
01-054 &  RX J0422.1+1934   & ...       &   ...     &         ... &   ... &         ... &                 ... &   ...          \\
01-062 &  2M J04221332+1934 & M8        &   66      &        1.02 &   ... &        2713 &               0.017 &   19           \\
11-079 &  CFHT-Tau 21       & M1.25     &   18      &        6.60 &   ... &        3665 &               0.381 &   18           \\
02-013 &  FV Tau AB         & K5/K6     &   58      &  5.33/ 5.33 &  1.15 &   4395/4130 &    0.98/ 0.44/ 1.20 &    5, 58       \\
02-000 &  FV Tau/c AB       & M2.5/3.5  &   20      &  3.25/ 7.00 &  0.49 &   3412/3155 &    0.18/ 0.06/ 0.21 &    5, 20, 58   \\
02-016 &  KPNO-Tau 13       & M5        &   34      &         ... &  0.70 &        3125 &               0.150 &   34           \\
02-000 &  DG Tau B          & ...       &  ...      &         ... &   ... &        4000 &               5.500 &   10           \\
02-022 &  DG Tau A          & K6        &    5      &        1.41 &  0.36 &        4205 &               1.700 &    5, 58       \\
02-000 &  KPNO-Tau 4        & M9.5      &    5, 18  &        2.45 &  0.00 &        2500 &               0.004 &    5, 18       \\
02-000 &  IRAS 04248+2612AB & M2-5.5    &    5, 59  &        7.02 &  1.51 &        2845 &               0.270 &    5, 59       \\
15-020 &  JH 507            & M4        &    5      &        0.76 &  0.29 &        3270 &               0.420 &    5, 27       \\
13-004 &  GV Tau AB         & K3-7      &   27, 59  &       12.10 &   ... &        4000 &               1.820 &   59           \\
13-000 &  IRAS 04264+2433   & M1        &   59      &       10.40 &   ... &        3605 &               0.140 &   59           \\
15-040 &  DH Tau AB         & M1        &    5      &        1.25 &  0.32 &        3705 &               0.560 &    5, 27       \\
15-042 &  DI Tau AB         & M0        &    5      &        0.76 &  0.43 &        3850 &               0.990 &    5, 27       \\
15-044 &  KPNO-Tau 5        & M7.5      &    5, 19  &        0.00 &  0.00 &        2783 &               0.023 &   19           \\
14-006 &  IQ Tau A          & M0.5      &    5      &        1.25 &  0.45 &        3778 &               0.880 &    5, 27       \\
13-000 &  CFHT-Tau 20       & M5.5      &   18      &        3.60 &   ... &        3065 &               0.138 &   18           \\
14-000 &  KPNO-Tau 6        & M9.0      &   18      &        0.88 &  0.00 &        2571 &               0.003 &    5, 18       \\
13-035 &  FX Tau AB         & M1        &   27      &        1.08 &  0.31 &        3720 &               1.020 &   27           \\
14-057 &  DK Tau AB         & K7        &    5      &        0.76 &  0.28 &        4060 &               1.300 &    5, 27       \\
14-000 &  KPNO-Tau 7        & M8.25     &    5      &         ... &  0.00 &        2632 &               0.003 &    5           \\
22-013 &  MHO 9             & M4.25     &    5      &        1.73 &  0.56 &        3234 &               0.220 &    5,  4       \\
22-021 &  MHO 4             & M7.1      &    5, 18  &        1.10 &  0.14 &        2880 &               0.048 &    5, 58       \\
22-040 &  L1551 IRS5        & ...       &  ...      &         ... &   ... &        4800 &               2.600 &   10           \\
22-042 &  LkHa 358          & M5.5      &    5      &       13.60 &  3.84 &        3058 &               0.590 &    5, 45       \\
22-000 &  HH 30             & M0        &   59      &        2.96 &   ... &        3800 &               0.006 &   59           \\
22-043 &  HL Tau            & K5        &   59      &        7.43 &   ... &        4395 &               1.530 &   59           \\
22-047 &  XZ Tau AB         & M2/M3.5   &   20      &  1.40/ 1.35 &  0.11 &   3561/3345 &    0.17/ 0.31/ 0.33 &    5, 20       \\
22-056 &  L1551 NE          & ...       &  ...      &         ... &   ... &         ... &                 ... &   ...          \\
03-005 &  HK Tau AB         & M0.5/M2   &    5      &        2.32 &  0.97 &   3778/3560 &               0.560 &    5, 27       \\
22-070 &  V710 Tau BA       & M0.5/M2   &    5      &  1.80/ 1.82 &  0.62 &   3778/3560 &    0.58/ 0.52/ 1.10 &    5, 58       \\
19-009 &  JH 665            & M5.5      &    5      &        0.97 &  0.54 &        3058 &               0.260 &    5, 27       \\
22-089 &  L1551 51          & K7        &    5      &        0.00 &  0.00 &        4060 &               0.470 &    5, 27       \\
22-097 &  V827 Tau          & K7        &    5      &        0.28 &  0.31 &        4060 &               1.100 &    5, 27       \\
03-016 &  Haro 6-13         & M0        &   59      &       11.90 &   ... &        3800 &               2.110 &   59           \\
\hline
\end{tabular}
\normalsize
\end{table*}
 
\setcounter{table}{8}
\begin{table*}[h!]
\caption{(Continued)}
\begin{tabular}{llrrrrrrr}
\hline
\hline
XEST & Name & Spec$^a$  & Refs & $A_{\rm V}^a$ & $A_{\rm J}$   & $T_{\rm eff}^a$ &  $L_*^b$        & Refs    \\
   &      &           &      & (mag)   & (mag)   &  (K)            & ($L_{\odot})$     &        \\
\hline
22-100 &  V826 Tau          & K7        &    5      &        0.28 &  0.19 &        4060 &               0.940 &    5, 27       \\
22-101 &  MHO 5             & M6        &    5      &        0.23 &  0.17 &        2990 &               0.110 &    5, 58       \\
03-017 &  CFHT-Tau 7        & M5.75     &   66      &        0.00 &   ... &        3030 &               0.061 &   18, 19       \\
03-019 &  V928 Tau AB       & M0.5      &    5      &        1.87 &  1.04 &        3778 &               1.400 &    5, 27       \\
03-022 &  FY Tau            & K5        &    5      &        3.47 &  1.07 &        4350 &               1.000 &    5, 27       \\
03-023 &  FZ Tau            & M0        &    5      &        2.72 &  0.99 &        3850 &               0.980 &    5, 58       \\
17-002 &  IRAS 04295+2251   & ...       &  ...      &         ... &   ... &        3400 &               1.200 &   10           \\
19-049 &  UZ Tau E+W(AB)    & M1/2/3    &   27, 20  &  1.49/ 0.83 &  0.26 &   3705/3560 &    0.40/ 0.49/ 0.89 &    5, 27       \\
17-009 &  JH 112            & K6        &   27      &        3.23 &  0.93 &        4205 &               0.740 &   27           \\
03-031 &  CFHT-Tau 5        & M7.5      &   18      &        9.22 &   ... &        2783 &               0.075 &   18           \\
04-003 &  CFHT-Tau 5        & M7.5      &   18      &        9.22 &   ... &        2783 &               0.075 &   18           \\
03-035 &  MHO 8             & M6        &    5      &        0.19 &  0.28 &        2990 &               0.170 &    5, 4        \\
04-009 &  MHO 8             & M6        &    5      &        0.19 &  0.28 &        2990 &               0.170 &    5, 4        \\
04-010 &  GH Tau AB         & M1.5/M2   &   58      &  0.69/ 0.64 &  0.11 &   3631/3556 &    0.38/ 0.34/ 0.81 &    5, 58       \\
04-012 &  V807 Tau  SNab    & K7/M3     &   58      &  0.36/ 0.36 &  0.04 &   3999/3388 &    1.07/ 0.32/ 2.10 &    5, 58       \\
18-004 &  KPNO-Tau 14       & M6        &   34      &         ... &  0.85 &        2990 &               0.110 &   34           \\
17-000 &  CFHT-Tau 12       & M6        &   66      &        3.44 &   ... &        2995 &               0.036 &   18, 19       \\
04-016 &  V830 Tau          & K7        &    5      &        0.28 &  0.13 &        4060 &               0.780 &    5, 27       \\
18-000 &  IRAS S04301+261   & M0        &    5      &         ... &  1.76 &        3850 &               0.025 &    5           \\
17-000 &  IRAS 04302+2247   & ...       & ...       &         ... &   ... &         ... &               0.300 &   13           \\
17-027 &  IRAS 04303+2240   & M0.5      &   59      &       11.70 &   ... &        3700 &               2.200 &   59           \\
04-034 &  GI Tau            & K7        &    5      &        0.87 &  0.43 &        4060 &               1.000 &    5, 27       \\
04-035 &  GK Tau AB         & K7        &    5      &        0.87 &  0.46 &        4060 &               1.400 &    5, 27       \\
18-019 &  IS Tau AB         & K7/M4.5   &   58      &  3.67/ 3.67 &  0.95 &   3999/3221 &    0.50/ 0.19/ 0.66 &    5, 58       \\
17-058 &  CI Tau            & K7        &   27      &        1.77 &  0.51 &        4060 &               0.870 &   27           \\
18-030 &  IT Tau AB         & K2        &    5      &        4.09 &  1.86 &        4900 &               2.400 &    5, 27       \\
17-066 &  JH 108            & M1        &   27      &        1.46 &  0.42 &        3720 &               0.300 &   27           \\
17-068 &  CFHT-BD Tau 1     & M7.1      &   39, 19  &        3.10 &   ... &        2853 &               0.017 &   39, 19       \\
25-026 &  AA Tau            & K7        &    5      &        0.49 &  0.27 &        4060 &               0.800 &    5, 27       \\
09-010 &  HO Tau AB         & M0.5      &    5      &        1.11 &  0.46 &        3778 &               0.170 &    5, 27       \\
08-019 &  FF Tau AB         & K7        &    5      &        2.22 &  0.51 &        4060 &               0.690 &    5, 27       \\
12-040 &  DN Tau            & M0        &    5      &        0.49 &  0.34 &        3850 &               1.000 &    5, 27       \\
12-000 &  IRAS 04325+2402AB & ...       &  ...      &         ... &   ... &         ... &               0.720 &   28           \\
12-059 &  CoKu Tau 3 AB     & M1        &    5      &        3.26 &  2.15 &        3705 &               0.980 &    5, 27       \\
09-022 &  KPNO-Tau 8        & M5.75     &    5      &         ... &  0.14 &        3024 &               0.021 &    5           \\
08-037 &  HQ Tau AB         & ...       &  ...      &         ... &   ... &         ... &                 ... &  ...           \\
09-026 &  HQ Tau AB         & ...       &  ...      &         ... &   ... &         ... &                 ... &  ...           \\
08-043 &  KPNO-Tau 15       & M2.75     &   34      &         ... &  0.56 &        3451 &               0.140 &   34           \\
09-031 &  KPNO-Tau 15       & M2.75     &   34      &         ... &  0.56 &        3451 &               0.140 &   34           \\
08-000 &  KPNO-Tau 9        & M8.5      &    5      &         ... &  0.00 &        2555 &               0.001 &    5           \\
09-000 &  KPNO-Tau 9        & M8.5      &    5      &         ... &  0.00 &        2555 &               0.001 &    5           \\
08-048 &  HP Tau AB         & K3        &    5      &        2.26 &  0.91 &        4730 &               1.400 &    5, 27       \\
08-051a&  HP Tau/G3 AB      & K7        &    5      &        2.32 &  0.77 &        4060 &               0.710 &    5, 27       \\
08-051 &  HP Tau/G2         & G0        &    5      &        2.08 &  0.66 &        6030 &               6.500 &    5, 27       \\
08-058 &  Haro 6-28 AB      & M2/M3.5   &   20      &  2.30/ 1.90 &  1.34 &   3556/3342 &    0.09/ 0.03/ 0.12 &    5, 20       \\
08-000 &  CFHT-BD Tau 2     & M7.5      &    5      &        0.00 &  0.56 &        2795 &               0.011 &    5, 39       \\
08-080 &  CFHT-BD Tau 3     & M7.75     &    5      &        0.00 &  0.28 &        2747 &               0.007 &    5, 39       \\
05-005 &  CFHT-Tau 6        & M7.25     &   33, 18  &        0.41 &  0.07 &        2818 &               0.024 &   33, 18       \\
05-000 &  IRAS 04361+2547   & ...       &  ...      &         ... &   ... &         ... &               3.700 &   13           \\
05-013 &  GN Tau AB         & M2.5      &   33      &         ... &  1.17 &        3488 &               0.720 &   33           \\
05-017 &  IRAS 04365+2535   & ...       &  ...      &         ... &   ... &         ... &               2.200 &   27(Lb)$^c$   \\
05-024 &  IRAS 04369+2539   & K4        &   59      &       18.10 &   ... &        4580 &              20.700 &   59           \\
07-011 &  JH 223            & M2        &   27      &         ... &  0.14 &        3560 &               0.180 &   33           \\
07-022 &  Haro 6-32         & M5        &   33      &         ... &  0.17 &        3125 &               0.120 &   33           \\
07-000 &  ITG 33 A          & M3        &   33      &        3.50 &  1.89 &        3415 &               0.051 &   33, 35       \\
07-000 &  CFHT-Tau 8        & M5.5      &   33      &        1.77 &  0.70 &        3058 &               0.024 &   33, 18, 19   \\
07-000 &  IRAS 04381+2540   & ...       &  ...      &         ... &   ... &         ... &               0.700 &   13           \\
07-041 &  IRAS 04385+2550AB & M0        &   33      &       10.20 &  1.13 &        3850 &               0.180 &   33, 59       \\
10-017 &  CoKuLk332/G2 AB   & M0.5/2.5  &   20      &  2.00/ 3.30 &  0.99 &   3778/3485 &    0.24/ 0.15/ 1.10 &   33, 20       \\
\hline
\end{tabular}
\normalsize
\end{table*}
 
\setcounter{table}{8}
\begin{table*}[h!]
\caption{(Continued)}
\begin{tabular}{llrrrrrrr}
\hline
\hline
XEST & Name & Spec$^a$  & Refs & $A_{\rm V}^a$ & $A_{\rm J}$   & $T_{\rm eff}^a$ &  $L_*^b$        & Refs    \\
   &      &           &      & (mag)   & (mag)   &  (K)            & ($L_{\odot})$     &        \\
\hline
10-018 &  CoKuLk332/G1 AB   & K7/M1     &   58      &  4.13/ 4.13 &  1.33 &   3707/3945 &    0.85/ 0.58/ 1.70 &   33, 58       \\
10-020 &  V955 Tau AB       & K5/M1     &   58      &  3.72/ 3.72 &  0.90 &   4395/3715 &    0.32/ 0.18/ 1.00 &   33, 58       \\
10-034 &  CIDA 7            & M4.75     &   66      &         ... &   ... &         ... &               0.050 &   27(Lb)$^c$   \\
10-045 &  DP Tau            & M0.5      &   27      &        1.46 &  0.41 &        3778 &               0.200 &   33, 27       \\
10-060 &  GO Tau            & M0        &   27      &        1.18 &  0.77 &        3850 &               0.370 &   33, 27       \\
26-012 &  2M J04552333+30   & M6.25     &   33, 19  &        0.00 &  0.00 &        2959 &               0.015 &   33, 19       \\
26-034 &  2M J04554046+30   & M5.25     &   33      &         ... &  0.07 &        3091 &               0.021 &   33           \\
26-043 &  AB Aur            & B9.5-A0   &   27, 7   &        0.25 &  0.24 &       10050 &              49.000 &   33, 27, 8, 9 \\
26-050 &  2MJ04554757/801   & M4.75/5.6 &   33      &         ... &  0.00 &   3161/3044 &    0.10/ 0.01/ 0.11 &   33           \\
26-067 &  SU Aur            & G2        &   27      &        0.90 &  0.21 &        5860 &               9.900 &   33, 27       \\
26-072 &  HBC 427           & K7        &   27      &        0.00 &  0.17 &        4350 &               1.100 &   33, 27       \\
\hline
\multicolumn{9}{l}{Additional sources from Chandra}\\
\hline
C1-0   &  KPNO-Tau 10       & M5        &   34      &         ... &  0.14 &        3125 &               0.052 &   34           \\
C1-1   &  IRAS 04158+2805   & M3        &    5      &        8.63 &   ... &        2760 &               0.050 &   59           \\
C2-1   &  Haro 6-5 B        & K5        &   59      &        9.96 &   ... &        4395 &               0.047 &   59           \\
C2-2   &  FS Tau AC         & M0/M3.5   &   20      &  4.95/ 5.15 &  0.53 &   3876/3345 &    0.15/ 0.17/ 0.32 &   27, 20       \\
C3-1   &  FV Tau/c AB       & M2.5/3.5  &   20      &  3.25/ 7.00 &  0.49 &   3412/3155 &    0.18/ 0.06/ 0.21 &    5, 20, 58   \\
C3-2   &  DG Tau B          & ...       &  ...      &         ... &   ... &        4000 &               5.500 &   10           \\
C4-1   &  GV Tau AB         & K3-7      &   27, 59  &       12.10 &   ... &        4000 &               1.820 &   59           \\
C5-2   &  HN Tau AB         & K5/M4     &    5      &  1.18/ 0.91 &  0.12 &   4395/3273 &    0.22/ 0.03/ 0.25 &    5, 58       \\
C5-1   &  L1551 55          & K7        &    5      &        0.69 &  0.20 &        4060 &               0.370 &    5, 27       \\
C5-4   &  HD 28867          & B9IVn     &   65      &        0.25 &   ... &       10500 &  64.90/62.10/127.00 &   65           \\
C5-3   &  DM Tau            & M1        &    5      &        0.59 &  0.31 &        3705 &               0.300 &    5, 58       \\
C6-1   &  CFHT-BD Tau 4     & M7        &   33      &        2.60 &  0.70 &        2853 &               0.062 &   33, 39, 19   \\
C6-0   &  L1527 IRS         & ...       &  ...      &         ... &   ... &         ... &               1.300 &   27(Lb)$^c$   \\
C6-0   &  CFHT-Tau 17       & M5.5      &   66      &        6.50 &   ... &        3030 &               0.068 &   18           \\
C6-2   &  IRAS 04370+2559   & ...       & ...       &         ... &   ... &         ... &               0.210 &   27(Lb)$^c$   \\
\hline
\end{tabular}
\begin{minipage}{0.89\textwidth}
\footnotetext{
\hskip -0.5truecm $^a$ For multiples, first number or spectral type refers to primary, second to secondary component\\
$^b$ For multiples, three numbers give primary/secondary/total system luminosity\\
$^c$ Referring to $L_{\rm bol}$ as derived from integration of the optical and infrared spectrum\\
NOTES on individual objects: 
\begin{itemize}
\item V773 Tau = XEST-20-042: Multiple entries refer to A and C, respectively
\item HD~28867 = C5-4: total $L_*$ \citep{berghoefer96} scaled to $d=$ 140~pc, split to components according to $V$ mag \citep{walter03} 
\item FQ Tau = XEST-23-067 = XEST-24-058, FS Tau = XEST-11-057, XZ Tau = XEST-22-047, CoKuLk332/G2 = XEST-10-017: $T_{\rm eff}$ of components read from Fig.~10 in Ref.~20 
\end{itemize}
}
\end{minipage}
\label{tab9}
\normalsize
\end{table*}

\clearpage

\setcounter{table}{9}
\begin{table*}[h!]
\caption{Fundamental parameters of targets in XEST (4): Age, mass, radius, rotation}
\begin{tabular}{llrlrlrlrl}
\hline
\hline
XEST & Name & Age$^{a,b}$   &  Mass$^{a,c}$       & Refs & Radius$^d$    &  $P$ & Refs & $v\sin i$         & Refs  \\
    &       & (Myr)         & ($M_{\odot}$)       &      & ($R_{\odot}$) &  (d) &      & (km~s$^{-1}$)     &        \\
\hline
27-115 &  HBC 352           &       ... &        1.05 &         2 &  0.79 & $<$ 0.53 & C          & $>$ 75.00 &   48       \\
27-000 &  HBC 353           &       ... &         ... &       ... &  0.71 & $<$ 4.08 & C          &      8.80 &    2       \\
06-005 &  HBC 358 AB        &      3.26 &        0.41 &        27 &  1.38 &      ... & C          & $<$ 10.00 &    2       \\
06-007 &  HBC 359           &      3.54 &        0.41 &        27 &  1.33 &      ... & C          & $<$ 10.00 &   48       \\
06-059 &  L1489 IRS         &      0.80 &        1.45 &        10 &  3.65 & $<$ 4.02 & C          &     46.00 &   10       \\
20-001 &  LkCa 1            &      0.87 &        0.27 &         5 &  1.93 & $<$ 4.18 & C          &     23.30 &   48       \\
20-005 &  Anon 1            &      0.50 &        0.56 &         5 &  3.63 &      ... & ...        &       ... &   ...      \\
20-000 &  IRAS 04108+2803 A &       ... &         ... &       ... &   ... &      ... & ...        &       ... &   ...      \\
20-022 &  IRAS 04108+2803 B &      1.60 &        0.36 &        10 &  1.72 & $<$ 6.23 & C          &     14.00 &   10       \\
20-000 &  2M J04141188+28   &       ... &        0.08 &        46 &  0.47 &      ... & ...        &       ... &   ...      \\
20-042 &  V773 Tau ABC      & 6.35/0.93 &  1.53/ 0.57 &        58 &  1.91 &     3.43 & 48         &     55.00 &   48       \\
20-043 &  FM Tau            &      2.76 &        0.57 &         5 &  1.53 &      ... & ...        &       ... &   ...      \\
20-046 &  CW Tau            &      6.97 &        1.40 &        27 &  1.57 &     8.25 & 48         &     27.40 &   48       \\
20-047 &  CIDA 1            &       ... &         ... &       ... &   ... &      ... & ...        &      5.30 &   48       \\
20-056 &  MHO 2/1           & 1.54/4.85 &  0.36/ 0.34 &         5 &  1.78 & $<$ 4.19 & C          &     21.50 &   59       \\
20-058 &  MHO 3             &       2.0 &        0.75 &         5 &  1.93 &      ... & ...        &       ... &   ...      \\
20-069 &  FO Tau AB         & 1.45/1.54 &  0.38/ 0.38 &        58 &  1.83 &      ... & ...        &       ... &   ...      \\
20-073 &  CIDA 2            &      0.39 &        0.20 &         5 &  2.02 &      ... & ...        &       ... &   ...      \\
23-002 &  CY Tau            &      1.53 &        0.42 &         5 &  1.79 &     7.50 & 48         &     10.00 &   48       \\
24-002 &  CY Tau            &      1.53 &        0.42 &         5 &  1.79 &     7.50 & 48         &     10.00 &   48       \\
23-004 &  LkCa 5            &      2.31 &        0.40 &         5 &  1.60 & $<$ 2.19 & C          &     37.00 &   48       \\
24-004 &  LkCa 5            &      2.31 &        0.40 &         5 &  1.60 & $<$ 2.19 & C          &     37.00 &   48       \\
23-008 &  CIDA 3            &      3.34 &        0.24 &         5 &  1.17 &      ... & ...        &       ... &   ...      \\
24-008 &  CIDA 3            &      3.34 &        0.24 &         5 &  1.17 &      ... & ...        &       ... &   ...      \\
23-015 &  V410 X3           &      2.68 &        0.12 &         5 &  1.08 & $<$ 3.79 & C          &     14.10 &   48       \\
24-015 &  V410 X3           &      2.68 &        0.12 &         5 &  1.08 & $<$ 3.79 & C          &     14.10 &   48       \\
23-018 &  V410 A13          &       ... &        0.10 &        58 &  0.72 & $<$ 3.72 & C          &      9.80 &   45       \\
24-000 &  V410 A13          &      ...  &        0.10 &        58 &  0.72 & $<$ 3.72 & C          &      9.80 &   45       \\
23-000 &  V410 A24          &      18.7 &        1.22 &         5 &  1.58 &      ... & ...        &       ... &   ...      \\
24-000 &  V410 A24          &      18.7 &        1.22 &         5 &  1.58 &      ... & ...        &       ... &   ...      \\
23-029 &  V410 A25          &      0.64 &        0.46 &         5 &  3.08 &      ... & ...        &       ... &   ...      \\
24-027 &  V410 A25          &      0.64 &        0.46 &         5 &  3.08 &      ... & ...        &       ... &   ...      \\
23-032 &  V410 Tau ABC      & 2.74/5.22 &  1.51/ 0.12 &        58 &  2.31 &     1.94 & 48         &     71.00 &   48       \\
24-028 &  V410 Tau ABC      & 2.74/5.22 &  1.51/ 0.12 &        58 &  2.31 &     1.94 & 48         &     71.00 &   48       \\
23-033 &  DD Tau AB         & 4.47/4.47 &  0.30/ 0.30 &        58 &  1.03 &      ... & ...        &       ... &   ...      \\
24-029 &  DD Tau AB         & 4.47/4.47 &  0.30/ 0.30 &        58 &  1.03 &      ... & ...        &       ... &   ...      \\
23-035 &  CZ Tau AB         &      2.10 &        0.32 &         5 &  1.49 &      ... & ...        &       ... &   ...      \\
24-030 &  CZ Tau AB         &      2.10 &        0.32 &         5 &  1.49 &      ... & ...        &       ... &   ...      \\
23-036 &  IRAS 04154+2823   &      5.30 &        0.33 &         5 &  0.99 &      ... & ...        &       ... &   ...      \\
24-031 &  IRAS 04154+2823   &      5.30 &        0.33 &         5 &  0.99 &      ... & ...        &       ... &   ...      \\
23-037 &  V410 X2           &      0.44 &        0.56 &         5 &  3.90 &      ... & ...        &       ... &   ...      \\
24-032 &  V410 X2           &      0.44 &        0.56 &         5 &  3.90 &      ... & ...        &       ... &   ...      \\
23-045 &  V410 X4           &      0.06 &        0.27 &         5 &  3.56 &      ... & ...        &       ... &   ...      \\
24-038 &  V410 X4           &      0.06 &        0.27 &         5 &  3.56 &      ... & ...        &       ... &   ...      \\
23-047 &  V892 Tau          &      2.97 &        2.89 &         5 &  2.66 &      ... & ...        &       ... &   ...      \\
24-040 &  V892 Tau          &      2.97 &        2.89 &         5 &  2.66 &      ... & ...        &       ... &   ...      \\
23-048 &  LR 1              &      17.5 &        1.00 &         5 &  1.15 &      ... & ...        &       ... &   ...      \\
24-000 &  LR 1              &      17.5 &        1.00 &         5 &  1.15 &      ... & ...        &       ... &   ...      \\
23-050 &  V410 X7           &      1.90 &        0.50 &         5 &  1.69 &      ... & ...        &       ... &   ...      \\
24-042 &  V410 X7           &      1.90 &        0.50 &         5 &  1.69 &      ... & ...        &       ... &   ...      \\
23-000 &  V410 A20          &      20.7 &        1.01 &         5 &  1.06 &      ... & ...        &       ... &   ...      \\
24-000 &  V410 A20          &      20.7 &        1.01 &         5 &  1.06 &      ... & ...        &       ... &   ...      \\
23-056 &  Hubble 4          &      0.68 &        0.74 &         5 &  3.33 & $<$13.17 & C          &     12.80 &   48       \\
24-047 &  Hubble 4          &      0.68 &        0.74 &         5 &  3.33 & $<$13.17 & C          &     12.80 &   48       \\
23-000 &  KPNO-Tau 2        &       ... &        0.05 &        46 &  0.34 &      ... & ...        &       ... &   ...      \\
24-000 &  KPNO-Tau 2        &       ... &        0.05 &        46 &  0.34 &      ... & ...        &       ... &   ...      \\
23-000 &  CoKu Tau 1        &      29.1 &        0.70 &        59 &  0.81 & $<$ 2.67 & C          &     15.30 &   59       \\
24-000 &  CoKu Tau 1        &      29.1 &        0.70 &        59 &  0.81 & $<$ 2.67 & C          &     15.30 &   59       \\
23-061 &  V410 X6           &      0.71 &        0.18 &         5 &  1.60 &      ... & ...        &       ... &   ...      \\
\hline
\end{tabular}
\normalsize
\end{table*}
 
 \setcounter{table}{9}
\begin{table*}[h!]
\caption{(Continued)}
\begin{tabular}{llrlrlrlrl}
\hline
\hline
XEST & Name & Age$^{a,b}$   &  Mass$^{a,c}$      & Refs & Radius$^d$    &  $P$ & Refs & $v\sin i$     & Refs  \\
     &      & (Myr)         & ($M_{\odot}$)      &      & ($R_{\odot}$) &  (d) &      & (km~s$^{-1}$) &            \\
\hline
24-054 &  V410 X6           &      0.71 &        0.18 &         5 &  1.60 &      ... & ...        &       ... &   ...      \\
23-063 &  V410 X5           &      4.21 &        0.14 &         5 &  1.03 &      ... & ...        &       ... &   ...      \\
24-055 &  V410 X5           &      4.21 &        0.14 &         5 &  1.03 &      ... & ...        &       ... &   ...      \\
23-067 &  FQ Tau AB         & 2.82/1.89 &  0.31/ 0.29 &        20 &  1.40 &      ... & ...        &       ... &   ...      \\
24-058 &  FQ Tau AB         & 2.82/1.89 &  0.31/ 0.29 &        20 &  1.40 &      ... & ...        &       ... &   ...      \\
28-100 &  BP Tau            &      1.91 &        0.75 &        27 &  1.97 &     7.60 & 48         &      7.80 &   48       \\
23-074 &  V819 Tau AB       &      2.02 &        0.76 &         5 &  1.93 &     5.60 & 48         & $<$ 15.00 &   48       \\
24-061 &  V819 Tau AB       &      2.02 &        0.76 &         5 &  1.93 &     5.60 & 48         & $<$ 15.00 &   48       \\
16-000 &  IRAS 04166+2706   &       ... &         ... &       ... &   ... &      ... & ...        &       ... &   ...      \\
16-000 &  IRAS 04169+2702   &       ... &         ... &       ... &   ... &      ... & ...        &       ... &   ...      \\
11-000 &  CFHT-Tau 19       &       ... &         ... &       ... &  0.93 &      ... & ...        &       ... &   ...      \\
11-000 &  IRAS 04181+2655   &      0.90 &        0.68 &        10 &  2.80 & $<$ 4.29 & C          &     33.00 &   10       \\
11-000 &  IRAS 04181+2654AB &       ... &         ... &       ... &   ... &      ... & ...        &       ... &   ...      \\
11-023 &  2M J04213459      &      4.56 &        0.13 &        33 &  0.91 &      ... & ...        &       ... &   ...      \\
01-028 &  IRAS 04187+1927   &       ... &         ... &       ... &   ... &      ... & ...        &       ... &   ...      \\
11-037 &  CFHT-Tau 10       &       ... &         ... &       ... &  0.52 &      ... & ...        &       ... &   ...      \\
11-000 &  2M J04215450+2652 &       ... &         ... &       ... &  0.26 &      ... & ...        &       ... &   ...      \\
21-038 &  RY Tau            &      2.11 &        2.37 &        27 &  3.57 &     5.60 & 48         &     52.20 &   48       \\
21-039 &  HD 283572         &      7.92 &        1.70 &        27 &  2.56 &     1.55 & 48         &     95.00 &   48       \\
01-045 &  T Tau N(+Sab)     &      2.67 &        2.41 &        27 &  3.62 &     2.80 & 48         &     20.10 &   48       \\
11-054 &  Haro 6-5 B        &       ... &         ... &       ... &  0.37 & $<$ 0.91 & C          &     20.70 &   59       \\
11-057 &  FS Tau AC         & 17.2/3.06 &  0.61/ 0.28 &        20 &  0.93 &      ... & ...        &       ... &   ...      \\
21-044 &  LkCa 21           &      1.16 &        0.35 &        27 &  2.18 &     8.80 & 48         &     60.00 &   48       \\
01-054 &  RX J0422.1+1934   &       ... &         ... &       ... &   ... &      ... & ...        &       ... &   ...      \\
01-062 &  2M J04221332+1934 &       ... &         ... &       ... &  0.59 &      ... & ...        &       ... &   ...      \\
11-079 &  CFHT-Tau 21       &       ... &         ... &       ... &  1.53 &      ... & ...        &       ... &   ...      \\
02-013 &  FV Tau AB         & 4.75/7.65 &  1.19/ 0.88 &        58 &  1.71 &      ... & ...        &       ... &   ...      \\
02-000 &  FV Tau/c AB       & 3.14/5.40 &  0.31/ 0.16 &    20, 58 &  1.22 &      ... & ...        &       ... &   ...      \\
02-016 &  KPNO-Tau 13       &      2.66 &        0.19 &        34 &  1.32 &      ... & ...        &       ... &   ...      \\
02-000 &  DG Tau B          &      0.30 &        0.72 &        10 &  4.90 &      ... & ...        &       ... &   ...      \\
02-022 &  DG Tau A          &      1.32 &        0.91 &         5 &  2.46 &     6.30 & 48         &     21.70 &   48       \\
02-000 &  KPNO-Tau 4        &       ... &        0.01 &        46 &  0.35 &      ... & ...        &       ... &   ...      \\
02-000 &  IRAS 04248+2612AB &      16.5 &        0.34 &         5 &  2.14 & $<$ 6.61 & C          &     16.40 &   59       \\
15-020 &  JH 507            &      1.33 &        0.27 &         5 &  2.02 &      ... & ...        &       ... &   ...      \\
13-004 &  GV Tau AB         &      0.89 &        0.68 &        59 &  2.82 & $<$ 5.63 & C          &     25.30 &   59       \\
13-000 &  IRAS 04264+2433   &      7.95 &        0.41 &        59 &  0.96 & $<$ 1.52 & C          &     32.00 &   10       \\
15-040 &  DH Tau AB         &      1.53 &        0.47 &         5 &  1.82 &     7.00 & 48         &     10.00 &   48       \\
15-042 &  DI Tau AB         &      1.07 &        0.56 &         5 &  2.24 &     7.70 & 48         &     10.50 &   48       \\
15-044 &  KPNO-Tau 5        &       ... &        0.04 &        46 &  0.65 &      ... & ...        &       ... &   ...      \\
14-006 &  IQ Tau A          &      1.06 &        0.51 &         5 &  2.20 &     6.25 & 48         &     11.50 &   48       \\
13-000 &  CFHT-Tau 20       &       ... &         ... &       ... &  1.31 &      ... & ...        &       ... &   ...      \\
14-000 &  KPNO-Tau 6        &       ... &        0.03 &        46 &  0.28 &      ... & ...        &       ... &   ...      \\
13-035 &  FX Tau AB         &      0.90 &        0.47 &        27 &  2.44 & $<$12.34 & C          &     10.00 &   48       \\
14-057 &  DK Tau AB         &      1.32 &        0.74 &         5 &  2.31 &     8.40 & 48         &     11.40 &   48       \\
14-000 &  KPNO-Tau 7        &       ... &        0.03 &        46 &  0.28 &      ... & ...        &       ... &   ...      \\
22-013 &  MHO 9             &      2.33 &        0.24 &         5 &  1.50 & $<$ 7.66 & C          &      9.90 &   48       \\
22-021 &  MHO 4             &       ... &        0.10 &        58 &  0.88 & $<$ 6.29 & C          &      7.10 &   48       \\
22-040 &  L1551 IRS5        &      4.80 &        1.58 &        10 &  2.34 & $<$ 3.82 & C          &     31.00 &   10       \\
22-042 &  LkHa 358          &      0.12 &        0.21 &         5 &  2.74 & $<$ 6.80 & C          &     20.40 &   45       \\
22-000 &  HH 30             &       ... &        0.52 &        59 &  0.19 &      ... & ...        & $<$ 12.00 &   59       \\
22-043 &  HL Tau            &      2.45 &        1.20 &        59 &  2.13 & $<$ 7.21 & C          &     15.00 &   48       \\
22-047 &  XZ Tau AB         & 4.58/1.79 &  0.37/ 0.29 &        20 &  1.18 &     2.60 & 48         &       ... &   ...      \\
22-056 &  L1551 NE          &       ... &         ... &       ... &   ... &      ... & ...        &       ... &   ...      \\
03-005 &  HK Tau AB         &      1.76 &        0.51 &         5 &  1.75 & $<$ 8.86 & C          &     10.00 &   48       \\
22-070 &  V710 Tau BA       & 1.69/1.69 &  0.51/ 0.40 &         5 &  1.78 & $<$ 5.67 & C          &     15.90 &   48       \\
19-009 &  JH 665            &      0.60 &        0.19 &         5 &  1.82 &      ... & ...        &       ... &   ...      \\
22-089 &  L1551 51          &      5.42 &        0.80 &         5 &  1.39 &     2.43 & 48         &     27.00 &   48       \\
22-097 &  V827 Tau          &      1.59 &        0.75 &         5 &  2.13 &     3.75 & 48         &     18.50 &   48       \\
03-016 &  Haro 6-13         &      0.57 &        0.52 &        59 &  3.36 & $<$ 7.30 & C          &     23.30 &   59       \\
\hline
\end{tabular}
\normalsize
\end{table*}
 
 \setcounter{table}{9}
\begin{table*}[h!]
\caption{(Continued)}
\begin{tabular}{llrlrlrlrl}
\hline
\hline
XEST & Name & Age$^{a,b}$   &  Mass$^{a,c}$      & Refs & Radius$^d$    &  $P$ & Refs & $v\sin i$     & Refs  \\
     &      & (Myr)         & ($M_{\odot}$)      &      & ($R_{\odot}$) &  (d) &      & (km~s$^{-1}$) &            \\
\hline
22-100 &  V826 Tau          &      1.94 &        0.75 &         5 &  1.96 &     3.70 & 48         &      4.20 &   48       \\
22-101 &  MHO 5             &       ... &        0.13 &        58 &  1.24 & $<$ 7.84 & C          &      8.00 &   48       \\
03-017 &  CFHT-Tau 7        &       ... &         ... &       ... &  0.89 &      ... & ...        &       ... &   ...      \\
03-019 &  V928 Tau AB       &      0.73 &        0.50 &         5 &  2.77 & $<$ 7.45 & C          &     18.80 &   48       \\
03-022 &  FY Tau            &      3.98 &        1.13 &         5 &  1.77 &      ... & ...        &       ... &   ...      \\
03-023 &  FZ Tau            &      1.08 &        0.56 &         5 &  2.23 &      ... & ...        &       ... &   ...      \\
17-002 &  IRAS 04295+2251   &      1.12 &        0.56 &        27 &  1.49 & $<$ 1.48 & C          &     51.00 &   10       \\
19-049 &  UZ Tau E+W(AB)    & 2.26/1.43 &  0.47/ 0.39 &         5 &  1.54 & $<$ 4.90 & C          &     15.90 &   48       \\
17-009 &  JH 112            &      4.06 &        0.95 &        27 &  1.62 &      ... & ...        &       ... &   ...      \\
03-031 &  CFHT-Tau 5        &       ... &         ... &       ... &  1.18 &      ... & ...        &       ... &   ...      \\
04-003 &  CFHT-Tau 5        &       ... &         ... &       ... &  1.18 &      ... & ...        &       ... &   ...      \\
03-035 &  MHO 8             &      0.47 &        0.15 &         5 &  1.54 & $<$ 4.67 & C          &     16.70 &   45       \\
04-009 &  MHO 8             &      0.47 &        0.15 &         5 &  1.54 & $<$ 4.67 & C          &     16.70 &   45       \\
04-010 &  GH Tau AB         & 2.04/2.03 &  0.42/ 0.38 &        58 &  1.56 & $<$ 3.57 & C          &     22.10 &   48       \\
04-012 &  V807 Tau  SNab    & 1.51/1.74 &  0.69/ 0.31 &        58 &  2.16 &      ... & ...        &       ... &   ...      \\
18-004 &  KPNO-Tau 14       &      1.09 &        0.13 &        34 &  1.24 &      ... & ...        &       ... &   ...      \\
17-000 &  CFHT-Tau 12       &       ... &         ... &       ... &  0.70 &      ... & ...        &       ... &   ...      \\
04-016 &  V830 Tau          &      2.48 &        0.76 &         5 &  1.79 &     2.75 & 48         &     29.10 &   48       \\
18-000 &  IRAS S04301+261   &       ... &         ... &       ... &  0.36 &      ... & ...        &       ... &   ...      \\
17-000 &  IRAS 04302+2247   &       ... &        0.34 &        59 &   ... &      ... & ...        &       ... &   ...      \\
17-027 &  IRAS 04303+2240   &      0.50 &        0.51 &        59 &  3.62 & $<$ 5.23 & C          &     35.00 &   59       \\
04-034 &  GI Tau            &      1.79 &        0.75 &         5 &  2.03 &     7.20 & 48         &     11.20 &   48       \\
04-035 &  GK Tau AB         &      1.22 &        0.74 &         5 &  1.97 &     4.60 & 48         &     18.70 &   48       \\
18-019 &  IS Tau AB         & 4.15/2.52 &  0.72/ 0.24 &        58 &  1.48 &      ... & ...        &       ... &   ...      \\
17-058 &  CI Tau            &      2.15 &        0.76 &        27 &  1.89 & $<$ 9.20 & C          &     10.40 &   48       \\
18-030 &  IT Tau AB         &      4.75 &        1.67 &         5 &  2.16 &      ... & ...        &       ... &   ...      \\
17-066 &  JH 108            &      3.40 &        0.48 &        27 &  1.32 &      ... & ...        &       ... &   ...      \\
17-068 &  CFHT-BD Tau 1     &       ... &         ... &      ...  &  0.53 &      ... & ...        &       ... &   ...      \\
25-026 &  AA Tau            &      2.40 &        0.76 &         5 &  1.81 &     8.22 & 48         &     11.40 &   48       \\
09-010 &  HO Tau AB         &      9.13 &        0.52 &         5 &  0.96 &      ... & ...        &       ... &   ...      \\
08-019 &  FF Tau AB         &      2.95 &        0.77 &         5 &  1.68 &      ... & ...        &       ... &   ...      \\
12-040 &  DN Tau            &      1.05 &        0.56 &         5 &  2.25 &     6.30 & 48         &      8.10 &   48       \\
12-000 &  IRAS 04325+2402AB &       ... &         ... &      ...  &   ... &      ... & ...        &       ... &   ...      \\
12-059 &  CoKu Tau 3 AB     &      0.92 &        0.46 &         5 &  2.41 &      ... & ...        &       ... &   ...      \\
09-022 &  KPNO-Tau 8        &       ... &         ... &      ...  &  0.52 &      ... & ...        &       ... &   ...      \\
08-037 &  HQ Tau AB         &       ... &         ... &      ...  &   ... &      ... & ...        &       ... &   ...      \\
09-026 &  HQ Tau AB         &       ... &         ... &      ...  &   ... &      ... & ...        &       ... &   ...      \\
08-043 &  KPNO-Tau 15       &      3.86 &        0.32 &        34 &  1.05 &      ... & ...        &       ... &   ...      \\
09-031 &  KPNO-Tau 15       &      3.86 &        0.32 &        34 &  1.05 &      ... & ...        &       ... &   ...      \\
08-000 &  KPNO-Tau 9        &       ... &         ... &      ...  &  0.19 &      ... & ...        &       ... &   ...      \\
09-000 &  KPNO-Tau 9        &       ... &         ... &      ...  &  0.19 &      ... & ...        &       ... &   ...      \\
08-048 &  HP Tau AB         &      6.90 &        1.39 &         5 &  1.77 &     5.90 & 48         &     15.40 &   48       \\
08-051a&  HP Tau/G3 AB      &      2.83 &        0.77 &         5 &  1.71 &      ... & ...        &       ... &   ...      \\
08-051 &  HP Tau/G2         &      10.5 &        1.58 &         5 &  2.34 &     1.20 & 48         &    100.00 &   48       \\
08-058 &  Haro 6-28 AB      & 10.0/18.5 &  0.35/ 0.21 &        20 &  0.79 & $<$ 3.98 & C          &     10.10 &   59       \\
08-000 &  CFHT-BD Tau 2     &       ... &         ... &       ... &  0.45 &      ... & ...        &       ... &   ...      \\
08-080 &  CFHT-BD Tau 3     &       ... &        0.04 &        46 &  0.38 &      ... & ...        &       ... &   ...      \\
05-005 &  CFHT-Tau 6        &       ... &         ... &       ... &  0.65 &      ... & ...        &       ... &   ...      \\
05-000 &  IRAS 04361+2547   &       ... &         ... &       ... &   ... &      ... & ...        &       ... &   ...      \\
05-013 &  GN Tau AB         &      1.03 &        0.36 &        33 &  2.33 & $<$11.80 & C          &     10.00 &   48       \\
05-017 &  IRAS 04365+2535   &       ... &         ... &       ... &   ... &      ... & ...        &       ... &   ...      \\
05-024 &  IRAS 04369+2539   &       ... &         ... &       ... &  7.24 &      ... & ...        & $<$ 15.00 &   59       \\
07-011 &  JH 223            &      4.27 &        0.37 &        33 &  1.12 &      ... & ...        &       ... &   ...      \\
07-022 &  Haro 6-32         &      3.33 &        0.09 &        33 &  1.18 &      ... & ...        &       ... &   ...      \\
07-000 &  ITG 33 A          &      12.9 &        0.26 &        33 &  0.65 &      ... & ...        &       ... &   ...      \\
07-000 &  CFHT-Tau 8        &      6.93 &        0.10 &        33 &  0.55 &      ... & ...        &       ... &   ...      \\
07-000 &  IRAS 04381+2540   &       ... &         ... &       ... &   ... &      ... & ...        &       ... &   ...      \\
07-041 &  IRAS 04385+2550AB &      12.2 &        0.59 &        33 &  0.96 & $<$ 2.13 & C          &     22.80 &   59       \\
10-017 &  CoKuLk332/G2 AB   & 5.60/4.50 &  0.52/ 0.34 &        20 &  0.99 & $<$ 2.25 & C          &     22.30 &   48       \\
\hline
\end{tabular}
\normalsize
\end{table*}
 
 \setcounter{table}{9}
\begin{table*}[h!]
\caption{(Continued)}
\begin{tabular}{llrlrlrlrl}
\hline
\hline
XEST & Name & Age$^{a,b}$   &  Mass$^{a,c}$      & Refs & Radius$^d$    &  $P$ & Refs & $v\sin i$     & Refs  \\
     &      & (Myr)         & ($M_{\odot}$)      &      & ($R_{\odot}$) &  (d) &      & (km~s$^{-1}$) &            \\
\hline
10-018 &  CoKuLk332/G1 AB   & 1.02/2.76 &  0.46/ 0.66 &        58 &  2.24 &      ... & ...        &       ... &   ...      \\
10-020 &  V955 Tau AB       & 24.1/6.70 &  0.90/ 0.47 &        58 &  0.98 & $<$ 4.90 & C          &     10.10 &   48       \\
10-034 &  CIDA 7            &       ... &         ... &       ... &   ... &      ... & ...        &       ... &   ...      \\
10-045 &  DP Tau            &      7.24 &        0.52 &        33 &  1.05 & $<$ 2.76 & C          &     19.20 &   59       \\
10-060 &  GO Tau            &      3.78 &        0.58 &        33 &  1.37 & $<$ 3.96 & C          &     17.50 &   48       \\
26-012 &  2M J04552333+30   &       ... &         ... &       ... &  0.47 &      ... & ...        &       ... &   ...      \\
26-034 &  2M J04554046+30   &      9.39 &        0.11 &        33 &  0.51 &      ... & ...        &       ... &   ...      \\
26-043 &  AB Aur            &      4.00 &        2.70 &         9 &  2.31 & $<$ 1.46 & C          &     80.00 &    6       \\
26-050 &  2MJ04554757/801   &      3.82 &        0.07 &        33 &  1.05 &      ... &  ...       &       ... &    ...     \\
26-067 &  SU Aur            &      6.02 &        1.91 &        33 &  3.06 &     1.70 & 48         &     65.00 &   48       \\
26-072 &  HBC 427           &      3.43 &        1.13 &        33 &  1.85 &     9.30 & 48         &     10.40 &    2       \\
\hline
\multicolumn{10}{l}{Additional sources from Chandra}\\
\hline
C1-0   &  KPNO-Tau 10       &       5.7 &        0.14 &        34 &  0.78 &      ... & ...        &       ... &   ...      \\
C1-1   &  IRAS 04158+2805   &      1.55 &        0.35 &        27 &  0.98 & $<$ 2.15 & C          &     23.10 &   59       \\
C2-1   &  Haro 6-5 B        &       ... &         ... &       ... &  0.37 & $<$ 0.91 & C          &     20.70 &   59       \\
C2-2   &  FS Tau AC         & 17.2/3.06 &  0.61/ 0.28 &        20 &  0.93 &      ... & ...        &       ... &   ...      \\
C3-1   &  FV Tau/c AB       & 3.14/5.40 &  0.31/ 0.16 &    20, 58 &  1.22 &      ... & ...        &       ... &   ...      \\
C3-2   &  DG Tau B          &      0.30 &        0.72 &        10 &  4.90 &      ... & ...        &       ... &   ...      \\
C4-1   &  GV Tau AB         &      0.89 &        0.68 &        59 &  2.82 & $<$ 5.63 & C          &     25.30 &   59       \\
C5-2   &  HN Tau AB         &     41/14 &  0.78/ 0.07 &        58 &  0.81 & $<$ 0.78 & C          &     52.80 &   48       \\
C5-1   &  L1551 55          &       8.2 &        0.81 &         5 &  1.23 &     6.20 & 48         & $<$ 10.00 &   48       \\
C5-4   &  HD 28867          & 3.71/3.76 &  2.85/ 2.82 &        65 &  2.60 & $<$ 2.00 & 65         &     65.00 &   65       \\
C5-3   &  DM Tau            &      3.26 &        0.47 &         5 &  1.33 & $<$ 6.74 & C          &     10.00 &   48       \\
C6-1   &  CFHT-BD Tau 4     &       ... &        0.06 &        46 &  1.02 &      ... & ...        &       ... &   ...      \\
C6-0   &  L1527 IRS         &       ... &         ... &       ... &   ... &      ... & ...        &       ... &   ...      \\
C6-0   &  CFHT-Tau 17       &       ... &         ... &       ... &  0.94 &      ... & ...        &       ... &   ...      \\
C6-2   &  IRAS 04370+2559   &       ... &         ... &       ... &   ... &      ... & ...        &       ... &   ...      \\
\hline
\end{tabular}
\begin{minipage}{0.82\textwidth}
\footnotetext{
\hskip -0.5truecm $^a$ For binaries, first number refers to primary, second to secondary component (calculated from $L_*$ and $T_{\rm eff}$).\\
$^b$ Ages derived after \citet{siess00} using the same principal parameters as for
masses, quoted in Table~\ref{tab9})\\
$^c$ Masses derived after \citet{siess00} using principal parameters quoted in Table~\ref{tab9}.\\
Exceptions quoted directly from literature: HBC 352 (XEST-27-115; ref. 2), 2M~J04141188+28 (XEST-20-000; ref. 46), V410 A13 (XEST23-018 = XEST-24-000; ref. 58),
KPNO-Tau 2 (XEST-23-000 = XEST-24-000; ref. 46), KPNO-Tau 4 (XEST-02-000; ref. 46), KPNO-Tau 5 (XEST-15-044; ref. 46), KPNO-Tau 6 (XEST-14-000; ref. 46), KPNO-Tau 7 (XEST-14-000; ref. 46), 
MHO 4 (XEST-22-021; ref. 58),  HH~30 (XEST-22-000; ref. 59), IRAS 04302+2247 (XEST-17-000; ref. 59), CFHT-BD Tau 3 (XEST-08-080; ref. 46), 
CFHT-BD Tau 4 (C6-1; ref. 46).\\
$^d$ For multiples, radius is given only for primary if luminosity of primary is explicitly known\\
NOTES on individual objects:
\begin{itemize}
\item Mass/age calculations for references different from those of $T_{\rm eff}$ and $L_*$ for CW Tau (XEST-20-046; ref. 27 instead of 5),
      and IRAS 04158+2805 (C1-1; ref. 27 instead of 59)
\item V773 Tau = XEST-20-042: Multiple entries refer to A and C, respectively
\item V410 Tau = XEST-23-032 = XEST-24-028: Multiple entries refer to A and C, respectively
\item HD~28867 = C5-4: radius and $P$ of G-type companion after \citet{walter03}
\end{itemize}
}
\end{minipage}
\label{tab10}
\normalsize
\end{table*}

\clearpage
 
\setcounter{table}{10}
\begin{table*}[h!]
\caption{Fundamental parameters of targets in XEST (5); Accretion and evolution}
\begin{tabular}{llrlrllrll}
\hline
\hline
XEST & Name & $\dot{M}$ (min/max)$^a$  &  Refs  &  EW(H$\alpha$)$^b$  & TTS    & Refs$^c$  & IR$^d$& Refs$^d$ & Type  \\
   &      & ($M_{\odot}$yr$^{-1}$) &        & (\AA)                   & type   &  	     & class  &          &         \\
\hline
27-115 &  HBC 352           &                 ... & ...       &          0 & W     &  29            & III     &  1, 27     & 3 \\
27-000 &  HBC 353           &                 ... & ...       &          0 & W     &  29            & III     & 27         & 3 \\
06-005 &  HBC 358 AB        & $<$  8.97           &   20      &      4- 10 & W/W   &  37, 29        & III     &  1, 27     & 3 \\
06-007 &  HBC 359           &                 ... & ...       &      2-  9 & W     &  56, 37        & III     & 27         & 3 \\
06-059 &  L1489 IRS         &     -7.15           &   59      &     41- 56 & C*    &  59, 29        & I       & 1, 59, 27  & 1 \\
20-001 &  LkCa 1            & $<$ -9.72           &   58      &      3-  4 & W     &  29, 45        & III     &  1, 27     & 3 \\
20-005 &  Anon 1            & $<$ -8.94           &   58      &      1-  3 & W     &  37, 29        & III     &  1, 27     & 3 \\
20-000 &  IRAS 04108+2803 A &                 ... &  ...      &         37 & C     &  29            & II      & 27         & 2 \\
20-022 &  IRAS 04108+2803 B &                 ... &  ...      &        ... & ...   &  ...           & I       & 59, 27     & 1 \\
20-000 &  2M J04141188+28   &    -10.00           &   46      &        250 & C     &  46            & ...     &  ...       & 4 \\
20-042 &  V773 Tau ABC      & $<$-10.00           &   58      &      2-  4 & W     &  21, 24        & II      &  1, 27     & 3 \\
20-043 &  FM Tau            &     -8.87/    -8.45 &   58, 44  &     51-101 & C     &  21, 29        & II      &  1, 27     & 2 \\
20-046 &  CW Tau            &     -7.99           &   58      &    135-140 & C     &   7, 29        & II      & 1, 59, 27  & 2 \\
20-047 &  CIDA 1            &     -8.50           &   57      &    112-149 & C     &  41, 29        & II      & 27         & 2 \\
20-056 &  MHO 2/1           & $<$ -8.48           &   59      &     58- 88 & C/C   &  59, 4         & I;II    & 59; 22     & 2 \\
20-058 &  MHO 3             &                 ... &  ...      &     16- 21 & C     &   4            & II      & 22         & 2 \\
20-069 &  FO Tau AB         &     -7.90/    -7.58 &   58, 20  &    116-137 & C     &  29, 20        & II      &  1, 27     & 2 \\
20-073 &  CIDA 2            &                 ... &  ...      &      5-  7 & W     &   3, 29        & III     &  1, 27     & 3 \\
23-002 &  CY Tau            &     -8.86/    -8.12 &   58      &     55- 70 & C     &  58, 7         & II      &  1, 27     & 2 \\
24-002 &  CY Tau            &     -8.86/    -8.12 &   58, 44  &     55- 70 & C     &  58, 7         & II      &  1, 27     & 2 \\
23-004 &  LkCa 5            & $<$-10.00           &   58      &          4 & W     &  29            & III     &  1, 27     & 3 \\
24-004 &  LkCa 5            & $<$-10.00           &   58      &          4 & W     &  29            & III     &  1, 27     & 3 \\
23-008 &  CIDA 3            &                 ... &  ...      &          6 & W     &   3            & II      &  1, 27     & 3 \\
24-008 &  CIDA 3            &                 ... &  ...      &          6 & W     &   3            & II      &  1, 27     & 3 \\
23-015 &  V410 X3           & $<$ -9.30           &   57      &     14- 30 & W?    &  41, 57        & ...     &  ...       & 3 \\
24-015 &  V410 X3           & $<$ -9.30           &   57      &     14- 30 & W?    &  41, 57        & ...     &  ...       & 3 \\
23-018 &  V410 A13          &    -11.30           &   45      &     27- 41 & C?    &  45, 58        & II      & 22         & 2 \\
24-000 &  V410 A13          &    -11.30           &   45      &     27- 41 & C?    &  45, 58        & II      & 22         & 2 \\
23-000 &  V410 A24          &                 ... &  ...      &        ... & ...   &  ...           & ...     &  ...       & 9 \\
24-000 &  V410 A24          &                 ... &  ...      &        ... & ...   &  ...           & ...     &  ...       & 9 \\
23-029 &  V410 A25          &                 ... &  ...      &        ... & ...   &  ...           & ...     &  ...       & 9 \\
24-027 &  V410 A25          &                 ... &  ...      &        ... & ...   &  ...           & ...     &  ...       & 9 \\
23-032 &  V410 Tau ABC      & $<$ -8.80           &   58      &      2-  3 & W     &  29, 7         & III     &  1, 27     & 3 \\
24-028 &  V410 Tau ABC      & $<$ -8.80           &   58      &      2-  3 & W     &  29, 7         & III     &  1, 27     & 3 \\
23-033 &  DD Tau AB         &     -9.10/    -7.21 &   58, 20  &     90-206 & C     &  37, 20        & II      &  1, 27     & 2 \\
24-029 &  DD Tau AB         &     -9.10/    -7.21 &   58, 20  &     90-206 & C     &  37, 20        & II      &  1, 27     & 2 \\
23-035 &  CZ Tau AB         &                 ... &  ...      &      4-  7 & W     &   7, 29        & II      &  1, 27     & 3 \\
24-030 &  CZ Tau AB         &                 ... &  ...      &      4-  7 & W     &   7, 29        & II      &  1, 27     & 3 \\
23-036 &  IRAS 04154+2823   &                 ... &  ...      & $>$     17 & C     &  59            & FS;II   & 1; 59, 27  & 2 \\
24-031 &  IRAS 04154+2823   &                 ... &  ...      & $>$     17 & C     &  59            & FS;II   & 1; 59, 27  & 2 \\
23-037 &  V410 X2           &                 ... &  ...      &        ... & ...   &  ...           & ...     &  ...       & 9 \\
24-032 &  V410 X2           &                 ... &  ...      &        ... & ...   &  ...           & ...     &  ...       & 9 \\
23-045 &  V410 X4           &                 ... &  ...      &        ... & ...   &  ...           & ...     &  ...       & 9 \\
24-038 &  V410 X4           &                 ... &  ...      &        ... & ...   &  ...           & ...     &  ...       & 9 \\
23-047 &  V892 Tau          &                 ... &  ...      &      7- 13 & Ae    &  29, 7         & II      &  1, 27     & 5 \\
24-040 &  V892 Tau          &                 ... &  ...      &      7- 13 & Ae    &  29, 7         & II      &  1, 27     & 5 \\
23-048 &  LR 1              &                 ... &  ...      &        ... & ...   &  ...           & ...     & ...        & 9 \\
24-000 &  LR 1              &                 ... &  ...      &        ... & ...   &  ...           & ...     & ...        & 9 \\
23-050 &  V410 X7           &                 ... &  ...      &          3 & W     &  58            & III     & 22         & 3 \\
24-042 &  V410 X7           &                 ... &  ...      &          3 & W     &  58            & III     & 22         & 3 \\
23-000 &  V410 A20          &                 ... &  ...      &        ... & ...   &  ...           & ...     & ...        & 9 \\
24-000 &  V410 A20          &                 ... &  ...      &        ... & ...   &  ...           & ...     & ...        & 9 \\
23-056 &  Hubble 4          & $<$ -8.16           &   58      &      3-  4 & W     &   7, 29        & III     &  1, 27     & 3 \\
24-047 &  Hubble 4          & $<$ -8.16           &   58      &      3-  4 & W     &   7, 29        & III     &  1, 27     & 3 \\
23-000 &  KPNO-Tau 2        & $<$-12.00           &   46      &      4- 12 & W     &  46, 5         & ...     & ...        & 4 \\
24-000 &  KPNO-Tau 2        & $<$-12.00           &   46      &      4- 12 & W     &  46, 5         & ...     & ...        & 4 \\
23-000 &  CoKu Tau 1        &     -7.36           &   59      &     70-111 & C     &  59, 29        & II      & 59, 1, 27  & 2 \\
24-000 &  CoKu Tau 1        &     -7.36           &   59      &     70-111 & C     &  59, 29        & II      & 59, 1, 27  & 2 \\
23-061 &  V410 X6           &                 ... &  ...      &         13 & W     &  38            &         & ...        & 3 \\
\hline
\end{tabular}
\normalsize
\end{table*}
 
 \setcounter{table}{10}
\begin{table*}[h!]
\caption{(Continued)}
\begin{tabular}{llrlrllrll}
\hline
\hline
XEST & Name & $\dot{M}$ (min/max)$^a$    &  Refs  &  EW(H$\alpha$)$^b$  & TTS    & Refs$^c$  & IR$^d$  & Refs$^d$  & Type  \\
   &      & ($M_{\odot}$yr$^{-1}$) &        & (\AA)                     & type   &  	    & class     &          &               \\
\hline
24-054 &  V410 X6           &                 ... &  ...      &         13 & W     &  38            & ...     & ...        & 3 \\
23-063 &  V410 X5           &                 ... &  ...      &     11- 19 & W?    &  58, 45        & II      & 22         & 3 \\
24-055 &  V410 X5           &                 ... &  ...      &     11- 19 & W?    &  58, 45        & II      & 22         & 3 \\
23-067 &  FQ Tau AB         &     -7.98/    -6.45 &   20, 44  &     81-114 & C/C   &  29, 7         & II      &  1, 27     & 2 \\
24-058 &  FQ Tau AB         &     -7.98/    -6.45 &   20, 44  &     81-114 & C/C   &  29, 7         & II      &  1, 27     & 2 \\
28-100 &  BP Tau            &     -7.88/    -7.54 &   58, 44  &     40- 92 & C     &  55, 29        & II      &  1, 27     & 2 \\
23-074 &  V819 Tau AB       & $<$ -8.86           &   58      &      2-  3 & W     &  60, 29        & III     &  1, 27     & 3 \\
24-061 &  V819 Tau AB       & $<$ -8.86           &   58      &      2-  3 & W     &  60, 29        & III     &  1, 27     & 3 \\
16-000 &  IRAS 04166+2706   &                 ... &  ...      &        ... & ...   &  ...           & I       & 59, 1, 27  & 1 \\
16-000 &  IRAS 04169+2702   &                 ... &  ...      &        ... & ...   &  ...           & I       & 59, 1, 27  & 1 \\
11-000 &  CFHT-Tau 19       &                 ... &  ...      &        442 &  C    &  18            &  ...    &  ...       & 2 \\
11-000 &  IRAS 04181+2655   &                 ... &  ...      &        ... & ...   &  ...           & I       & 59, 27     & 1 \\
11-000 &  IRAS 04181+2654AB &                 ... &  ...      &        ... & ...   &  ...           & I       & 59, 27     & 1 \\
11-023 &  2M J04213459      &                 ... &  ...      &        ... & W     &  33            & ...     & ...        & 3 \\
01-028 &  IRAS 04187+1927   &                 ... &  ...      &        ... & ...   &  ...           & II      &  27        & 2 \\
11-037 &  CFHT-Tau 10       &                 ... &  ...      &         17 &  W    &  18            &  ...    &  ...       & 3 \\
11-000 &  2M J04215450+2652 &                 ... &  ...      &        ... & ...   &  ...           & ...     &  ...       & 4 \\
21-038 &  RY Tau            &     -7.19/    -7.04 &   63      &         13 & C     &  29            & II      &  1, 27     & 2 \\
21-039 &  HD 283572         &                 ... &  ...      &          0 & W     &  29            & III     &  1, 27     & 3 \\
01-045 &  T Tau N(+Sab)     &     -7.50/    -7.24 &   58, 63  &         41 & C     &  29            & II      & 59, 1, 27  & 2 \\
11-054 &  Haro 6-5 B        &     -6.76           &   59      &         91 & C*    &  59            & (II)    & (59)       & 1 \\
11-057 &  FS Tau AC         &     -9.50/    -8.09 &   58, 44  &     57- 81 & C/C   &   7, 58        & FS;II   &  1; 27     & 2 \\
21-044 &  LkCa 21           &                 ... &  ...      &          6 & W     &  37            & III     &  1, 27     & 3 \\
01-054 &  RX J0422.1+1934   &                 ... &  ...      &         19 & W     &  36            & ...     & ...        & 3 \\
01-062 &  2M J04221332+1934 &                 ... &  ...      &        ... & ...   &  ...           & ...     &  ...       & 4 \\
11-079 &  CFHT-Tau 21       &                 ... &  ...      &         45 & C     &  18            & ...     & ...        & 2 \\
02-013 &  FV Tau AB         &     -7.70/    -6.23 &   58, 44  &      6- 23 & C/C   &  58, 7         & II      & 23, 1, 27  & 2 \\
02-000 &  FV Tau/c AB       & $<$ -8.70/$<$ -8.60 &   20, 58  &     17- 29 & C/C   &  58, 21        & II      & 23, 1, 27  & 2 \\
02-016 &  KPNO-Tau 13       &                 ... &  ...      &      8- 10 & W     &  41, 34        & ...     & ...        & 3 \\
02-000 &  DG Tau B          &                 ... &  ...      & $>$    276 & C*    &  59            & I/II;II & 23; 59     & 1 \\
02-022 &  DG Tau A          &     -7.34/    -6.13 &   58, 59  &     63-125 & C*    &  59, 29        & I/II;II & 23;59, 27  & 2 \\
02-000 &  KPNO-Tau 4        & $<$-12.00           &   46      &     38-158 & C     &  46, 18        & III     & 23         & 4 \\
02-000 &  IRAS 04248+2612AB &     -8.97           &   59      &        163 & C*    &  59            & I       & 59, 1, 27  & 1 \\
15-020 &  JH 507            &                 ... &  ...      &        ... & ...   &  ...           & III     & 23, 1, 27  & 3 \\
13-004 &  GV Tau AB         &     -6.71           &   59      &         86 & C*    &  59            & I       & 59, 1, 27  & 1 \\
13-000 &  IRAS 04264+2433   &     -7.11           &   59      &     69- 96 & C*    &  29, 59        & I       & 59, 27     & 1 \\
15-040 &  DH Tau AB         &     -8.95/    -8.30 &   58, 44  &     39- 72 & C     &  29, 58        & II      & 23, 1, 27  & 2 \\
15-042 &  DI Tau AB         &                 ... &  ...      &      1-  2 & W     &  14, 29        & III;II  & 23; 27     & 3 \\
15-044 &  KPNO-Tau 5        & $<$-12.00           &   46      &      4- 30 & W     &  46, 5         & III     & 23         & 4 \\
14-006 &  IQ Tau A          & $<$ -8.32/    -7.55 &   58, 44  &      8- 17 & C     &  29, 21        & II      & 27         & 2 \\
13-000 &  CFHT-Tau 20       &                 ... &  ...      &        114 & C     &  18            &  ...    &  ...       & 2 \\
14-000 &  KPNO-Tau 6        &    -11.40           &   46      &     41-350 & C     &  46, 5         & II      & 23         & 4 \\
13-035 &  FX Tau AB         &     -8.65           &   44      &     10- 15 & C/W   &   7, 29        & II      &  1, 27     & 2 \\
14-057 &  DK Tau AB         &     -7.42           &   44      &     31- 50 & C/C   &  26, 29        & II      &  1, 27     & 2 \\
14-000 &  KPNO-Tau 7        &    -11.40           &   46      &     31-300 &  C    &  46, 5         & II      & 23         & 4 \\
22-013 &  MHO 9             & $<$ -9.70           &   57      &      3-  6 & W     &  41, 4         & ...     & ...        & 3 \\
22-021 &  MHO 4             & $<$ -9.50           &   57      &     28- 43 & W?    &  57, 45        & ...     & ...        & 4 \\
22-040 &  L1551 IRS5        &                 ... &  ...      &     83-412 & C*    &  29, 59        & I       & 59, 1, 27  & 1 \\
22-042 &  LkHa 358          &     -8.50           &   45      &     47- 87 & C     &   7, 29        & II      &  27        & 2 \\
22-000 &  HH 30             &     -6.45           &   59      &    185-199 & C     &  29, 59        & II      & 59         & 2 \\
22-043 &  HL Tau            &     -8.83/    -6.80 &   58, 59  &     43- 55 & C*    &  59, 7         & I;II    &  1,59; 27  & 1 \\
22-047 &  XZ Tau AB         &     -8.90/    -7.00 &   58, 20  &     62-274 & C/C   &  58, 7         & II      & 59, 27     & 2 \\
22-056 &  L1551 NE          &                 ... &  ...      &        ... & ...   &  ...           & I       & 59, 1, 27  & 1 \\
03-005 &  HK Tau AB         &     -7.65           &   59      &     29- 54 & C/C   &   7, 29        & FS;I/II & 1; 27, 59  & 2 \\
22-070 &  V710 Tau BA       &                 ... &  ...      &     34- 89 & C/W   &  37, 26        & II+II   &  1, 27     & 2 \\
19-009 &  JH 665            &                 ... &  ...      &        ... & ...   &  ...           & III     &  27        & 3 \\
22-089 &  L1551 51          & $<$ -9.50           &   58      &      1-  2 & W     &  37, 29        & III     &  1, 27     & 3 \\
22-097 &  V827 Tau          & $<$ -8.53           &   58      &      2-  4 & W     &   7, 29        & III     &  1, 27     & 3 \\
03-016 &  Haro 6-13         &     -7.54           &   59      &     34- 88 & C     &  59, 7         & FS;I/II & 1; 27, 59  & 2 \\
\hline
\end{tabular}
\normalsize
\end{table*}
 
 \setcounter{table}{10}
\begin{table*}[h!]
\caption{(Continued)}
\begin{tabular}{llrlrllrll}
\hline
\hline
XEST & Name & $\dot{M}$ (min/max)$^a$    &  Refs  &  EW(H$\alpha$)$^b$  & TTS    & Refs$^c$  & IR$^d$  & Refs$^d$  & Type  \\
   &      & ($M_{\odot}$yr$^{-1}$) &        & (\AA)                     & type   &  	    & class     &          &               \\
\hline
22-100 &  V826 Tau          &                 ... &  ...      &      2-  4 & W     &   7, 29        & III     &  1, 27     & 3 \\
22-101 &  MHO 5             &    -10.80/$<$ -9.70 &   45, 57  &     36- 60 & C?    &  57, 45        & ...     & ...        & 2 \\
03-017 &  CFHT-Tau 7        &                 ... &  ...      &          9 & W     &  18            & ...     & ...        & 3 \\
03-019 &  V928 Tau AB       &                 ... &  ...      &      1-  2 & W     &  37, 29        & III     &  1, 27     & 3 \\
03-022 &  FY Tau            &     -7.48           &   58      &     48- 73 & C     &  58, 29        & II      &  1, 27     & 2 \\
03-023 &  FZ Tau            &     -7.70           &   58      &    181-204 & C     &  29, 7         & II      &  1, 27     & 2 \\
17-002 &  IRAS 04295+2251   &                 ... &  ...      & $>$ 11- 66 & C*    &  59, 29        & I;FS    & 59, 27; 1  & 1 \\
19-049 &  UZ Tau E+W(AB)    &     -8.70/    -8.01 &   58, 20  &     65- 82 & C/C/C &  29, 7         & II+II   & 23         & 2 \\
17-009 &  JH 112            &                 ... &  ...      &         16 & C     &  29            & II      &  1, 27     & 2 \\
03-031 &  CFHT-Tau 5        &                 ... &  ...      &         30 & W     &  18            & ...     & ...        & 4 \\
04-003 &  CFHT-Tau 5        &                 ... &  ...      &         30 & W     &  18            & ...     & ...        & 4 \\
03-035 &  MHO 8             &                 ... &  ...      &     14- 21 & W     &  45, 4         & II      &  22        & 3 \\
04-009 &  MHO 8             &                 ... &  ...      &     14- 21 & W     &  45, 4         & II      &  22        & 3 \\
04-010 &  GH Tau AB         &     -8.90/    -7.92 &   58, 44  &     10- 31 & C/C   &  20, 58        & II      &  1, 27     & 2 \\
04-012 &  V807 Tau  SNab    & $<$ -8.68/    -8.40 &   20, 58  &      5- 16 & C/W   &  21, 37        & III;II  &  1; 27     & 2 \\
18-004 &  KPNO-Tau 14       & $<$-12.00           &   46      &     14- 40 & W?    &  46, 34        & ...     & ...        & 3 \\
17-000 &  CFHT-Tau 12       &                 ... &  ...      &         80 & C     &  18            & ...     & ...        & 2 \\
04-016 &  V830 Tau          & $<$ -8.97           &   58      &      2-  3 & W     &  29, 60        & III     &  1, 27     & 3 \\
18-000 &  IRAS S04301+261   &                 ... &  ...      &        ... & ...   &  ...           & II      & 23, 27     & 2 \\
17-000 &  IRAS 04302+2247   &                 ... &  ...      &         67 & C*    &  59            & I       & 59, 1, 27  & 1 \\
17-027 &  IRAS 04303+2240   &     -6.63/    -6.05 &   59      &     67-137 & C     &  59            & II      & 59, 27     & 2 \\
04-034 &  GI Tau            &     -8.08           &   58, 44  &     15- 21 & C     &  29, 21        & II      & 27         & 2 \\
04-035 &  GK Tau AB         &     -8.19           &   58      &     15- 35 & C/C   &  21, 11        & II      & 59, 1, 27  & 2 \\
18-019 &  IS Tau AB         &     -8.10/    -7.91 &   58, 20  &     10- 26 & C/C   &  20, 58        & II      & 1, 23, 27  & 2 \\
17-058 &  CI Tau            &     -7.59/    -7.19 &   58, 44  &     77-102 & C     &  29, 7         & II      & 1, 27      & 2 \\
18-030 &  IT Tau AB         &                 ... &  ...      &     10- 22 & C/C   &  29, 11        & II      & 1, 23, 27  & 2 \\
17-066 &  JH 108            &                 ... &  ...      &          3 & W     &  29            & III     & 1, 27      & 3 \\
17-068 &  CFHT-BD Tau 1     &                 ... &  ...      &      7- 19 & W     &  41, 39        & ...     & ...        & 4 \\
25-026 &  AA Tau            &     -8.48/    -8.19 &   44, 58  &     37- 46 & C     &   7, 29        & II      & 1, 27      & 2 \\
09-010 &  HO Tau AB         &     -8.87           &   58      &    102-115 & C     &  29, 7         & II      & 1, 27      & 2 \\
08-019 &  FF Tau AB         &                 ... &  ...      &      1-  3 & W     &  24, 29        & III     & 1, 27      & 3 \\
12-040 &  DN Tau            &     -8.73/    -7.79 &   58, 59  &     12- 87 & C     &   7, 14        & II      & 59, 1, 27  & 2 \\
12-000 &  IRAS 04325+2402AB &                 ... &  ...      &        ... & ...   &  ...           & I       & 59, 1, 27  & 1 \\
12-059 &  CoKu Tau 3 AB     &                 ... &  ...      &          5 & W     &   7            & II      &  1, 27     & 3 \\
09-022 &  KPNO-Tau 8        &                 ... &  ...      &     15- 18 & W     &  41, 5         & ...     & ...        & 3 \\
08-037 &  HQ Tau AB         &                 ... &  ...      &        ... & ...   &  ...           & III     &  1         & 3 \\
09-026 &  HQ Tau AB         &                 ... &  ...      &        ... & ...   &  ...           & III     &  1         & 3 \\
08-043 &  KPNO-Tau 15       &                 ... &  ...      &          6 & W     &  34            & ...     & ...        & 3 \\
09-031 &  KPNO-Tau 15       &                 ... &  ...      &          6 & W     &  34            & ...     & ...        & 3 \\
08-000 &  KPNO-Tau 9        &                 ... &  ...      &          1 & W     &  41            & ...     & ...        & 4 \\
09-000 &  KPNO-Tau 9        &                 ... &  ...      &          1 & W     &  41            & ...     & ...        & 4 \\
08-048 &  HP Tau AB         &                 ... &  ...      &     20- 35 & C     &  29, 7         & FS;II   &  1; 27     & 2 \\
08-051a&  HP Tau/G3 AB      &                 ... &  ...      &      1-  2 & W     &  21, 29        & III     & 27         & 3 \\
08-051 &  HP Tau/G2         &                 ... &  ...      &      0-  5 & W     &  21, 7         & III     & 27         & 3 \\
08-058 &  Haro 6-28 AB      &     -8.70/    -8.10 &   45, 20  &     48- 92 & C/C   &  45, 7         & FS;I    & 1; 59, 27  & 2 \\
08-000 &  CFHT-BD Tau 2     &                 ... &  ...      &      7- 13 & W     &  41, 39        & ...     & ...        & 4 \\
08-080 &  CFHT-BD Tau 3     & $<$-12.00           &   46      &     11- 55 & W?    &  46, 39        & ...     & ...        & 4 \\
05-005 &  CFHT-Tau 6        &    -11.30           &   46      &     64-102 & C     &  18, 46        & ...     & ...        & 4 \\
05-000 &  IRAS 04361+2547   &                 ... &  ...      &        ... & ...   &  ...           & I       & 59, 1, 27  & 1 \\
05-013 &  GN Tau AB         &     -7.90           &   57      &     59- 62 & C     &  57, 3         & II      & 1, 23, 27  & 2 \\
05-017 &  IRAS 04365+2535   &                 ... &  ...      &        ... & ...   &  ...           & I       & 1,23,59,27 & 1 \\
05-024 &  IRAS 04369+2539   &     -6.20           &   59      &         22 & C     &  59            & II      &  1,59, 27  & 2 \\
07-011 &  JH 223            &                 ... &  ...      &          4 & W     &  29            & II      &  1,23, 27  & 3 \\
07-022 &  Haro 6-32         &                 ... &  ...      &        ... & W     &  33            & ...     & ...        & 3 \\
07-000 &  ITG 33 A          &                 ... &  ...      &         53 & C     &  35            & ...     & ...        & 2 \\
07-000 &  CFHT-Tau 8        &                 ... &  ...      &         52 & C     &  18, 33        & ...     & ...        & 2 \\
07-000 &  IRAS 04381+2540   &                 ... &  ...      &        ... & ...   &  ...           & I       & 23,1,59,27 & 1 \\
07-041 &  IRAS 04385+2550AB &     -8.11           &   59      &     15- 20 & C     &  59, 29        & II;I    & 23,27; 59  & 2 \\
10-017 &  CoKuLk332/G2 AB   & $<$ -8.58           &   20      &      2-  3 & W/W   &  20, 7         & III     & 1, 23, 27  & 3 \\
\hline
\end{tabular}
\normalsize
\end{table*}
 
 \setcounter{table}{10}
\begin{table*}[h!]
\caption{(Continued)}
\begin{tabular}{llrlrllrll}
\hline
\hline
XEST & Name & $\dot{M}$ (min/max)$^a$    &  Refs  &  EW(H$\alpha$)$^b$  & TTS    & Refs$^c$  & IR$^d$  & Refs$^d$  & Type  \\
   &      & ($M_{\odot}$yr$^{-1}$) &        & (\AA)                     & type   &  	    & class     &          &               \\
\hline
10-018 &  CoKuLk332/G1 AB   & $<$  8.20           &   58      &      0-  5 & W/W   &  58, 29        & III;II  & 23, 1; 27  & 3 \\
10-020 &  V955 Tau AB       & $<$ -8.97/    -8.50 &   20, 58  &     11- 45 & C/W   &  20, 58        & II      & 23, 1, 27  & 2 \\
10-034 &  CIDA 7            &                 ... &  ...      &         79 & C     &   3            & II      & 23, 1      & 2 \\
10-045 &  DP Tau            &     -8.50/    -6.92 &   58, 59  &     74-102 & C     &  59, 29        & II      & 23,1,59,27 & 2 \\
10-060 &  GO Tau            &     -8.42           &   58      &     78- 81 & C     &  29, 7         & II      & 23, 1, 27  & 2 \\
26-012 &  2M J04552333+30   &                 ... &  ...      &        ... & W     &  33            & ...     & ...        & 4 \\
26-034 &  2M J04554046+30   &                 ... &  ...      &        ... & W     &  33            & ...     & ...        & 3 \\
26-043 &  AB Aur            &                 ... &  ...      &     22- 44 & Ae    &  15, 29        & II      &  1, 27     & 5 \\
26-050 &  2MJ04554757/801   &                 ... &  ...      &         25 & C     &  46            & ...     & ...        & 2 \\
26-067 &  SU Aur            &     -8.30/    -8.20 &   63      &      2-  6 & C     &  29, 14        & II      &  1, 27     & 2 \\
26-072 &  HBC 427           &                 ... &  ...      &          1 & W     &  29            & III     & 23, 1, 27  & 3 \\
\hline
\multicolumn{10}{l}{Additional sources from Chandra}\\
\hline
C1-0   &  KPNO-Tau 10       &                 ... &  ...      &         36 & C     &  34            & ...     & ...        & 2 \\
C1-1   &  IRAS 04158+2805   & $<$ -9.50           &   59      &        ... & ...   &  ...           & I;II    & 59; 27     & 2 \\
C2-1   &  Haro 6-5 B        &     -6.76           &   59      &         91 & C*    &  59            & II      & 59         & 1 \\
C2-2   &  FS Tau AC         &     -9.50/    -8.28 &   58, 20  &     57- 81 & C/C   &   7, 58        & FS;II   &  1; 27     & 2 \\
C3-1   &  FV Tau/c AB       & $<$ -8.70/$<$ -8.60 &   20, 58  &     17- 21 & C/C   &  58, 20        & II      & 23, 1, 27  & 2 \\
C3-2   &  DG Tau B          &                 ... &  ...      & $>$    276 & C*    &  59            & I/II;II & 23; 59     & 1 \\
C4-1   &  GV Tau AB         &     -6.71           &   59      &         86 & C*    &  59            & I       & 59, 1, 27  & 1 \\
C5-2   &  HN Tau AB         &                 ... &  ...      &    138-163 & C     &  21, 58        & II      &  1, 27     & 2 \\
C5-1   &  L1551 55          & $<$ -9.70           &   58      &          1 & W     &  58            & III     &  1, 27     & 3 \\
C5-4   &  HD 28867          &                 ... &  ...      &        ... & ...   &  ...           & ...     & ...        & 9 \\
C5-3   &  DM Tau            &     -8.67           &   58      &        ... & ...   &  ...           & II      &  1, 27     & 2 \\
C6-1   &  CFHT-BD Tau 4     &    -11.30           &   46      &    129-340 & C     &  46, 39        & ...     & ...        & 4 \\
C6-0   &  L1527 IRS         &                 ... &  ...      &        ... & ...   &  ...           & 0       & 59, 23     & 0 \\
C6-0   &  CFHT-Tau 17       &                 ... &  ...      &          7 & W     &  18            & ...     & ...        & 3 \\
C6-2   &  IRAS 04370+2559   &                 ... &  ...      &        ... & ...   &  ...           & II      & 27         & 2 \\
\hline
\end{tabular}
\begin{minipage}{0.84\textwidth}
\footnotetext{
\hskip -0.5truecm $^a$ Range of $\dot{M}$ reported in literature given. For multiple systems, numbers refer to primary or integrated system\\
$^b$ Range of EW reported in literature given. For multiple systems, numbers refer to primary or integrated system\\
$^c$ For EW range, first reference for minimum, second for maximum reported.\\
$^d$ Infrared classification; double entries: '/' for transition objects, '+' for components, ';' for different types, ',' for  different references. FS = flat-spectrum source\\
}
\end{minipage}
\label{tab11}
\normalsize
\end{table*}

\clearpage

\end{document}